\documentclass[a4paper, 12pt, twoside]{Thesis}
\usepackage{graphicx}
\usepackage[dvipsnames]{xcolor}
\usepackage{epsfig}
\usepackage[justification=justified,singlelinecheck=false]{caption}
\usepackage{pdflscape}
\usepackage{enumerate}
\usepackage{rotating}
\usepackage{mathrsfs}
\usepackage{addfont}
\usepackage{wasysym}
\usepackage{natbib}
\bibpunct{(}{)}{;}{a}{}{,}
\usepackage{threeparttable}
\usepackage{etoolbox}
\usepackage{lipsum}

\newcommand{\zap}{{\it Z. Astrophys.}}
\newcommand{\araa}{{\it Ann. Rev. Astron. Astrophys.}}

\newcommand{\aap}{{\it Astron. Astrophys.}}
\newcommand{\aaps}{{\it Astron. Astrophys. Suppl.}}
\newcommand{\aapr}{{\it Astron. Astrophys. Rev.}}

\newcommand{\apj}{{\it Astrophys. J.}}
\newcommand{\apjl}{{\it Astrophys. J. Lett.}}
\newcommand{\apjs}{{\it Astrophys. J. Suppl.}}

\newcommand{\jgr}{{\it J. Geophys. Res.}}
\newcommand{\mnras}{{\it Mon. Not. Roy. Astron. Soc.}}
\newcommand{\nat}{{\it Nature}}
\newcommand{\pasp}{{\it Pub. Astron. Soc. Pac.}}

\newcommand\prl{{\it Phys. Rev.~Lett.}}    \newcommand{\solphys}{{\it Solar Phys.}}
 
\newcommand{\ssr}{{\it Space Sci. Rev.}}

\newcommand{\lrsp}{  {\it Living Rev. Sol. Phys.}}
\newcommand{\aspcs}{  {\it Astr. Soc. Pacific Conf. Ser.}}
\newcommand{\al}{     {\it Astron. Lett.}}
\newcommand{\qjras}{     {\it Quarterly Journal of the RAS}}
\newcommand{\planss}{     {\it Planet. Space Sci.}}\usepackage[unicode]{hyperref}
\usepackage{epigraph}
\usepackage{multicol}
\usepackage{physics}
\usepackage{bm}
\usepackage{bibentry}
\usepackage{color, soul}
\setstcolor{red}
\setulcolor{red}
\usepackage{pdfpages}

\addfont[1.25]{OT1}{hge}{\hge}
\graphicspath{{Figures/}}
\appto\TPTnoteSettings{\footnotesize}

\def\myfnt{\ifx\protect\@typeset@protect\expandafter\footnote\else\expandafter\@gobble\fi}

\def\hyp{--}
\newcommand{\lt}{long--term}
\newcommand{\tobs}{${\rm T}_{\rm OBS}$}
\newcommand{\q}{\mathit q}
\renewcommand{\degree}{^\circ}
\newcommand{\pa}{\partial}
\newcommand{\Rsun}{${\rm R}_\odot$}
\newcommand{\Bmax}{$B_{\rm max}$}
\newcommand{\angles}[1]{\ensuremath{\left\langle #1 \right\rangle}}
\def\bl{Babcock--Leighton}
\newcommand{\aname}[1]{\textcolor{RoyalBlue}{\textbf{#1}}}

\def\ndash{\,--\,}

\begin{document}

\frontmatter                            \pagenumbering{alph}

\thispagestyle{empty}
\par
\parskip 0.3in
\begin{center}
\noindent
\vspace{0.1cm}
\sffamily 
\noindent\rule[1.5ex]{\linewidth}{2pt}
\\
\vspace{0.30cm}
\textbf{\Huge{Long--term Study of the Sun and Its Implications to Solar Dynamo Models}}\\
\vspace{0.45cm}
\noindent\rule[1.5ex]{\linewidth}{2pt}
		 \textbf{A Thesis \\
		 Submitted for the Degree of} \\
\normalfont\hge
		 \Large{{Doctor of Philosophy}}\\ 
\normalfont\sffamily 
		 
\vspace{0.1in}
		\textbf{\large{in}}\\
\normalfont\sffamily 
\vspace{0.1in}
\textbf{\large{
The Department of Physics, \\ 
Pondicherry University, \\ Puducherry - 605 014, India}} \\
\vspace{0.15in}
	\includegraphics[width=0.175\textwidth]{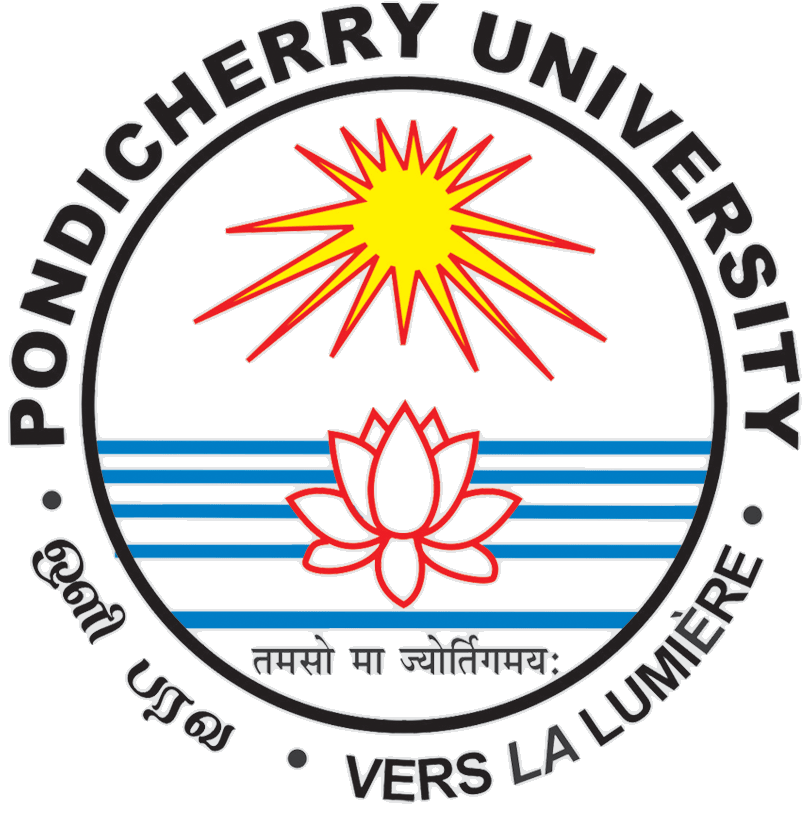}\\
\vspace{0.15in}
		\textbf{\large{by}}\\
\normalfont\sffamily 
\vspace{0.25in}
		{\huge{\textbf{Bibhuti Kumar Jha}}} \\
		        \large \textbf{Indian Institute of Astrophysics, \\  Bangalore - 560 034, India}\\
        \vspace{0.4cm}
        \hspace{0.05cm}        \includegraphics[width=0.170\textwidth]{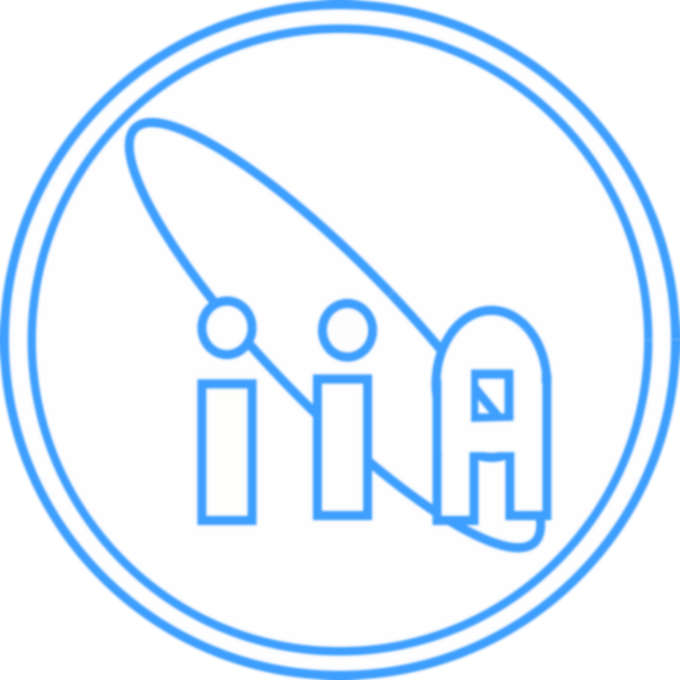} 
        \vspace{0.4cm}
        \large
	 \\

         \textbf{Jan 2022}
    \end{center}

\newpage

\newpage
 \pagestyle{empty}
\clearpage\null\newpage
\setboolean{@twoside}{true}

\thispagestyle{empty}
\par
\begin{center}
\noindent
\sffamily 
\textbf {\Huge {Long--term Study of the Sun and Its Implications to Solar Dynamo Models}}
\\
\normalfont
\vspace{4.0 cm}
{\huge\textbf{Bibhuti Kumar Jha}}\\ 
\textit{\textbf{Indian Institute of Astrophysics }} \\
\vspace{11.0cm}
        \hspace{0.02cm}   \includegraphics[width=0.170\textwidth]{logo_IIA} 
\vspace{-1.0cm}
\begin{center}\large \textbf{Indian Institute of Astrophysics \\  Bangalore - 560 034, India}\\ \end{center}
\large
\end{center}
\newpage

\newpage
 \pagestyle{empty}                        \thispagestyle{plain}
\onehalfspacing

\thispagestyle{plain}
\onehalfspacing
\noindent
        \vspace{2.8cm}
        \\
\noindent\rule[0.5ex]{\linewidth}{1.5pt}
\\
\\
\normalfont\sffamily 
    \begin{tabular}{ p{5cm} p{.2cm} p{7.8cm} l }

Title of the thesis  & : & \textbf{\large{Long--term Study of the Sun and Its Implications to Solar Dynamo Models
}} \\ \\
   
Name of the author & : &\textbf{\large{Bibhuti Kumar Jha}} \\[.15cm]
Address & : & Indian Institute of Astrophysics \\
         &   &      II Block, Koramangala \\
          &  &     Bangalore - 560 034, India \\ [.15cm]        
Email    & : &             bibhuti.kj@iiap.res.in \\ \\

Name of the supervisor & : & \textbf{Prof. Dipankar Banerjee} \\[.15cm]  
Address & : &  Indian Institute of Astrophysics \\
              &  &     II Block, Koramangala \\
            &  &      Bangalore - 560 034, India \\ [.15cm]  
Email               & : &     dipu@iiap.res.in \\
    \end{tabular}
\noindent
\\
        \\
\noindent\rule[0.5ex]{\linewidth}{1.5pt}

\thispagestyle{empty}

\cleardoublepage
\phantomsection
\normalfont
\setstretch{1.3}

\fancyhead{}  
\rhead{\thepage}
\lhead{}

\pagestyle{fancy}

\setboolean{@twoside}{false}
\noindent
\vspace{2.5cm}
\begin{center}{\huge\bf Declaration of Authorship}\end{center}
\thispagestyle{empty}
I hereby declare that the matter contained in this thesis is the result of the investigations carried out by me at the Indian Institute of Astrophysics, Bangalore, under the supervision of Prof. Dipankar Banerjee. This work has not been submitted for the award of any other degree, diploma, associateship, fellowship, etc. of any other university or institute.

Signed:\\
\rule[1em]{25em}{1.0pt}  

Date:\\
\rule[1em]{25em}{1.0pt}  

\clearpage  \thispagestyle{empty}
 \noindent
\vspace{2.5cm}
\newpage
\begin{center}{\huge\bf Certificate}\end{center}
\thispagestyle{empty}
This is to certify that the thesis entitled \textbf {``Long--term study of the Sun and its implications to solar dynamo models''} submitted to the Pondicherry University by Mr. Bibhuti Kumar Jha for the award of the degree of Doctor of Philosophy, is based on the results of the investigations carried out by him under my supervision and guidance, at the Indian Institute of Astrophysics. This thesis has not been submitted for the award of any other degree, diploma, associateship, fellowship, etc. of any other university or institute.

Signed:\\
\rule[1em]{25em}{1.0pt}  

Date:\\
\rule[1em]{25em}{1.0pt}  

\clearpage
 
\clearpage
\thispagestyle{empty}
\noindent
\vspace{2.5cm}
\newpage 
\pagestyle{empty}

\begin{center}{\huge\bf List of Publications}\end{center}
\begin{enumerate}

    \item \footnote{Presented in \cref{Chap3}.}\aname{Bibhuti Kumar Jha}, Sudip Mandal, \& Dipankar Banerjee, ``Study of Sunspot Penumbra to Umbra Area Ratio Using Kodaikanal White-light Digitised Data" \href{https://doi.org/10.1007/s11207-019-1462-2}{ Sol Phys (2019) 294: 72}
    
    \item \footnote{Presented in \cref{Chap6}.}\aname{Bibhuti Kumar Jha}, Bidya Binay Karak, Sudip Mandal \& Dipankar Banerjee; ``Magnetic field dependence of bipolar magnetic region tilts on the Sun: Indication of tilt quenching" \href{https://ui.adsabs.harvard.edu/abs/2019arXiv191213223J/abstract}{APjL (2020) 889:L19}

    \item \footnote{Presented in \cref{Chap4}.}\aname{Bibhuti Kumar Jha}, Aditya Priyadarshi, Sudip Mandal, Subhamoy Chatterjee \& Dipankar Banerjee; ``Measurements of Solar Differential Rotation Using the Century Long Kodaikanal Sunspot Data"; \href{https://ui.adsabs.harvard.edu/abs/2021SoPh..296...25J/abstract}{Sol Phys (2021) 296: 25}
    
    \item  \footnote{Presented in \cref{Chap5}.}\aname{ Bibhuti Kumar Jha} \& Arnab Rai Choudhuri; ``A theoretical model of the near-surface shear layer of the Sun"; \href{https://ui.adsabs.harvard.edu/abs/2021MNRAS.tmp.1557J/abstract}{MNRAS (2021) 506:2 (2189)}
    
    \item \footnote{Briefly Presented in \cref{Chap2}.}\aname{Bibhuti Kumar Jha}, Theodosios Chatzistergos, Dipankar Banerjee, Ilaria Ermolli, Natalie A. Krivova, Sami K. Solanki \& Aditya Priyadarshi; ``Revisiting the Ca-K data from Kodaikanal Solar Observatory: An Update on butterfly diagram"; \textcolor{blue}{\bf Under Preparation; {\it Sol. Phys.} (2022)}
    
    \item \footnotemark[\value{footnote}]\aname{Bibhuti Kumar Jha}, Manjunath Hegade, Aditya Priyadarshi, Sudip Mandal \& Dipankar Banerjee; ``Extending the Sunspot Area Series from Kodaikanal Solar Observatory"; \textcolor{blue}{\bf Under Preparation; {\it Frontiers In Astronomy And Space Sciences} (2022)}
    
    \item \footnote{Not included in thesis.}Theodosios Chatzistergos, Ilaria Ermolli, Sami K. Solanki, Natalie A. Krivova, Dipankar Banerjee, \aname{ Bibhuti Kumar Jha} \& Subhamoy Chatterjee; ``Delving into the Historical Ca II K Archive from the Kodaikanal Observatory: the Potential of the Most Recent Digitised Series"  \href{https://doi.org/10.1007/s11207-019-1532-5}{ Sol Phys (2019) 294: 145}

    \item \footnotemark[\value{footnote}]Aditya Priyadarshi, Manjunath Hegde, \aname{Bibhuti Kumar Jha}, Subhamoy Chatterjee, Sudip Mandal, Mayukh Chowdhury \& Dipankar Banerjee; ``A Machine Learning Approach towards Segmentation and Analysis of Solar Filaments from Kodaikanal Solar Observatory Hand-drawn Archive"; \textcolor{blue}{\bf Under Preparation; {\it ApJS} (2021)}
    
\end{enumerate}

\clearpage 
\pagestyle{empty}
 
\begin{center}{\huge\bf Conference Proceedings}\end{center}
\begin{enumerate}
    \item \aname{Bibhuti Kumar Jha}, Sudip Mandal, \& Dipankar Banerjee; ``Long-term variation of sunspot penumbra to umbra ratio: A study using Kodaikanal white-light digitized data"; \href{https://doi.org/10.1017/S1743921318001989}{ Proceedings of the International Astronomical Union (2018), 13, 185–186}
\end{enumerate}

\clearpage 
\pagestyle{empty}
 
\clearpage
\thispagestyle{empty}
\noindent
\vspace{2.5cm}
\newpage 
\pagestyle{empty}

\begin{center}{\huge\bf Presentations}\end{center}
\begin{enumerate}

    \item Presented a poster titled {\bf ``Long-term variation of sunspot penumbra to umbra ratio: A study using Kodaikanal white-light digitized data.''}, IAUS340,19 - 24 February, 2018, Jaipur, India
    
    \item Presented an oral talk titled {\bf ``Magnetic field dependency of Bipolar magnetic region tilt angle: A study from SOHO/MDI data''}, Young Astronomers Meet, 24-28 September, 2018, PRL, Ahmadabad, India
    
    \item Presented an oral talk titled {\bf ``An update on Kodaikanal Digital Archived Data''} in a meeting entitled \href{http://www.issibern.ch/teams/solheliomagnet/}{``Reconstructing Solar and Heliospheric Magnetic Field Evolution Over the Past Century''}, ISSI Team led by Alexei Pevtsov; 12 - 15 February, 2019
    
    \item Presented a poster titled {\bf ``Solar Differential Rotation in last century: A study from Kodaikanal white light digitised data''}, Young Astronomers Meet, 23-27 September, 2019, Kodaikanal Solar Observatory, IIA Kodaikanal, India
    
    \item Presented a poster titled {\bf ``Magnetic field dependency of bipolar magnetic region tilt angle: A study using MDI and HMI data sets''}, IRIS-10, 4-8 November, 2019, Christ University Bangalore, India
    
    \item Presented a poster titled {\bf ``Solar differential rotation as measured from century long Kodaikanal white light digitized data''}, 5th Asia Pacific Solar Physics Meeting (APSPM), 3-7 February, IUCAA, Pune, India
    
    \item Presented a talk titled {\bf ``Magnetic field dependence of bipolar magnetic region tilts on the Sun: Evidence of tilt quenching''}, Astronomical Society of India Meeting 2020, 13-17 February, 2020, IISER Tirupati, India

    \item Presented a talk titled {\bf ``Signature of quenching from observation of tilted bipolar magnetic regions on the Sun''}, IIA-50 Conference - Advances in Observations and Modelling of Solar Magnetism and Variability, 1-4 March, 2021, IIA, Bangalore, India
    
    \item Presented a e-poster titled {\bf ``A Theoretical Model of the Near-Surface Shear Layer of the Sun''}, The 16th European Solar Physics Meetings (ESPM-16), 6-10 September, 2021, {\bf Online}
    
    \item Presented a talk titled {\bf ``A Theoretical Model of the Near-Surface Shear Layer of the Sun''}, The 15th Quadrennial Solar-Terrestrial Physics (STP-15) symposium, 21-25 February, 2022, {\bf Online}
    
    \item Presented a poster titled {\bf ``A Theoretical Model of the Near-Surface Shear Layer of the Sun''}, The 40th Astronomical Society of India Meeting, 25-29 March, 2022
    
    \item Presented a talk titled {\bf ``Update on Ca-K data from Kodaikanal Solar Ob- servatory''}, Workshop on “Long-term study of the solar activity” in the 40th Astronomical Society of India Meeting, 25-29 March, 2022
    
    \item Presented an e-talk titled {\bf ``Signature of tilt quenching from observation of tilted bipolar magnetic regions on the Sun''} in International Astronomical Union General Assembly (IAUGA) 2022, 2\,--\,11 August 2022, Busan, South Korea
    
\end{enumerate}
\clearpage

\setboolean{@twoside}{true} 
\setstretch{1} 
\pagestyle{empty}
\addtocontents{toc}{\protect\setcounter{tocdepth}{-1}}

\acknowledgements{

\thispagestyle{empty}

I still cherish the memory of my {\it Baba} (grandfather) telling me the story of {\it Dhurv Tara} (the pole star), {\it Akashganga} (Milkyway), {\it Surya} (the Sun) and many more. Those stories had a very significant impact on me, and because of that, as a very young child, I started wondering about the night sky, much before I learned the words ``Astronomy'' or ``Astrophysics''. Today, I am delighted to say that my {\it Baba} was the one who sowed the seed in my mind, and today, that seed is flourishing as my interest in the subject. I am also thankful to all my teachers who taught me since childhood, kept enlightening my path, and believed in me. I would also like to pass my gratitude to the former and current directors of the Indian Institute of Astrophysics for providing me with the excellent opportunity to pursue my career in Astronomy and Astrophysics. Many thanks to all the faculty members of IIA, RRI and IISc who taught me during the course\hyp work which served as the foundation for rest of my research.

I am sure I will be short of words when I thank my mentor Dipu Da (Prof Dipankar Banerjee) because he has always been more than a thesis supervisor. I will thank him by saying that ``the Sun will never be just a mere star for me'' because of the insight he has given me to look at the Sun. I am beyond grateful to Bidya Da (Dr Bidya Binay Karak) for his constant guidance and support, like an elder brother, during this period. It has been such a blessing for me to get a chance to work with Arnab Da (Prof Arnab Rai Choudhuri), and I would like to pass my deepest gratitude to him for the opportunity. I would also like to thank Prof Sami Solanki and Prof Natalie Krivova for their guidance during my visit to MPS, Germany. I am also grateful to the DC members Prof B. Ravindra and Dr Alok Sharan for their valuable inputs and suggestions for my thesis works.

I want to convey my gratitude to Sudip Da, Subhamoy Da and Vaibhav for patiently teaching me the essential things and bearing all my ridiculous questions, without which I could not have done any of my work. It is impossible to put aside the contribution of Ritesh and Satabdwa in this journey. They have always been colleagues, friends and critics, which helped me to shape my works and made this journey memorable. I would also like to thank Manju Da, Rakesh Da, Nancy, and my juniors Aditya, Arpit, Anu, Dibya, Jyoti, Nitin, Tanushree and Upasna for the valuable input in scientific discussions and for the fun we had together.

Thanks a lot to all the administrative staff, library and computer support team, particularly Fayaz, Anish and Ashok, for their help during my stay at IIA. I would like to convey my sincere gratitude to Sankar Sir for his help in completing all the administrative processes. 
I sincerely applaud the COE, HOD Physics Department and administration staff of Pondicherry University; without their constant support, it was not possible for me to concentrate on my work without worrying about the different non-academic pieces of stuff. 

I started at IIA in 2017 and then moved to ARIES in 2020; after that, I worked at ARIES, Nainital, for the last two years. It was not possible without the support of all the members of ARIES, including faculty members, administrative staff and the computer support team. Therefore, I am greatly obliged to all the ARIES members.

No one can work on a scientific problem without a good environment. I am fortunate to have such a wonderful group of people as Partha, Athira, Soumya, Sharmila, Manika and Suman, who created such an atmosphere for me. I will always cherish the memory of the time that I spent with Priyanka Di, Megha Di, Dipanweeta Di, Bhoomika Di, Rubi Di, Snehlata Di, Priya, Prerna, Chayan Da, Anirban Da, Samrat Da, Sandeep, Avinash, Panini, Varun, Raghu, Deepak, Manoj, Aritra, Ankit, Vikrant, Jyoti, Indrani, Anirban, Deepthi, Fazlu, Sonith, Pallavi and Swastik. I would also like to thank Ankur, Vivek, Amar, Rahul, Dimple, Pankaj, Varun, Arpan, Priyanka S, Sadhana, and Priyanka J for making life easier at ARIES during my stay here in the last two years. I would also like to sincerely thank all the canteen staff at Bhaskra (IIA) and Rohini (ARIES) for making my stay comfortable.

This journey could not have been completed without the constant support of my family {\it Maa}, {\it Nanimaa} and Prabhujee in all my ups and downs.

\pagestyle{empty} \vspace{4 cm}
\hspace{12 cm}  \large{\textbf{--Bibhu}}}

\newpage
\textit{\huge Data usage}\\
\\
During different studies, presented in this thesis, I have used telescopic data from various space-based facilities. I duly acknowledge the data usage. I also want to thank the members and associates of these facilities for providing the data. 

Kodaikanal Solar Observatory (KoSO) is a facility of Indian Institute of Astrophysics (IIA), Bangalore, India. This data is now available for public use at \url{http://kso.iiap.res.in} through a service developed at IUCAA under the Data Driven Initiatives project funded by the National Knowledge Network (NKN).

SOHO is a project of international cooperation between ESA and NASA. SDO Data supplied courtesy of the SDO/HMI and SDO/AIA consortia. SDO is the first mission to be launched for NASA's Living With a Star (LWS) Program.

\newpage
\thispagestyle{empty}
\noindent
\vspace{2.5cm}

\newpage
\pagestyle{empty}
\null\vfill\vfill\vfill
\begin{center}
    \textit{\textbf{\Large In loving memory of}}\\
    \vskip 0.2in
\textit{\textbf{\Large my Father}}
\end{center}
    
\vfill\vfill\vfill\vfill\vfill\vfill\null
\clearpage  \clearpage\null\newpage

\addtocontents{toc}{\protect\setcounter{tocdepth}{2}}
\clearpage

\pagenumbering{roman}
\frontmatter

\Abstract{
\addtocontents{toc}{}
{\huge \textbf{Abstract}}

The Sun shows a wide range of temporal variations, from a few seconds to decades and even centuries, broadly classified into two classes short\hyp term and \lt. The solar dynamo mechanism is believed to be responsible for these global changes happening in the Sun. Hence, many dynamo models have been proposed to explain the observed behaviour of the Sun. This thesis is primarily focused on studying the \lt\ variation of the Sun and provides  various inputs to the solar dynamo models.

With a renewed interest on the subject several automatic techniques have been developed for extensive data analysis as applied to the long term datasets and presented in this thesis.  This approach provides better consistency and  eliminate human subjectivity, which has been the normal practice in the past. These newly developed techniques are used to extract the umbra and penumbra from sunspots to study area ratio ($q$) and track the sunspot for the measurement of solar differential rotation in the white-light data from the Kodaikanal Solar Observatory (KoSO; 1923\ndash2011), Michelson Doppler Imager (MDI; 1996\ndash2011) and Helioseismic and Magnetic Imager (HMI; 2010\ndash2018). In addition, a completely automatic method to detect the bipolar magnetic regions (BMRs) from the line of sight magnetogram observed using MDI and HMI is also developed. Furthermore, the sunspot area series has been updated for the observation period (1906\ndash2017). We also discovered an issue with the incorrect time stamp in Ca-K data as recorded in the Kodaikanal archive, which we corrected  and oriented the images appropriately.

The ratio of penumbra to umbra area show an increase from 5.5 to 6 as the size of sunspot increases from 100~$\mu$Hem to 2000~$\mu$Hem, it does not show any \lt\ systematic trend as found in Royal Observatory, Greenwich (RGO) photographic results. The solar differential rotation, a  vital parameter in the solar dynamo models, measured in this study is, $\Omega (\theta)= (14.381\pm0.004)-(2.72\pm0.04)\cos^2\theta$, where $\theta$ is the co-latitude. While this study does not show any significant variation on $\Omega$ with time and between the maximum and minimum of activity, it reveals that the bigger sunspots (area$>400~\mu$Hem) give a relatively slower rotation rate than the smaller ones (area$<200~\mu$Hem). The measurement of solar differential rotation using sunspots compared to the spectroscopic observations indicates the existence of an intriguing layer known as the near-surface shear layer (NSSL), which was further verified by a more accurate measurement by helioseismology. The theoretical explanation for NSSL is presented based on the thermal wind balance equation. The idea is based on the fact that, in the top layer of the solar convection zone the temperature profile of the Sun is  independent of latitude; and in addition to that, the sharp fall in temperature causes the thermal wind term to grow. To compensate for the growth of the thermal wind term--the centrifugal term has to increase, leading to the observed NSSL. Based on the observed value of $\Omega (r,\theta)$, $r_c=0.96$~\Rsun\ shows the good agreement with observed NSSL, and it is also hand in hand with the observed estimate of pole equator temperature difference (PETD). Another crucial input for the dynamo model is the observed systematic tilt of the BMRs, particularly in \bl\ types. Based on the thin flux tube models, the tilt induced in BMR depends on the flux tube's rise time, which is a function of the magnetic field it carries. The magnetic field dependence can provide the non-linearity required in kinematic dynamo models to suppress the magnetic field's growth. The distribution of maximum magnetic field (\Bmax) in BMRs is found to be bimodal in both the data sets, peaking at 600~G and 2.2~kG, which turns out to be for BMRs of two classes. The results show an increase in the amplitude of Joy's law for \Bmax$<2$~kG; however, for \Bmax$>2$~kG, it decreases, indicating tilt quenching routinely used in \bl\ models.

The variation of penumbra to umbra area ratio, $q$, observed here will provide constraints in sunspot simulations. In addition, the absence of any difference in the behaviour of small and big spots does not support the idea of the global and local dynamo. Two classes of BMRs observed in the magnetograms further verify this behaviour. The importance of the NSSL is not studied so well in the context of solar dynamo models, but it will be worth waiting to see its significance for understanding solar dynamo. Finally, the indication of tilt quenching presented here needs to be further verified using the more comprehensive data set, including stronger cycles.

\pagebreak
\newpage }
\clearpage
\pagestyle{fancy}

\lhead{\emph{Contents}}
\tableofcontents

\lhead{\emph{List of Figures}}
\listoffigures 

\lhead{\emph{List of Tables}}
\listoftables

\lhead{\emph{Abbreviations}}  \listofsymbols{ll}  {
\textbf{NASA} & \textbf{N}ational \textbf{A}eronautics and \textbf{S}pace \textbf{A}dministration \\
\textbf{SOHO} & \textbf{SO}lar and \textbf{H}eliospheric \textbf{O}bservatory\\
\textbf{MDI} & \textbf{M}ichelson \textbf{D}oppler \textbf{I}mager \\
\textbf{KoSO} & \textbf{Ko}daikanal \textbf{S}olar \textbf{O}servatory \\
\textbf{WSO} & \textbf{W}ilcox \textbf{S}olar \textbf{O}servatory \\
\textbf{RGO} & \textbf{R}oyal \textbf{O}servatory of \textbf{G}reenwich \\
\textbf{SDO} & \textbf{S}olar \textbf{D}ynamics \textbf{O}bservatory \\
\textbf{AIA} & \textbf{A}tmospheric \textbf{I}maging \textbf{A}ssembly \\
\textbf{HMI} & \textbf{H}elioseismic and \textbf{M}agnetic \textbf{I}mager\\
\textbf{MHD} & \textbf{M}agneto\textbf{H}ydro\textbf{D}ynamics \\
\textbf{CCD} & \textbf{C}harge \textbf{C}oupled \textbf{D}evice \\
\textbf{UTC} & \textbf{C}oordinated \textbf{U}niversal \textbf{T}ime \\
\textbf{ISN} & \textbf{I}nternational \textbf{S}unspot \textbf{N}umber \\
\textbf{ISSN} & \textbf{I}nternational \textbf{S}moothed \textbf{S}unspot \textbf{N}umber \\
\textbf{BL} & \textbf{B}abcock--\textbf{L}eighton\\
\textbf{NSSL} & \textbf{N}ear--\textbf{S}urface \textbf{S}hear \textbf{L}ayer\\
}

\addtocontents{toc}{\vspace{2em}}
\setboolean{@twoside}{true}
\mainmatter	
\pagestyle{fancy}

\clearpage{}\chapter{Introduction}
\label{Chap1}
\lhead{\emph{Chapter 1: Introduction}} 

\epigraph{\itshape ``The purpose of life is the investigation of the Sun, the Moon, and the heavens.''}{--Anaxagoras}

\noindent
The Sun, our backyard star, is far more fascinating than it may appear in its first impression. Most of us believe it to be uninteresting because it is just an ordinary star among the billions of stars that we see in the night sky. However, the question is, if it is such a typical star, what makes it so fascinating? The brief answer is that it is the closest star to our home, i.e., the Earth, and it is the primary source of energy for the survival of living beings on this blue planet. The dependency of our life on this star for energy makes it necessary to study and at the same time exciting to know the future of this ordinary star. When we hear or see the word ``Sun'', the first thing that pops up in our mind is a yellow star that rises in the East and sets in the West every day. Even though the Sun looks the same in the sky every day, but the importance of this yellow star did not remain hidden for long, and the curious minds started realizing it very early. Starting from the mythological story to the scientific quest, we have expanded our horizons beyond the limit to understand the Sun. Being the closest star to us provides an excellent opportunity to study it with enormous details. Moreover, it also serve as a laboratory for stellar physics to test their hypotheses and models. Throughout this thesis, we will try to get an even more comprehensive answer to what makes the Sun even more exciting.

The study of the Sun is broadly classified into two classes, the short\hyp term, where one studies the changes in Sun happening with the time scale of seconds to days, and the long\hyp term, where the changes happening in the Sun over the years is studied. This thesis aims to understand the long\hyp term behaviour of the Sun and how it can help our understanding of physics underneath. In this thesis, I will use the historical archival data as obtained from the Kodaikanal Solar Observatory (KoSO) and others, along with space-based data from Solar and Heliospheric Observatory (SOHO) and Solar Dynamic Observatory (SDO). Before I explain the details, let me start with our fundamental understanding of the Sun.

\section{The Sun}
 The Sun is young but around five billion-year-old, residing at around 150~Mkm from us. It has a mass of $\approx2\times10^{30}$~kg and is primarily made up of  Hydrogen ($\approx74.9\%$), Helium ($\approx23.8\%$) and some other metals \citep{Asplund2009}. It is a GII~V star burning hydrogen in its core via nuclear fusion, and the energy produced in this process provides support against its gravity. The structure of the Sun is broadly classified into two classes (i) internal structure and (ii) solar atmosphere. I am going to discuss these two classes briefly in the following subsections.
 
\begin{figure}[!htbp]
\centering
\centerline{\includegraphics[width=0.6\textwidth,clip=]{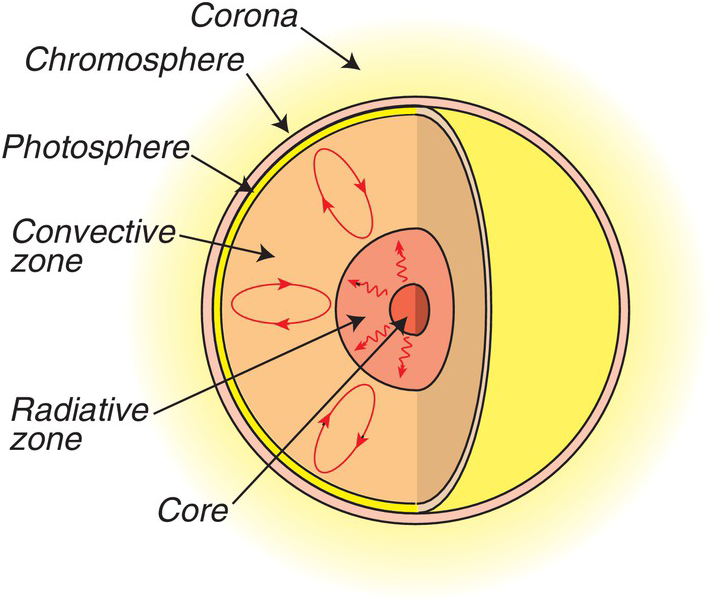}}
\caption{The anatomy of the Sun labeling the different internal and external structure of it \citep[\textit{Image courtesy:}][]{Christian2017}.}
\label{fig1.1}
\end{figure} 
 
\subsection{Internal Structure}
Being a black body makes the internal structure of the Sun invisible to all the wavelengths, and it was almost impossible to peep inside the Sun. The advent of a technique called helioseismology \citep{Dalsgaard2002} gives us an opportunity to indirectly look inside the Sun to confirm its internal structures known from the standard models \citep{Stix2012book}, which comprises of: 
\begin{itemize}
    \item {\bf Core:} This is the powerhouse of the Sun, which spans around $\approx$20\% its radius. This is the place where He, along with photons and neutrinos, are produced in {\it p-p} chain reaction. In this region of the Sun, temperature varies from $\approx$15~MK at its centre to $\approx$10~MK at its edge. The energy produced in this reaction not only provides support against gravity but also gives the energy that we receive on the Earth. 
    
    \item {\bf Radiative Zone:} The layer next to the core, which extends from 0.2~\Rsun\ to 0.7~\Rsun, where radiation is the energy transport mechanism, is called the radiative zone. The gamma photons produced in the {\it p-p} reaction have to go through a random walk in this layer, and when it comes out of the Sun, it comes out as black body radiation peaking in the visible band. It takes around $1.7\times 10^5$ years for a photon to come out of the Sun and only around 8 minutes to reach us.
    
    \item {\bf Tachocline:}  In helioseismology observations, a very thin layer of around 4\% of the solar radius, at 0.7~\Rsun, has been discovered, which is called Tachocline \citep{Charbonneau1999}. In this region as we move from radiative zone to convection zone, the rotation profile of the Sun changes from almost solid body to differential. Now, it is believed that the rapid change in the rotation could possibly has important implication to solar dynamo for the generation of a large scale magnetic field in the Sun \citep{Miesch2005}, see  \citet{Wright2016} for the more discussion about this layer.

    \item {\bf Convection Zone:} This is the region of Sun above 0.7~\Rsun and extends up to the visible surface. Here, the temperature is such that the electrons and nuclei start forming ions, leading to increased opacity and sufficient temperature gradient to start the convection. Here, the energy is transported via mass motion, which can be seen on the visible surface as granules. 
\end{itemize}
\subsection{Solar Atmosphere}

The modern space-based instruments, on-board  SOHO, SDO etc., have taken our capability to look at the Sun like never before. They enabled us to image the different layers of the solar atmosphere by observing the Sun in many wavelengths most of which are impossible to observe from the ground-based instruments. These atmospheric layers are discussed below:

\begin{figure}[!htbp]
\centering
\centerline{\includegraphics[width=\textwidth,clip=]{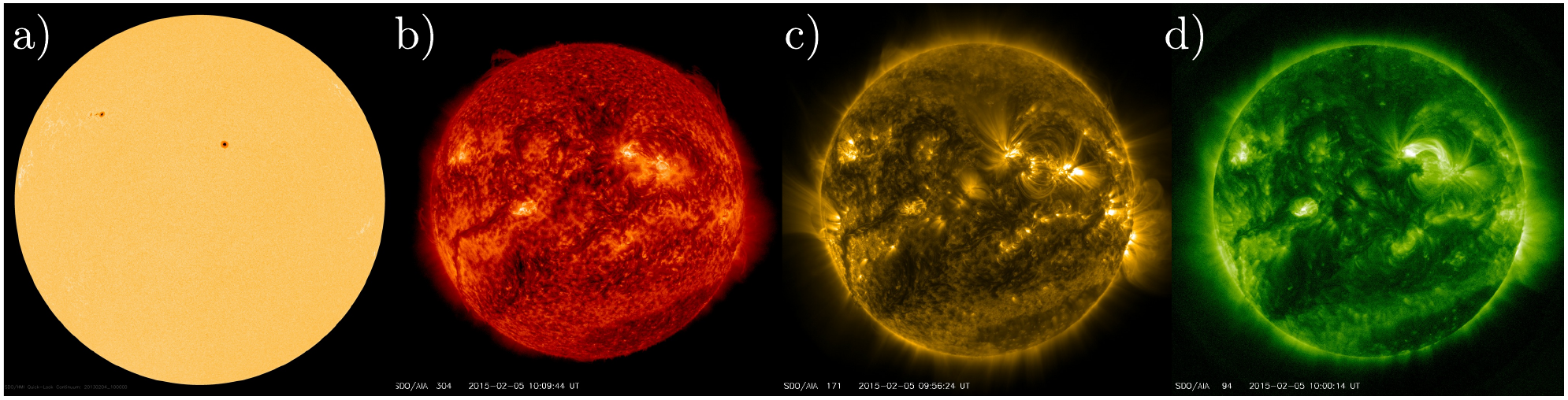}}
\caption{Image of the Sun as observed in different wavelength showing different atmospheric layers (a) Photosphere in HMI Intensity Continuum; (b) Chromosphere in AIA 304~\AA; (c) Upper Transition region in AIA 171~\AA; (d) Corona in AIA 0.94~\AA\ ({\it Image courtesy: SDO/AIA}).}
\label{fig1.2}
\end{figure} 

\begin{itemize}
    
    \item {\bf Photosphere:} It is the visible surface of the Sun (\autoref{fig1.2}a) and plays the role of the boundary between the solar interior and atmosphere. We notice a significant change in density in this region, making the solar plasma opaque to transparent. The temperature of this layer is around 5778~K; hence, plasma is weakly ionized here.
    
    \item {\bf Chromosphere:} It is derived from the Greek word meaning `` sphere of colour'' because of its colourful appearance during the total solar eclipse. In this region, temperature initially falls for around 500~km and then starts increasing as the distance from the Sun increases. Even though this region of the solar atmosphere is primarily transparent in the visible band, it can be still observed in the H-alpha and He II lines. The \autoref{fig1.2}b show the image of Sun observed He II line by SDO/AIA. 
    
    \item {\bf Transition Region:} This is an extremely thin region at the boundary of the chromosphere and corona. This region of the solar atmosphere has a large temperature gradient from thousands to million kelvin in a very short distance. \autoref{fig1.2}c show the upper transition region as observed in AIA~171~\AA.
    
    \item {\bf Corona:} This is the outermost region of the Sun, which extend from the transition region to the interplanetary medium. This region of the Sun can only be naturally seen during the total solar eclipse because of its extremely low intensity compared to the photospheric intensity. The advancement of instruments has enabled us to observe it even without the total solar eclipse, \autoref{fig1.2}d is one such example (AIA~94~\AA). The extreme temperature, varying from a million kelvin to 10 MK, makes this region very interesting. The reason behind this unusual high temperature is still one of the outstanding problems of solar physics and known as the Coronal Heating problem. Apart from that, a frequent change in the magnetic topology makes this region very dynamic.
\end{itemize}

Now, since we got an idea about the atmospheric layer of the Sun, it's time to talk about different features such as sunspots, active regions, plages, filaments, prominences etc., which are seen in these layers. The outstanding fact about these features is that they carry a lot of information about the physical conditions where they originate, and hence, the Sun. Therefore, by observing these features, one can enrich the understanding of the Sun, and the development of modern instruments has pushed our observing capability even further. One of the most prolonged observed solar features is the sunspots which have a significant place in this thesis and will be discussed in detail.

\section{Sunspot: Observers' Perspective}
\label{ch1-sunspot}

\begin{figure}[!htbp]
\centering
\centerline{\includegraphics[width=0.5\textwidth,clip=]{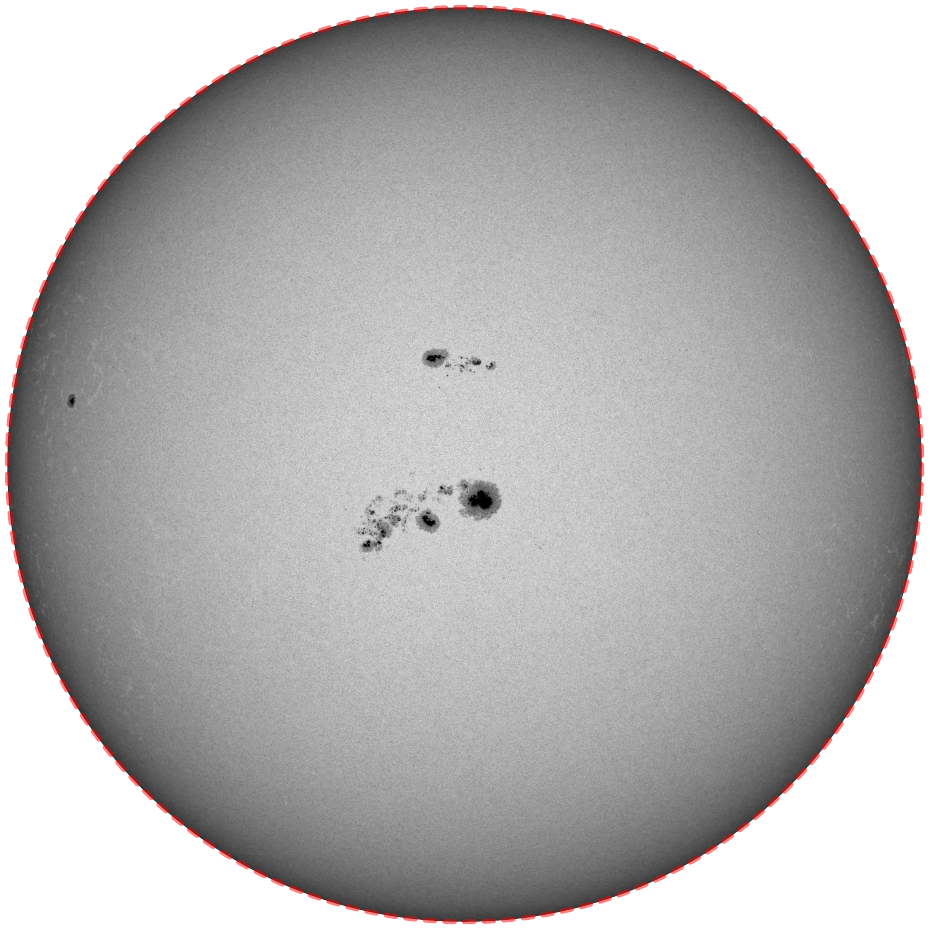}}
\caption{Full disk image of the Sun showing a few sunspots as observed from HMI on 2014-04-07 ({\it Image courtesy: SDO/HMI}).}
\label{fig1.22}
\end{figure}

On the visible surface of the Sun, the photosphere, sometime we see a few dark spots (\autoref{fig1.22}) which are called sunspots. It has been reported that the Chinese were the ones who first recorded these spots \citep{Clark1978, Wittmann1987}, but the regular observation of these dark spots only started at the beginning of the 17th century by Galilee Galileo and Christoph Scheiner, just after Galileo's advent of the telescope. Even though the systematic observation of sunspots started so early, their physical nature was hidden to us till 1908. \citet{Hale1908}, while working at Mount Wilson Observatory, by studying the Zeeman splitting of a spectral line \citep{Zeeman1896}, discovered that these sunspots are the regions of the strong and concentrated magnetic field (see \autoref{fig1.6}) with a typical field strength of $\sim 10^3$~G, which can go up to 2000--3700~G \citep{Solanki2003}. This is the first time the magnetic field was observed outside the Earth.
The presence of a strong magnetic field suppress the convection which leads to the inefficient transfer of heat to the photosphere \citep{Biermann1941}. The undersupply of heat result in lower temperature ($\approx 4500$~K) in sunspots relative to their surrounding ($5778$~K) and hence, they appear dark in the visible band. 

Sunspots as seen in the visible band (white light images) show a wide range in their morphology, size (area on disk), lifetime etc., going to be discussed briefly.

\subsection{Morphology of Sunspots}

\begin{figure}[!htbp]
\centering
\centerline{\includegraphics[width=0.7\textwidth,clip=]{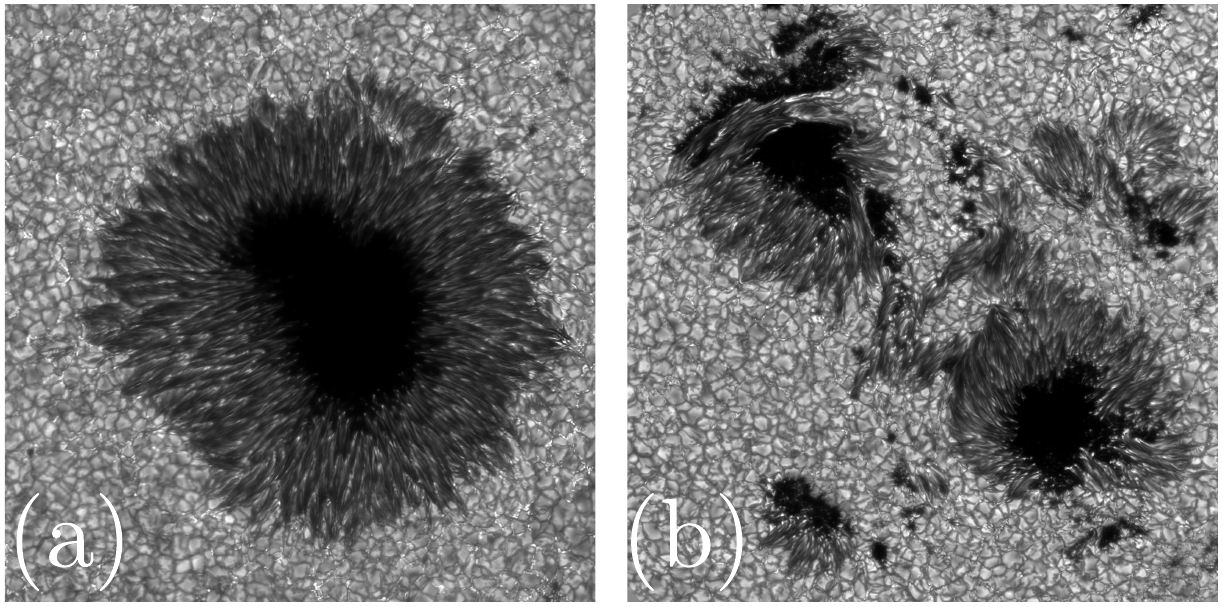}}
\caption{(a) An isolated sunspot (AR 397) and (b) a group of sunspots (AR 431), observed using Swedish Solar Telescope on July 3, 2003 and Aug 14, 2003 respectively ({\it Image courtesy: SST}).}
\label{fig1.17}
\end{figure}

Sunspots are formed isolated (\autoref{fig1.17}a) as well as in the groups (sunspot group; \autoref{fig1.17}b), consists of many sunspots. Most of the time, a typical sunspot has a two-part structure, a central darker region called umbra surrounded by a region lighter
than umbra but darker than photosphere is called penumbra (shown in \autoref{fig1.17}).
The spot that lacks penumbra is referred as the pores and believe to be the initial stage of sunspot formation \cite[see][and references therein]{Sobotka1999}. Sunspots show a wide range of shape and size; hence, based on their morphology, they are classified into different categories such as A (uni-polar with no penumbra), B (bipolar without penumbra) etc., \citep[][and references therein]{McIntosh1990, McIntosh2000}. After the 70s, the magnetic field measurement of the Sun led to another classification scheme known as Mount Wilson Magnetic classification scheme \citep{Smith1968} based on their magnetic complexity, e.g. $\alpha$, $\beta$, $\delta$ etc.  

\subsection{Size of Sunspots}

Sunspots show a broad spectrum in their sizes, starting from a few thousand km to 60,000~km and even more \citep{Solanki2003}. Sometimes, they could be so big that they can be seen with naked eyes provided the favourable weather condition. The distribution of sunspot area shows a log normal distribution which implies that the smaller sunspots are more common than, the larger ones. This observation was first reported by \citet{Bogdan1988} based on the sunspot observation at Mt. Wilson Observatory. 

The two-part structure of sunspots also encouraged to measure the ratio of penumbra to umbra area, and plenty of effort has been done to look at this ratio \citep{Antalova1971, Hathaway2013, Jha2018, Jha2019}. These studies show that the ratio increases initially with the size of the sunspot, i.e., the whole sunspot area. A detailed discussion on this ratio will be presented in \autoref{Chap3}.

\subsection{Lifetime of Sunspots}

Similar to the sizes, sunspots also show a large variation in their lifetime, it can vary from a few hours to months. Statistically, the lifetime of a sunspot show linear dependence on its maximum size \citep{Gnevyshev1938, Waldmeier1955}. The relation is given by
\begin{equation}
    a_{\rm max} = W\tau,
    \label{eq1.25}
\end{equation}
where $a_0$ and $\tau$ are the maximum areas and lifetime of the sunspot and, $W$ is proportionality constant having value $10~\mu$Hem/day ($\mu$Hem is millionth of the solar hemisphere). The value of $W$ has been further updated to $W=10.89\pm0.18~\mu$Hem/day \citep{Petrovay1997}.

\subsection{Bipolar Magnetic Region}
\begin{figure}[!htbp]
\centering
\centerline{\includegraphics[width=0.7\textwidth,clip=]{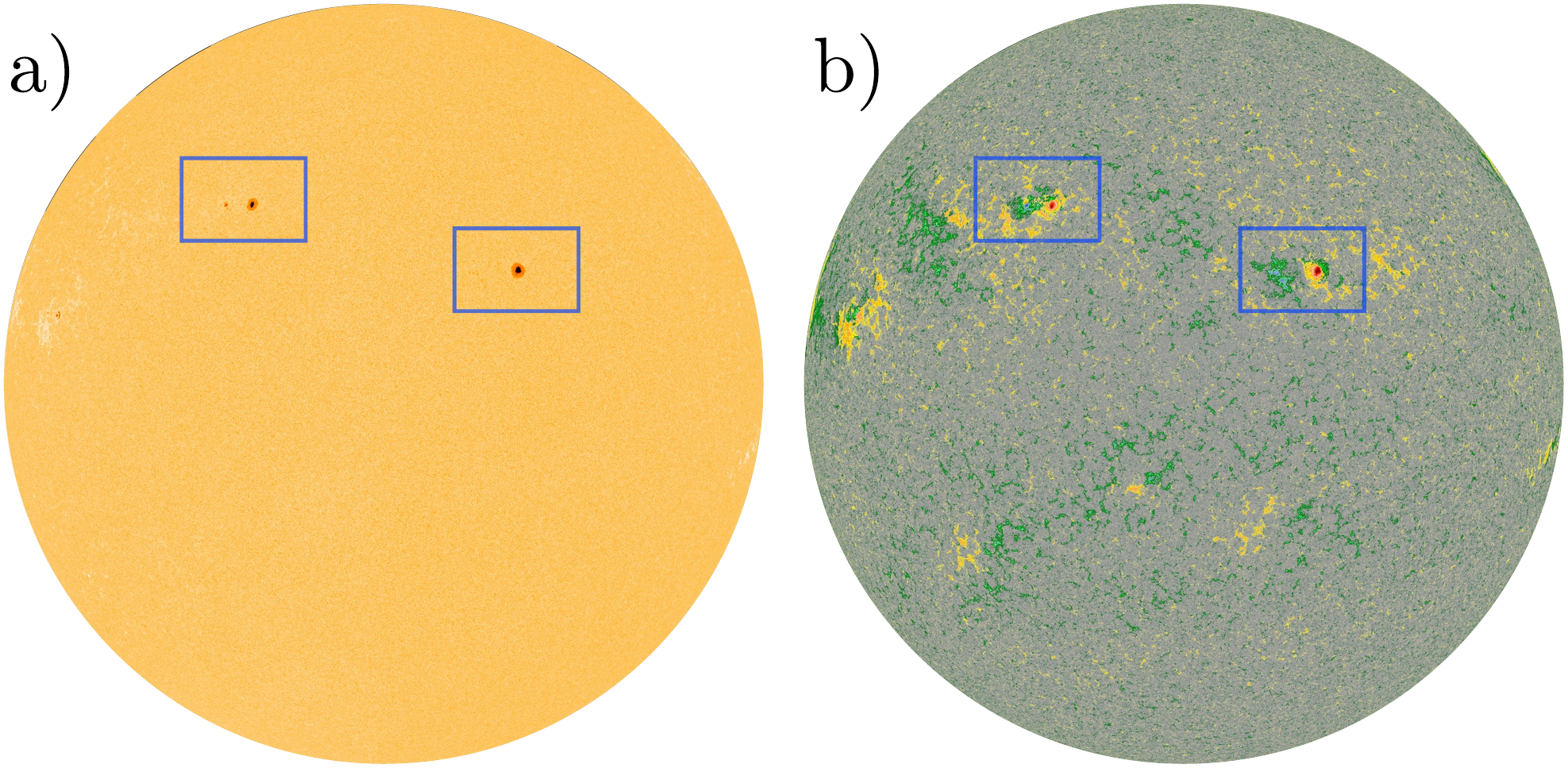}}
\caption{(a) Intensity continuum (white light) image and (b) Line of Sight Magnetogram as observed from SDO/HMI on 2013-02-05. The boxes represent two same BMR regions in both images ({\it Image courtesy: SDO/HMI}).}
\label{fig1.6}
\end{figure}

Hale's discovery of magnetic field in the sunspots \citep{Hale1908} also reveals that a group of sunspots are actually the regions of opposite magnetic polarities residing very close to each other (see \autoref{fig1.6}). Due to this reason, they are called Bipolar Magnetic Region (BMR), which is a more general feature of sunspots. The two polarities if these BMRs are named as leading and following polarities based on their movement on the visible disk. Since the Sun is rotating from East to West, it appears that one polarity of BMR is following the other. Hence, the polarity (spot) towards the West is leading, and the one towards the East is following polarity. 

\begin{figure}[!htbp]
    \centering
    \centerline{\includegraphics[width=0.7\textwidth,clip=]{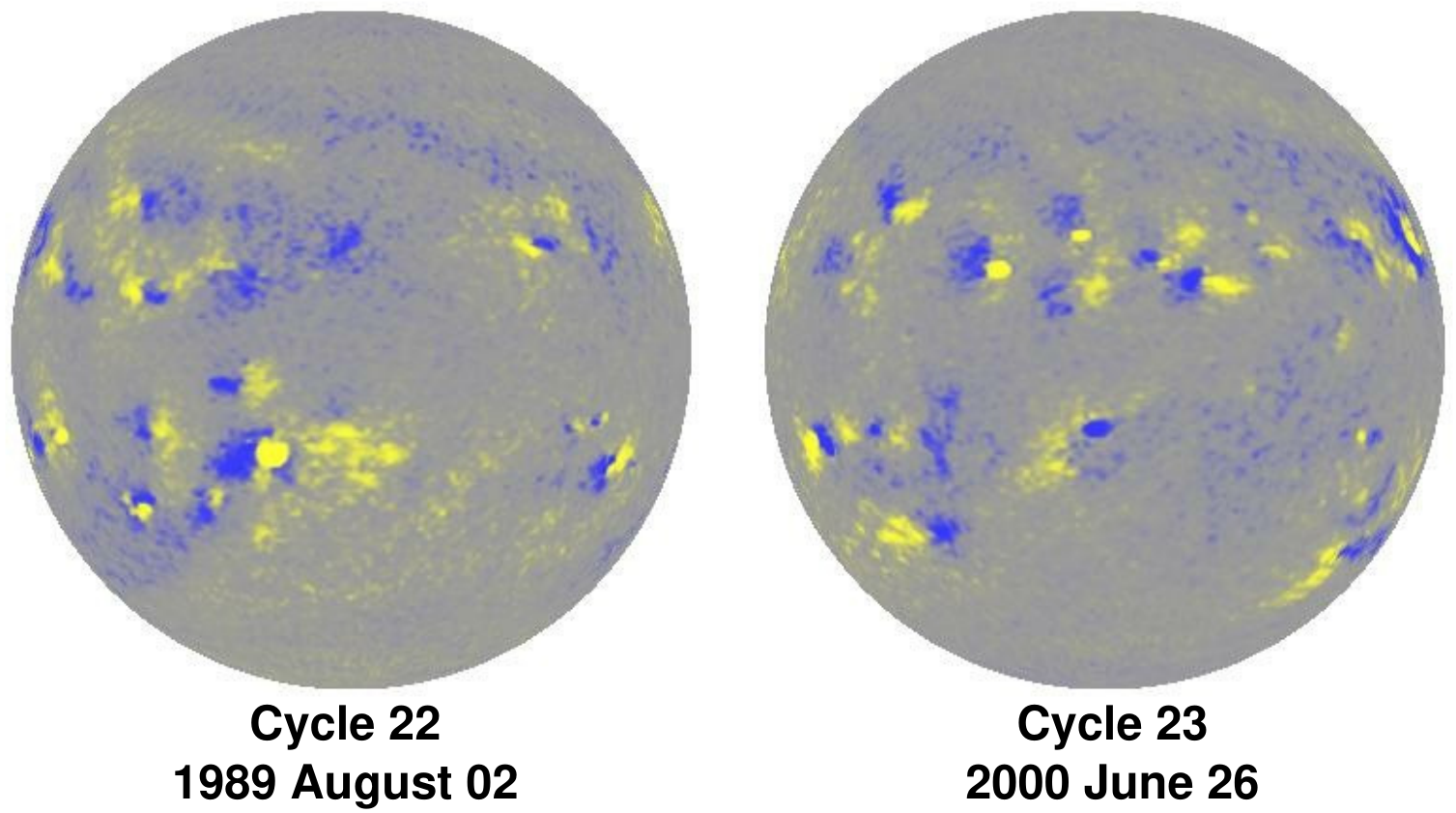}}
    \caption{The magnetogram for Cycle-22 (left) and Cycle-23 (right), with yellow and blue representing positive and negative polarity, respectively. In Cycle-22, the leading and following polarity show the opposite sign in the northern and southern hemispheres, and they flip their signs in the next cycle, Cycle-23 \citep[\textit{Image courtesy:}][]{Hathaway2015}.}
    \label{fig1.8}
\end{figure}

 The careful observation of these BMRs reveals that all BMRs in one hemisphere have the same orientation (with few exceptions), and this orientation is opposite in another hemisphere \citep{Hale1919}. This interesting characteristic of BMRs is called ``Hale's Polarity Law'' after \citet{Hale1919}. Furthermore, he also discovered that the leading and the following spots reverse their sign in each cycle as shown in \autoref{fig1.8}. But sometimes, we also notice that a few BMRs do not obey the Hale's polarity law and are hence called ``Anti Hale Sunspots''. Even though such BMRs are very few, they play a crucial role in our understanding of the Sun and solar cycle models \citep{Karak2017, Karak2018} to be discussed later.
 
\begin{figure}
\centering
\centerline{\includegraphics[width=0.7\textwidth,clip=]{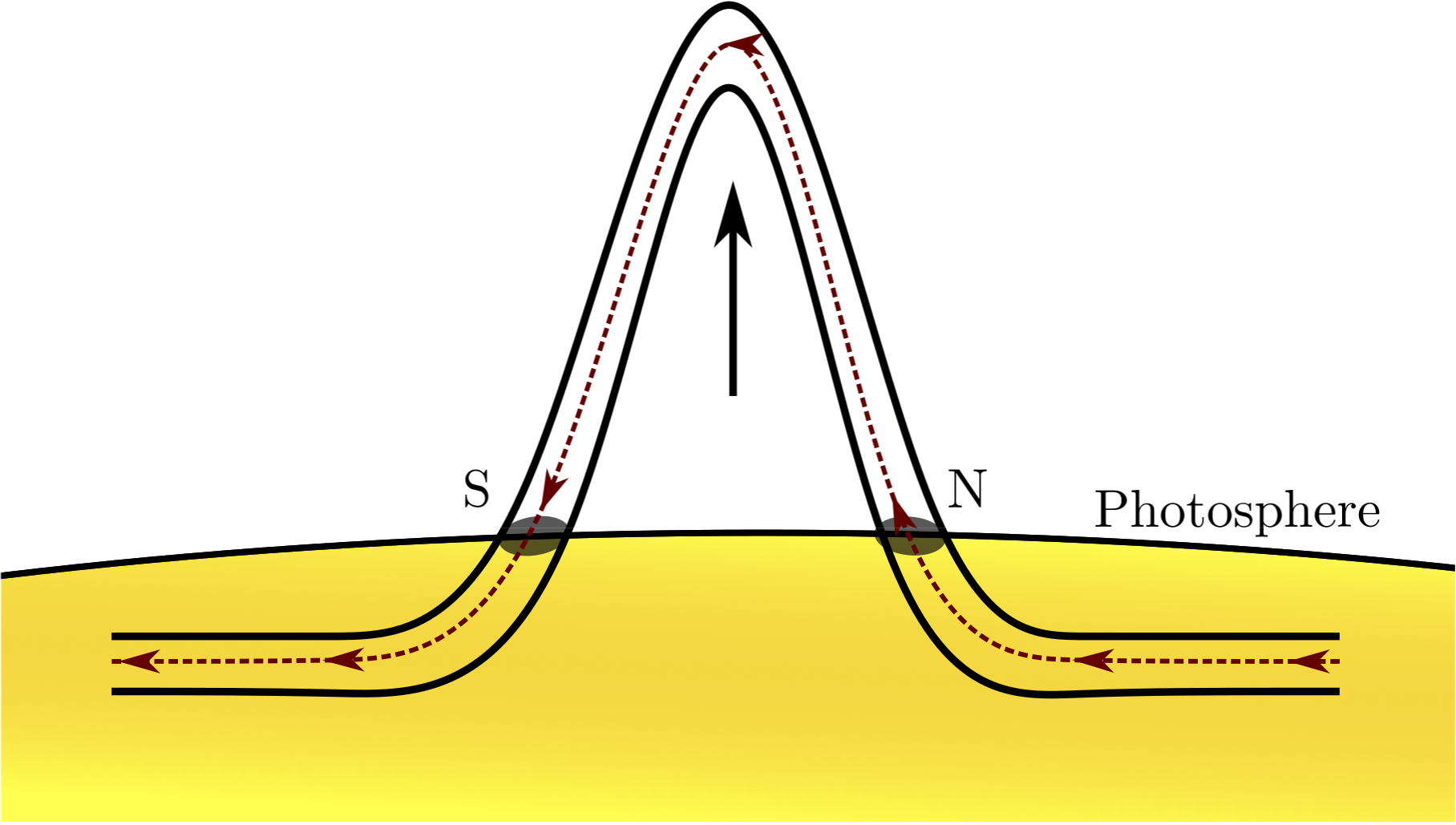}}
\caption{The rise of flux tube due to magnetic buoyancy  which leads to the formation of magnetic bipolar magnetic region (BMR).}
\label{fig1.18}
\end{figure}

Now, the question arises, how are these BMRs or sunspots formed? To understand that, let us consider a bundle of magnetic field lines which constitute a flux-tube. Now, due to the presence of a magnetic field in the flux tube, there will be an additional pressure inside called magnetic pressure ($P_{\rm mag}$) along with gas pressure ($P_{\rm gas}$), which must be balanced by only gas pressure ($P_{\rm gas}$) outside. Hence, in equilibrium,
\begin{align}
    P^{\rm out}=& P^{\rm in} \nonumber\\
    P_{\rm gas}^{\rm out}=&P_{\rm gas}^{\rm in}+P_{\rm mag}^{\rm in}, ~~~\because~~P_{\rm mag}^{\rm in}=\frac{B^2}{8\pi}\nonumber\\
    \therefore~~P_{\rm gas}^{\rm out} > &P_{\rm gas}^{\rm in};
\end{align}
which is not always but usually leads to,
\begin{equation}
    \rho^{\rm in}<\rho^{\rm out}.
\end{equation}
Lower density inside the flux tube will make it buoyant \citep{Parker1955a}, and hence it will come out of the photosphere and appear as a BMR (shown in \autoref{fig1.18}).

\subsection{Tilt of Bipolar Magnetic Region}
\begin{figure}[!htbp]
\centering
\centerline{\includegraphics[width=0.8\textwidth,clip=]{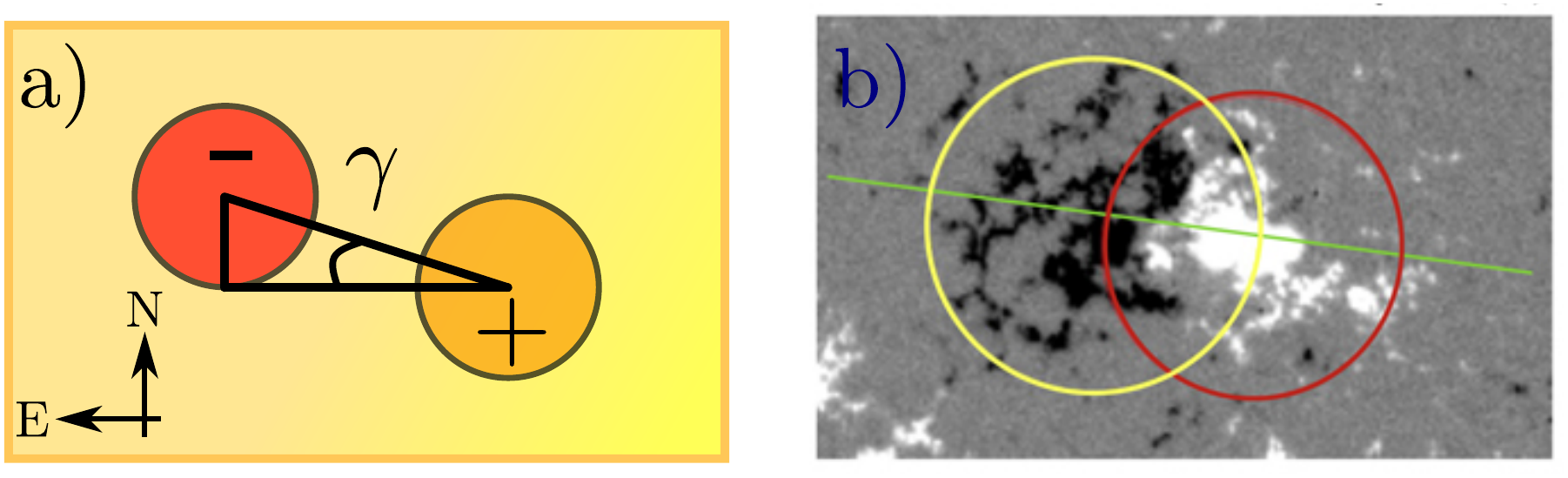}}
\caption{(a) Cartoon diagram, (b) magnetogram image \citep[taken from][]{Li2018} , showing the position of leading and following polarity of BMR with respect to solar equator.}
\label{fig1.9}
\end{figure}

 A careful examination of the BMRs (or sunspot group) show that the line joining the centre of leading and following polarity, is tilted with respect the solar East--West direction, as shown in cartoon diagram (\autoref{fig1.9}a) and in magnetogram data (\autoref{fig1.9}b).  This is called tilt ($\gamma$) of BMR, and it systematically increases with the latitude of the BMR in both hemispheres \citep{Hale1919}. The relation between tilt ($\gamma$) and co-latitude ($\theta$) is called Joy`s law and given as:
 \begin{equation}
     \gamma = \gamma_0\cos{\theta}
     \label{eq1.2}
 \end{equation}
where, $\gamma_0$ is a constant and called amplitude of Joy's law.

The tilt of BMR is induced by the effect of Coriolis force acting during the rise of flux tube due to the rotation of the Sun. The Coriolis force acting on the rising loop apexes has opposite direction and will try to twist it, bringing one polarity closer to the equator and the other away from it \citep{DSilva1993}. The systematic tilt observed in BMRs give rise to net dipole moment, which have a significant impact on the flux transport dynamo models (going to be discussed in the upcoming section), particularly in Babcock-Leighton type dynamo models \citep{Babcock1961, Leighton1964, Charbonneau2010}.

The theoretical work of \citet{DSilva1993} which explain the observed tilt in the BMRs and its dependency on the latitude, hence unfold the physics behind Joy`s Law. In their work, \citet{DSilva1993} have also elaborated the role of the strength of the magnetic field, flux for the observed tilt of BMRs.

\section{Solar Activity Cycle}
The discovery of sunspot has significant influence on the observers around the globe and prompt the systematic observation of these dark spots. Even though Christian Horrebow gave the first indication of variation in the appearance of these spots but it was not really discovered until 1844 when Heinrich Schwabe \citep{Schwabe1844} reported the existence of $\approx 10$ years periodicity in the number of appearance of these spots. This turns out to be the discovery of solar activity cycle or solar cycle. The discovery of the solar cycle has changed our perspective about the Sun, and astronomers started turning their telescopes towards it. Sunspots show a large variation in their shape and size, and hence an obvious question raised is that how quantify it. Here I discuss a few of them  that are widely used as measures for solar activity.

\subsection{Sunspot Number}
\begin{figure}[!htbp]
\centering
\centerline{\includegraphics[width=\textwidth,clip=]{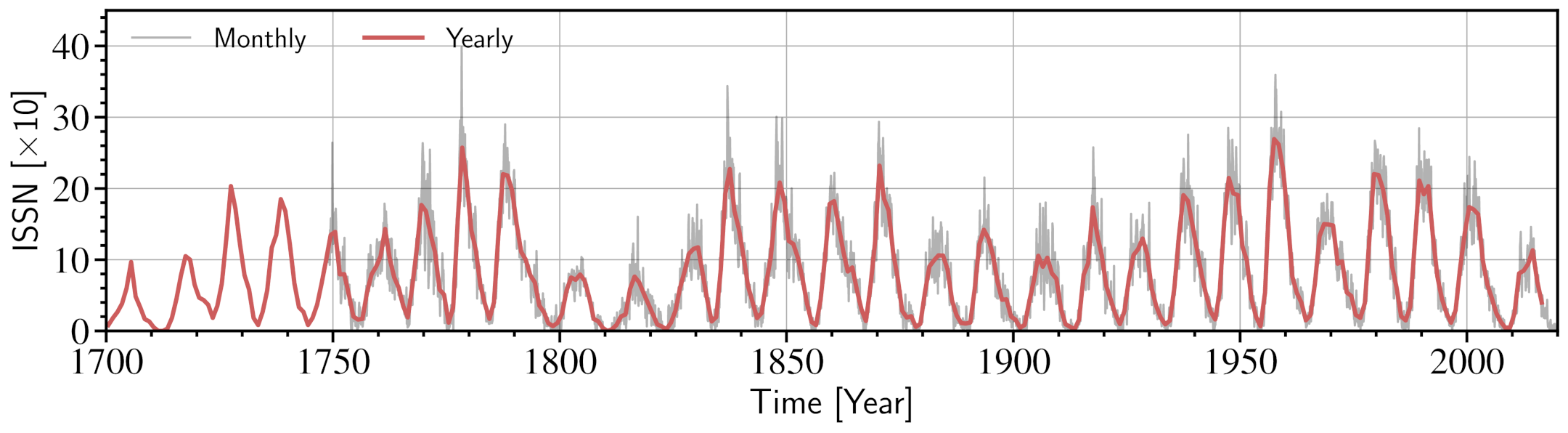}}
\caption{Variation of monthly and yearly sunspot number with time ({\it Source: WDC-SILSO, Royal Observatory of Belgium, Brussels}).}
\label{fig1.3}
\end{figure}

Encouraged by the discovery of the solar cycle by Schwabe, Rudolf Wolf, working at Bern Observatory, initiated the regular observation of these spots. He also devised the way to calculate the number of sunspots on the solar disk, called ``relative sunspot number (R)'' or ``International Sunspot Number (ISN)'' or ``Zurich Number'', defined as
\begin{equation}
    R=k(10g+n)
    \label{eq1.1}
\end{equation}
where $k$ is the correction factor, $g$ is the number of groups and, n is the number of isolated sunspots. If we look carefully at the \autoref{eq1.1}, we notice that it gives higher priority to the groups which is because they are easy to identify on the solar disk. Later in 1981, the Belgium Observatory also started the regular observation of sunspots which continued till date, and these data are freely available to the community via the Solar Influences Data Analysis Center (SIDC)\footnote{\url{https://wwwbis.sidc.be/silso/home}}. Today, the international sunspot number has been treated as one of the best proxies for solar activity; it is not because everyone agrees to it, but because of its availability. \autoref{fig1.3} shows the variation of monthly and yearly sunspot numbers taken from WDC-SILSO. ISN is the one, which is mainly used in the community as the best proxy for solar activity, there are few others, such as sunspot area, 10.7~cm Solar Flux \citep{Tapping1994}, etc., that have been also regularly used for the same.

\subsection{Sunspot Area}
\begin{figure}[!htbp]
\centering
\centerline{\includegraphics[width=\textwidth,clip=]{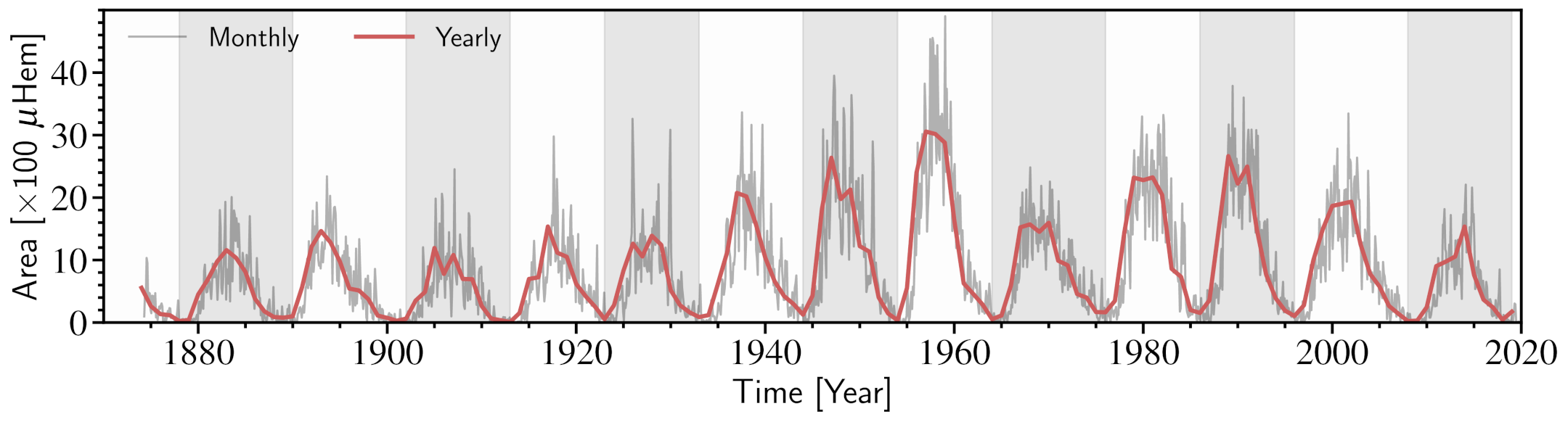}}
\caption{Monthly and yearly averaged, cross-calibrated sunspot area series as reported in \citet{Mandal2020}.}
\label{fig1.4}
\end{figure}

It is advocated that the area covered by the sunspots have greater physical significance than the sunspot number. Therefore, in 1874, Royal Observatory, Greenwich (RGO) started compiling the sunspot area and its position on the disk from the photographic plates. These plates were collected from different observatories around the globe, including Kodaikanal Solar Observatory, India. RGO continued reporting the sunspot area till 1976, and then  Debrecen Observatory \citep{Gyori2010, Baranyi2016} joined the campaign and started updating the area series. Realising the importance of this series, recently \citet{Mandal2020} has cross-calibrated the sunspot area data taken from different observatory. The cross-calibrated monthly, and yearly, sunspot area data from RGO and NOAA is shown in \autoref{fig1.4}.

In India, Kodaikanla Solar Observatory (KoSO) has also started the regular observation of the Sun from the beginning of 20th century and the area time series from the digitized data have been reported in \citet{Ravindra2013} and \cite{Mandal2017a}.

\subsection{Other Proxies}
Apart from ISN and Sunspot Area there are multiple physical quantities such as 10.7 cm Solar Flux, Total Solar Irradiance, Geomagnetic Activity, Cosmic Ray Modulation and etc. that show a good correlation with the ISN \citep[see][and references therein for details]{Hathaway2015}. Here I discuss a few of them.

\begin{itemize}
    \item {\bf 10.7 cm Solar Flux:} Another measure of solar activity period is the disc integrated radio flux observed at 10.7~cm \citep[2800~MHz;][]{Tapping1994}. Starting from 1946 Canadian Solar Radio Monitoring Program and then in 1990 new observation program at Penticton British Columbia regularly measures radio flux from the Sun. The radio flux measurements show a good correlation with the ISN \citep{Holland1984}.

    \item {\bf Total Irradiance:} The total energy from the Sun integrated over all wavelengths per unit area per second outside the Earth's atmosphere is called Total Solar Irradiance (TSI). The accurate measurement of TSI was not possible before the development of space-borne instruments such as Nimbus-7 (1978--1993), Solar Maximum Mission (SMM), ACRIM-I (1980--1989), Earth Radiation Budget Satelite (ERBS; 1984--1995) etc. This observation of the TSI suggests that it also varies with solar activity and shows a good correlation with ISN and, hence, a possible candidate for activity measurement \citep[and references therein]{Hathaway2015}.  
\item {\bf Cosmic Ray Modulation:} A cosmic ray contains high energy (GeV) particles produced in violent cosmic phenomena. When a positively charged one reaches the Earth atmosphere, it makes a cascading shower of particles. These could be measured by monitoring neutrons, and the number of neutron particles shows anti-correlation with ISN. \citet{Parker1965} explained that during high activity, these particles get scattered by the magnetic structure carried by the solar wind responsible for reduction in neutron number.
\end{itemize}

\subsection{Characteristics of Solar Cycle}

The consistent observation of the sunspots over many cycles uncovered so many solar cycle characteristics. Many of them are common, but still, a few show differences, making the sun even more interesting. A few aspects which will be frequently used in this thesis are discussed below.    

\begin{enumerate}
    \item {\bf Spoerer's Law of Zones:} The latitude time diagram of these spots show a beautiful pattern like butterfly (as seen in \autoref{fig1.4}) and called ``Butterfly Diagram'' or ``Spoerer`s Law of Zones'' first reported by \citep{Maunder1903, Maunder1904}. At the beginning of the solar cycle we observe the sunspots at higher latitude around $30^\circ -35^\circ$ in both the hemisphere but as the cycle progress sunspots start appearing towards the equator (\autoref{fig1.5}). 

\begin{figure}[!htbp]
    \centering
    \centerline{\includegraphics[width=0.9\textwidth,clip=]{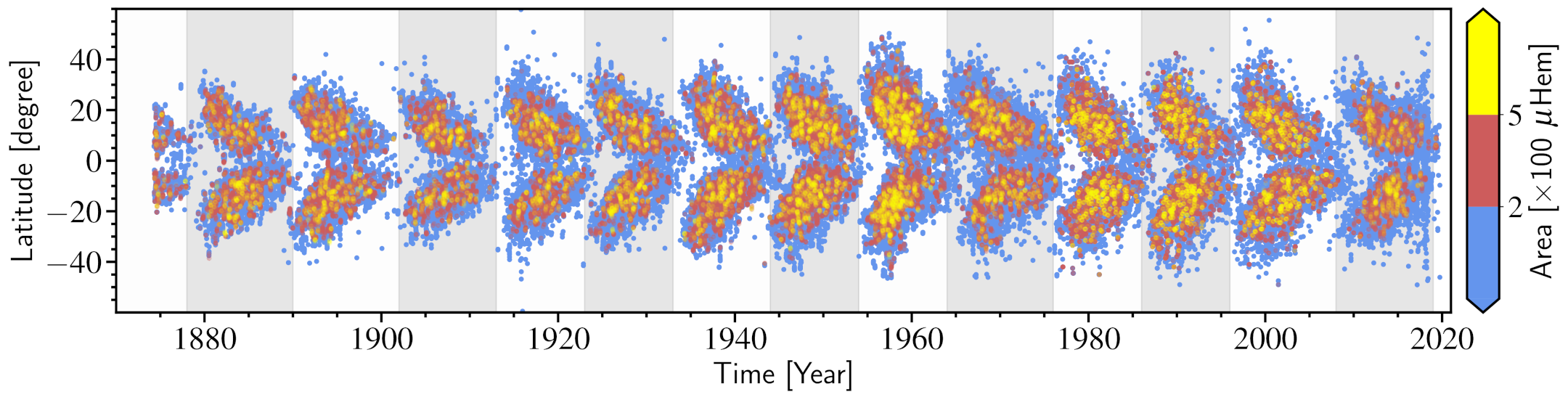}}
    \caption{A composite Latitude-Time diagram of sunspot position on the solar disk is known as the ``Butterfly Diagram''. This plot is generated using cross-calibrated data reported in \citet{Mandal2020}.}
    \label{fig1.5}
\end{figure}

    \item {\bf Cycle Maxima and Minima:}The solar maxima and minima times play a crucial role in understanding the solar activity cycle. Getting an exact time of maxima and minima is a little tricky due to inherent noise in the data. It is necessary to smooth the ISN in some way to get the proper time of occurrence of maxima and minima. Currently, these times are calculated from the 13-month running average, centred at the month of interest, over monthly mean sunspot number data \citep{Clette2016, Hathaway2015}. This is similar to the method used earlier by \citet{Waldmeier1961, McKinnon1987}.

\item {\bf Cycle Period:} Like any other periodic function, the solar cycle period is also defined as the time between two consecutive minima accompanying a maximum (sometimes more than one). In the case of the solar cycle, this definition has one issue, as each solar cycle starts a little earlier than the preceding minima and stays a little longer than the following maxima. Apart from that, the accuracy in the time calculation of minima also impacts the measurement of the activity period. Hence, the cycle period depends on the cycle of interest and the method used to determine it. Usually, the average cycle period is calculated as the time difference between two minima divided by the cycles it includes. Based on that, the average cycle period is around 131.7 months which is $\approx$~11~years \citep[][and references therein]{Hathaway2015}.

\item {\bf Cycle Amplitude or Strength:} The solar cycle amplitude or strength is the maximum number of smoothed ISN or any other proxy. It has been noticed that the cycle's amplitude depends on the proxy and method of smoothing used before determining it. Therefore, the amplitude of the solar cycle differ from each other when different proxies are used, e.g., Cycle-15 and 16 have similar amplitude when sunspot area is used as a proxy but differ significantly in the case of ISN \citep[][and references therein]{Hathaway2015}. This becomes even more complicated in the case of double peak cycles \citep[e.g., Cycle-20 \& 22;][]{Karak2018b}.
\item {\bf The Waldmeier Effect:} Most of the time, the rise time of a solar cycle is shorter than the time it takes to reach maximum to minimum \citet{Waldmeier1935}. In addition, it has also been observed that the rise time of a cycle is anti\hyp correlated with the amplitude of the cycle, and this effect is called  ``Waldmeir Effect'' after \citet{Waldmeier1935, Waldmeier1939}. The theoretical explanation for the observed effect was given by \citet{Karak2011}.

\end{enumerate}

\section{Long--term Variability}
 Apart from the dominant 11~years solar activity cycle, the historical observation of the Sun also show long\hyp term variation in solar activity \citep{Maunder1890, Eddy1976, Hoyt1998, Solanki2004, Usoskin2013}. In \autoref{fig1.3} and \autoref{fig1.4}, we see that all the cycles are not the same; they differ from each other in amplitude as well as period. This significant variation that we see in solar activity might play an essential role in understanding and predicting future cycles \citep{Petrovay2020}. There are a few major aspects of long\hyp term variation of the solar cycle, which are discussed below.
 
 \begin{figure}[!htbp]
\centering
\centerline{\includegraphics[width=0.8\textwidth,clip=]{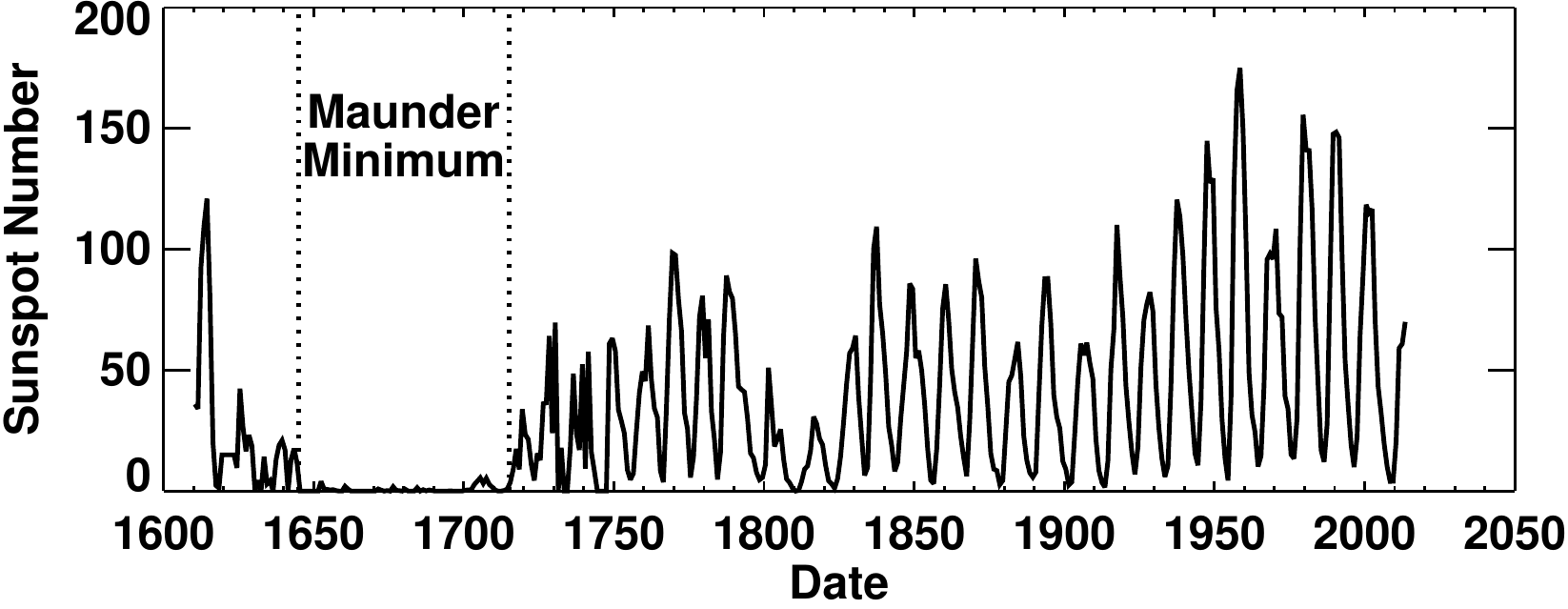}}
\caption{The yearly averaged sunspot number for the last 400 years as a function of time \citep{Hoyt1998,Hathaway2015}.}
\label{fig1.5_1}
\end{figure}

\subsection{The Maunder Minimum}
In \autoref{fig1.5_1}, during the period 1645 to 1715, we notice very little activity and this is called ``The Maunder Minimum''. This curious behaviour of the Sun was first reported by \cite{Maunder1890} while compiling the work of Spoerer`s and later by \citet{Eddy1976} with more observational support. \citet{Hoyt1998} reported the complete observation during this period depicting the drought of activity in that period. \cite{NesmeRibes1993} results also indicated the presence of a weak magnetic field in this period. The other proxies for solar activity, such as $^{10}$Be \citep{Beer1998} and $^{14}$C \citep{Stuiver1980} revels that the solar activity was so weak during that period it produced very small or no sunspots in the period. These observation also indicate the existence of many such extreme events in the history of Sun \citep{Usoskin2013}, and it is still an open problem of solar physics.

\subsection{The Secular Trend}

Just after the Maunder Minimum, we notice the slight but steady increase in the solar cycle amplitude, which was earlier reported by \citet{Wilson1988} and recently by \citet{Svalgaard2012}. \citet{Hathaway2002} also noted the positive correlation coefficient (0.7) between the amplitude and number of cycles. Furthermore,  A recent result based on radioisotope shows many such upward and a downward trends in the solar activity period in the $11,000$~years \citep{Solanki2004}. Although the trend is minimal, it might still significantly impact the understanding of the Sun. 

\subsection{Even--Odd Effect}

There is an interesting characteristic of inter cycle variation of cycle amplitude called ``Even--Odd rule'' or ``Gnevyshev–Ohl Rule'' after its discovery by \citet{Gnevyshev1948}. They found that the sum of the number of sunspots in even numbered cycle is lower than the odd numbered cycle, with few exceptions \citep[see][Figure~40]{Hathaway2015}. Our current understanding of the Sun cannot answer why such a pair exists at all? However, the observational support to this inter cycle property impacts our solar activity models \citep{Sheeley2005}.  

\subsection{Other Long--term Trends}
There are few others observed periodic variation in cycle amplitude such as a period of $\approx9.1$-cycles called ``The Gleissberg Cycle'' \citep{Gleissberg1939, Hathaway2015, Usoskin2013}, and $\approx$210~years, reported based on the radiocarbon studies often refered as the Suess or de Veies cycle \citep{Suess1980}.

\section{Large--scale Flows in the Sun}
\label{ch1-largemotion}
The observations of sunspots have not only helped us to understand the magnetic nature of the Sun but it has taken us even beyond that. One can infer that the Sun is rotating by looking at the position of sunspots on the visible disk. The methodical measurement of sunspots position on the visible disk has enabled us to measure the solar differential rotation. Based on these measurements, we know that there are two major large\hyp scale flows in the Sun (i) differential rotation and (ii) meridional circulation, which I will discuss in the following sections.

\subsection{Differential Rotation}
\label{s-differential}

\begin{figure}[!htbp]
\centering
\centerline{\includegraphics[width=\textwidth,clip=]{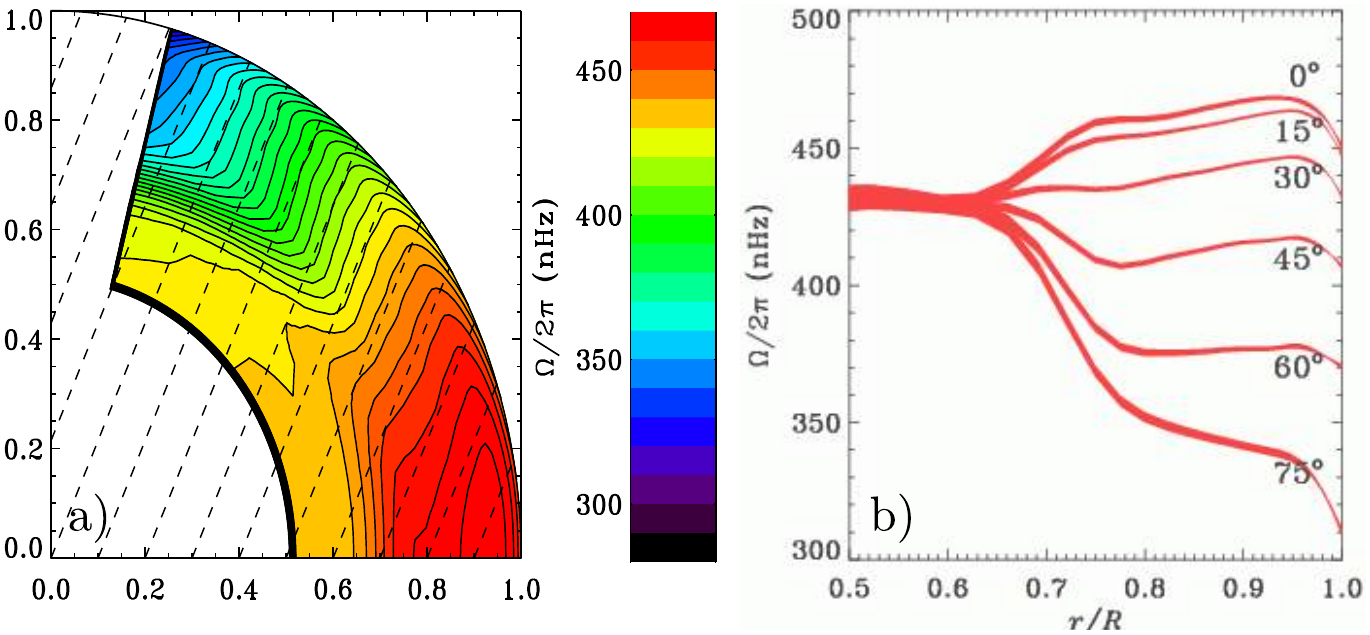}}
\caption{(a) Differential rotation map (b) latitudinal average rotation profile as a function of radius of the Sun as infered from helioseismology observation \citep[GONG data; ][]{Howe2000, Howe2005}}
\label{fig1.12}
\end{figure}

In the past, historical observations of sunspots and other solar features have been utilized (as a tracer) so many times to measure the photospheric solar differential rotation \citep[see a review by][and references therein]{Beck2000}. Based on these observation the solar differential rotation can be expressed as a function of co-latitude~($\theta$)
\begin{equation}
    \Omega(\theta) = A+B\cos^2{\theta}+B\cos^4{\theta},
    \label{eq1.3}
\end{equation}
where $A$ is equatorial rotation rate and $B$, and $C$ are called latitudinal gradients \citep[and reference therein]{Beck2000}. The issue with this method is that we see very few or no features during the minimum activity period, hence making the measurement difficult. Therefore, another method, such as spectroscopy, is also used to measure the solar differential rotation \citep{Howard1970}. A major breakthrough came in understanding solar differential rotation after the helioseismology era. Helioseismology uses the different modes of oscillation in the Sun to map the solar differential rotation profile to the surface as shown in \autoref{fig1.12}a (from the 0.5\Rsun\ to \Rsun). Variation of latitudinally averaged rotation profile with solar radius is shown in \autoref{fig1.12}b. In both the figure a sharp change in rotation profile is noticed at 0.7\Rsun. The detail discussion about the solar differential rotation will be presented in \cref{Chap4}.

\subsection{Meridional Circulation}

\begin{figure}[!htbp]
\centering
\centerline{\includegraphics[width=0.4\textwidth,clip=]{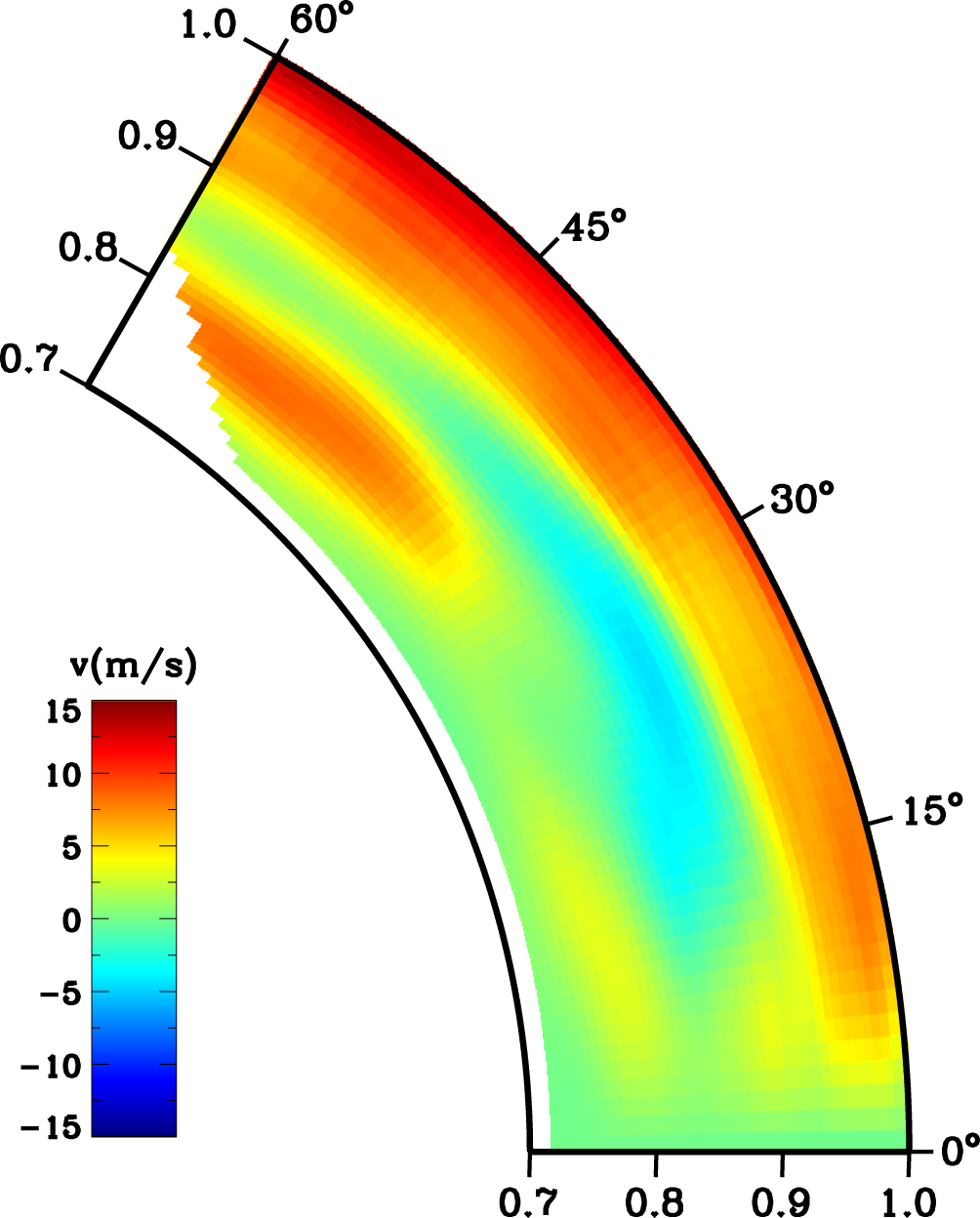}}
\caption{The flow pattern in a quadrant of the meridional plane representing meridional circulation \citep[\textit{Image courtesy:}][]{Chen2017}.}
\label{fig1.13}
\end{figure}

Another aspect of large scale flow in the Sun is a flow confined in the meridional plane, a plane that passes through the rotation axis and perpendicular to the equatorial plane, called meridional circulation. Meridional circulation is the motion of plasma from the equator towards the pole (shown in \autoref{fig1.13}) ; since we do not expect plasma to pile up near the pole, it must return to the equator at some depth below the solar surface, in both the hemisphere  \citep{Rajaguru2015, Chen2017}. It takes a little while to infer this motion in the Sun due to its very low magnitude \citep[20~ms$^{-1}$;][]{Choudhuri2021b}. The meridional circulation is far weaker than the other motion on the surface, and that is the reason it is so difficult to measure. There is something very interesting about the meridional circulation; it transports various solar features with it to the pole, making it possible to infer at least. If we look at the magnetic butterfly diagram (\autoref{fig1.10}), we notice the poleward transport of magnetic flux, which is responsible for polarity reversal, is believed to be carried by meridional circulation \citep{Charbonneau2010, Choudhuri2021b}.

\section{Global Magnetic Field}

So far, I have only talked about the magnetic field in the sunspots i.e., the magnetic field only localized in some regions, now I will, talk about the global magnetic field of the Sun.

\begin{figure}[!htbp]
\centering
\centerline{\includegraphics[width=\textwidth,clip=]{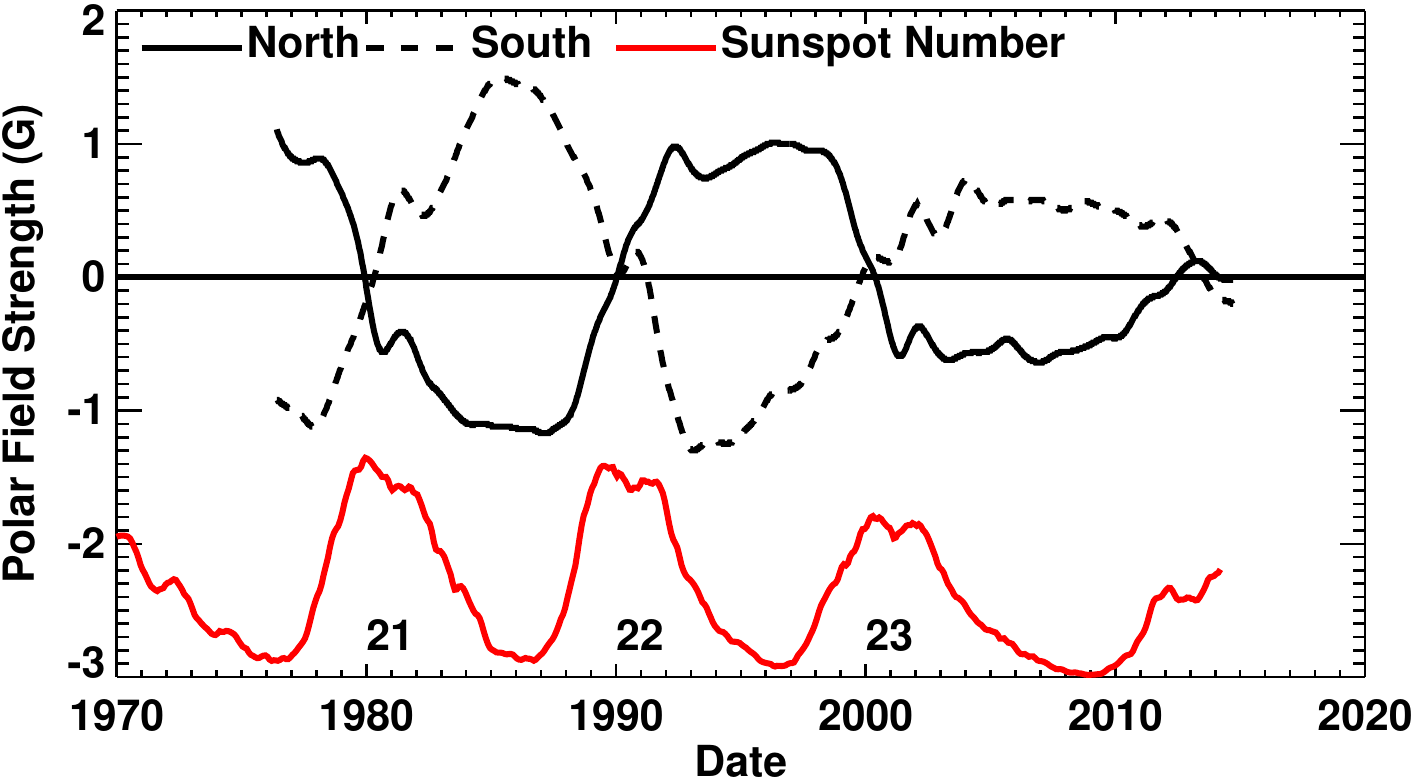}}
\caption{Polar field measurement of the Sun at WSO. The solid and dotted lines represent the polar field for the Northern and Southern hemispheres. The solid red curve at the bottom represents the scaled smoothed ISN. \citep[\textit{Image courtesy:}][]{Hathaway2015}.}
\label{fig1.21}
\end{figure}

In the second half of the 20th century, the works of \citet{Babcock1958, Babcock1959} suggests that the polar field of the Sun reverses its polarity systematically every 11~years, and this polarity reversal happens during the time of maximum solar activity. After this discovery around a decade later, Wilcox Solar Observatory \citep[WSO;][]{Scherrer1977} started to measure the polar field of the Sun regularly. This observation confirms the findings of \citet{Babcock1958, Babcock1959} and it also highlights that these polarity reversals do not need to occur at the same time in both the hemisphere\hyp it might have delay of months to years. Another interesting fact about the polar field is its strength, which is of the order of a few Gauss contrary to the field measured in sunspots is of kilo Gauss \citep{Hale1908}. The smoothed polar field is shown in \autoref{fig1.21} as measured at WSO.

\begin{figure}[!htbp]
\centering
\centerline{\includegraphics[width=\textwidth,clip=]{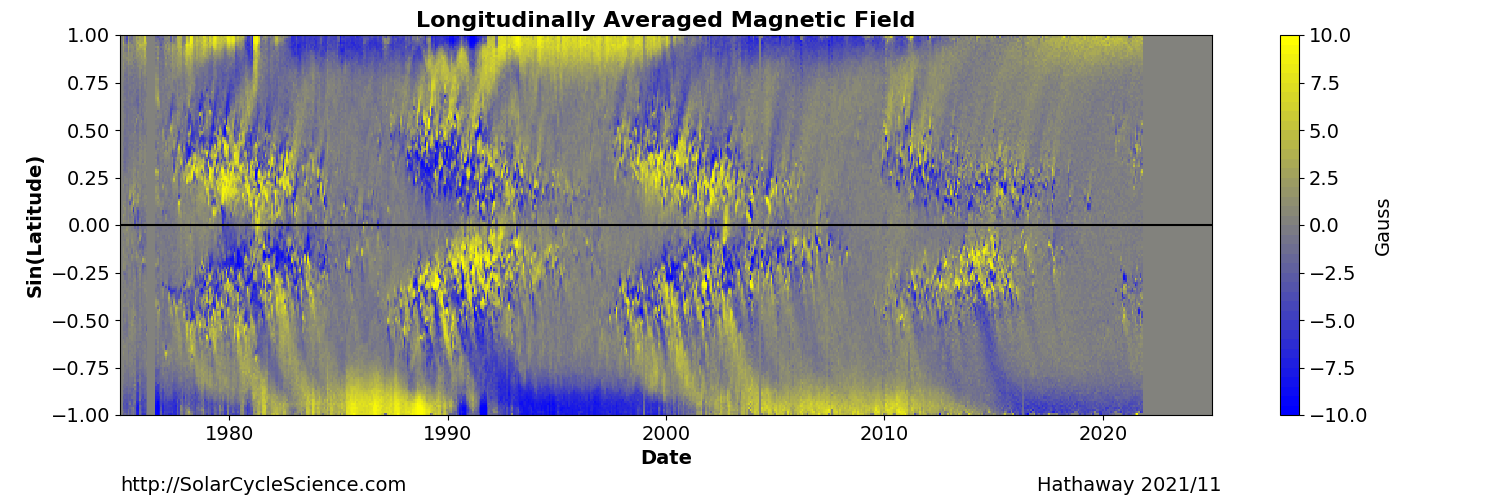}}
\caption{An azimuthally averaged magnetic map of the Sun stitched together to form a magnetic butterfly diagram. The map is constructed using the data taken from Kitt Peak, SOHO/MDI and SDO/MHI (\textit{Image courtesy: David H. Hathaway}).}
\label{fig1.10}
\end{figure}

In the early 1970s the Kitt Peak National Observatory (KPNO) started measuring line of Sight (LOS) magnetic field map (magnetogram) of the Sun. Later, National Solar Observatory \citep[NSO- SOLIS;][]{Keller1998} with slightly better resolution joied the hands and started observing the full disk magnetogram. A substantial change happened after the development of space based instruments like Michelson Doppler Imager (MDI) on the Solar and Heliospheric Observatory \citep[SOHO, 1996;][]{Scherrer1995}, and Helioseismic and Magnetic Imager (HMI) on-board Solar Dynamic Observatory \citep[SDO, 2010;][]{Schou2012}. In \autoref{fig1.10}, we show the longitudinally averaged radial magnetic field of map of the Sun, which is called magnetic butterfly diagram, for the last four-cycle. We notice in the diagram that the butterfly wings are dominated by opposite polarities in the northern and southern hemispheres, near the equator and the pole, as expected based on Hale's Polarity law. Moreover, we also observe that higher latitude field is transported to the pole, where they are eventually responsible for the polarity reversal during the maximum time \citep{Mordvinov2022}.  

\begin{figure}[!htbp]
\centering
\centerline{\includegraphics[width=0.95\textwidth,clip=]{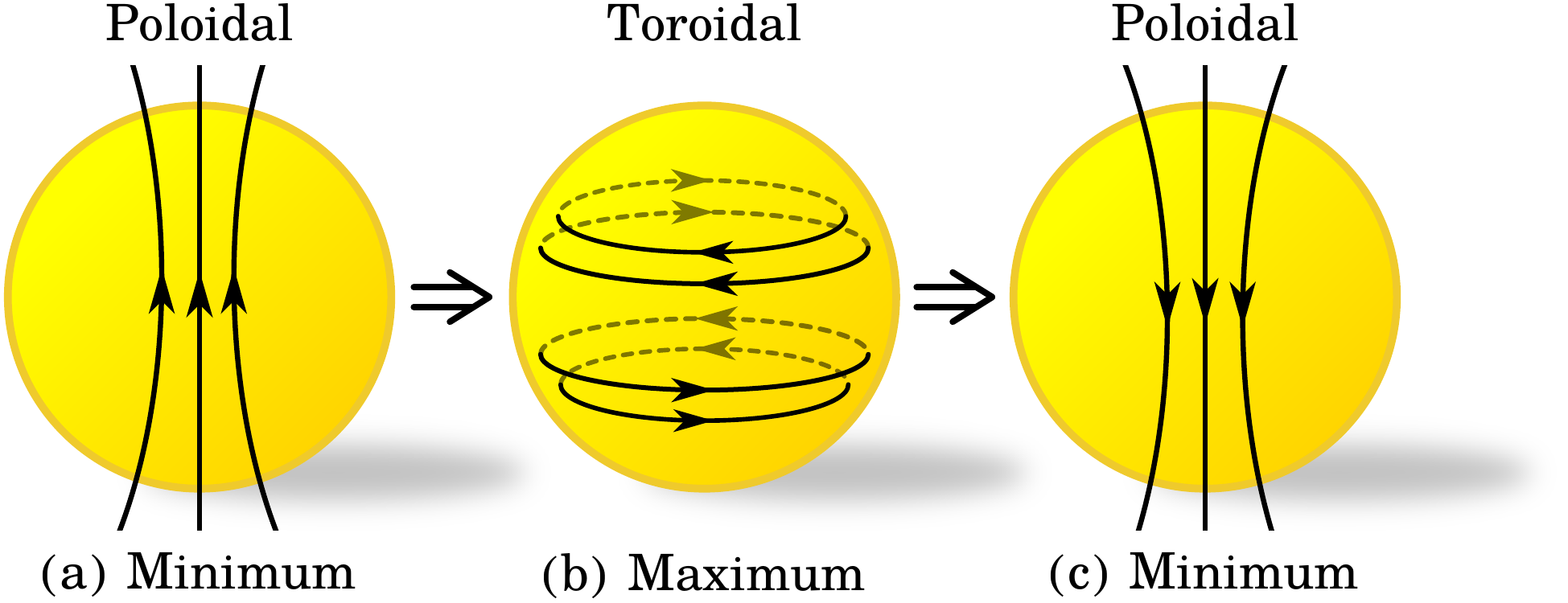}}
\caption{A flow diagram showing the cyclic nature of solar magnetism based on the observation.}
\label{fig1.10_1}
\end{figure}

These observations of global solar activity not only reveal the cyclic nature of solar magnetism it also helps us to understand the change in magnetic field configuration during the solar cycle. During the minimum activity period, the magnetic field configuration of the Sun is like a bar magnet and called a poloidal field (\autoref{fig1.10_1}a); on the other hand, the field configuration is like a toroid in the plane parallel to the equator and called a toroidal field (\autoref{fig1.10_1}b). The poloidal field of a few Gauss is getting converted to the toroidal field and amplified to the order of kilo Gauss \citep[as observed in the form of sunspots;][]{Choudhuri2003}, which is responsible for the generation of sunspots (\sref{ch1-sunspot}). After that, the diffused magnetic flux is transported to the pole by meridional circulation, which is responsible for polarity reversal (\autoref{fig1.10}) and generation of poloidal field with opposite orientation (\autoref{fig1.10_1}b). The conversion of poloidal to toroidal and back to poloidal with opposite polarity takes around 11~years (solar cycle). In contrast, it takes around 22~years to get the same global magnetic field configuration, known as the magnetic cycle.

Now, it is essential to understand the physics behind all these observed behaviour of the Sun and here comes the solar dynamo theory which is discussed in \sref{ch1-dynamo}.

\section{Solar Dynamo}
\label{ch1-dynamo}

Hitherto, we have only discussed about the observation aspects of the Sun, the solar activity cycle, solar magnetism, and large\hyp scale flows. This section will discuss the causes of the observed aspects of solar magnetism based on our current understanding of the solar dynamo, which is believe to operating in the convection zone of the Sun.

The ``Solar Dynamo'' is the mechanism that helps the Sun amplify and sustain its magnetic field. Joseph Larmer was the first person who came up with the idea about the inductive action of the fluid, and it turned out to be the foundation stone for the solar dynamo. In his idea, the axisymmetric solar differential rotation was an obvious choice responsible for amplification and conversion of poloidal (configuration of the magnetic field during solar minimum) field to toroidal field (configuration of the magnetic field during solar maximum). Moreover, it also nicely fit Hale's polarity law \citep{Hale1919}. A major hurdle came into the path of this idea as Cowling's anti-dynamo theorem \citep{Cowling1933}. This theorem says that axisymmetric flows could not, in themselves, maintain the self-sustaining axisymmetric magnetic field, against the Ohmic dissipation \citep{Cowling1933, Charbonneau2010}. \citet{Parker1955} came with the idea that can bypass this problem. He pointed out that the Coriolis force will twist the rising fluid elements in the solar convection zone due to the rotation of the Sun and so break the axisymmetry. This groundbreaking idea led to the development of mean-field electrodynamics, the theory of solar dynamo modelling. Before I discuss the mean-field dynamo models, let's briefly discuss the fundamental equations of magnetohydrodynamics (MHD), which is at the crust of solar dynamo models.

\subsection{MHD Equations to Solar Dynamo}
In a magnetised medium, there are four set of equations (eight scalar) that describe the complete dynamical system \citep{Choudhuri1998book}. These equations are as follow and all the terms have their standard meaning \citep[see][for details about these equation]{Priest2014book}.

\begin{enumerate}
    \item The most important equation of MHD, which talk about the inductive effect of the flow and called induction equation is given as,
\begin{equation}
\frac{\pa\vb{B}}{\pa t}=\bm{\nabla}\times[\vb{v}\times\vb{B}-\lambda\bm{\nabla}\times\vb{B}]. \label{eq1.6}
\end{equation}
    \item The second in the list is called momentum equation and it talk about the conservation of momentum and written as,
\begin{equation}
\rho\frac{d\vb{v}}{dt}=-\bm{\nabla}p+\frac{1}{4\pi}(\bm{\nabla}\times\vb{B})\times \vb{B}+\vb{F}+\nu\bm{\nabla}^2\vb{v}.   \label{eq1.5}
\end{equation}
    \item The third equation in the set is the continuity equation which describe the conservation of mass and given as,
\begin{equation}
\frac{d\rho}{dt}+ \rho\bm{\nabla}\cdot\vb{v}=0.
    \label{eq1.4}
\end{equation}
    \item The fourth and final equation in this set is the energy conservation equation, which is
\begin{equation}
\frac{\rho^\gamma}{\gamma-1}\frac{d}{dt}\left( \frac{p}{p^\gamma}\right)=-\bm{\nabla}\cdot\vb{q}-L_r+\frac{j^2}{\sigma}+F_H.
    \label{eq1.7}
\end{equation}
\end{enumerate}
If we consider in-compressible fluid for simplicity, the continuity equation (\autoref{eq1.4}) and energy equation (\autoref{eq1.7}) become redundant and we only left with momentum and induction equations (\autoref{eq1.5} and \ref{eq1.6}). Let`s have a detailed look at the induction equation first (\autoref{eq1.6}).
\begin{equation}
    \frac{\pa\vb{B}}{\pa t}=\underbrace{\bm{\nabla}\times(\vb{v}\times\vb{B})}_{\text{Inductive }}-\underbrace{\bm{\nabla}\times(\lambda\bm{\nabla}\times\vb{B})}_{\text{Dissipative}}, \label{eq1.8}
\end{equation}
the first term in this equation is a source term, and another is Ohmic dissipation. The relative significance of these two terms is measured in terms of magnetic Reynolds Number (${\rm R_m}$), which is the ratio of two terms,
\begin{equation}
{\rm R_m}=   \frac{vL}{\lambda}=\frac{L^2}{\lambda \tau}.
\label{eq1.9}
\end{equation}
In the case of the Sun, this number is pretty huge and one can ignore the second term, which makes it little simpler. A further simplistic approach is used by assuming a given velocity field which again makes it unnecessary to solve the momentum equation (\autoref{eq1.5}), and this approach is known as the ``kinematic approach'' or ``kinematic dynamo problem'' \citep{Karak2014}. Therefore, in this approach, only the induction equation is sufficient to understand the magnetic field's evolution for a given velocity field. In the first impression, this assumption may be ad-hock, but the recent observation of solar differential rotation and meridional circulation \citep{Dalsgaard2002} have shown that this assumption is not so inappropriate. 

Now, once we have the velocity field ($\vb{v}$) in such a way that it can sustain the magnetic field in the Sun against Ohmic Dissipation, the source term in induction equation will amplify the magnetic field by shearing, compression and transport as seen in the equation
\begin{equation}
    \bm{\nabla}\times(\vb{v}\times\vb{B})=\underbrace{(\vb{B}\cdot\bm{\nabla})\vb{v}}_{\text{Shearing}}-\underbrace{\vb{B}(\bm{\nabla}\cdot\vb{v})}_{\text{compression}}-\underbrace{(\vb{v}\cdot\bm{\nabla})\vb{B}}_{\text{Transport}}.
    \label{eq1.10}
\end{equation}
One point is worth noticing that the first term (shearing) itself of this equation will give rise to the required amplification of the magnetic fields.

\subsubsection*{The Solar Dynamo Problem}
\vspace{-0.7cm}

Although the \autoref{eq1.11} is sufficient enough to sustain and amplify the magnetic field, the dynamo problem in case of the Sun is more profound, and it is much more than sustaining and amplifying the field.  We need a cyclic generation of the magnetic field along with other observed solar cycle characteristics. The axisymmetric solar differential rotation can amplify the poloidal field but the conversion of poloidal back to toroidal is ruled out by anti-dynamo theorem in case of axisymmetric flows. This leads to the development of mean field dynamo model which was qualitatively given by \citet{Parker1955} and the mathematical formulation for the same was demonstrated by \citet{Krause1980}, keeping the following shopping list in mind \citep{Charbonneau2010}.
\begin{itemize}
    \item A polarity reversal with a periodicity of 11~years.
    \item Equatorial migration of sunspot generating magnetic field i.e., butterfly diagram.
    \item Migration of diffused magnetic flux towards the pole.
    \item Anti-symmetric with respect to equator, i.e., Hales`s polarity law,
    \item Poloidal and toroidal field strength in agreement with the observation.
    \item Last but not the least, the model should also be able to produce a fluctuation in the cycle amplitude.
\end{itemize}

Now I will briefly discuss the different dynamo models while  keeping our shopping list in mind.

\subsection{Mean Field Dynamo Model}
A simple and axisymmetric solution for a solar dynamo models has been already ruled out by Cowling's anti dynamo theorem \citep{Cowling1933}. Eugen Parker's qualitative idea about the effect of Coriolis force \citep{Parker1955} came as a savior and further leads to the mean field electrodynamics or mean field dynamo model.

In mean field dynamo model, since we are only interested in the evolution of the system only in the large scale hence a physical quantity can be written as the sum of its large scale mean and small scale fluctuation. Therefore, in this context velocity and magnetic field could be written as
\begin{equation}
    \vb{v}=\vb{\langle v\rangle}+\vb{v^\prime}~~\&~~\vb{B}=\vb{\langle B\rangle}+\vb{B^\prime}.
    \label{eq1.11}
\end{equation}
Where, $\angles{\vb{v}^\prime}=0$ and $\angles{\vb{B}^\prime}=0$. If we use these assumption in induction equation (\autoref{eq1.6})
\begin{equation}
    \frac{\pa}{\pa t}(\angles{\vb{B}}+\vb{B^\prime})=\bm{\nabla}\times\left[ (\angles{\vb{v}}+ \vb{v^\prime})\times (\angles{\vb{B}}+ \vb{B^\prime)} -\lambda\bm{\nabla}\times(\angles{\vb{B}}+\vb{B^\prime})\right].
    \label{eq1.12}
\end{equation}
After simplification and taking the average we will get
\begin{equation}
    \frac{\pa}{\pa t}\angles{\vb{B}}=\bm{\nabla}\times\left[\angles{\vb{v}}\times \angles{\vb{B}}+\angles{\vb{v}^\prime\times\vb{B}^\prime} -\lambda\bm{\nabla}\times\angles{\vb{B}}\right].
    \label{eq1.13}
\end{equation}
Now since I am going do deal with only average quantity so I can drop the average sign and replace $\angles{\vb{B}}$ by $\vb{B}$, and hence
\begin{equation}
    \frac{\pa \vb{B}}{\pa t}=\bm{\nabla}\times\left(\vb{v}\times \vb{B}+ \bm{\mathcal{E}} -\lambda\bm{\nabla}\times\vb{B}\right).
    \label{eq1.14}
\end{equation}
 where 
$\mathcal{E}=\angles{\vb{v}^\prime\times\vb{B}^\prime}$ and called ``mean turbulant electromotive force''. In case of homogeneous and isotropic turbulence $\mathcal{E}$ could be written as \citep{Choudhuri1998book, Charbonneau2010}
\begin{equation}
    \bm{\mathcal{E}}=\alpha \vb{B}-\beta\bm{\nabla}\times \vb{B},
    \label{eq1.15}
\end{equation}
where $\alpha$ and $\beta$ are tensors, which depend on the statistical properties of the flow \citep{Krause1980}. Using \autoref{eq1.15} in \autoref{eq1.14} we get
\begin{equation}
    \frac{\pa \vb{B}}{\pa t}=\bm{\nabla}\times\left[\vb{v}\times \vb{B}+ \alpha \vb{B} -(\lambda+\beta)\bm{\nabla}\times\vb{B}\right],
    \label{eq1.16}
\end{equation}
where, $\lambda+\beta$ is called magnetic diffusivity \citep{Charbonneau2010} and this is the called the basic dynamo equation \citep{Choudhuri1998book}.

Now, let`s consider the magnetic and velocity field of the form of axisymmetric poloidal and azimuthal (toroidal) components
\begin{align}
    \vb{B}(r,\theta,t)&=\underbrace{\bm{\nabla}\times (A(r, \theta, t)\bm{\hat{e}}_\phi)}_{\text{Poloidal}}~+~\underbrace{B(r,\theta,t)\bm{\hat{e}}_\phi}_{\text{Toroidal}}, ~~\text{and} \label{eq1.17}\\
     \vb{v}(r,\theta)&=\underbrace{v_r(r,\theta)\bm{\hat{e}}_r+v_\theta(r,\theta)\bm{\hat{e}}_\theta}_{\text{Meridional  Circulation}~(\vb{v}_{\rm p})}~~+~~\underbrace{\bar{\omega}\Omega(r,\theta)\bm{\hat{e}}_\phi}_{\text{Diff. Rotation}}, \label{eq1.18}
\end{align}
where $\bar{\omega}=r\sin{\theta}$ and $\Omega$ is the angular velocity. Substituting \autoref{eq1.17} and \ref{eq1.18} in \autoref{eq1.16} and using $\bm{\nabla}\cdot\vb{v}_{\rm p}$ (incompressiblilty). We can separate \autoref{eq1.16} into two equations of $B$ and $A$ \citep[for derivation see ][]{Choudhuri1998book, Charbonneau2010, Priest2014book}.
\begin{align}
    \frac{\pa B}{\pa t}=&(\lambda+\beta)\left( \bm{\nabla}^2-\frac{1}{\bar{\omega}^2}\right)B+\frac{1}{\omega}\frac{\pa(\bar{\omega}B)}{\pa r}\frac{\pa(\lambda+\beta)}{\pa r}-\bar{\omega}\vb{v}_{\rm p}\cdot\left( \frac{B}{\bar{\omega}}\right)\nonumber\\
    &+\underbrace{\bar{\omega}[\bm{\nabla}\times(A\bm{\hat{e}}_\phi)]\cdot\bm{\nabla}\Omega}_{\text{shear}}~+~\underbrace{\bm{\nabla}\times[\alpha\bm{\nabla}\times(A\bm{\hat{e}}_\phi)]}_{\text{MFT source}}\label{eq1.19}\\
\frac{\pa A}{\pa t}=&(\lambda+\beta)\left( \bm{\nabla}^2-\frac{1}{\bar{\omega}^2}\right)A-\frac{\vb{v}_{\rm p}}{\bar{\omega}}\cdot\bm{\nabla}(\bar{\omega}A)+\underbrace{\alpha B}_{\text{MFT source}} \label{eq1.20}
\end{align}

\begin{figure}[!htbp]
\centering
\centerline{\includegraphics[width=0.9\textwidth,clip=]{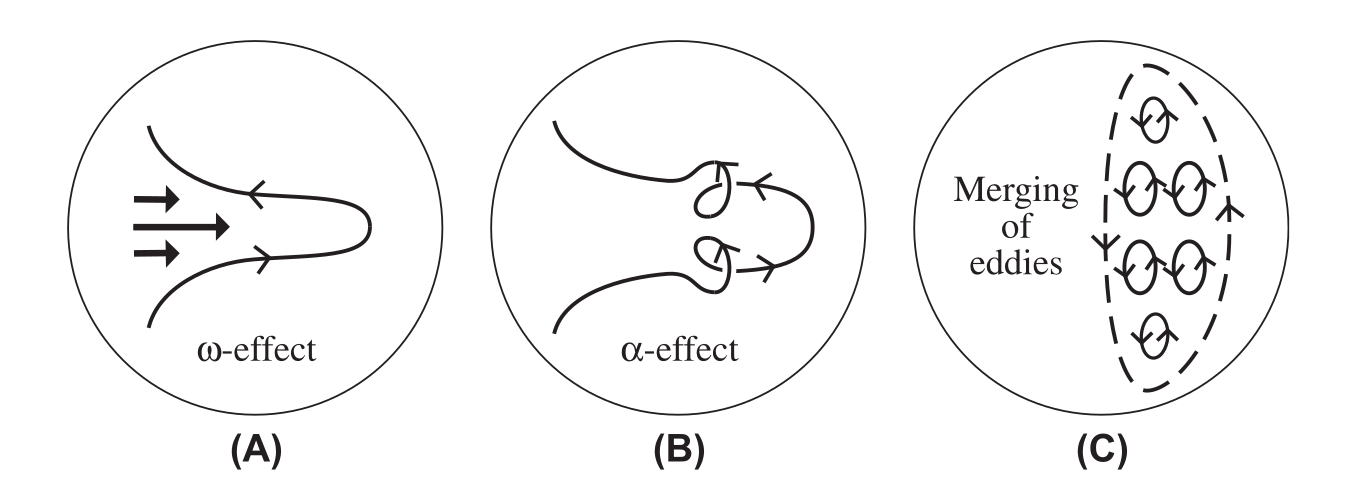}}
\caption{A cartoon diagram showing (A) $\Omega$-process, (B) helical twist (eddy) of rising flux tube due to Coriolis force and, (C) the resultant effect of many such eddies; in the mean-field solar dynamo model \citep[\textit{Image courtesy: }][]{Engvold2019book}.}
\label{fig1.14}
\end{figure}

If we carefully look at these equations, we see the additional source term coming due to the mean-field theory (MFT), and these two terms play a significant role in circumventing Cowling's anti dynamo theorem \citep{Cowling1933}. In the case of the Sun, the shear term dominates over the MFT term \citep{Charbonneau2010} in \autoref{eq1.19} hence, one can ignore it, and this leads to so-called $\alpha\Omega$-dynamo. If for a time being we put aside the other terms (turbulent diffusion and transport) in \autoref{eq1.19} we see that the shear term will leads to the amplification of poloidal field $(\vb{B}_{\rm p}=\bm{\nabla}\times(A\bm{\hat{e}}_\phi))$ proportional to the gradient of differential rotation ($\Omega$-effect, see \autoref{fig1.14}a).
\begin{equation}
    B(r,\theta,t) \approx (\vb{B}_{\rm p}\cdot\bm{\nabla}\Omega)t
    \label{eq1.21}
\end{equation}
Now, we come to another equation (\autoref{eq1.20}) that tells the evolution of the poloidal field. The first two-term in the equation are diffusion and transport, the third one is the source term, and the effect of this term is known as $\alpha$-effect (\autoref{fig1.14}b). This $\alpha$-effect is nothing but the quantitative form of the helical twist suggested by \citet{Parker1955}. Based on these equations, one can say that the $\alpha$-effect is a viable mechanism for converting the toroidal field to a poloidal field (\autoref{fig1.14}c). Another remarkable characteristic of this $\alpha\Omega$ is that it gives the travelling wave solution, as proposed by \citet{Parker1955}, in agreement with the observed butterfly diagram.

Another possibility has been suggested by \cite{Babcock1961} and \citet{Leighton1964, Leighton1969}, which is called ``Babcock-Lighton Mechanism''. It provides another possible mechanism for the conversion of toroidal to the poloidal field, which I will discuss in the next section.

\subsection{Babcock--Leighton Dynamo Model}
The process that we called Babcock and Leighton (BL) mechanism was proposed by \citet{Babcock1961} and \citet{Leighton1964, Leighton1969} in the 60s. Initially, the solar cycle models based on this process were shadowed by mean-field models, but the synoptic observation of solar magnetic field in the latter part of the 20th century has given new life to this model.

In the BL mechanism, the diffused flux resulting from the decay and dispersal of the BMRs is transported to the pole via surface flow, particularly the meridional circulation \citep{Charbonneau2010}. This transported flux first cancels out the existing opposite flux and finally leads to the polarity reversal. In other words, the systematic tilt observed in BMRs has the net dipole moment. The fraction of this dipole moment resulting from BMRs' dispersal is transported to the pole, resulting in the global dipole moment. This process can be clearly seen in the magnetic butterfly diagram (figure).

\begin{figure}[!htbp]
\centering
\centerline{\includegraphics[width=0.9\textwidth,clip=]{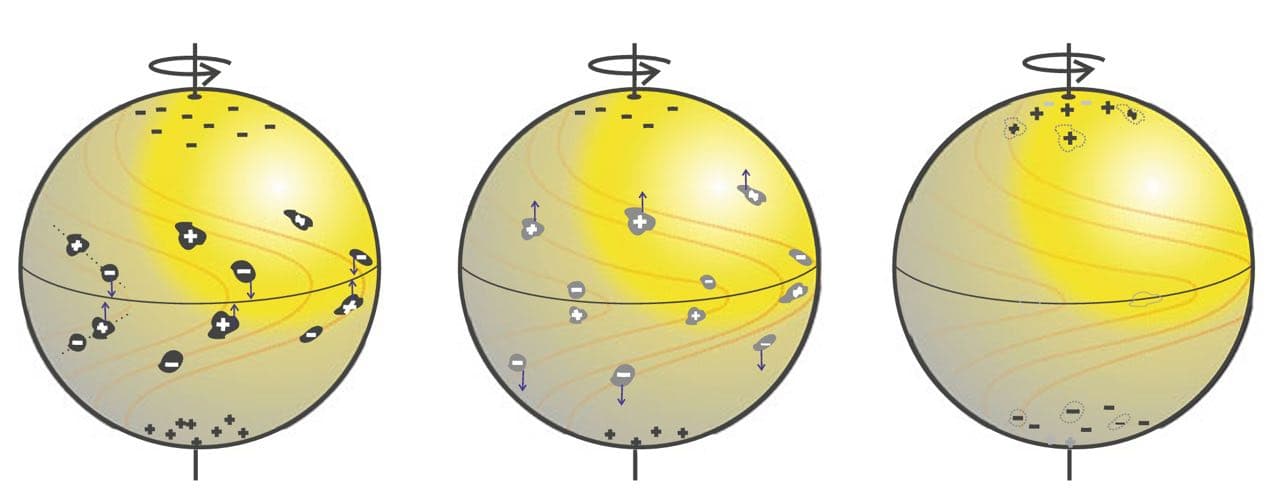}}
\caption{A representative diagram showing the Babcock--Leighton Process (left) the emergence of tilted BMRs, (middle) transportation of diffused opposite flux to the equator and pole (right) resultant effect of such process ({\it Image courtesy:} D. Passos}.
\label{fig1.15}
\end{figure}

The resultant effect of this process ultimately leads to the conversion of the toroidal field into the poloidal field. These poloidal fields are then carried to the bottom of the convection zone by in-surface meridional flows. Again, it is amplified by differential rotation and converted back to the toroidal field, completing the dynamo loop. Hence, the BL mechanism can provide a sustainable dynamo loop. A number of dynamo models have been constructed based on the BL process \citep{Leighton1969, Wang1991, Karak2014, Choudhuri2018}.

\subsection{$\alpha$-quenching and Tilt-quenching}

In all types of kinematic dynamo models, such as mean-field and BL type models, the dynamo generated magnetic field will keep on growing in time (see \autoref{eq1.21}). With the increase in strength of the magnetic field, the magnetic tension will try to resist the change caused by differential rotation as an effect of Lorentz force and hence try to reduce it. The long\hyp term measurement of the solar differential rotation does not show any significant variation in it \citep{Howard1984, Javaraiah2005a, Jha2021}. Therefore, it could not be the possible answer to suppress the growth of the magnetic field. 

An alternate way is routinely used in solar dynamo models is called ``$\alpha$-quenching''. This idea is based on the fact that the increased magnetic tension will try to resist small scale turbulent motion as the magnetic field grows. This will happen when the magnetic energy per unit volume reaches close to the kinetic energy of turbulent motion,
\begin{equation}
    \frac{B_{\rm eq}^2}{8\pi}=\frac{1}{2}\rho(\vb{v}^\prime)^2.
    \label{eq1.22}
\end{equation}
Based on this idea, an ad hoc nonlinear dependency of $\alpha$ on $\vb{B}$ is used \citep[][and references therein]{Charbonneau2010}, which is given by
\begin{equation}
    \alpha \rightarrow \alpha(\vb{B})=\frac{\alpha_0}{1+\left(\frac{B}{B_{\rm eq}}\right)^2}.
\end{equation}
This equation tells that as $\vb{B}$ start increasing and exceed $B_{\rm eq}$, the $\alpha$ approaches zero, which limits the amplification of the magnetic field. It is pointed out by \citep{Charbonneau2010} that  ``This is an oversimplification of the complex interaction between flow and field, but it does the right thing'' Hence, it is widely used in solar dynamo models.

The problem of the growth of the magnetic field is even more severe in the \bl\ type dynamo models due to the absence of $\alpha$-effect. Since in the \bl\ process, the tilt of BMRs play a significant role in the conversion of the toroidal field to poloidal, and it encouraged the modellers to use tilt quenching to suppress the growth of the magnetic field in their models \citep{Lemerle2017, Karak2017, Karak2018}. The idea behind the tilt quenching is as follows: BMRs are formed due to the rise of the flux tube due to magnetic buoyancy, and, during its rise, it will experience the Coriolis force and hence appear as tilted BMR on the Surface. The time it takes to come out of the photosphere depends on the magnetic field's strength, and therefore higher the magnetic field lowers the rise time \citep{DSilva1993}. If the rise time is less, the Coriolis force will not get sufficient time to tilt the BMR. Therefore a tilt quenching of the form
\begin{equation}
    \gamma = \frac{\gamma_0\cos{\theta}+\delta}{1+\left(\frac{B}{B_{\rm satu}}\right)^2},
    \label{eq1.24}
\end{equation}
is routinely used in the solar cycle models. Here, $\theta$ and $\delta$ are co-latitude and scatter around Joy's law respectively \citep{Karak2017, Karak2018}. \citet{Lemerle2017} and \citet{Karak2017, Karak2018} have used this type of tilt quenching in their models and successfully produced the many observed features of the solar activity cycle. Recently, a few observations also 
supported the idea of tilt quenching \citep{DasiEspuig2010, Jha2020}. The detailed discussion on tilt quenching will be presented in \autoref{Chap6}. An alternative mechanism has also been proposed in \bl\ process to limit the growth of magnetic field based on the latitudinal quenching \citep{Jiang2014, Jiang2020, Karak2020}; but it needs more sophisticated simulations and observations to support this idea.  

\begin{figure}[!htbp]
\centering
\centerline{\includegraphics[width=0.6\textwidth,clip=]{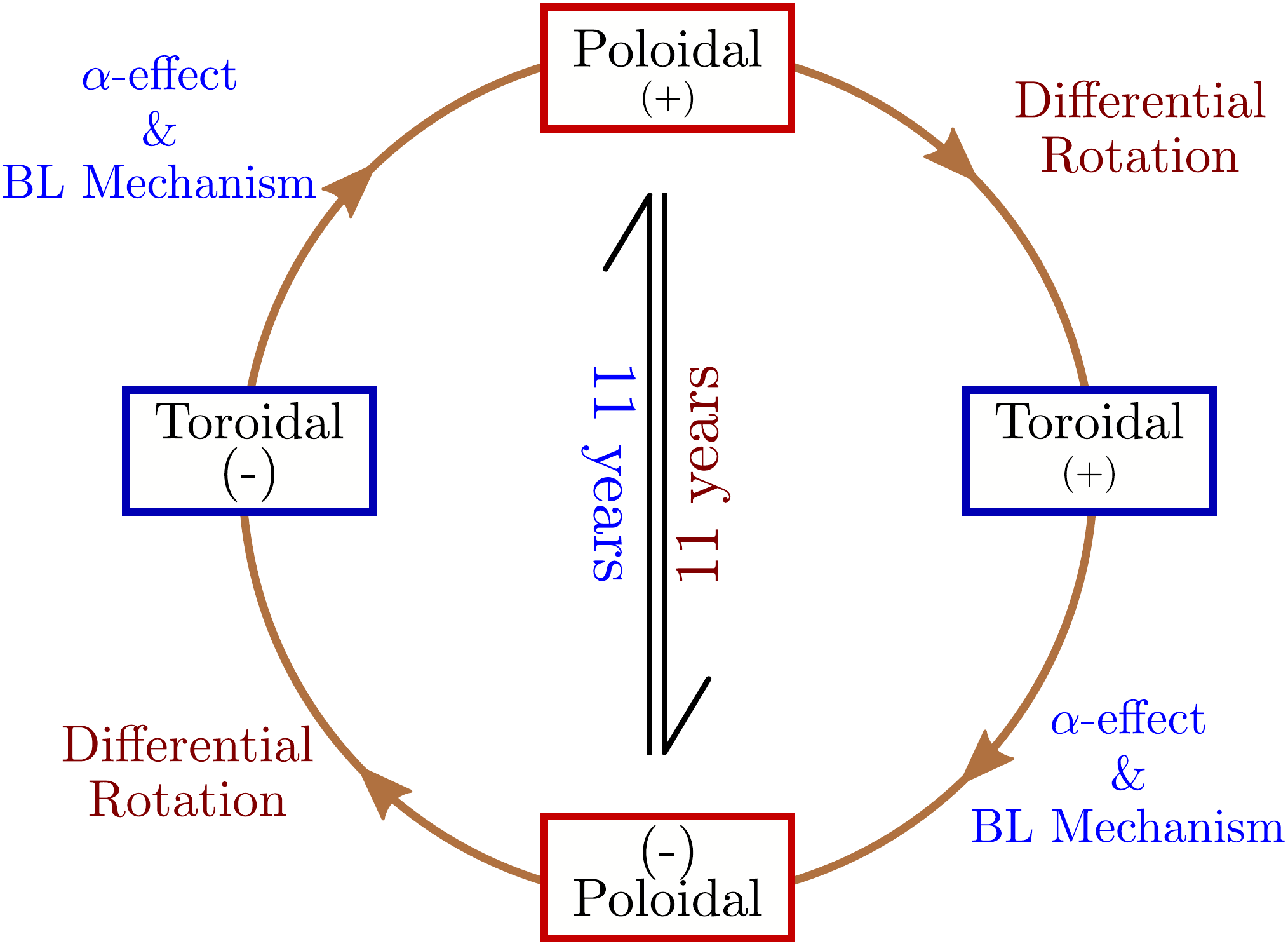}}
\caption{Similar to \autoref{fig1.10_1} along with the different processes responsible for the conversion of these fields.}
\label{fig1.16}
\end{figure}

The summary of this section is that in the solar dynamo, the poloidal field is amplified and converted to the toroidal field by the shearing effect of solar differential rotation and then via $\alpha$ process and BL process, they get converted back to toroidal field completing the one activity cycle (11~years) and half of the magnetic cycle (22~years). This cyclic process is also presented in \autoref{fig1.16}.

\section{Motivation Behind this Thesis}
Theory and observation have an intimate relationship in science, and one can not keep aside any of them. There are many branches where theory is far ahead of our observational capability, and there are others where observation is at an advantageous place than our understanding of physics underneath. Solar astrophysics is one such branch of science where modern and sophisticated instruments have taken our observational capability at unbelievable height, lagging in theoretical understanding. This thesis is motivated to provide the theoretician inputs, which will help them reduce the gap between these two aspects of the science, particularly for our understanding of solar dynamo models.

The KoSO digital archive, primarily used data in this thesis, provides one of the most consistent data series taken from the same instrument and at the exact location for such a long period. It puts the KoSO at an advantageous position unlike other observatories, e.g., RGO, which collects data from different places and observatories, including KoSO. The homogeneity in KoSO data makes it appropriate for the long\hyp term study. Other than KoSO, data from modern space-based instruments such as SOHO and SDO have also been used to extract the different solar magnetic features using automatic algorithms. The development of automated algorithms is very important as it helps us get rid of human biasing and provides a consistent approach throughout the data, which is really important for archival data. Most of these features are magnetic and proxies for the magnetic field's evolution in the Sun. The long\hyp term variation in the properties of these proxies provides valuable input to the solar dynamo models.

Understanding the importance of sunspots for solar magnetism, \citet{Hathaway2013} has studied the penumbra to umbra area ratio of the sunspot (RGO, 1874\ndash1976). He found that this ratio does not have any significant long\hyp term change for bigger spots or spot groups, whereas, interestingly, the smaller ones show a systematic trend. He was not able to conclude anything due to a lack of data. The white light digitized data at Kodaikanal Solar Observatory (KoSO) provides the data till 2011, which allow us to extend this series and look for the systematic trend speculated by \citet{Hathaway2013}. If such a trend exists and the ratio shows some difference in bigger and smaller sunspots, it raises the question: Are they coming from two different origins? Hence, it will be vital for the dynamo models as they treat both of them similarly \citep{Charbonneau2010, Hathaway2015}. Motivated by the work, I have developed an automated algorithm to extract the umbra and penumbra from the sunspots and studied the long\hyp term variation of their ratio. This work is presented in \autoref{Chap3} and published as \citet{Jha2018, Jha2019}.

In \sref{ch1-dynamo}, I have discussed that solar differential rotation is responsible for the conversion of the poloidal field to toroidal field. In the past, several studies \citep[see review by][]{Beck2000} of this subject show either no or minor variation in solar differential rotation. Since most of these studies have used manually tracked sunspots (or other solar feature) data and may have suffered from human subjectivity, it is necessary to develop an automatic algorithm to track the sunspots. In addition, earlier studies of the differential rotation also show the slight difference in measurement when larger and smaller sunspots are used to measure it. This difference again raises questions about their origin; hence, KoSO data becomes even more important for this study. The study of solar differential rotation is presented in \autoref{Chap4} and published as \citet{Jha2021}.

The differential rotation rate measured based on sunspots \citep{Jha2021} show a higher rotation rate when compared with Doppler measurement \citep{Howard1970}. This observation indicates the sharp change of rotation profile near the surface, and the helioseismology measurements verified it \citep{Schou1998, Howe2005}. This intriguing feature in the solar rotation profile is called the ``Near-Surface Shear Layer'' of the Sun. Even though there are multiple attempts have been made to explain this layer, none of them has successfully produced the observed NSSL \citep{Foukal1975, Gilman1979, Hotta2015}. The missing theoretical understanding encouraged us to make the theoretical model of NSSL \citep{Jha2021a}, which is presented in \autoref{Chap5}.

Another essential process in solar dynamo models based on the BL process is the tilt quenching, which limits the growth of the magnetic field in kinematic dynamo models. Although an ad hoc tilt quenching (\autoref{eq1.24}) is routinely used in solar dynamo models to produce the observed cycle behaviours \citep{Charbonneau2010, Karak2017, Karak2018}. What was lacking is the observational support for this idea? The indirect evidence for this idea has presented \citet{DasiEspuig2010} but the availability of space-based high-resolution magnetogram from SOHO and SDO make it possible to look for the direct observation support of this idea. I have used these data to demonstrate this idea in \autoref{Chap6}, and this work is published in \citet{Jha2020}.

\section{Organization of the Thesis}

\begin{figure}[!htbp]
\centering
\centerline{\includegraphics[width=0.9\textwidth,clip=]{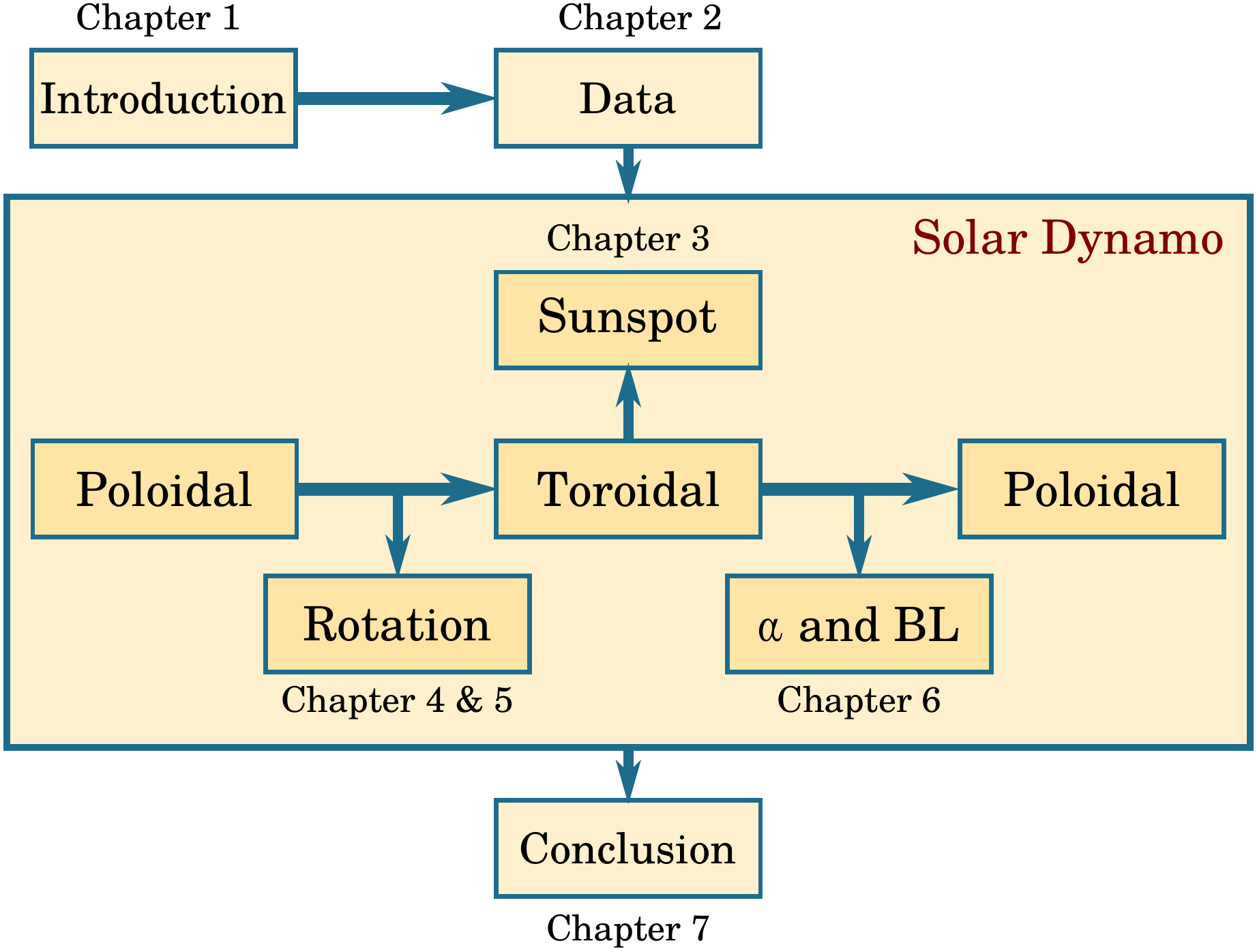}}
\caption{A representative outline of the thesis.}
\label{fig1.19}
\end{figure}

This thesis consists of 7 chapters; the first two chapters deal with the introduction and data. In \cref{Chap2} I have discussed the different data sources that have been analyzed in this thesis. In this chapter, I will also briefly discuss the updated white light area series from the KoSO, along with a short update on Ca-K data from KoSO. \cref{Chap3} and \cref{Chap4} deals with the long\hyp term variation of sunspot penumbra to umbra area ratio and solar differential rotation, respectively, using white-light digitized data from KOSO. In \cref{Chap5} I have given a theoretical model for the observed NSSL in the Sun. \cref{Chap6} presents the indication of tilt quenching as observed from the SOHO and SDO data. Finally, in \cref{Chap7} we discuss the conclusion, nobility of this thesis and the future aspects of the works presented in the thesis. A representative flow diagram of thesis outline is shown in \autoref{fig1.19}.

\clearpage{}

\clearpage{}\chapter{Data}
\label{Chap2}
\lhead{\emph{Chapter 2: Data}} 

\epigraph{\itshape ``The goal is to turn data into information, and information into insight.''}{--Carly Fiorina
}
\noindent

The importance of historical archival data is immeasurable for the long--term study of the Sun. The significance of these data was not out of the site, and hence observatories around the globe started preserving them. Most of these data were taken on photographic plates or films, which are now either digitized or in the process of digitization. Among all the observatories, the Royal Observatory of Greenwich has one of the most extended data series from 1874 to 1976, collected and compiled from its different observing stations. Apart from RGO, Kodaikanal Solar Observatory, Kodaikanal (1906 onward), Debrecen Photographic Data (1974--2018), Pulkovo (1932--1991) etc., have systematic sunspot observation.

\begin{figure}[!htbp]
\centering
\centerline{\includegraphics[width=0.8\textwidth,clip=]{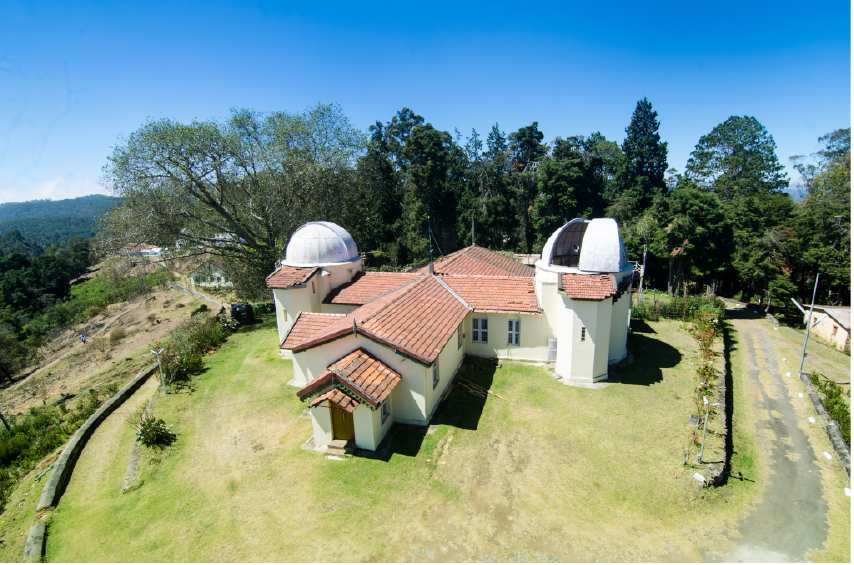}}
\caption{A fish eyes view of Kodaikanal Solar Observatory ({\it Image courtesy: KoSO}).}
\label{fig2.1}
\end{figure}

\section{The KoSO Digital Archive}
\label{ch1-koso}
\begin{figure}[!t]
\centering
\centerline{\includegraphics[width=0.95\textwidth,clip=]{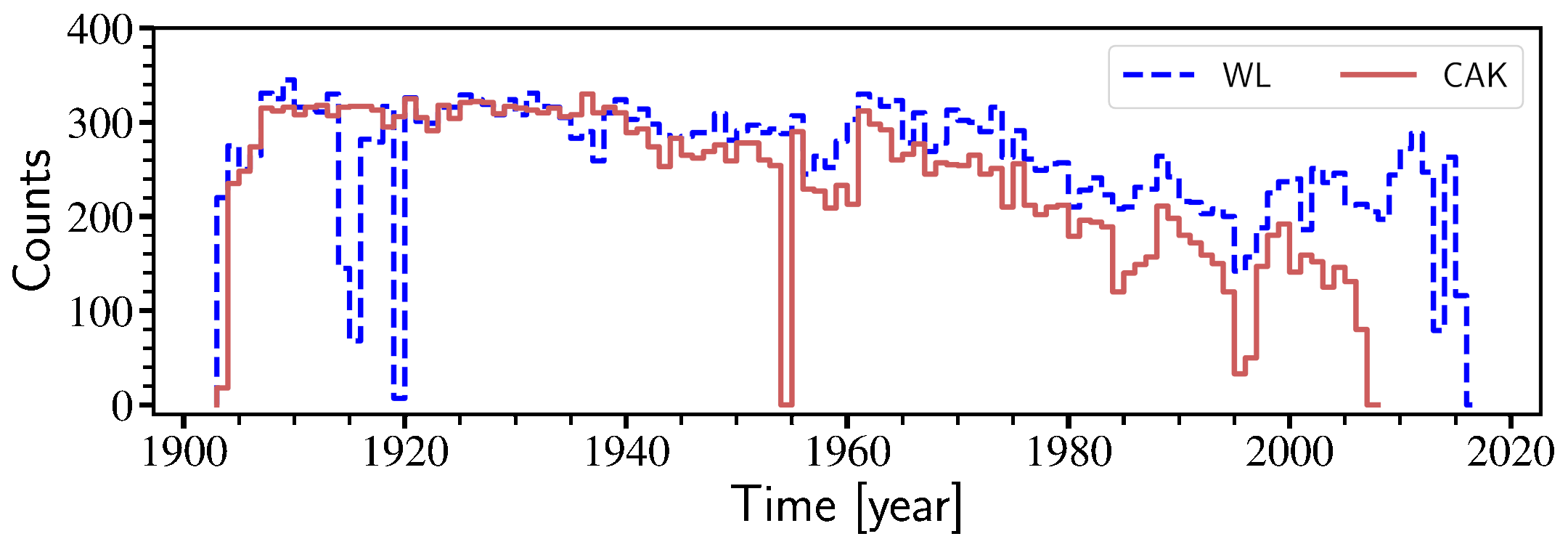}}
\caption{No of observation at KoSO (red) white-light and (blue) Ca-K.}
\label{fig2.2}
\end{figure}

Kodaikanal Solar Observatory (KoSO), established in 1899, is one of the oldest observatories in the world observing the Sun in multi-wavelengths (white-light, H-alpha, Ca-K) from the beginning of the 20th century \citep{Hasan2010}. Earlier, these observations were taken on photographic plates or films and preserved in the humidity controlled environment at Kodaikanal. Recently these data have been digitized using a digitizer unit and made public for the scientific community\footnote{\url{https://kso.iiap.res.in/new/data}}. The statistics of the digitized data is shown in \autoref{fig2.2}. First, I will briefly discuss the telescope used to acquire these observations and the digitizer unit used to make it digital before I get into the details the data products.

\subsection{The Telescope}
\label{ch2-tele}
The white light observation at Kodaikanal has been performed using an equatorially mounted 10~cm, f/15 telescope from the begging of 1904. With the help of additional optics, this telescope produces a solar image of a diameter of 20.4~cm, which is acquired on a photographic plate or film. In 1918 this 10~cm lens was replaced by a 15~cm achromatic lens, and then onward, the same setup is used for the observation \citep{Sivaraman1993}. 

The Ca-K (393.37~nm) and H-alpha (656.28~nm) observations were taken using a spectroheliograph. To fed the reflected sunlight to the spectrograph, an 18~inches single mirror siderostat is used along with the additional optics. The siderostat also take care of the Earth's rotation and reflect the sunlight in a fixed direction which is fed to the spectroheliograph. In the spectrograph a two prisms/mirrors along with CA-K/H-alpha filter are used to observe the Sun in these two wavelengths.

\subsection{The Digitizer Unit}
\label{ch2-digi}
The digitizer unit available at KoSO is used to digitize the observation taken on plates/films into 4k$\times$4k resolution. This unit consists a 1~m diameter spherical uniform light source with a small opening at the top where plates/films are placed in a holder for digitization. Light coming from the sphere passes through the plate/film and is captured by a cryogenic cooled (-100$^\circ$~C) CCD camera to reduce the dark noise. This camera takes the image with resolution of 4k$\times$4k and with a bit depth of 32-bit and stores it in Flexible Image Transport System (FITS) format for the use of scientific community \citep[for further details about the digitization see][]{Ravindra2013}. This digitizer unit has been extensively used for digitizing all the photographic plates and films.

\section{White\hyp light Data}
\label{ch2-wl}
The white--light photographic data for the period 1906--2011 has been digitized recently using the digitizer discussed in \sref{ch2-digi}. Barring the period 1906--1920, the white--light data for 1921--2011 has been calibrated and reported in \citet{Ravindra2013}. These calibration steps include (i) correction for flat field, (ii) disk identification using Hough transform, (iii) conversion of intensity to relative plate density and (iv) addition of proper header information as per FITS standard. After that, a semi--automated sunspot detection technique was developed based on the already existing sunspot detection algorithm such as STARA \citep{Watson2009} to identify the sunspots from these images and these informations are stored in the form of binary images. The sunspots area time series extracted from these data have been reported in \citet{Ravindra2013} and \citet{Mandal2017a}. The stored binary mask keeping the information about sunspots are used in \cref{Chap3} \citep{Jha2018, Jha2019} and \cref{Chap4}\citep{Jha2021} and could be further used for subsequent studies.

\subsection{Updated Sunspots Area Time Series}
\label{ch2-upwl}

The white\hyp light observation at KoSO started in 1904, but due to difficulties in calibration for the period 1904\ndash1920, the sunspot area series has been only reported for the period 1921\ndash2011 \citep{Ravindra2013, Mandal2017a}. The initial two years of data (1904 and 1905) have difficulty identifying reference lines that could be used to get the correct orientation of the images, as they do not contain any. Therefore, barring the initial two years here, I will briefly discuss the updated area series, which includes the data for the period 1906\ndash1920 and extends the series till 2018. The detail report on the update is under preparation and will be presented in in \citet[in preparation]{Jha2022}.

\begin{figure}[!htbp]
\centering
\centerline{\includegraphics[width=\textwidth,clip=]{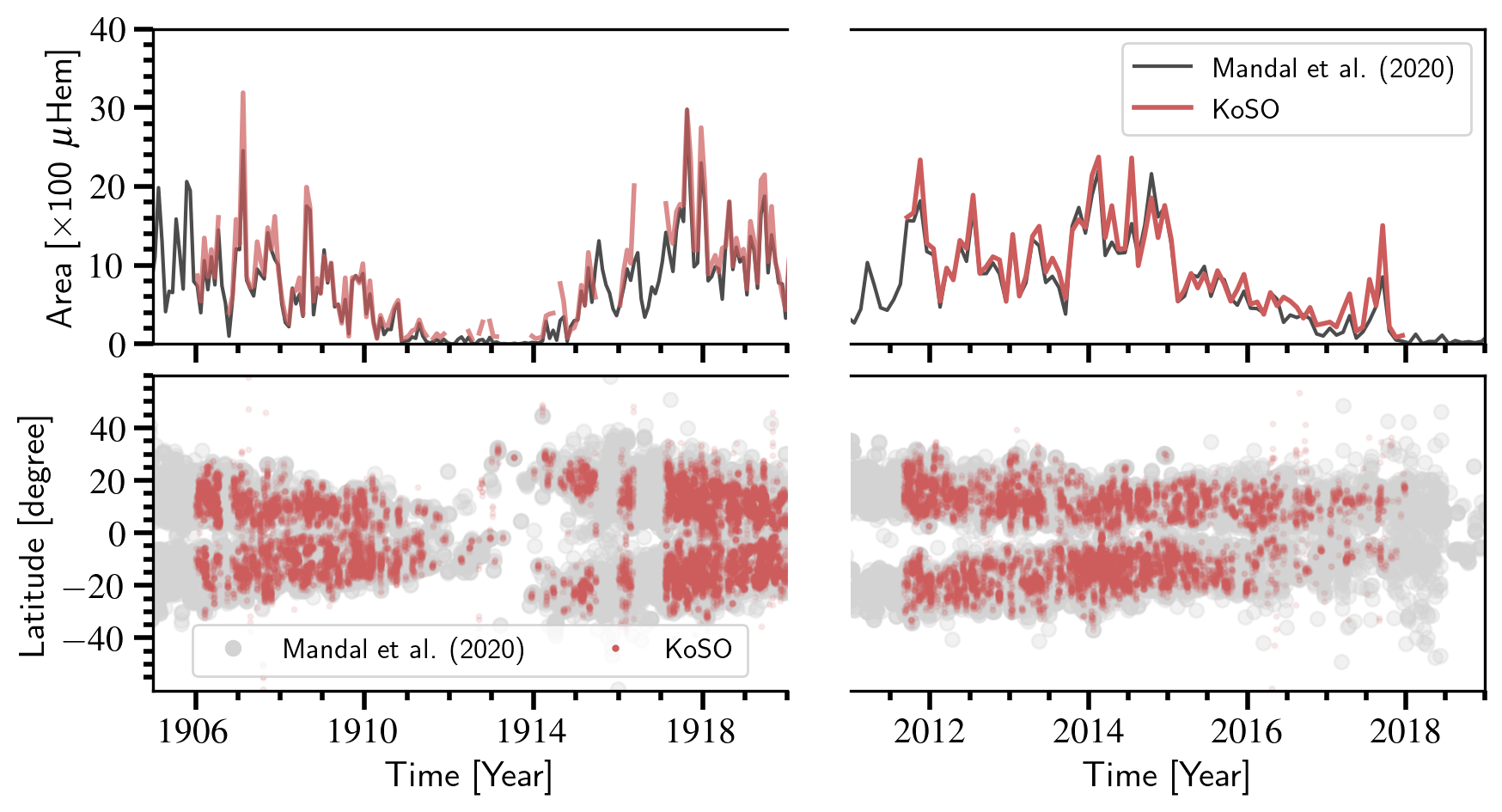}}
\caption{ Top: Monthly averaged sunspot area from KoSO with recently cross-calibrated area series \citep{Mandal2020} and bottom: the latitude time diagram, for the two different epochs.}
\label{fig2.3}
\end{figure}

It has been already pointed out by \citet{Ravindra2013} that the configuration of the telescope has been changed several times before 1918. These changes result in the varying image quality and disk size on the photographic plates of films. Along with the East\hyp West line, which is used to orient the images correctly, these observations contain an additional North\hyp South line. Due to the presence of two lines, it is not clear which of these lines correspond to North\hyp South and  East\hyp West, and hence, the older version of codes was unable to deal with these observations. Now, a semi-automated method is used use manually verify the orientation of the images based on the movement of features on the surface. The steps of calibration and sunspot detection is the same as discussed and also reported in \citet{Ravindra2013}. Furthermore, here I also present the extended area series from the recent  period 2011\ndash2018, which are also extracted in the similar way as discussed in \citet{Ravindra2013} and \citet{Mandal2017a}.

\begin{figure}[!t]
\centering
\centerline{\includegraphics[width=\textwidth,clip=]{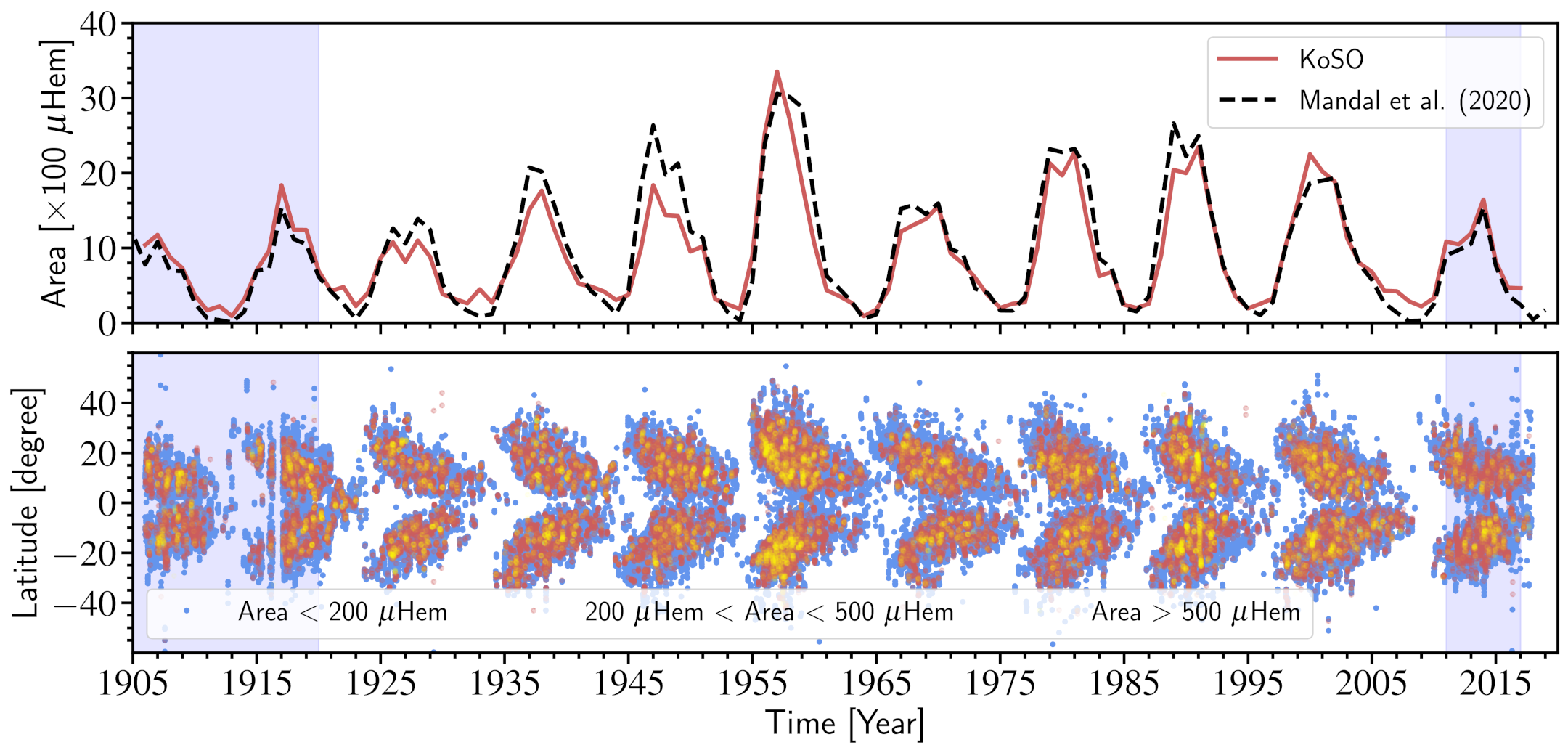}}
\caption{The updated area time series from the KoSO digital archive along with the location of sunspots shown in the form of butterfly diagram.}
\label{fig2.4}
\end{figure}

After the detection of sunspots the position of these spots are extracted. Top panel of \autoref{fig2.3} shows the monthly averaged sunspot area for the two different periods (1906\ndash1920 \& 2011\ndash2018). The comparison of these monthly averaged area show a good agreement with the recently cros-calibrated area series \citep{Mandal2020}. The bottom panel of \autoref{fig2.3} also show the comparison of the position of the detected spots which also in accordance with \citet{Mandal2020}. \autoref{fig2.4} shows the (top) updated area series from the KoSO along with the cross-calibrated yearly averaged area from \citet{Mandal2020}, and (bottom) location of sunspots on the surface in form of latitude time diagram.

\section{Ca-K Data}
\label{ch2-cak}

The Ca-K observation captures the chromosphere and carries an important information about the magnetic nature of the Sun. Ca-K data is used as an important proxy for the magnetic field measurement. Moreover, this data also plays a crucial role in the historical reconstruction of solar irradiance. Therefore, analogous to white light, the Ca-K data has also been digitized for the period 1904--2007 and calibrated by following the similar steps as mentioned in \sref{ch2-wl}. These data have been taken with the help of siderostat, a single mirror system, and suffers from rotation of the field of view (FOV). It is important to correctly locate the solar north in the image for the scientific use of the data. Therefore, after the observations, the position of solar North/South has been marked by double/single dot based on the tabulated value of rotation calculated from observation time \citep{Priyal2014b}. It has been recently discovered that there are a few issues with this marking, and it will be addressed in the next section.

\subsection{Pole Angle Calculation}
\label{ch2-pole}

As mentioned already the siderostat cause the rotation of FOV and hence the position of solar north rotates 360$^\circ$  in one complete day. Now, instead of tabulated value I suggest to use the mathematical formula to calculate the rotation caused by siderostat ($\Theta$), which is given by
\begin{equation}
    \Theta = 2\tan^{-1}{\left[ K\tan{\left( \frac{{\rm HA}}{2}\right)}\right]};
    \label{eq2.1}
\end{equation}
where,
\begin{equation*}
    K \equiv \frac{\sin{\left( \frac{L-\delta}{2}\right)}}{\sin{\left( \frac{L+\delta}{2}\right)}},
\end{equation*}
    
{\rm HA} is the hour angle\footnote{To calculate hour angle of the sun we have used a IDL routine in SolaSoftWare (SSW) {\it sunpos.pro} and {\it eq2hor.pro} along with the Longitude of the Observatory as 77.46$^\circ$E.} of the Sun at the time of observation, $L$ (10.23$^\circ$N for KoSO) is the latitude of the observatory and, $\delta$ ($=90^o - {\rm declination~of~Sun}$) is the pole angle of the Sun \citep{Cornu1900}. The mathematical relation given in \autoref{eq2.1} depends on HA, which need a correct time of observation (\tobs). 
    
\subsection{Issues with Timestamps}

The \tobs\ are usually written on plates/films, envelops containing the plate and in the logbook. During the digitization of these observations, the file name\footnote{filename is given as CAK\_19100114T084400\_Q1L0a.fits.} is given as par the \tobs. To use the digitized images and get the correct position of solar North, we need this time from the filenames and here come the problems. There are a few issues with the timestamp in the filename, which are following.

\begin{figure}[!t]
\centering
\centerline{\includegraphics[width=\textwidth,clip=]{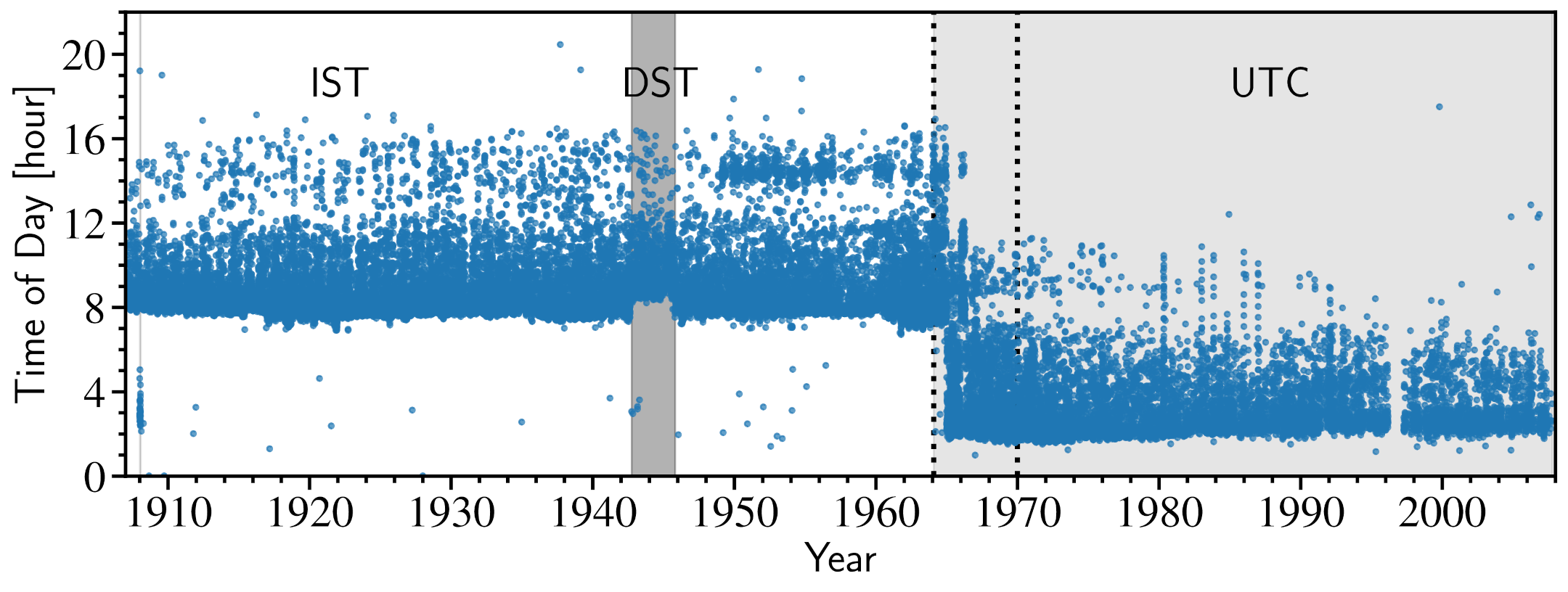}}
\caption{Time of observation of the day as extracted from the filename with \tobs.}
\label{fig2.5}
\end{figure}

\begin{enumerate}\item  A major change in time zone from IST (+05:30 GMT) to UTC (00:00 GMT) in February 1964 (see \autoref{fig2.5}) is noticed and it can straightly corrected.
    
    \item During Sep 1, 1942 to Oct 15, 1945 there is a systematic shift of a hour daylight saving time\footnote{\url{https://www.timeanddate.com/time/zone/india/kolkata}} (DST; +06:30 UTC) observed in India. It is also very easy to be taken care once known.
    
    \item A careful look at the \tobs\ in Jan 1908 (for a small period) suggests that the \tobs\ is in UTC, which is again easy to compensate.
    
    \item Scatters from 1964 to 1970 is mainly due to mixing time zones (IST and UTC) during digitization and renaming the files even though the \tobs\ on plates are in UTC. Because of its randomness it is little tricky to correct. 
    
    \item Scatters seen beyond the usual observation time (06:30 to 17:00 IST) is primarily due to the mistake in the name of files during digitization. 
\end{enumerate}
 The points raised in 1, 2 and 3 are straight forward to bring to either IST or GMT, whatever one may prefer. The scatter and the mistakes in renaming the file are really difficult to correct because of its random nature. It was unknown that how much data is effected by such mistakes and it becomes important to first identify that fraction of data and then try to correct them if possible.  

\subsection{Identification and Correction}
To identify the in-correct timestamp in the filenames (observations) these steps have been followed.

\begin{enumerate}
    \item Three consecutive observations (images) have been selected and rotated based on the \tobs\ as extracted from filename and using \autoref{eq2.1}.
    
    \item The preceding/following observations are differentially forward/back rotated to match with the \tobs\ of central observation.
    
    \item Using a fixed threshold, the contours (represents the chromospheric feature) of differentially forward/back rotated preceding/following observation are overlapped and compared manually with central observation.
    
    \item If either of the contours shows correct overlap, the central observation is marked as correct timestamp or \tobs\ and if non of them are overlapping it is marked as in-correct.
    
    \item Observation marked in-correct are again compared with the two correctly marked near-time observations and marked again as Step\hyp4. 
\end{enumerate}

I have gone through all the observations, and based on the method briefed above, the summary is presented in \autoref{tab2.1} 
\begin{table}[!htbp]
\centering
\begin{tabular}{lcc}
Number of Observations & & Number\\
\hline
Total (1907\ndash2007)&:& 47935\\
Correct timestamp (IST)&:& 31911 (66.6\%)\\
Correct timestamp (UTC)&:& 11654 (24.3\%)\\
In-Correct timestamp&:& 04370 (09.1\%)\\
\hline
\end{tabular}
\caption{A brief statistics of the number of images with correct and in-correct timestamps.}
\label{tab2.1}
\end{table}

To correct all those observations identified as in-correct timestamps, we check the plates and logbooks for the \tobs. After noting down \tobs\ for all the observations, I again check for their correctness using a similar approach explained earlier. Hence, the final statistics for the observation with correct and in-correct \tobs\ is as in \autoref{tab2.2}. Apart from these, there are around 720 observations during the period 1904\ndash1906, in which no \tobs\ information is available, and I am currently exploring the possiblity to get the \tobs\ in those observations. This complete work will be submitted soon to the Journal of Solar Physics as Jha et al. (2022).

\begin{table}[!htbp]
\centering
\begin{tabular}{lcc}
Number of Observations & & Number\\
\hline
Total (1907\ndash2007)&:& 47935\\
Correct \tobs &:& 44977 (93.8\%)\\
In-Correct \tobs &:& 01607 (03.4\%)\\
Bad Quality (visually) Images &:& 01150 (02.4\%)\\
Unknown \tobs &:& 00207 (00.4\%)\\
\hline
\end{tabular}
\caption{A brief statistics of the number of final images with correct and in-correct \tobs.}
\label{tab2.2}
\end{table}

\section{The Michelson Doppler Imager}
\label{ch2-mdi}

\begin{figure}[!htbp]
\centering
\centerline{\includegraphics[width=0.5\textwidth,clip=]{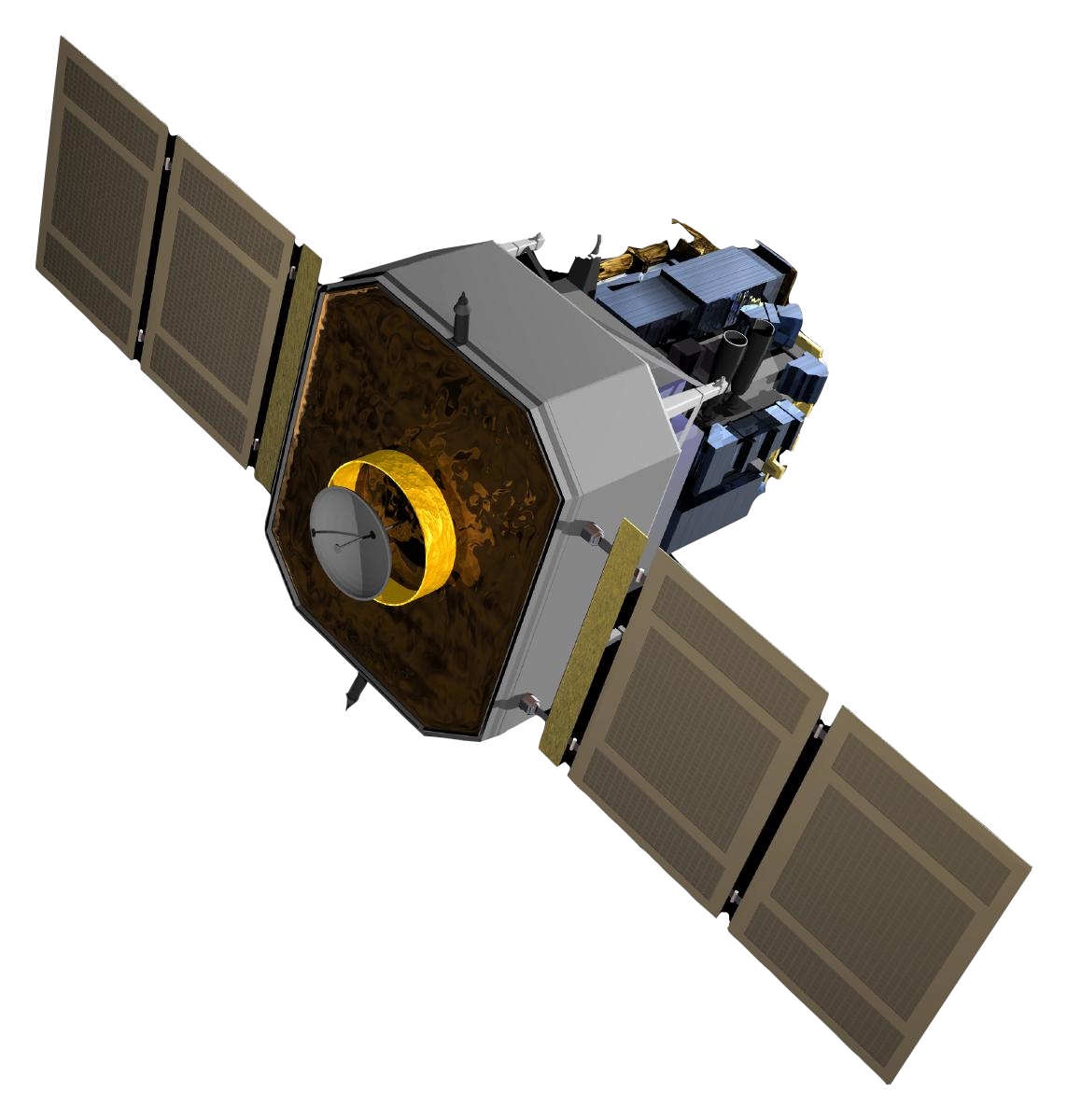}}
\caption{Image of SOHO ({\it Image Courtesy: ESA \& NASA}).}
\label{fig2.6}
\end{figure}

The Michelson Doppler Imager \citep[MDI;][]{Scherrer1995} is a space-based instrument onboard Solar and Heliospheric Observatory (SOHO, shown in \autoref{fig2.6}) launched in December 1995, which is a project of international collaboration between The National Aeronautics and Space Administration (NASA) and European Space Agency (ESA). MDI observes the Sun in visible light (6768~\AA; white\hyp light or intensity continuum) with a cadence of 6~hr and resolution of 4'' on 1k$\times$1k CCD. This provides the photospheric observation of the Sun, which is helpful for the study of sunspots and other photospheric activity. MDI also measures the line of sight (LOS) magnetic field component using Zeeman splitting of magnetically sensitive Ni I 6768~\AA line. These LOS magnetograms are available with the cadence of 96~m and have the exact resolution as white\hyp light. In this thesis work, the intensity continuum and magnetogram data have been used for 1996\ndash 2011.

\section{The Helioseismic and Magnetic Imager}
\label{ch2-hmi}

The Helioseismic and Magnetic Imager \citep[HMI;][]{Schou2012} onboard Solar Dynamic Observatory (SDO) is a succesor of MDI with better resolution and cadence. It has two 4k$\times$4k CCD cameras which observe the Sun with 45~sec and 720~sec cadence. HMI provides the full disk LOS and vector magnetic filed measurementsa and it also provides the full disk intensity continuum with the resolution of 1''. In this thesis the intensity continum and magnetogram data for the period 2010\ndash2018 has been used.

\begin{figure}[!htbp]
\centering
\centerline{\includegraphics[width=0.7\textwidth,clip=]{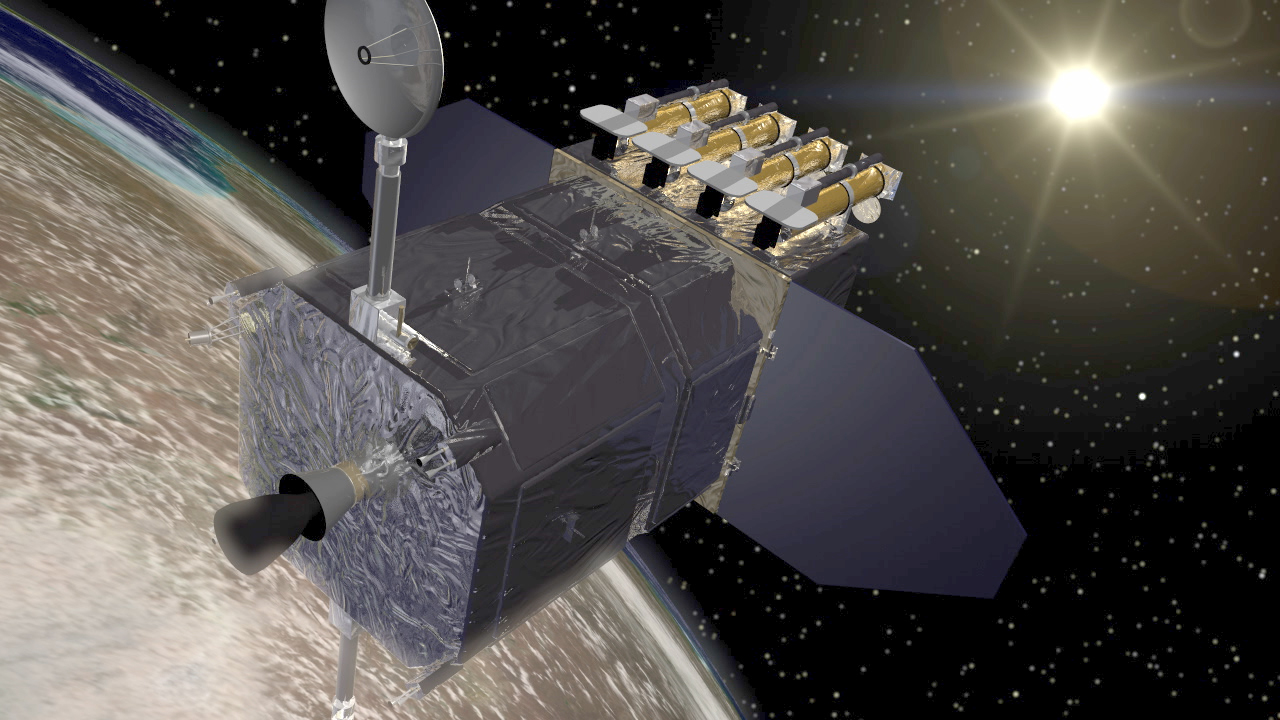}}
\caption{SDO ({\it Image Courtesy: NASA)})}
\label{fig2.7}
\end{figure}

\clearpage{}

\clearpage{}\chapter[Penumbra to Umbra Area Ratio from KoSO Data]{Study of Sunspot Penumbra to Umbra Area Ratio using Kodaikanal White\hyp light Digitized Data}
\label{Chap3}
\lhead{\emph{Chapter 3: PPenumbra to Umbra Area Ratio from KoSO Data}}

\noindent

\section{Introduction}
\label{ch3-intro}
Sunspots, the most prominent features on the solar photosphere, appear dark when observed in visible wavelengths. They also show periodic variations in properties over an $\approx$11 years time-scale, generally referred as the solar cycle \citep{Hathaway2015}. In fact, after the observations by \cite{Hale1908}, it became clear that sunspots are the locations of strong magnetic fields ($\approx$4 kG) which inhibit convection within them. Due to such suppression of energy, they appear as dark structures \citep{Solanki2003}. A closer inspection of sunspot images reveals that there are, actually, two different features within a spot: a darker (with respect to photospheric intensity) umbra surrounded by a lighter penumbra. This contrast in appearance is generally attributed to different strengths and orientations of the magnetic fields which are present in these two regions \citep{Mathew2003a}. Hence, area measurements of umbra and penumbra carry these magnetic field information too. The other importance of these measurements come from their application in calculating the Photometric Solar Index (PSI) values which quantize the decrement of Total Solar Irradiance (TSI) due to the presence of a spot on solar disc \citep{Froehlich1977,Hudson1982a}. Thus, a knowledge of long-term variations in the umbra and penumbra area will enhance our understanding of solar variability.

One of the earliest measurements of umbra and penumbra area values was reported by \cite{Nicholson1933} who studied almost one thousand unipolar or preceding member of bipolar sunspots from Royal Observatory, Greenwich (RGO) between 1917 to 1920. The average ratio ($ \q $), between the area of penumbra to that of umbra, was quoted as $\approx$4.7 and it was also found to be independent of sunspot sizes. However, examining the diameters of umbra and penumbra of 53 sunspots as photographed by Wolfer at Z\"urich, \cite{Waldmeier1939} noted that the $\q$ value decreases from 6.8 to 3.4 as the sunspot area increases from $100~\mu$Hem to $1000~\mu$Hem. The first investigation of the long-term evolution of this ratio was reported by \cite{Jensen1955,Jensen1956} where the authors analysed the RGO data from 1878 to 1945. Interestingly, they noted that the ratio is a decreasing function of sunspot size during cycle maxima but the variation is much lower than the values as reported in \cite{Waldmeier1939}. Several follow up studies by \cite{TandbergHanssen1956,Antalova1971,Beck1993} also confirmed similar results by including more complex sunspots and larger statistics.

Using the largest set of observations as recorded in RGO data (161839 sunspot groups between 1874-1976), \cite{Hathaway2013} calculated the $\q $ values for each of these cases and noted that it increases from 5 to 6 as sunspot group size increases from $100~\mu$Hem to $2000~\mu$Hem. However, the author did not find any dependency of $\q $ on the cycle phase or the locations of the spots. The most remarkable result of all was the behaviour of smaller sunspot groups (area $<100~\mu$Hem), for which the author found a substantial change in the $\q $ values within a relatively smaller timescale. The ratio decreased significantly from 7 to 3 during solar Cycles 14$-$16, however, it again increased to $>$7 in 1961 at the end of Cycle 19.

\section{Data}

\begin{figure}
\centering
\centerline{\includegraphics[width=\textwidth,clip=]{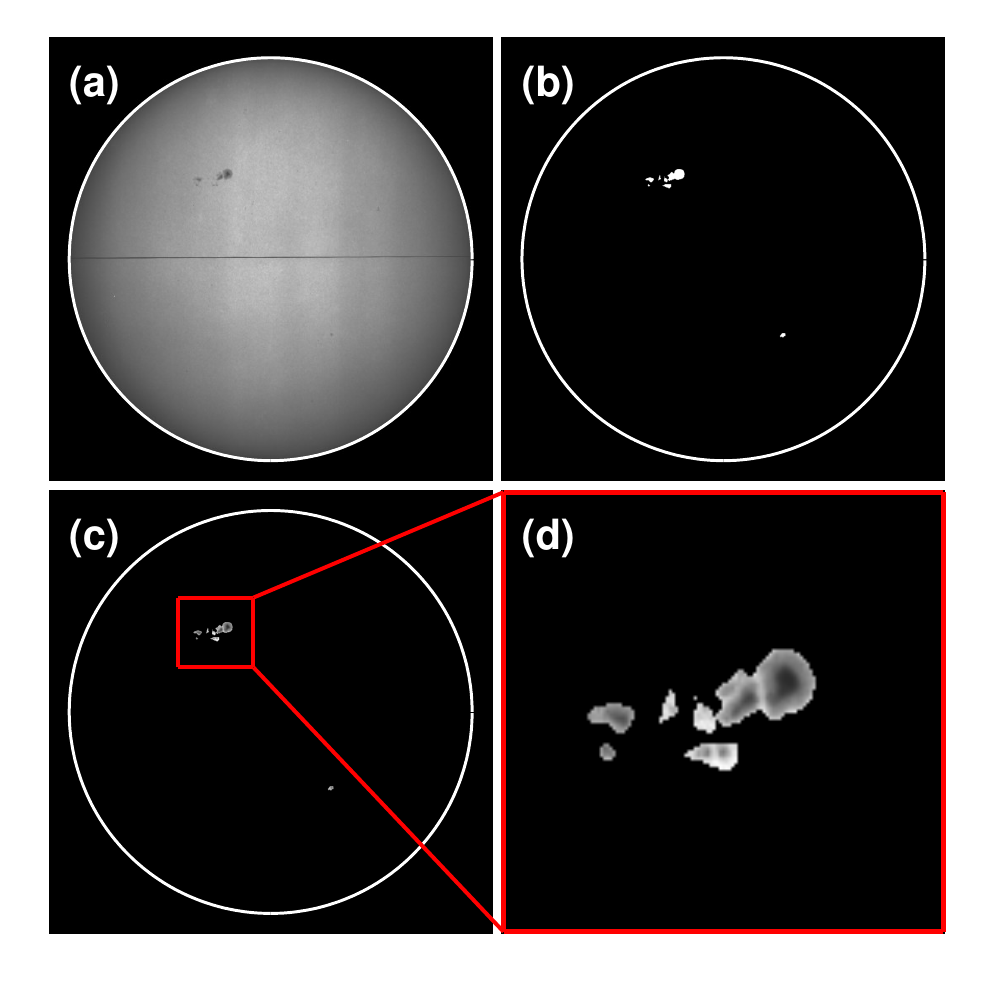}}
\caption{{\it Panel-(a):} A calibrated white-light image from Kodaikanal Observatory as recorded on 1955-01-07 08:15. {\it Panel-(b):} Binary image of the extracted sunspots. {\it Panel-(c):} Isolated spots in the original grey scale image produced by multiplying images on panel-(a) with panel-(b). A zoomed in view is presented in {\it Panel-(d).}}
\label{fig3.1}
\end{figure}

In this study, we have used the newly digitized and calibrated white-light full disk images (Figure~\ref{fig3.1}a) from Kodaikanal Solar Observatory (KoSO). Details of this digitization, including the various steps of calibration process, are reported in \cite{Ravindra2013}. Recently, \cite{Mandal2017a} catalogued the whole-spot area series\footnote{This catalogue is available online at \url{https://kso.iiap.res.in/new/white_light}.} (between 1921 and 2011) by using a semi-automated sunspot detection algorithm on this data. We start our analysis with these detected binary images of sunspots as shown in Figure~\ref{fig3.1}b. In order to isolate the spots, we multiply the binary mask with the limb-darkening corrected full disc images. The final results are displayed in Figure~\ref{fig3.1}c-\ref{fig3.1}d.

Considering the volume of the data to be processed, we opted for an automatic boundary detection algorithm. A number of methods have already been used in the past to automatically detect umbrae of sunspots: \cite{Brandt1990} \& \cite{Pucha2016} using fixed intensity threshold; \cite{Pettauer1997} using cumulative histogram method and \cite{Steinegger1997a} using the inflection method. Despite their successes on other data sets (mostly of smaller duration), we found that none of these methods actually produces a faithful result when applied on the entire Kodaikanal data. The main reasons behind this are the varying image quality over time, poor contrast, presence of artefacts etc. Keeping these limitations in mind, we select an adaptive umbra detection method based on the Otsu thresholding technique \citep{Otsu1979}.
This method finds the optimum threshold for an image which has a bimodal intensity distribution. In our case, the two different intensity levels of umbra and penumbra constitute a similar type of distribution which is suitable for such an application. Mathematically, to calculate the threshold, this method maximizes the between-class variance of the distribution. If $t$ is the threshold that separates ${\rm L}$ bins of histogram in background class ($C_{\rm b}$) and foreground class ($C_{\rm f}$), then the probability of occurrence of background ($\omega _{\rm b}$) and foreground classes ($\omega _{\rm f}$) are
\begin{eqnarray}
   \omega _{\rm b}&=&\sum _{i=1}^{{\rm t}}P(i)=\omega(t),\\
   \omega _{\rm f}&=&\sum _{{i=t+1}}^{{\rm L}}P(i)=1-\omega(t)
\end{eqnarray}
where $P(i)$ represents the probability of occurrence of the $i_{\rm th}$ bin.
The between-class variance {\bf($\sigma_{\rm B}$)} of the distribution for a particular $t$ can be written as
\begin{equation}
 \sigma _{\rm B}(t)^2=\omega _{\rm b}(\mu _{\rm b}-\mu)^2+\omega _{\rm f}(\mu _{\rm f}-\mu)^2.
 \label{eq3.1}
\end{equation}
In Equation \ref{eq3.1}, $\mu$ (the mean of the distribution) and $\mu _{\rm b}$ and $\mu _{\rm f}$ (the means of the background and foreground class) are defined as
\begin{eqnarray}
    \mu&=&\sum _{i=1}^{{\rm L}}iP(i),\\
    \mu _{\rm b}&=&\sum _{i=1}^{t}iP(i/C_{\rm b})=\frac{\mu(t)}{\omega(t)},\\
    \mu _{\rm f}&=&\sum _{i=t+1}^{\rm L}iP(i/C_{\rm f})=\frac{\mu-\mu(t)}{1-\omega(t)}.
\end{eqnarray}
    
In this work, we use the {\texttt cgotsu$\_$threshold.pro}\footnote{Description is available at \url{http://www.idlcoyote.com/idldoc/cg/cgotsu_threshold.html}.} routine, an IDL\footnote{For more details, visit \url{ https://www.harrisgeospatial.com/Software-Technology/IDL}.} implementation of the above concept.

\begin{figure}
\centering
\centerline{\includegraphics[width=\textwidth,clip=]{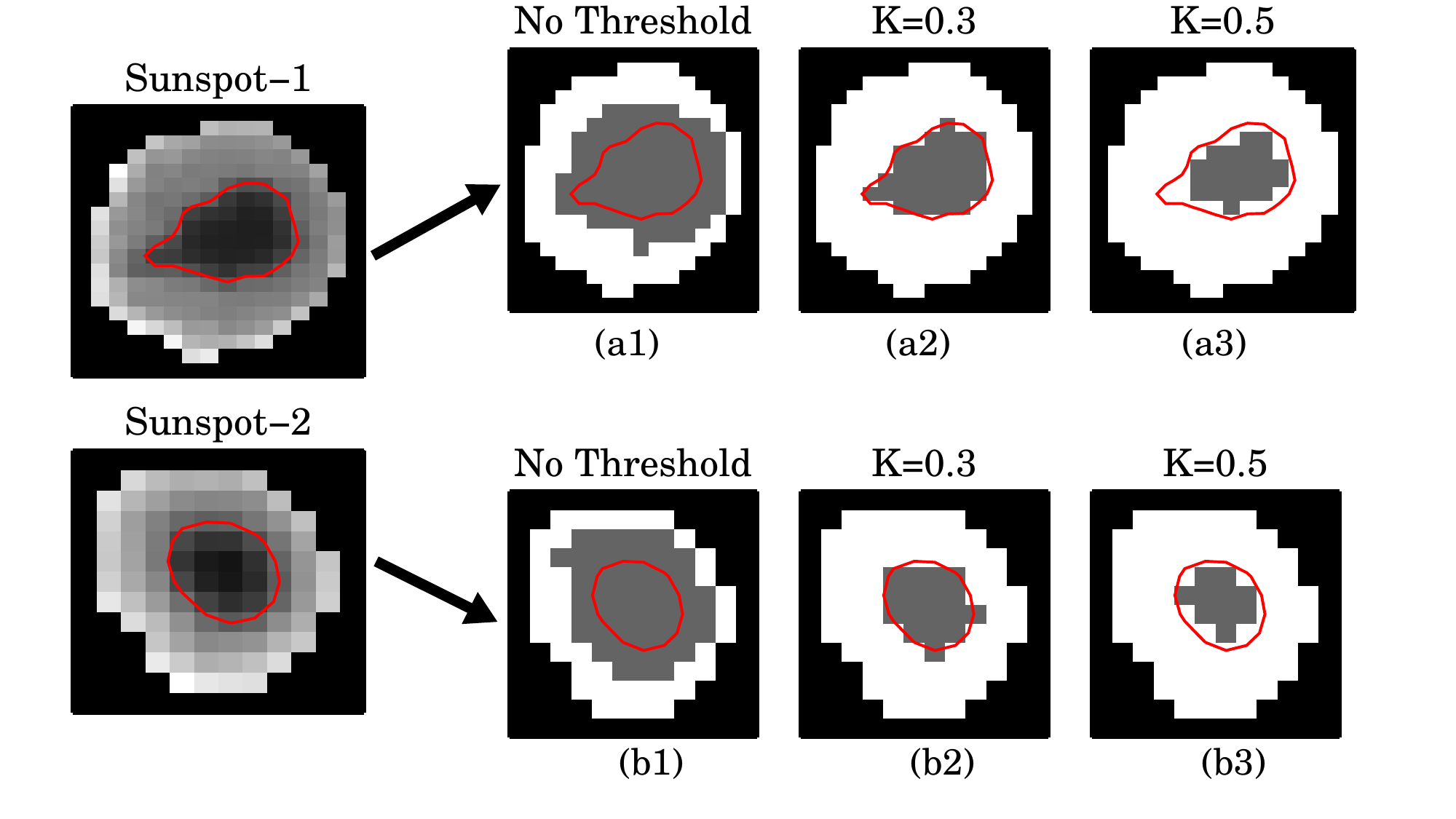}}
\caption{Two representative examples of our umbra detection technique on Kodaikanal sunspot data. The red contours (in panels (a1-b3)) highlight the umbra-penumbra boundaries as estimated by eye whereas the detected umbrae for different set of threshold values are shown as grey regions. See text for more details.}
\label{fig3.2}
\end{figure} 

 We demonstrate the application of this algorithm on our data with two representative examples as shown in Figure~\ref{fig3.2}. The red contours on the spots represent the umbra-penumbra boundary as estimated by visual inspections. We expect an umbral boundary, as detected by this Otsu method, to more-or-less coincide with this contour. When applied on the original image, the detected umbra comes out to be significantly larger in size as seen in panels (a1,b1) of Figure~\ref{fig3.2}. Upon investigation, we realize that this over-estimation occurs due to the presence of few brighter pixels on the edge of the detected spots. In fact, these bright pixels are originally a part of the quiet Sun region and got picked up during the sunspot detection procedure. Though the number of such pixels is very small compared to the total pixels of a typical sunspot, it seems to have a significant influence on the derived threshold value. To get rid of these ``rouge pixels", a pre-processing technique is applied before feeding the spots into the Otsu method. We set up an intensity filter which is based on a threshold defined as:
\begin{equation}
I_{\rm th}=\bar{I}-k\sigma
\label{eq3.2}
\end{equation}
where $\bar{I}$ and $\sigma$ are mean and standard deviation of spot region. With this criteria, a pixel with intensity ($ I_{\rm n}$) greater than $I_{\rm th}$ gets removed form that specific spot {\it i.e.} we set $I_{\rm n}=0$. Although, from Equation~\ref{eq3.2}, we note that {\it k} is a free parameter which needs to be optimized. We fix this issue by taking a large subset of randomly chosen sunspots (of different contrasts and morphologies) and repeating the above procedure with multiple values of {\it k}. After visual inspections of each of those results, we find that two values, $ k=0.3$ and $k=0.5$, produce the most accurate results as compared to other {\it k} values. However, most often or not, the umbra gets underestimated with $k = 0.5$ (Figure~\ref{fig3.2}a3, \ref{fig3.2}b3).

\begin{figure}
    \centering
    \centerline{\includegraphics[width=\textwidth,clip=]{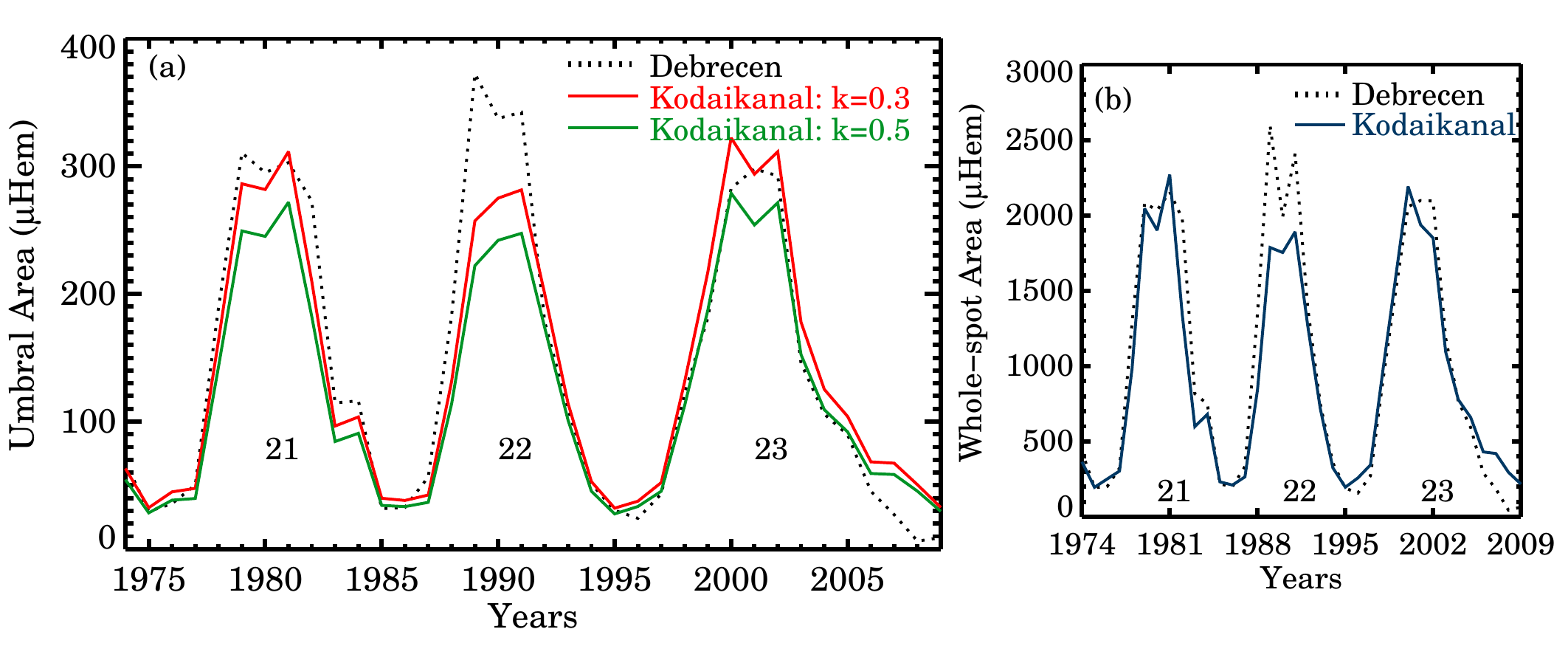}}
    \caption{{\it Panel(a):} Comparison of  yearly averaged umbral areas between Kodaikanal ($k=0.3$ (red) \& $k=0.5$ (green)) and Debrecen data (black). {\it Panel(b):} Same as before but for the whole-spot area.}
\label{fig3.3}
\end{figure}
To better visualize this effect, we compare our results with the umbra measurements from Debrecen Observatory\footnote{This data is downloaded from \url{http://fenyi.solarobs.csfk.mta.hu/ftp/pub/DPD/data/dailyDPD1974_2016.txt}.} \citep{Baranyi2016,Gyoeri2017} as shown in Figure~\ref{fig3.3}a. The plot highlights the fact that $k=0.3$ is indeed a better choice for our Kodaikanal data. However, there is a large discrepancy between the Kodaikanal values with that from Debrecen, near the Cycle 22 maxima. To eliminate the possibility of this being an artefact of our umbra detection technique, we also plot the whole spot area between the two observatories in Figure~\ref{fig3.3}b. Presence of a similar difference in this case too, indicates an underestimation of total sunspot area during the original spot detection procedure, as reported in \cite{Mandal2017a}.

Finally, we compute the penumbra to umbra area ratio as:
\begin{equation}
{\rm Ratio }= \q = \frac{A_{\rm W}}{A_{\rm U}}-1
\end{equation}
where $A_{\rm W}$ and $A_{\rm U}$ are whole spot area and umbra area. This definition is same as \cite{Antalova1971} and \cite{Hathaway2013}.

\section{Results}
We calculate the the ratio  $\q$ for the whole period of the currently available Kodaikanal data which covers Cycle 16 to Cycle 23. Different aspects of this ratio are discussed in this section.

\begin{figure}
\centerline{\includegraphics[width=\textwidth,clip=]{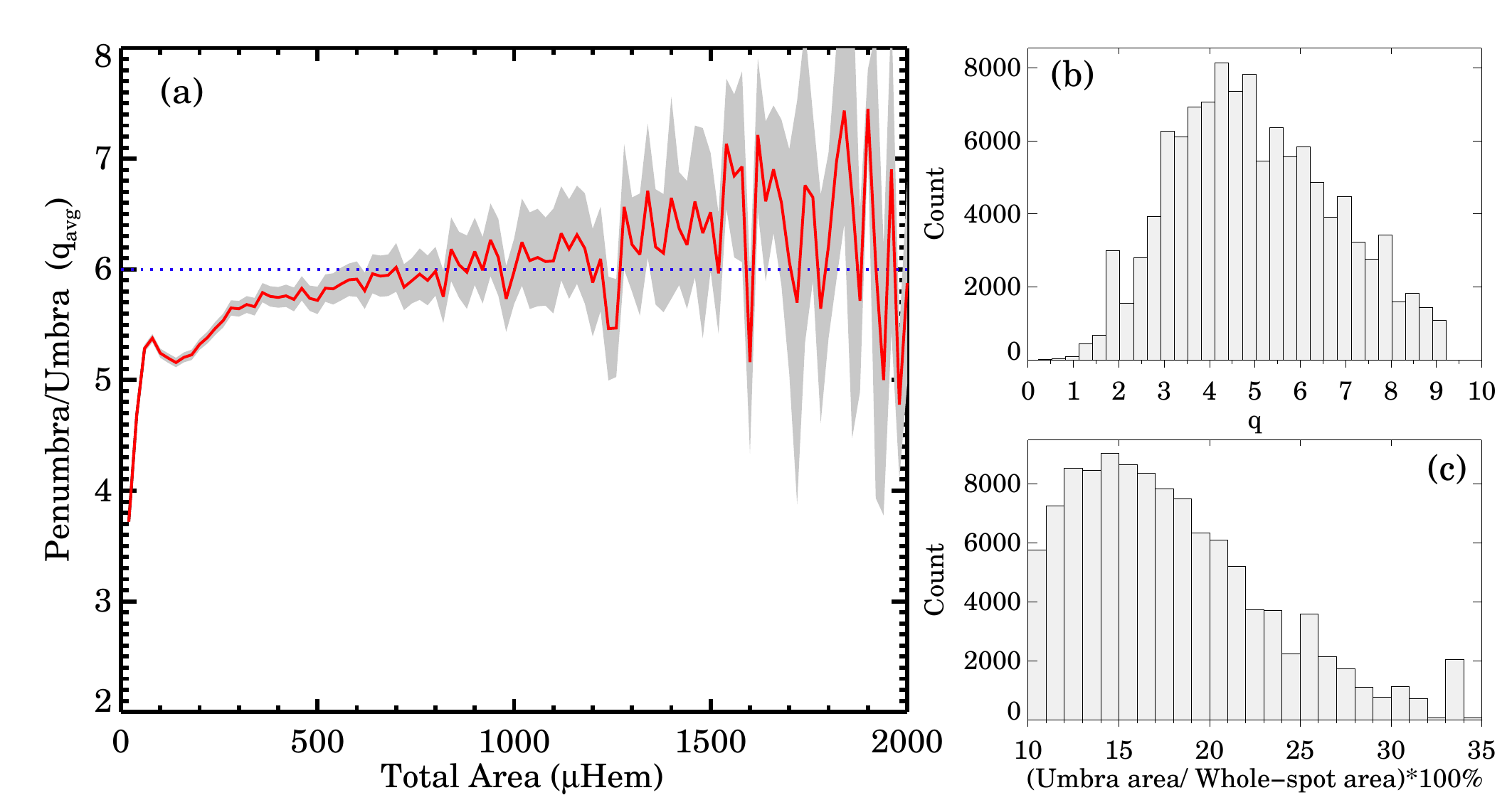}}
\caption{ {\it Panel(a):} Penumbra to umbra area ratio as function of total sunspot area binned over $20~\mu$Hem. Grey shaded region represent the $2\sigma$ errors. {\it Panel(b)} shows the distribution of individual ratio ($\q$) whereas the distribution of percentage coverage of umbral area over the whole-spot area is shown in {\it Panel(c)}.}
\label{fig3.4}
\end{figure}

\subsection{Individual Variations}

To investigate the overall behaviour of  $\q$, we group the sunspot areas into bin sizes of $20~\mu$Hem between $20-2000~\mu$Hem and calculate the average ratio ({$\q_{\rm avg}$}) for all the sunspots falling in that particular bin. Figure~\ref{fig3.4}a shows the quantity {$ \q_{\rm avg}$} as a function of total spot area. The shaded region represents the standard error of $2\sigma$ uncertainty. The error bars beyond area $>1500~\mu$Hem are considerably larger due to poor statistics in those bins. Initially, the ratio for smaller spots (area $<100~\mu$Hem), increases rapidly from 3.4 to 5.2. As the area increases further ($>100~\mu$Hem), {$ \q_{\rm avg}$} tends to settle down to a value of $\approx$6 \citep{Jha2018}. In fact, these results are consistent with the findings by \cite{Antalova1971} \& \cite{Hathaway2013}. Physically this means larger spots tend to have larger penumbra (the observed slow upward trend), however large uncertainties make this conclusion rather weak. In addition to this, we note that there is a local minima of {$\q_{\rm avg}$} around 150~$\mu$Hem which also needs further investigation and we do not have a convincing explanation for the same. Behavior of $\q$ for every detected sunspots is also analysed and presented in a histogram as shown in Figure~\ref{fig3.4}b. The distribution peaks $\approx$4.5 and falls rapidly on both sides from the peak. Another interesting aspect is the coverage of umbra with respect to the total area for any individual sunspot. Figure~\ref{fig3.4}c shows the distribution of this quantity (expressed in \%). The distribution peaks at 15\%, although there are significant number of cases between 15\% to 25\%. These properties are in good  agreement with previously measured values by \cite{Watson2011,Carrasco2018}.

\subsection{Dependency on Cycle Strength and its Phases}
During the onset of a solar cycle, we see very few spots present on the disc (mostly of smaller sizes \cite{Mandal2017a}). They are also located at higher latitudes and with the progress of the cycle, they move towards the equator to form the popular `sunspot butterfly diagram'. 

\begin{figure}
\centerline{\includegraphics[width=\textwidth,clip=]{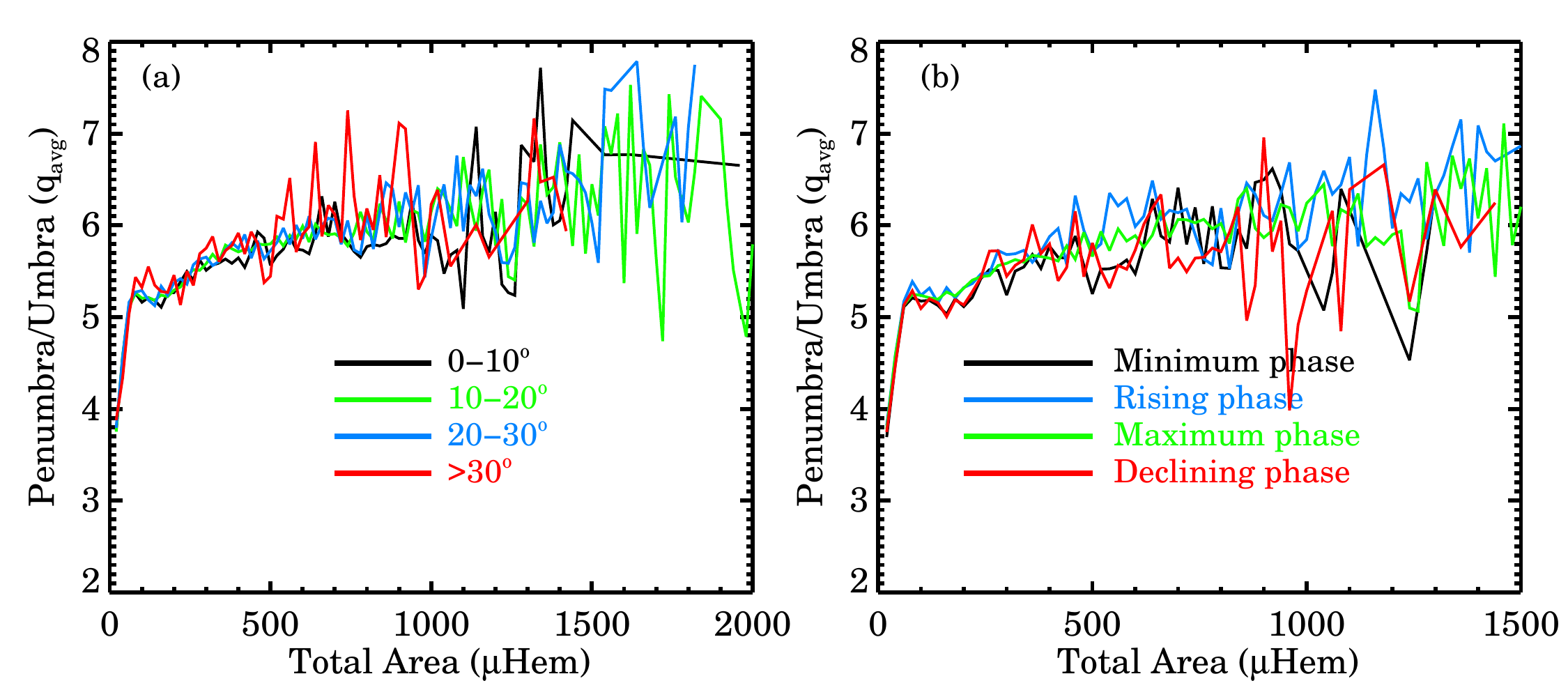}}
\caption{{\it Panel (a):} Variation of $\q_{\rm avg}$ as a function of total area in four different latitude bands as written on the panel; {\it Panel (b):} Same as previous but separated for four different activity phases of cycles.}
\label{fig3.5}
\end{figure}
We look for any such dependency of $\q_{\rm avg}$ by dividing the solar disc into several latitudinal bands. We fold the two hemispheres together and the results are plotted in Figure~\ref{fig3.5}a. As seen from the plot, we find that the ratio does not depend on the latitude of a spot \citep{Antalova1971,Hathaway2013}. In a slightly different representation of the same phenomena, we isolate the spots according to their appearances during a solar cycle. In fact, we are also motivated by some of the earlier studies by \cite{Jensen1955, TandbergHanssen1956,Antalova1971}, where these authors reported different values of $\q_{\rm avg}$ during a cycle maxima as opposed to a cycle minima. To check this, a cycle is divided into four phases: minimum phase, rising phase, maximum phase and declining phase. The definition of each of these phases is the same as described in \cite{Hathaway2013}. Considering all the cycles together, we generate a plot as shown in Figures~\ref{fig3.5}b. In this case, too, we do not notice any change for a given spot range in different phases of cycles. This is consistent with the RGO data as found by \cite{Hathaway2013}.
\begin{figure}
\centerline{\includegraphics[width=\textwidth,clip=]{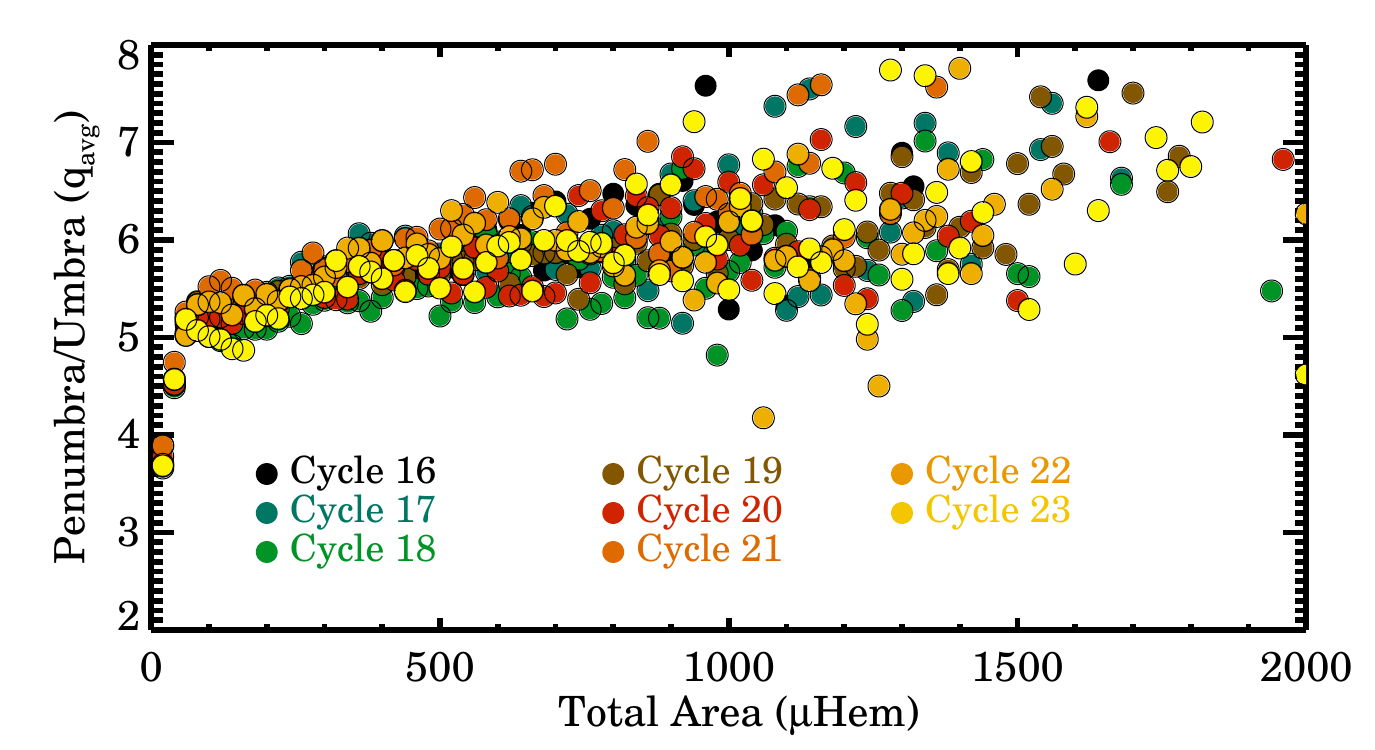}}
\caption{Figure above (a-h) shows the variations in $\q_{\rm avg}$ as recorded for each solar cycle (Cycles 16-23). Dashed red line is plotted just for reference.}
\label{fig3.6}
\end{figure}

 The other factor to potentially affect this ratio is the strength of a cycle. Similar spots in a weak cycle (Cycle 16) may have different $\q_{\rm avg}$ values than a strong cycle (Cycle 19). From Figure~\ref{fig3.6}a-\ref{fig3.6}h we note that there is absolutely no variation of $\q_{\rm avg}$ with cycles of different strengths. In a similar analysis by \cite{Hathaway2013} on RGO data, showed two different behaviours, specifically for the smaller spots (area $<100~\mu$Hem), between even and odd numbered cycles. However, we do not find any such relation in our data.

\begin{figure}
\centering
\centerline{\includegraphics[width=\textwidth,clip=]{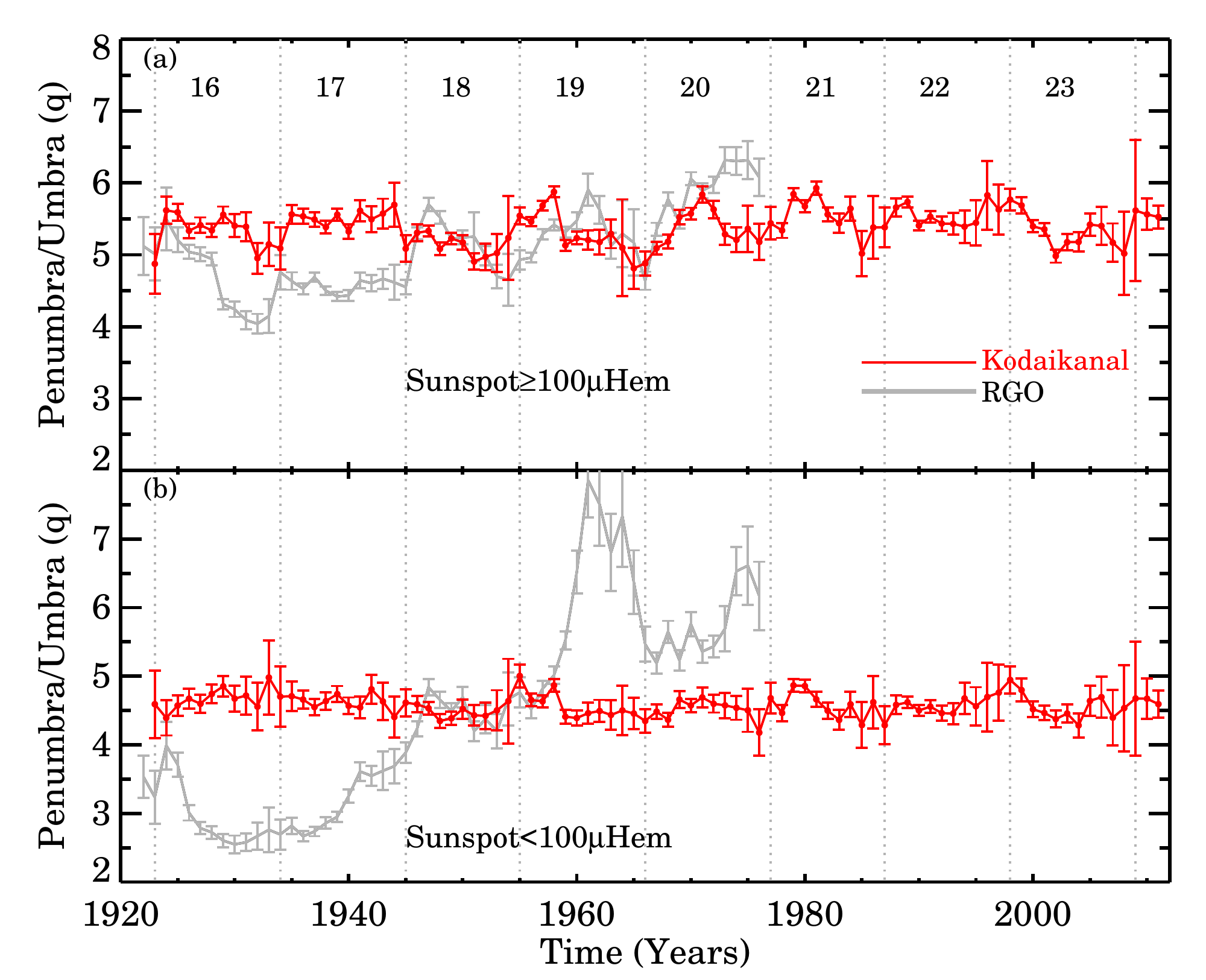}}
\caption{ Yearly averaged values of $\q$ as obtained from Kodaikanal data (red points) for two sunspot classes; for area $\geq$100~$\mu$Hem ({\it Panel-(a)}) and for area $<$100~$\mu$Hem ({\it Panel-(b)}). Similar values from RGO are also over plotted (grey points) for comparison. Error bars in each case represents the 2$\sigma$ uncertainties.}
\label{fig3.7}
\end{figure}

\subsection{Behaviour of Smaller and Larger Spots}

Sunspots of different sizes tend to show different behaviour \citep{Mandal2016}. In this section, we look for the temporal evolution of $\q $ from two class of sunspots: {\it i)} Sunspots with area $<$100~$\mu$Hem (Figure~\ref{fig3.7}a); {\it ii)} Sunspots with area $>$100$~\mu$Hem (Figure~\ref{fig3.7}b). The choice of this threshold at 100$\mu$Hem is primarily dictated by the fact that we see a jump in $\q$ value at this area value in Figure~\ref{fig3.4}a . In order to compare our results with \cite{Hathaway2013}, we over plot the $\q$ values for RGO data as shown in Figure~\ref{fig3.7}. For spots $>$100$~\mu$Hem, the ratio neither show any significant time variation, nor any tendency to follow the solar cycles. The over plotted RGO data is in accordance with our values, except some systematically lower values during Cycles 16 to 17. One of the highlights of the work by \cite{Hathaway2013}, was the large secular variation of the ratio for smaller spots which showed 300\% increment with time. However, this property is not visible from Kodaikanal data which shows the ratio remains constant at $\approx$4.5 throughout the duration. In fact, analysing the Coimbra Astronomical Observatory (COI) data, \cite{Carrasco2018} also reported the absence of any type of secular variation in smaller spots.

As mentioned in the introduction, differences in the derived $\q$ values largely depend on the methods that have been used to detect umbra-penumbra boundary \citep{Steinegger1997a}. Our method of Otsu thresholding has not been utilized in the literature before and thus, we feel the need of checking the robustness of this method on other independent datasets. The following section describes the application of the same on the space-based SOHO/MDI continuum images.

\subsection{Application on SOHO/MDI}
\begin{figure}
\centering
\centerline{\includegraphics[width=0.7\textwidth,clip=]{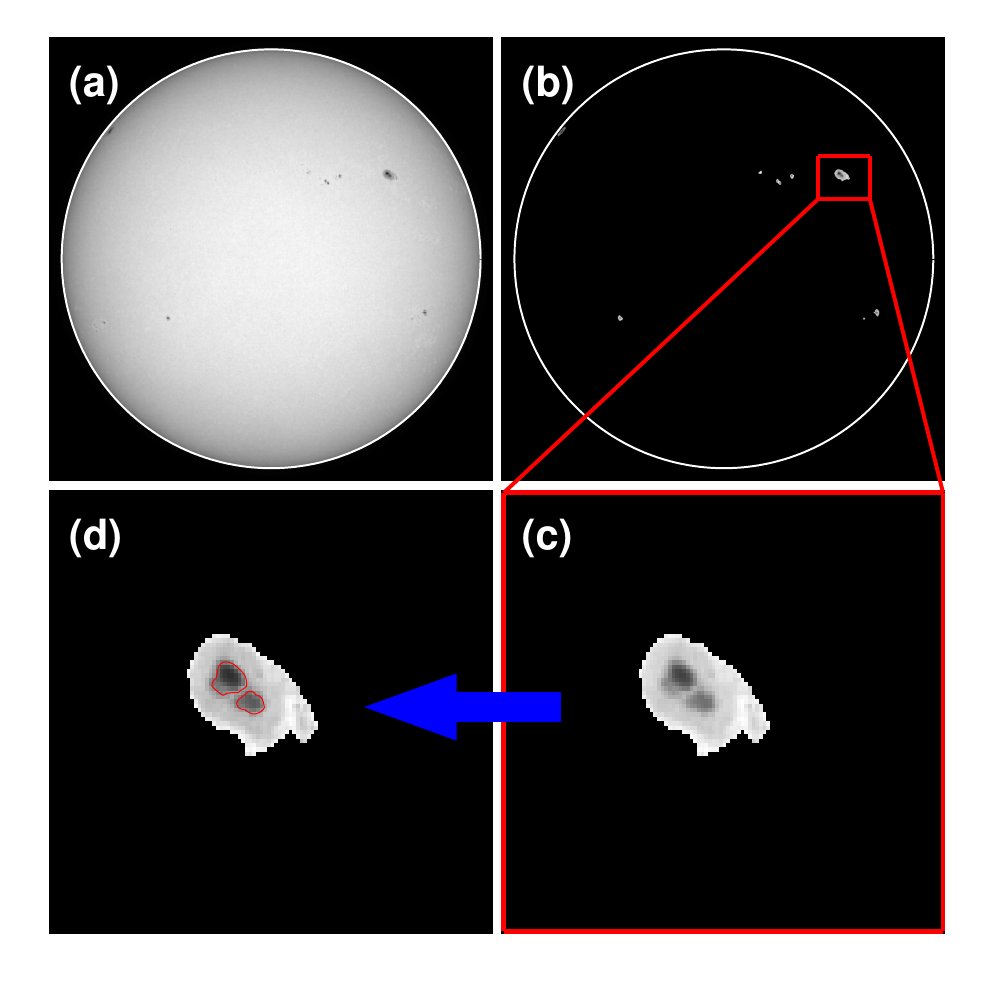}}
\caption{ Detection of umbra from SOHO/MDI data. {\it Panel-(a):} A representative continuum image as captured on 1999-05-14 23:59. {\it Panel-(b)} and {\t (c):} Detected sunspots and its zoomed in view respectively. {\it Panel-(d):} Contours of the umbrae over plotted onto the spot.}
\label{fig3.8}
\end{figure}

We analyse SOHO/MDI \citep{Scherrer1995} continuum images from 1996 to 2010 with a frequency of one image per day. First, we detect the sunspots using the same Sunspot Tracking and Recognition Algorithm (STARA: \cite{Watson2009}) as used on the Kodaikanal data. The detected spots are then fed to the Otsu algorithm for umbra detection. Figure~\ref{fig3.8} summarizes the whole procedure.

\begin{figure}
\centerline{\includegraphics[width=\textwidth,clip=]{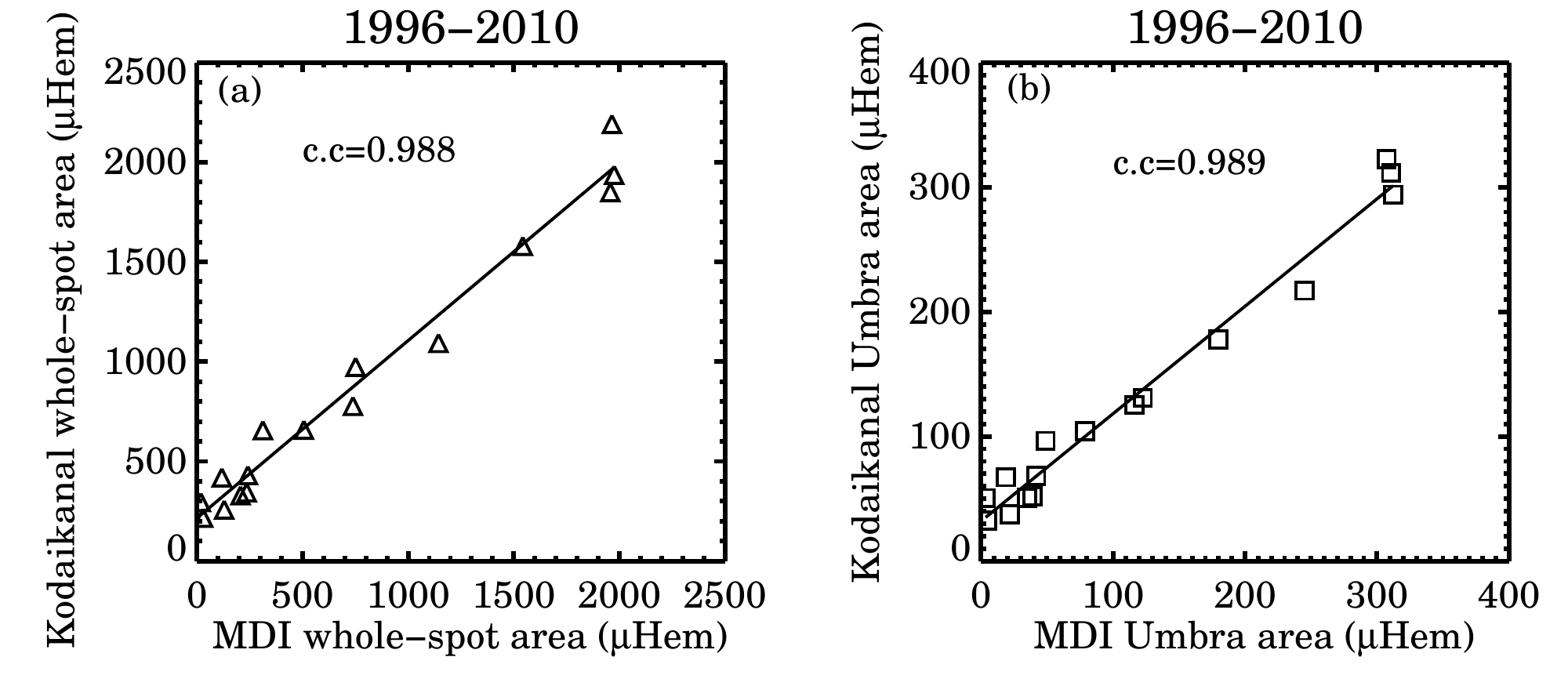}}
\caption{Comparison of yearly averaged whole sunspot area and umbra area as extracted from MDI and Kodaikanal.}
\label{fig3.9}
\end{figure}

We first compare the whole spot area values between Kodaikanal and MDI and the result is shown in Figure~\ref{fig3.9}a. Computed yearly averages of whole spot areas are very similar to each other ({\rm c.c}=0.99). A similar behaviour is found for the umbral areas too (Figure~\ref{fig3.9}b). Hence, the overall spot areas measured from these two observatories, show similar trends. However, our prime interest in this case, is to recover the behaviour of small ($<$100 $\mu$Hem) spots as seen from Kodaikanal. 

\begin{figure}
\centerline{\includegraphics[width=0.9\textwidth,clip=]{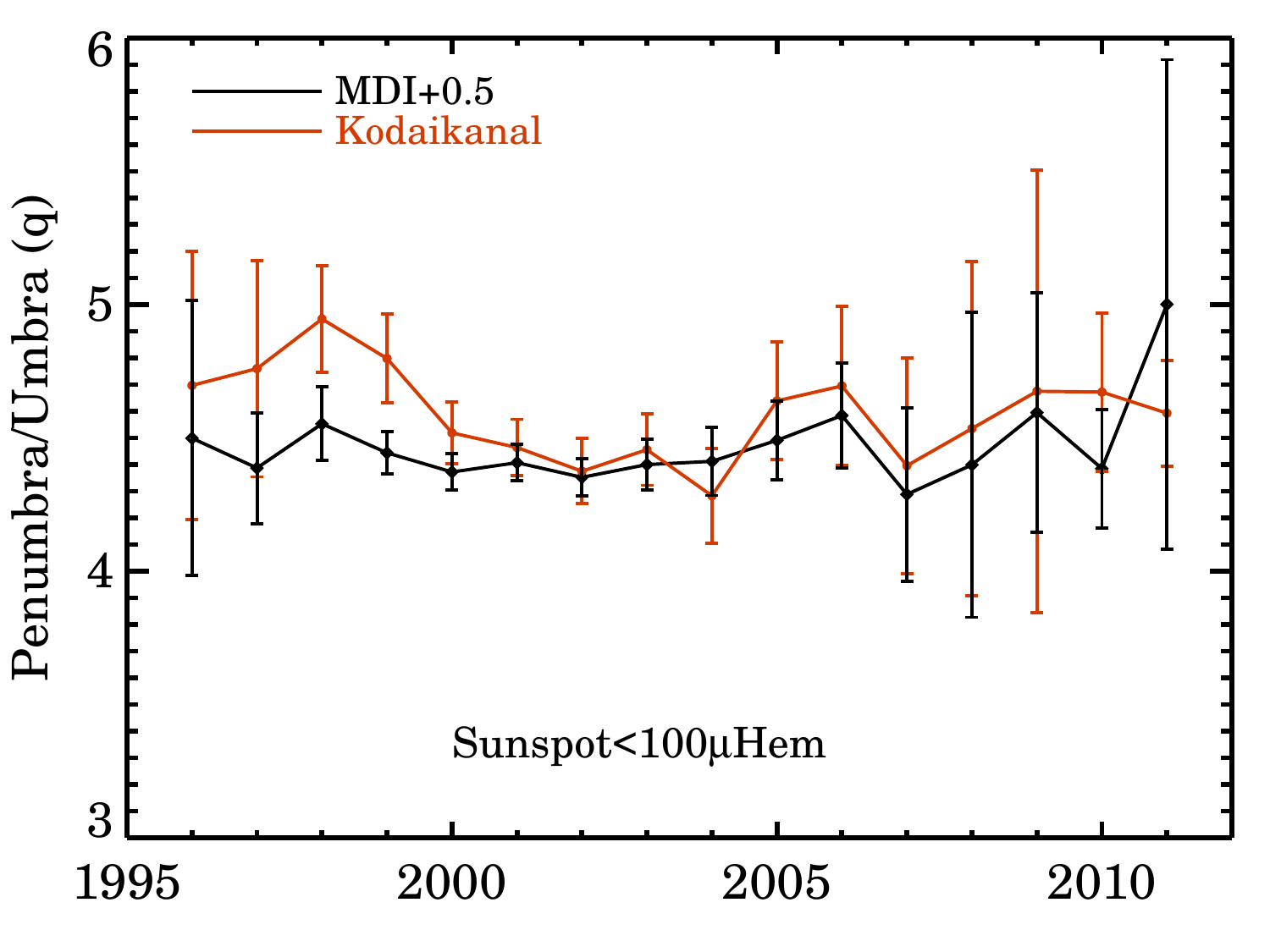}}
\caption{Ratio of areas of penumbra to umbra as a function of time for smaller sunspots.}
\label{fig3.10}
\end{figure}

In Figure~\ref{fig3.10} we plot the $\q$ values (black solid line) for spots with area $<$100$\mu$Hem as calculated from MDI. Kodaikanal values, for the overlapping period, are also over plotted (in red) for the ease of comparison. We see similar trends in both the curves, however the MDI values are needed to scale up by adding a constant factor of 0.5 to match the absolute values of Kodaikanal. This underestimation in MDI data is again, primarily due to the bright pixels present near the spot boundaries. During our analysis, we learnt that it is impossible to completely avoid these bright pixels while using the Sunspot Tracking and Recognition Algorithm (STARA) algorithm on large datasets. We can get around this problem by using a suitable {$k$} value as used in the earlier case. However, such a treatment only scales the absolute values, not the trend. Hence we present the results as it is.

\section{Conclusion}

In this paper, we investigated the long-term evolution of sunspot penumbra to umbra area ratio primarily using Kodaikanal white-light data. The main findings are summarized below:

$\bullet$ A total of 8 solar cycle (Cycles 16-23) data of Kodaikanal white-light digitized archive (1923-2011) and 15 years of MDI data (1996-2010) have been analysed in this work. We have used an automated umbra detection technique based on Otsu thresholding method and found that this method is efficient in isolating the umbra from variety of spots with different intensity contrasts.

$\bullet$ The penumbra to umbra ratio is found to be in the range of 5.5 to 6 for the spot range of 100~$\mu$Hem to 2000~$\mu$Hem. It is also found to be independent of cycle strengths, latitude zones and cycle phases. These results are in agreement with the previous reports in the literature.

$\bullet$ We segregated the spots according to their sizes and found that there is no signature of long-term secular variations for spots $<$100~$\mu$Hem. This results contradicts the observations made by \cite{Hathaway2013} using the RGO data. However, our results are in close agreement with a recent study by \cite{Carrasco2018}.

$\bullet$ To check the robustness of our umbra detection technique, we analysed SOHO/MDI continuum images. These results also confirmed our previous findings from Kodaikanal data including the absence of any trend for smaller spots. During this study, we realized that although the Otsu technique is robust and adaptive in determining the umbral boundaries, it is also sensitive to any presence of artefacts within the spots.

In future, we plan to continue our study using the Solar Dynamics Observatory (SDO)/Helioseismic and Magnetic Imager (HMI) \citep{Schou2012} data. This will not only extend the time series but will also allow us to study the effect of higher spatial resolution ({\it i.e.} more pixels within a spot) in determining the optimum threshold. We also plan to use the Debrecen sunspot images (which are available online) and repeat the measurements of this ratio using our method. Debrecen has more than fifty years of overlap with Kodaikanal which makes this data suitable for cross calibration too.

\clearpage{}

\clearpage{}\chapter[Differential Rotation of the Sun Using KoSO Data]{Measurements of Solar Differential Rotation Using the Century Long Kodaikanal Sunspot Data}
\label{Chap4}
\lhead{\emph{Chapter 4: Differential Rotation of the Sun Using KoSO Data}} \noindent
\section{Introduction}
\label{ch4-intro}
A sunspot, a dark photospheric feature, is widely considered to be a suitable proxy of solar surface magnetism and its long-term variability. As we now understand, a dynamo which operates beneath the photospheric layer, is responsible for generating the magnetic field which we observe on the surface \citep{Parker1955a}. Within this framework, a sufficiently strong and buoyant flux tube rises through the convection zone and forms a bipolar magnetic patch which often manifests itself as a spot pair on the visible solar surface \citep{Parker1955a, Solanki2003}. Thus, analysis of spot properties on the surface, presents a unique opportunity to sneak peek the sub-surface physical processes which are otherwise hidden from us. 

One of the key solar parameters that can be measured using sunspots, is the differential rotation profile of the Sun. In fact, it is this differential rotation in the solar dynamo theory which stretches the poloidal field and converts it into the spot-generating toroidal field \citep{Charbonneau2010}. Therefore, precise and long-term measurements of solar rotation rates are required to further help and improve the current dynamo models \citep{Badalyan2017}. Modern day space-borne high-resolution data from the Solar and Heliospheric Observatory \citep[SOHO:][]{Domingo1995} and the Solar Dynamic Observatory \citep[SDO:][]{Pesnell2012} offer significant improvements in measuring the rotation profile. However, such data are only limited to the last two solar cycles (1996 onward). On the other hand, regular sunspots measurements from a number of ground based observatories are available for a significantly longer time (more than a century) and can be used to determine the differential rotation profile of the past.

Being one of the oldest recorded solar parameters, sunspots have already been utilised many times in the past to derive the rotation profile. Earliest of such measurements were reported by Christoph Scheiner in 1630 and almost after 200 years, by \citet{Carrington1863}. These authors followed spots as they moved across the solar disc and reported a differential rotation profile in which the equator rotates faster than the poles. Later, as more sunspot data from different observatories became available, several follow-up studies on this subject were conducted. These include analyses using sunspot data from the Royal Greenwich Observatory \citep[][etc]{Newton1951, Ward1966, Balthasar1986, Javaraiah2005a}, Kenzelh\"ohe Observatory \citep{Lustig1983}, Mt. Wilson Observatory \citep[][etc.]{Howard1984, Gilman1984, Gilman1985}, Meudon Observatory \citep{Ribes1993}, Catania Observatory \citep{Ternullo1981}, Kodaikanal Solar Observatory \citep[KoSO:][]{Howard1999, Gupta1999}, etc. Overall, the observed rotation profile ($\Omega$) can be expressed by the following mathematical formula \citep{Newton1951, Howard1984}:
\begin{equation}
    \Omega~=~A~+~B\sin^2 {\theta}~+~C\sin^4{\theta}.
    \label{eq1}
\end{equation}
Since sunspots are mostly restricted within $\pm40 \degree$ latitudes, one can drop the $C$ term \citep{Newton1951, Howard1984, Woehl2010} and rewrite the equation as:
\begin{equation}
    \Omega=A+B\sin^2{\theta},
    \label{ch4:eq2}
\end{equation}
where $A$ and $B$ represent the equatorial rotation rate and the latitudinal rotation gradient, respectively. Here, $\theta$ is the heliographic latitude of the spot.

Despite the considerable amount of independent studies on this subject, there still remain certain systematic differences in the rotation rates derived from different sunspot catalogues (see Table~\ref{ch4:table1}). This is partly due to the use of different spot tracing methods \citep{Newton1951, Howard1984, PoljancicBeljan2017}, different trace subjects within the spot (such as only the umbra vs. the whole spot; \citealp{Howard1999, Gupta1999}) etc. In fact, in most of the older publications, authors had used manual inspections to identify the spots \citep{Newton1951, Gupta1999}. This, in turn, has introduced human errors and biases.

In this context, we present here the long-term measurements (1923\,--\,2011) of solar rotation rates using the newly digitised, high resolution KoSO sunspot observations. A fully automated sunspot tracing algorithm has been implemented to eliminate any detection bias and robustness of the method is also demonstrated by applying it on modern data from the Michelson Doppler Imager \citep[MDI:][]{Scherrer1995} onboard SOHO.

\section{Data and Method}
\label{ch4-method}

Kodaikanal Solar Observatory (KoSO) is recording regular white-light sunspot data since 1904. This data, originally stored in photographic plates and films, have now been digitised into high resolution images of 4k$\times$4k size (for more information about this digitisation, see \citealp{Ravindra2013}). Barring the initial 15 years (1904\,--\,1920), calibration as well as sunspot detection on the rest of the data (1921\,--\,2011) have also been completed recently \citep{Ravindra2013, Mandal2017a}. KoSO sunspot area catalogue\footnote{Available for download at \url{https://kso.iiap.res.in/new/white_light}.} provides area values (i.e umbra+penumbra) of individual spots\footnote{Individual spots are not yet classified into groups in the current KoSO catalogue.} along with their heliographic positions (longitude and latitude).

For the purpose of this work, we utilise the binary, full-disc images of detected sunspots as generated by \citet{Mandal2017a} (see Section~4 of \citet{Mandal2017a} for more details). In this work, we implement a fully automated algorithm to track the spots in successive sunspot images. The whole procedure can be summarised as follows.

\begin{figure}[!htbp]
\centering
\includegraphics[width=0.75\textwidth,clip=]{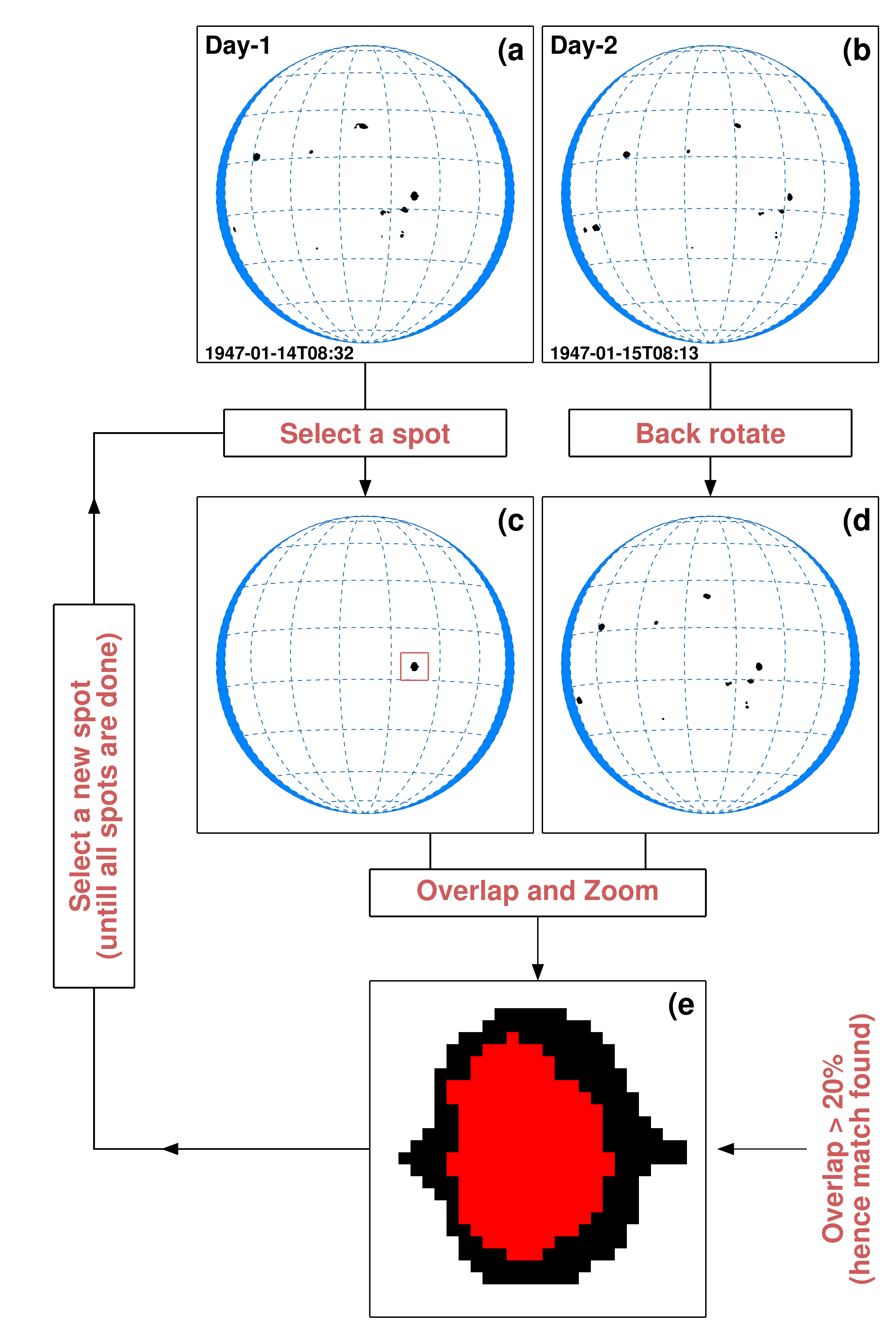}
\caption{Representative examples of our sunspot tracking algorithm. (a), (b): Show two consecutive observations which are one day apart. Blue shaded regions near the solar limb highlight areas beyond $\vert 70\vert\degree$ longitudes.
(c) Shows a selected spot from Day-1 and (d) shows a back rotated image from Day-2.
(e) Depicts a zoomed in view of the overlapped region of Image1 (with the selected spot) on to the differentially back rotated image of Day-2.}
\label{ch4:fig1}
\end{figure}
\begin{figure}[!htbp]
\centerline{\includegraphics[height=0.9\textheight,clip=]{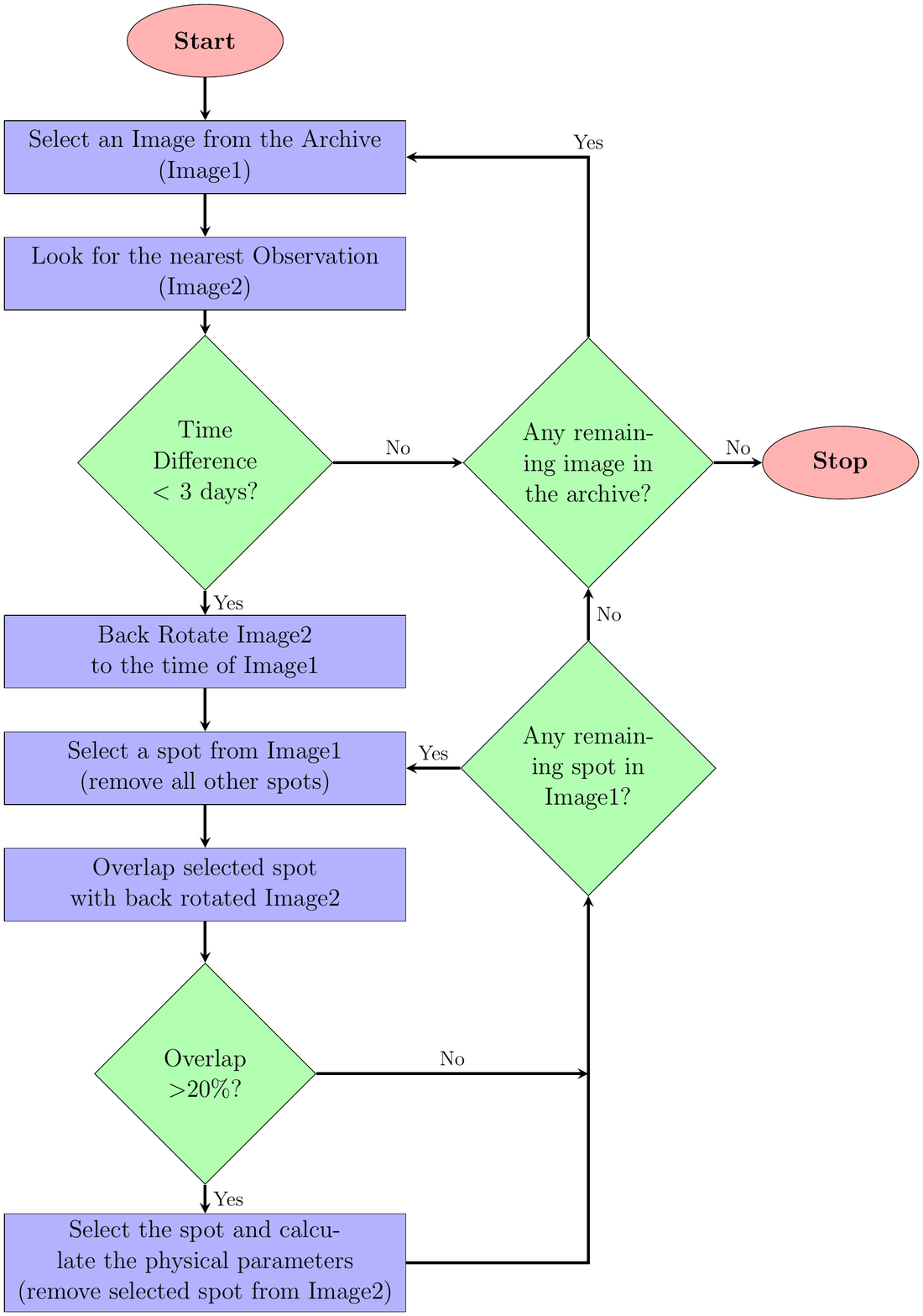}}
\caption{A flow chart describing the different steps of our spot tracking algorithm.}
\label{flow_chart}
\end{figure}

To follow the same sunspot in two consecutive observations, we first choose two images, Image1 and Image2, which are (preferably) taken on consecutive days. \autoref{ch4:fig1}a-b, show one such pair of observations. To incorporate the occasional `missing data' scenario, we also allow the time difference between Image1 and Image2 to be a maximum of up to three days. Additionally, to minimise the projection errors, we restrict our analysis to spots whose absolute heliographic longitude are $<\vert 70\vert\degree$.
In order to identify a selected spot of Image1 (e.g. \autoref{ch4:fig1}c) in a subsequent observation, we first differentially back rotate Image2 (using the IDL routine {\sf drot\_map.pro}\footnote{Detail of this function is available at \url{https://hesperia.gsfc.nasa.gov/ssw/gen/idl/maps/drot_map.pro}. This routine  uses the differential rotation parameter from \citet{Howard1990} to differentially rotate an input image.}) to the time of observation of Image1 (\autoref{ch4:fig1}d). In this context, it is also worth noting that the opposite sense of image rotation, i.e. forward-rotating Image1 to match the time of Image2 and repeat the same process also produces the same result.
Next, the selected spot from Image1 is overlapped onto the back rotated Image2 for the final step of the procedure (\autoref{ch4:fig1}e).
Basically, an overlap here is an indication of a potential match. However, group evolution makes this process somewhat complicated. To eliminate any false detection, we employ a two step verification method. Firstly, the overlap (by pixel count) has to be more than 20\%. This accounts for the uncertainties that are introduced during the back rotation of Image2. The example shown in \autoref{ch4:fig1}d has an overlap of 52\%. Next, there are situations when spots appear significantly different on Day 2 due to their rapid evolution (which are mostly mergers or bifurcations). To handle these cases, we invoke a cut-off on the fractional change in area as $\frac{|~a_2-a_1~|}{a_2+a_1} < 0.7$, where $a_1$ and $a_2$ are the areas of a spot in Image1 and Image2. This limit (of 0.7) has been decided after manually inspecting several randomly chosen image sequences. An Image1 spot which passes both these tests, is then flagged as the `same spot'. This whole procedure is repeated for all the spots in Image1 and for all images in our archive. 
In \autoref{flow_chart}, we present a summary of our spot tracking algorithm in the form of a flow chart.

If $\phi _1$ and $\phi _2$ represent the heliographic longitudes of a spot at times $t_1$ and $t_2$, then the synodic rotation rate
($\Omega _{\rm synodic}$) is calculated as
\begin{equation}
     \Omega _{\rm synodic}~=~\frac{\phi _2 - \phi _1}{t_2 - t_1},
\end{equation}
to convert the synodic values into sidereal rotation rate, we use the following relation \citep{Rosa1995, Wittmann1996, Skokic2014}:
\begin{equation}
    \Omega _{\rm sidereal}~=~\Omega _{\rm synodic}+ \frac{0.9856}{r^2}\left(\frac{\cos^2{\psi}}{\cos{i}}\right),
\end{equation}
where $i$ is the inclination of the solar equator to the ecliptic, $\psi$ is the angle between the pole of the ecliptic and the solar rotation axis orthographically projected on the solar disk and $r$ is the Sun-Earth distance in astronomical units \citep{Lamb2017}. We apply the above mentioned procedure on all the sunspots present in our data and calculated $\Omega _{\rm sidereal}$ (hereafter $\Omega$) values.

\section{Results}
\label{s-results}

\subsection{The Average Rotation Profile}

\begin{figure}[!htbp]
\centering
\includegraphics[width=0.8\textwidth,clip=]{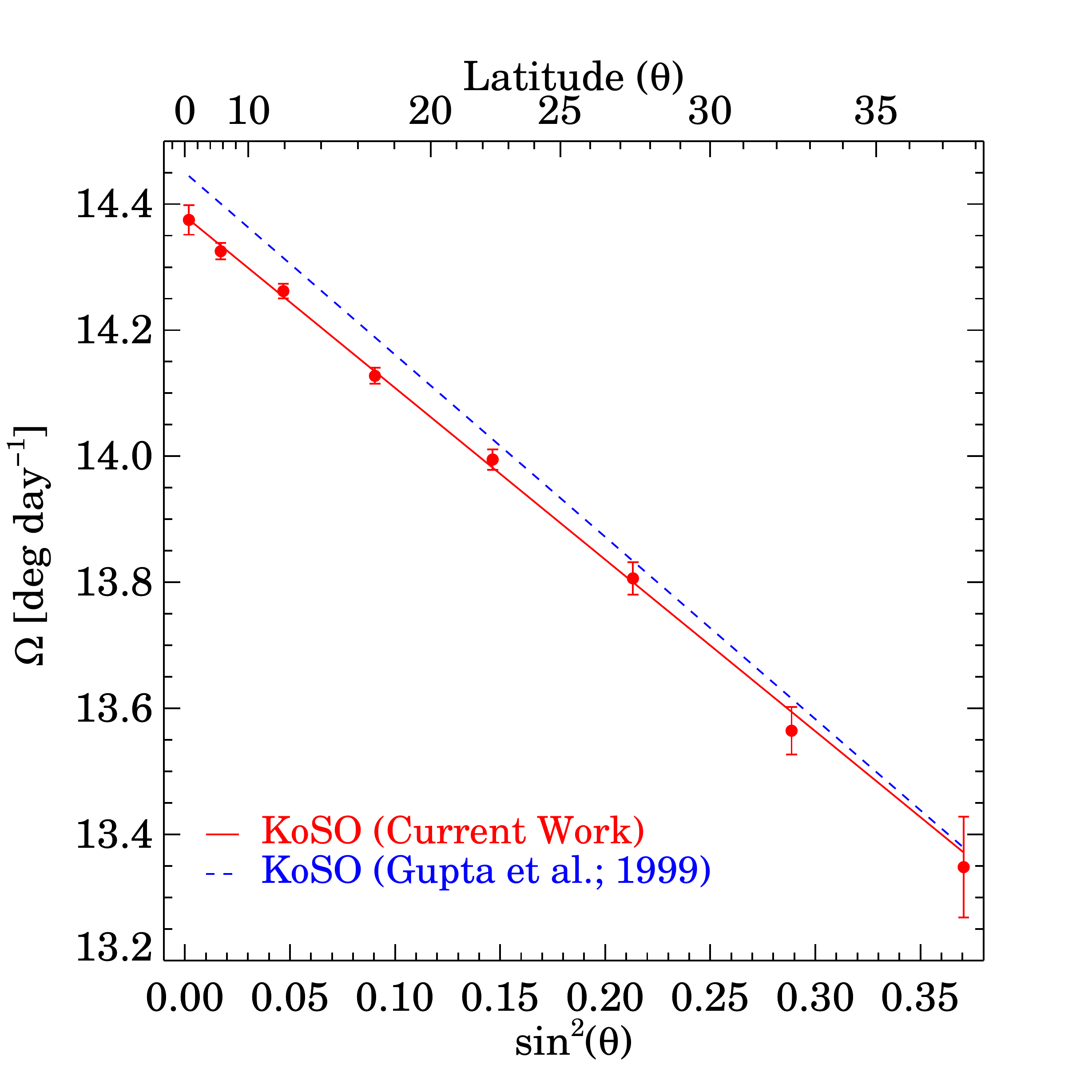}
\caption{Solar rotation profile (red solid line) measured using the KoSO sunspot data for the period 1923--2011. For a comparison, we overplot (blue dashed line) the results from \citet{Gupta1999}.}
\label{ch4:fig2}
\end{figure}

To study the latitudinal dependency of solar rotation rate, we first organise the spots according to their latitudes ($\theta$) in $5\degree$ bins and calculate the mean $\Omega$ in each of those bins. In \autoref{ch4:fig2}, these mean $\Omega$ values (indicated by filled red circles) are plotted as a function of $\sin^2\theta$. The error bars shown in the plot are the standard errors calculated for each bin. As seen from the figure, $\Omega$ seems to have a linear relation with $\sin^2\theta$ \citep{Howard1999}. To measure the equatorial rotation rate ($A$) and the latitudinal gradient of rotation ($B$), we fit \autoref{ch4:eq2} onto our data by using the Levenberg-Marquardt least squares (LMLS) method\footnote{We have used the {\sf mpfit\_fun.pro} function available in IDL for this purpose.} \citep{Markwardt2009} and, the obtained best fit is shown by the solid red line in \autoref{ch4:fig2}. The $A$ and $B$ values, returned from the fit, are $14.381\pm0.004$~deg/day and $-2.72\pm0.04$~deg/day respectively, and they compare well with the existing literature as shown in \autoref{ch4:table1}.
We directly compare our results with the values from \citet{Gupta1999} (shown by the blue dashed line in \autoref{ch4:fig2}) who had used an older low-resolution version of the Kodaikanal data, of limited period as well. Interestingly, our $A$ and $B$ values are slightly different from their measurements (see \autoref{ch4:table1}) and we attribute such differences to the following two reasons. Firstly, \citet{Gupta1999} used only the umbra to measure the rotation rate as opposed to the the whole-spot area (i.e. umbra and penumbra) as used in our study. Secondly, the epochs covered by these two datasets are different. As we will find in later sections not only the spatial extend of the tracer (i.e. sunspot) affects the derived $B$ value, but both parameters, $A$ and $B$, vary significantly with solar cycles. Thus, a combination of these two effects produces the observed differences.

\begin{table}
\centering
\begin{tabular}{ ccccc }
\hline
     References  & Period  & $A \pm \Delta A$ & $B \pm \Delta B$ \\
      & & ($\deg$$/$day) & ($\deg$$/$day) & \\
    \hline

    \citet{Howard1984}         & 1921--1982 & $14.522  \pm 0.004$   & $-2.84 \pm 0.04$    \\
    \citet{Howard1999}
                            & 1907--1987 & $14.547\pm 0.005$     & $-2.96\pm 0.05$     \\
    \citet{Howard1999}
                           & 1917--1985 & $14.459\pm 0.005$     & $-2.99\pm 0.06$     \\
    \citet{Gupta1999}
                          & 1906--1987  & $14.456 \pm 0.002$    & $-2.89 \pm 0.02$    \\
    \citet{Ruzdjak2017}
                          & 1874--2016  & $14.483\pm 0.005$     & $-2.67\pm 0.05$     \\

    Current Work
                           & 1923--2011 & $14.381\pm 0.004$        & $-2.72\pm0.04$  \\
\hline
\end{tabular}
\caption{Differential rotation parameters measured from different observations since \citeyear{Carrington1863}.}
\label{ch4:table1}
\end{table}

\subsection{Variations in Rotational Profile}

\subsubsection{Sunspot Sizes}
\label{sizes}
\begin{figure}[!htbp]
\centering
\includegraphics[width=0.7\textwidth,clip=]{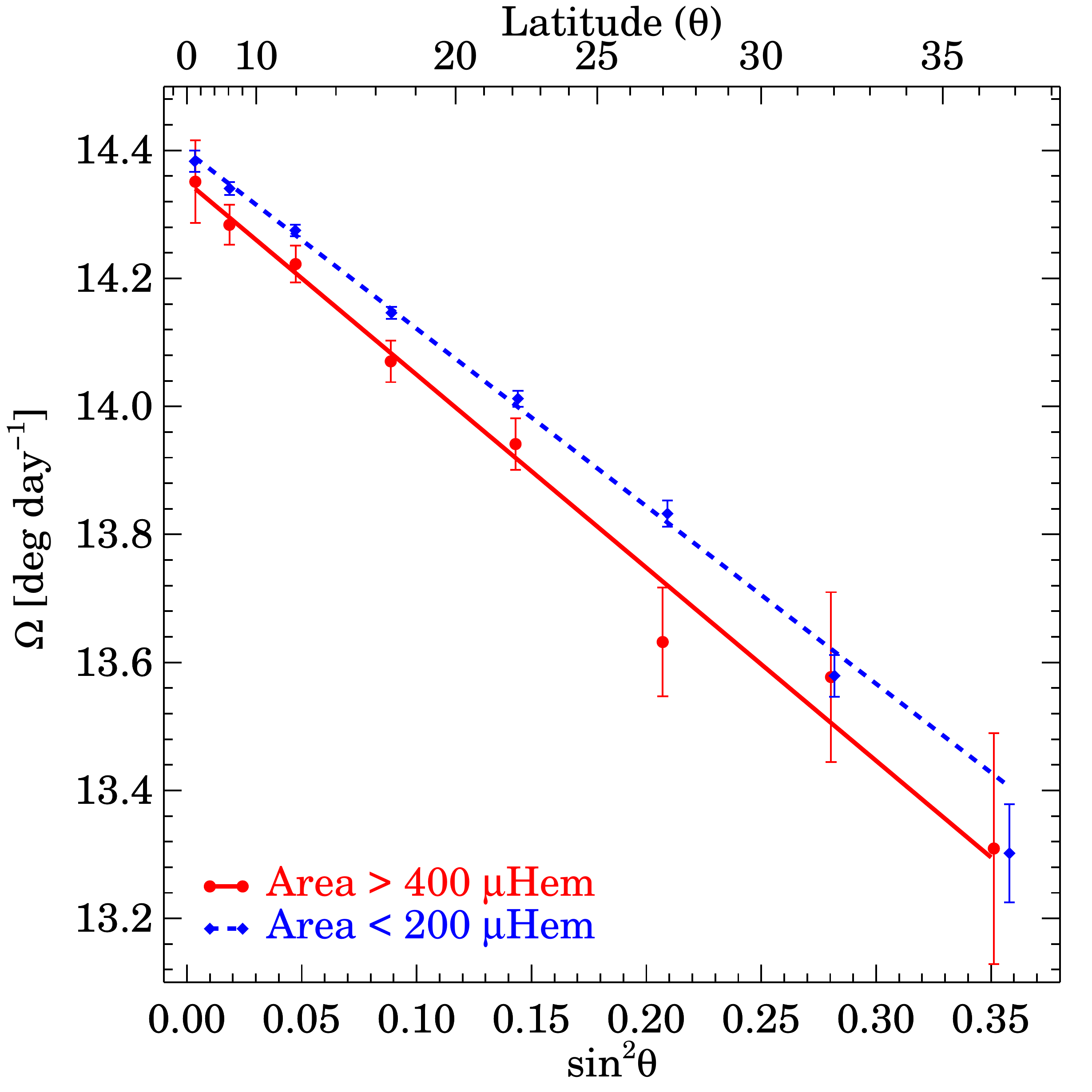}
\caption{Rotation profiles of sunspots with area $<$200~$\mu$Hem (blue dashed line) and with area $>$400~$\mu$Hem (red solid line).}
\label{ch4:fig3}
\end{figure}

Sunspots are observed in a variety of shapes and sizes and from past observations, we know that bigger spots typically host stronger magnetic field \citep{MunozJaramillo2015}. These strong fields, which are believed to be anchored deep in the convection zone, may have a possible effect on the spot rotation at the surface \citep{Ward1966, Gilman1983, Gupta1999}. Hence, we investigate the effect of stronger fields on the inferred solar rotation by dividing all the spots according to their areas. They are grouped into two area categories:  (i) small spots with area  $<200~\mu$Hem and, (ii) large spots with area  $>400~\mu$Hem. \autoref{ch4:fig3} shows the obtained rotation profiles for these two spot area classes.

Our results show that bigger spots rotate slower than smaller ones (see the $A$ values listed in \autoref{ch4:table2}). This is consistent with previous reports by \citet{Ward1966, Gilman1984, Gupta1999}; and \citet[][using magnetic field maps]{Kutsenko2020}. At the same time, we do not observe any significant change (considering the error limits) in $B$ values between the two area categories. Physically, the slower rotation rates for bigger spots (hence, stronger magnetic fields; \citealp{Livingston2006}) may hint towards deeper anchoring depths of the parent flux tubes \citep{Balthasar1986}. Additionally, it has been pointed out that spots with larger areas experience a greater drag which further affects their rotation rates \citep{Ward1966, Gilman1984}.

\begin{table}[!ht]
\centering
\begin{tabular}{ccccc}
\hline
     Area Group  &$A\pm \Delta A$ &$B \pm \Delta B$ \\\hline
    Small (Area $<200~\mu$Hem) &$14.399\pm 0.004$        &$-2.77 \pm 0.04$\\      Large (Area $>400~\mu$Hem) &$14.351\pm 0.011$        &$-3.01 \pm 0.12$   \\   \hline
\end{tabular}
\caption{Differential rotation parameters measured from different sizes of spots for Kodaikanal  Solar Observatory data (current work).}
\label{ch4:table2}
\end{table}

\subsubsection{Solar Activity Phase and Cycle Strength}
Properties of sunspots such as their areas, numbers, etc. vary with solar cycle phases \citep{Hathaway2015}. Hence, it is reasonable to look for signatures of any such variation in the $\Omega$ profile.

\begin{figure}[!htbp]
\centering
\includegraphics[width=\textwidth,clip=]{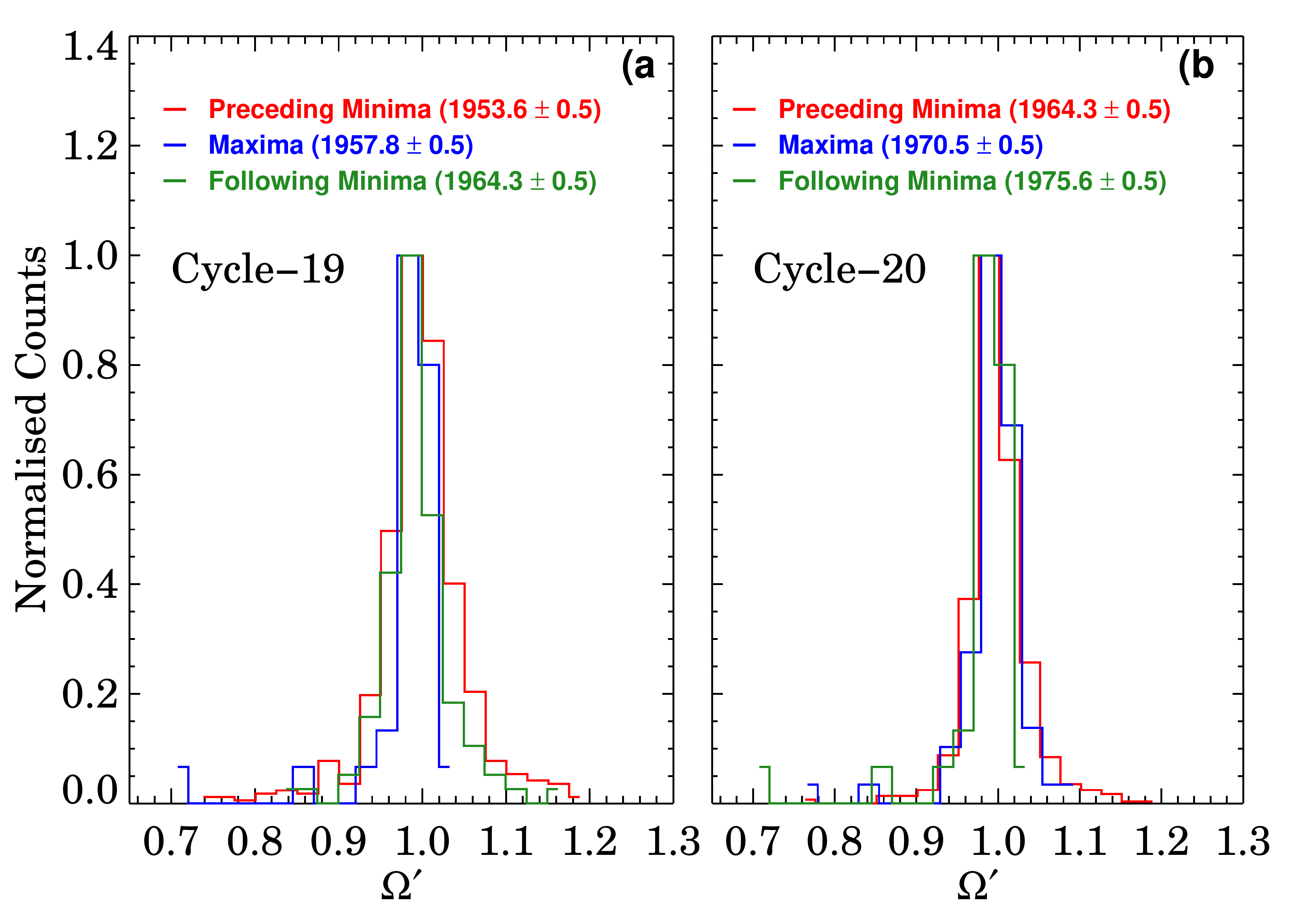}
\caption{(a): Distributions of $\Omega^\prime$, for Cycle 19, as calculated over the three activity phases: preceding minima (red), maxima (blue), and following minima (green). See text for more details. (b): Same as in (a), but for cycle 20.}
\label{ch4:fig4}
\end{figure}

For every solar cycle, we first isolate the data into three activity phases: preceding minimum, cycle maximum, and following minimum. These phases (each with one year duration) are identified after performing a 13 month running average on the KoSO area data. Next, we choose a 30$\degree$ latitude band centred around the solar equator. Choice of such a bandwidth ensures that we have enough statistics to work with during each of these phases. Since the average size of spots varies with the phase of the solar cycle and as shown in Subsection~\ref{sizes}, spots with different sizes rotate with different rates, it is important to remove this effect from the measured $\Omega$ before we look for its temporal variation. We achieve this by dividing the calculated $\Omega$ values of every spot by the $\Omega$ derived using the rotation parameters $A$ and $B$ from \autoref{ch4:table2}. This newly normalised quantity is denoted by $\Omega^\prime$ and the distributions of $\Omega^\prime$, for all three activity phases and for all eight solar cycles, are then examined thoroughly. \autoref{ch4:fig4} shows the results for two representative cases, i.e for Cycle 19 and Cycle 20. As noted from the plot, distribution of $\Omega^\prime$ shows no change, either in shape or location, with activity phases. \autoref{ch4:table3} lists the statistical parameters (such as mean, skewness, etc.) of each of these distributions which further confirm our previous observation. Overall, our results are in accordance with the findings of \citet{Gilman1984}, \citet{Ruzdjak2017} and \citet{Javaraiah2020}, who either found none or statistically insignificant correlation of sunspot rotation with the activity phase of the solar cycle.

\begin{table}[!htbp]
\centering
\begin{tabular}{crrrrrr}
\hline
    \multicolumn{1}{c}{}&
    \multicolumn{3}{c}{Maximum}&
    \multicolumn{3}{c}{Following Minimum}\\
\hline
Cycle &Mean &Median &Skewness &Mean &Median &Skewness\\
\hline

    16 &$     1.000$ &$     0.998$ &$     0.190$ &$     1.000$ &$     0.998$ &$     0.010$ \\
    17 &$     1.002$ &$     0.999$ &$     0.279$ &$     0.993$ &$     0.991$ &$     0.521$ \\
    18 &$     0.996$ &$     0.995$ &$    -0.030$ &$     0.991$ &$     0.992$ &$    -0.042$ \\
    19 &$     1.000$ &$     0.998$ &$     0.035$ &$     0.993$ &$     0.993$ &$    -0.027$ \\
    20 &$     0.998$ &$     0.996$ &$     0.188$ &$     0.997$ &$     0.997$ &$    -0.272$ \\
    21 &$     1.003$ &$     1.000$ &$     0.275$ &$     0.995$ &$     0.992$ &$    -0.079$ \\
    22 &$     0.997$ &$     0.996$ &$    -0.053$ &$     0.998$ &$     0.997$ &$    -0.054$ \\
    23 &$     1.000$ &$     0.997$ &$     0.203$ &$     1.010$ &$     1.008$ &$     0.577$ \\

\hline
\end{tabular}
\caption{Statistical parameters of $\Omega^\prime$ distributions for each cycle.}
\label{ch4:table3}
\end{table}

\begin{figure}[!htbp]
\centering
\includegraphics[width=0.8\textwidth,clip=]{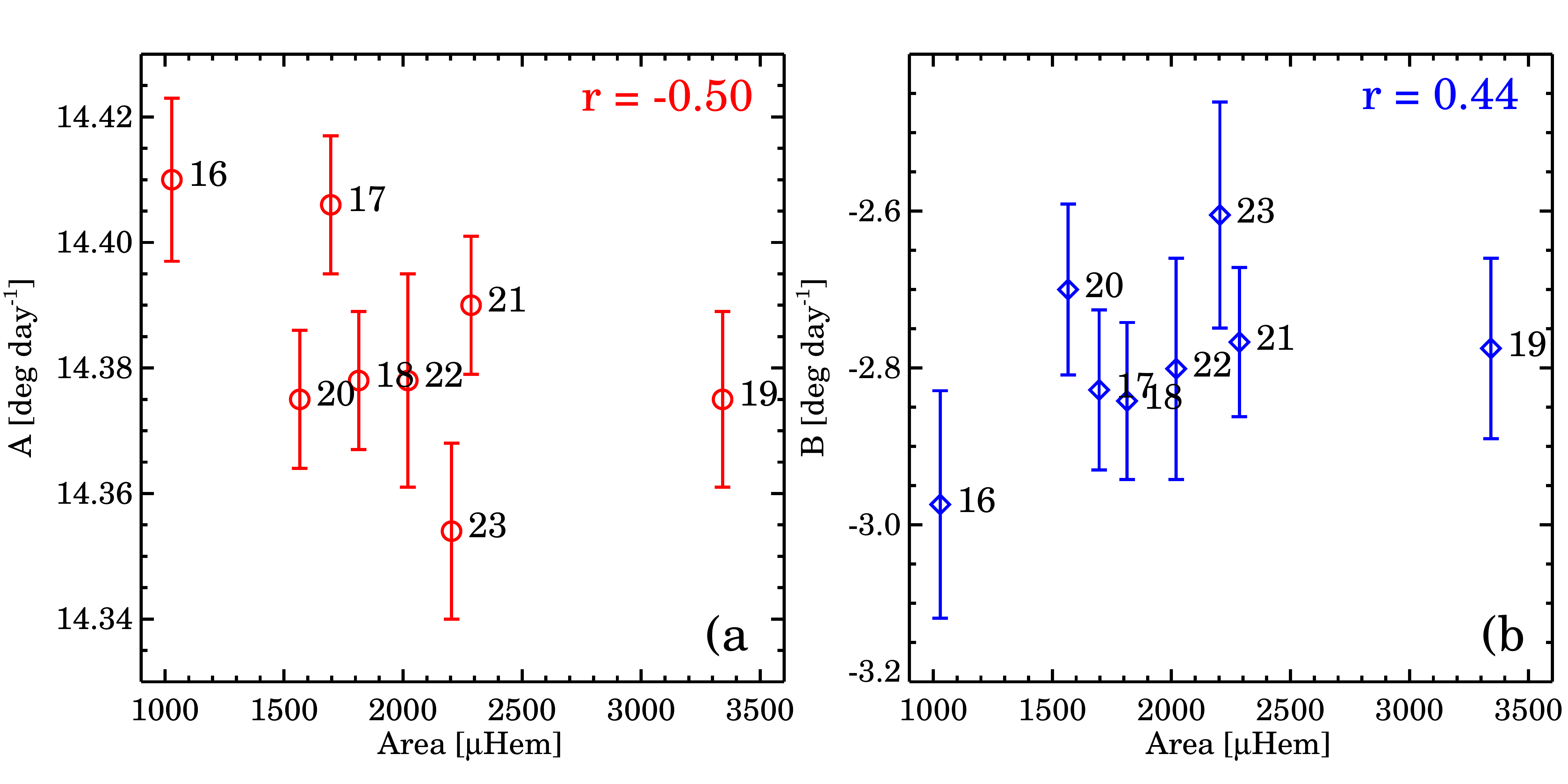}
\caption{(a): Scatter plots between $A$ and cycle strengths. (b): Scatter plots between $B$ and cycle strengths. Obtained Pearson correlations ($r$) are printed in the respective panels.
}
\label{ch4:fig5}
\end{figure}

Next, we explore how a rotational profile depends upon the strength of a solar cycle. We do this by analysing the rotation parameters $A$ and $B$ of each cycle. The peak area value within a cycle (of yearly averaged data) is assigned as the strength of that cycle. In \autoref{ch4:fig5}, we show the scatter plot of $A$ vs. cycle strength (\autoref{ch4:fig5}a), as well as the plot between $B$ vs. cycle strength (\autoref{ch4:fig5}b). Immediately we notice that $A$, which represents the equatorial rotation rate, decreases with increasing cycle strengths (Pearson correlation, $r=-0.50$). Interestingly, $B$, the latitudinal gradient of rotation, shows a positive correlation ($r=0.44$). This implies that, during a strong cycle, the Sun not only rotates slowly at the equator but the latitudinal gradient of rotation also gets reduced \citep{Gilman1984, Obridko2001, Javaraiah2005a, Obridko2016, Javaraiah2020}. It is important to highlight here that in both panels, we notice that the values for Cycle 19 seem to lie significantly far away from the trends. At this moment, we do not know any physical process that can explain this observed anomaly.

\subsubsection{Rotation Rates of Northern and Southern Hemispheres}

Almost all sunspot indices display profound differences in their properties, in the northern and southern hemisphere \citep{Hathaway2015}, including sunspot rotation rates \citep{Gilman1984}. However, in the case of rotation rates, the difference, as reported by many authors in the past, differs significantly from each other. For example, \citet{Gilman1984} used sunspots as tracers and noted that the variation in solar rotation is more profound in the southern hemisphere relative to the northern one. In a recent work, \citet{Xie2018} reported that the northern hemisphere rotates considerably faster than the southern one during Cycles 21\,--\,23. Similar findings, for the same period, have been also reported earlier \citep{Zhang2011, Zhang2013, Li2013a}. Thus, comparing the rotation rates in both hemispheres from our KoSO sunspot data is appropriate.

\begin{figure}[!htbp]
\centering
\includegraphics[width=0.8\textwidth,clip=]{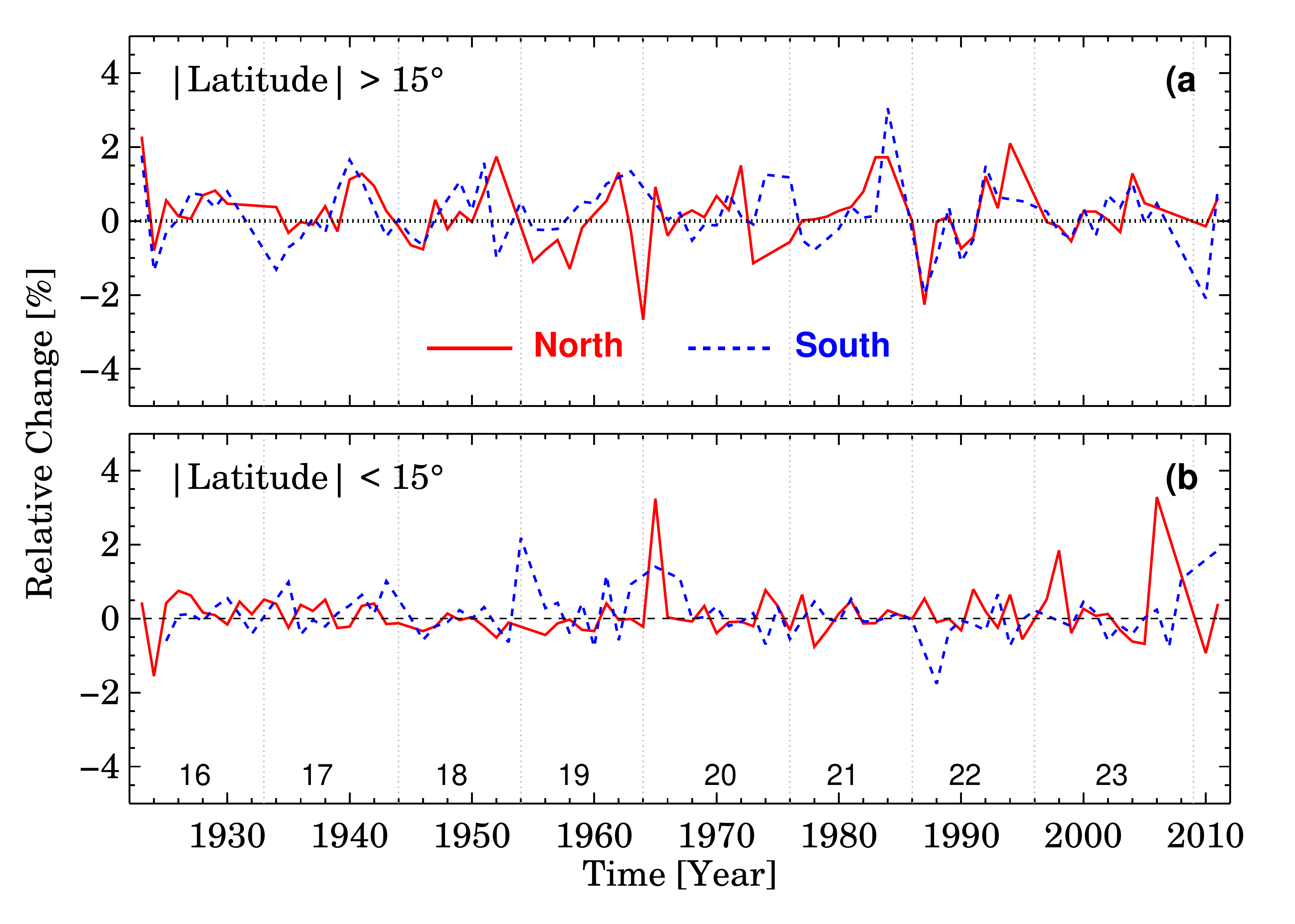}
\caption{Relative (\%) changes in $\Omega$, separately for northern and southern hemispheres, in latitude band~I (a) and band~II (b). See text for more details.}
\label{ch4:fig6}
\end{figure}

To achieve this, the data is first separated for the two hemispheres and then each hemispheric data is divided into two separate latitude bands: band~I, 0\,--\,15$\degree$ and band~II,  $\geq$15$\degree$ . In each case, we then calculate the relative change by first calculating the mean $\Omega$ in that band over the year ($\bar{\Omega}_{\rm Year}$) and then calculate the relative change as $\frac{\bar{\Omega}_{\rm Year}-\bar{\Omega}_{\rm All}}{\bar{\Omega}_{\rm All}}\times 100\%$, where $\bar{\Omega}_{\rm All}$ is the mean $\Omega$ calculated over the entire duration of the data (1923\,--\,2011) within our chosen band.
\autoref{ch4:fig6}a-b show these relative changes for band~II and band~I, respectively. In the case of band~I, the relative change is $<$0.5\% and we do not observe any signature of solar cycle like variation. On the other hand, in the case of band~II, there are clear signatures of cyclic changes (e.g. for Cycles 16, 17, 18, 21) along with a slightly larger variation ($\approx$ 1.5\%) as compared to band~I.

\section{Cross-Validation of Our Method Using Space Based Data}

As discussed in the introduction, rotational profiles that have been derived using sunspots, depend significantly (i) on the method used, and (ii) on the quality of the data. To examine the aforementioned effects onto our measurements of rotation rates from KoSO data, we use the space based sunspot data available from MDI onboard SOHO. These data cover a period of 15 years (1996\,--\,2011). 

\begin{figure}[!htbp]
\centering
\includegraphics[width=\textwidth,clip=]{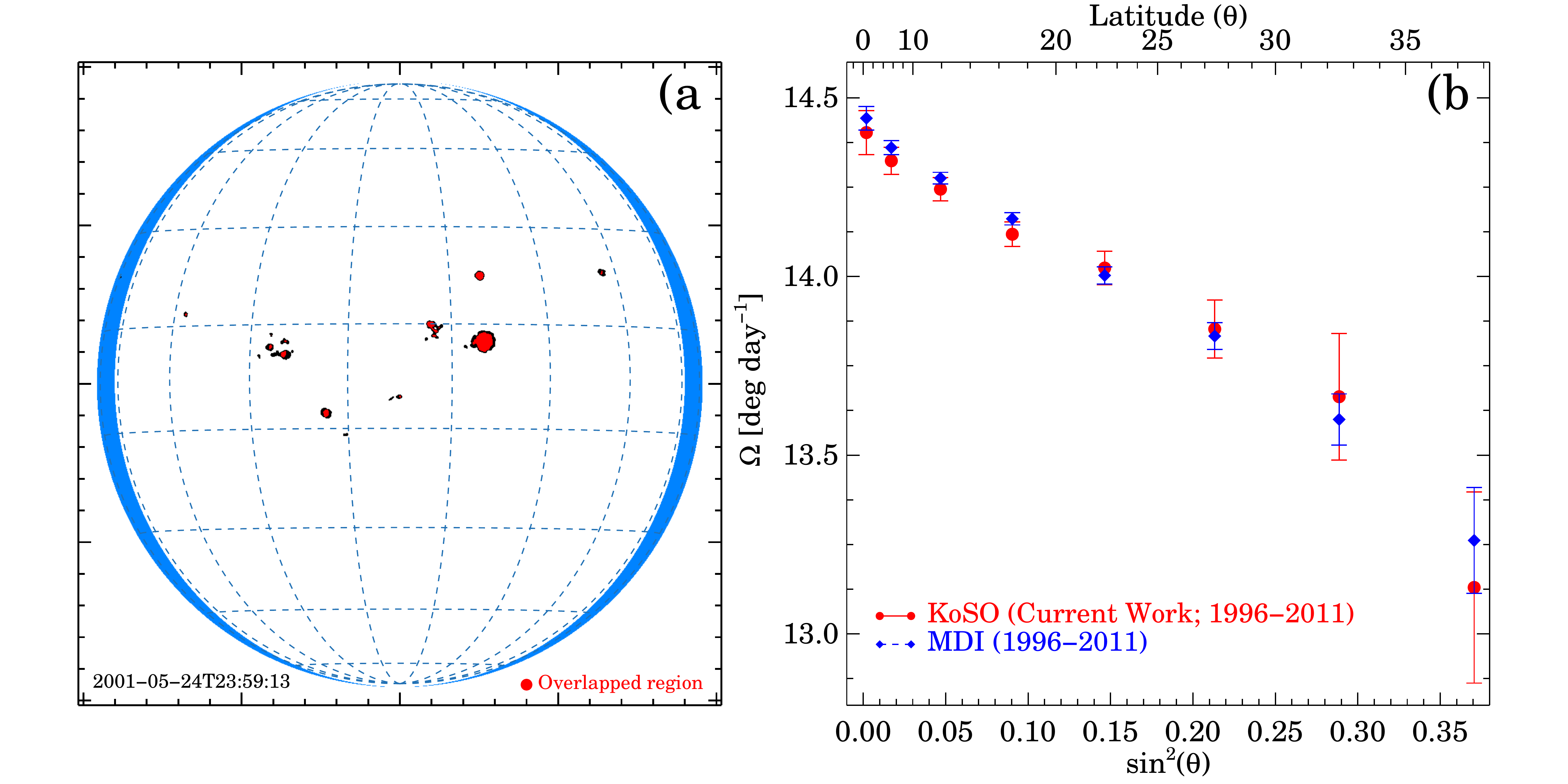}
\caption{(a): Representative MDI image after implementing our sunspot tracking algorithm. (b): Solar rotation profile measured using MDI data (blue circles) and the rotation profile derived from KoSO data for the overlapping period (red circles).
}
\label{ch4:fig7}
\end{figure}

Sunspot masks are generated using the same methods as used in KoSO data and likewise, the method implemented to track the spots (\autoref{ch4:fig7}a) and derive the rotation profile, is also the same as described in Subsection~\ref{ch4-method}. Results from MDI are shown in \autoref{ch4:fig7}b using blue circles. For easy comparisons, results from our KoSO data (corresponding to the same epoch) are also overplotted (via red circles) in both panels. As seen from the plots, the KoSO profile matches well with the values from MDI. At the same time, we also note that the error bars are substantially smaller in MDI which is mostly due to the better data availability in MDI as compared to KoSO within this period. Overall, these results not only show the effectiveness of our method on ground and space based data but also highlights the quality of the KoSO catalogue.

\section{Conclusion}
\label{ch4-con}

A better knowledge of solar differential rotation is key towards a deeper understanding of solar dynamo theory. In this work, we have derived the solar rotation profile using the white-light digitised data from Kodaikanal Solar Observatory (KoSO). These data cover a period of $\approx$90 years (1923\,--\,2011) i.e. Cycle~16 to Cycle~24 (ascending phase only). We have implemented a correlation-based automated sunspot tracking algorithm which correctly identifies the same sunspot in two consecutive observations. This positional information is then used to calculate the angular rotation rate for every individual spot. Furthermore, the derived rotation parameters $A$ and $B$ ($14.381\pm0.004$~deg/day,~$-2.72\pm0.04$~deg/day; \autoref{ch4:fig2}), which represent the equatorial rotation rate and the latitudinal gradient of rotation, respectively, are in good agreement with existing reports from other sunspot catalogues \citep{Javaraiah2005a, Ribes1993}. In fact, our results also compared well with results from the older, low-resolution, and manually processed KoSO sunspot data \citep{Howard1999, Gupta1999}.

During our analysis, we found that bigger spots rotate slower than smaller ones (\autoref{ch4:fig3}), which could well be a possible signature of deeper anchoring depths of bigger spots \citep{Balthasar1986}. We have also studied the cyclic variations in the rotation profile along with its modulation with cycle strengths. Results suggest that during a strong cycle, the Sun not only rotates slower at the equator but the latitudinal gradient of rotation also gets reduced (\autoref{ch4:fig5}). Another interesting finding from this study is the relative changes in the rotation rate in the northern and southern hemispheres, where high latitude ($>$15$\degree$) spots seem to show pronounced variations ($\approx$1.5\% as compared to low latitude ($<$15$\degree$) ones (\autoref{ch4:fig6}). Lastly, we check for the robustness of our tracking algorithm by implementing it onto the MDI data. Similar results from KoSO and MDI confirm the same.

In this study we have assumed that sunspots are basically embedded structures within the photosphere and their motions, such as the rotation about their own axis \citep{Brown2003,Svanda2009} is not accounted for in our study. The KoSO image database (with one day cadence and relatively low spatial resolution) is not really capable of taking care of these effects and we need modern day high resolution (spatial and temporal) data for such a study.
Furthermore, KoSO has historical data for the last century in two other wavelengths, i.e. H$\alpha$ and CaII K. These datasets can further be used in the future to investigate the change in solar rotation above the photospheric height and in turn, this will provide insight about the coupling between different solar atmospheric layers. 
\clearpage{}

\clearpage{}\chapter[Theory of Near--Surface Shear Layer]{A Theoretical Model of the Near--Surface Shear Layer of the Sun}
\label{Chap5}
\lhead{\emph{Chapter 5: A Theoretical Model of the Near--Surface Shear Layer of the Sun}} \noindent

\section{Introduction}
\label{ch5:intro}
\begin{figure}
\centering
	\includegraphics[width=\columnwidth]{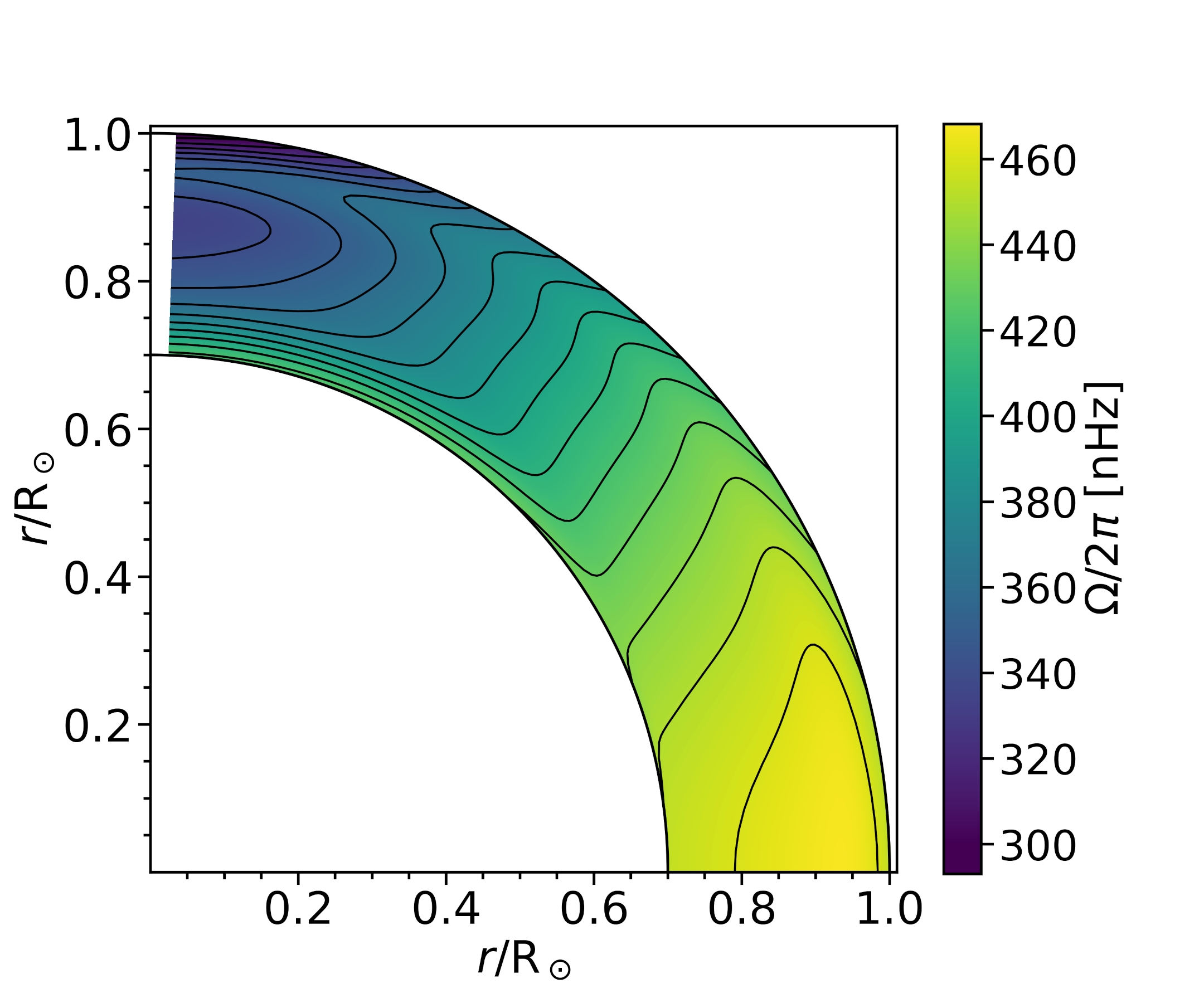}
    \caption{Rotation profile on the basis of helioseismology data. The contours correspond
    to values of $\Omega$ at the interval of 10 nHz,
    the extreme contours inside the main body
    of the convection zone being for the values 340 nHz
    and 460 nHz (apart from a few lower value contours
    near the surface at the polar region, where
    helioseismic inversions are not trustworthy).}
    \label{ch5:fig1}
\end{figure}

One of the intriguing features in the differential rotation map of the Sun, as seen, for example, in Figure~1 of \citet{Howe2009} or in Figure~26 of \citet{Basu2016}, is the existence of the near-surface shear layer (NSSL). This is a layer near the solar surface at the top of the convection zone, within which the angular velocity decreases sharply with increasing solar radius.  The first indication of the existence of such a layer came more than half a century ago, when it was noted that the rotation rate of the solar surface measured from the Doppler shifts of photospheric spectral lines was about 5\% lower than the rotation rate inferred from the positions of sunspots on the solar surface \citep{Howard1970}. While the depth at which sunspots are anchored remains unclear and probably changes with the age of a sunspot group \citep{Longcope2002}, the rotation rate inferred from the sunspots was assumed to correspond to a layer underneath the solar surface, implying that the angular velocity was higher in that layer. When helioseismology mapped the internal differential rotation of the Sun, the existence of this layer was fully established.  \autoref{ch5:fig1} shows the differential rotation map
of the Sun obtained by helioseismology (with contours of constant angular velocity) which we shall use later in our calculations in Section~\ref{ch5:helio}.  The contours of constant angular velocity, which are nearly radial within a large part of the body of the solar convection zone, bend towards the equator within a layer of thickness of order $\approx 0.05$\Rsun at the top of the convection zone \citep{Schou1998, Howe2005}.

What causes this NSSL is still not properly understood.  The first attempts to explain it \citep{Foukal1975, Gilman1979} were based on the idea that convection in the upper layers of the convection zone mixes angular momentum in such a manner that the angular momentum per unit mass tends to become constant in these layers, leading to a decrease of the angular velocity with radius.  Once the differential rotation of the Sun was properly mapped and no evidence was found for the constancy of the specific
angular momentum within the convection zone, it was realized that this
could not be the appropriate explanation. Within the last few years, there have been attempts to explain the NSSL on the basis of numerical simulations of the solar convection \citep{Guerrero2013, Hotta2015, Matilsky2019}. It has been argued by \citet{Hotta2015} that the Reynolds stresses play an important role in creating the NSSL, whereas \citet{Matilsky2019} suggested that the steep decrease in density in the top layers of the convection zone is crucial in giving rise to the NSSL. \citet{Choudhuri2021a} has recently proposed a possible alternative theoretical explanation of the NSSL based on order-of-magnitude estimates. The aim of the present paper is to substantiate the ideas proposed by \citet{Choudhuri2021a} through detailed calculations.

The theories of the two large-scale flow patterns within the convection zone of the Sun---the differential rotation and the meridional circulation---are intimately connected with each other \citep{Kitchatinov2013, Choudhuri2021b}.  The idea we wish to develop follows from the central equation in the theory of the meridional circulation: the thermal wind balance equation. Since the nature of the Coriolis force arising out of the solar rotation varies with latitude, the effect of this force on the convection is expected to vary with latitude \citep{Durney1971, Belvedere1976}. Since the Coriolis force provides the least hindrance to convective heat transport in the polar regions, the poles of the Sun are expected to be slightly hotter than the equator \citep{Kitchatinov1995}. There are some observational indications that this may indeed be the case  \citep{Kuhn1988, Rast2008}. Hotter poles would tend to drive what is called a thermal wind, i.e.\ a meridional circulation which would be equatorward at the solar surface. Since the observed meridional circulation is the opposite of that, we must have another effect which overpowers this and drives the meridional circulation in the poleward direction at the surface as observed. It is easy to show that the centrifugal force arising out of the observed differential rotation of the Sun can do this job: see Figure~8 and the accompanying text in \citet{Choudhuri2021b}. The term corresponding to the dissipation of the meridional circulation is found to be negligible compared to the driving terms within the bulk of the convention zone, as pointed out in the Appendix of \citet{Choudhuri2021b}.  As a result, we expect the two driving terms of the meridional circulation---the thermal wind term and the centrifugal term---to be comparable within the main body of the solar convection zone.  This is often referred to as the thermal wind balance condition.

It is generally believed that the thermal wind balance condition holds within the body of the convection zone
\citep{Kitchatinov2013, Karak2014a}, although different authors may not completely agree as to the extent to which 
it holds  \citep{Brun2010}.  
There is, however, not much agreement among different authors whether the thermal wind condition should
hold even within the top upper layer of the solar convection zone. A widely held view is that this upper
layer is a kind of boundary layer within which the dissipation term or Reynolds stresses become important, giving rise to a
violation of the thermal wind balance condition.  It is argued that the NSSL arises in some manner out of this
violation.  A completely opposite argument is given in the earlier paper by \citet{Choudhuri2021a} and in the
present paper.  We point out that the thermal wind term becomes very large in the top layer of the
solar convection zone due to a combination of two factors: (i) the temperature falls sharply as we move outward
through this layer, and (ii) the pole-equator temperature difference does not vary with depth in this
layer because of the reduced effect of the Coriolis force on convection in this layer, as explained
in the next section (Section~\ref{ch5:method}). The dissipation term is much smaller than the
thermal wind term even within the main body of the convection zone, as the order of magnitude
estimate in the Appendix of \citet{Choudhuri2021a} suggests. If the thermal wind term becomes
even much larger in the upper layers of the convection zone, then it appears unlikely to us that this large thermal wind term
can be balanced by the dissipation term.  The only possibility is that the centrifugal term also
has to become very large in the top layer to balance the thermal wind term.  This dictates that the
top layer has to be a region within which the angular velocity undergoes a large variation.  We show
through quantitative calculations that the structure of the NSSL calculated theoretically on the basis
of our ideas agrees with the observational data remarkably well.

We explain our basic methodology in the next section (Section~\ref{ch5:method}).  After that Section~\ref{ch5:empirical} is devoted to the applying
our methodology to an analytical expression of the differential rotation in the interior of the solar
convection zone.  Then the actual data of differential rotation obtained by helioseismology are applied to
calculate the structure of the NSSL in Section~\ref{ch5:helio}. Finally, our conclusions are summarized in Section~\ref{ch5:conclusion}.

\section{Basic Methodology}
\label{ch5:method}

The equation for thermal wind balance is
\begin{equation}
r \sin{\theta} \frac{\pa}{\pa z} \Omega^2=\frac{1}{r}\frac{g}{\gamma C_V} \frac{\pa S}{\pa \theta},
\label{ch5:eq1}
\end{equation}
where $\Omega$ is the angular velocity, $z$ is the distance from the equatorial plane
measured upward, $g$ is the acceleration due to gravity of the Sun at the point
under consideration and $\gamma$ is the adiabatic index, while $S$ and $C_V$ are respectively the entropy and the specific heat of the gas per
unit mass. See \citet{Choudhuri2021b} for the derivation and a discussion of this equation.

Over any isochoric surface,
the entropy and temperature differentials between two points are related by
\begin{equation}
\Delta S = C_V \frac{\Delta T}{T},
\label{ch5:eq2}
\end{equation}
We also point out in Appendix~\ref{ch5:appn} that 
Equation~\ref{ch5:eq1} and Equation~\ref{ch5:eq2} give the following equation for the temperature
differences on isochoric surfaces
\begin{equation}
r^2 \sin \theta \frac{\pa}{\pa z} \Omega^2 = \frac{g}{\gamma T} \left( \frac{\pa}{\pa \theta} \Delta T \right)_{\rm isochore}.
\label{ch5:eq3}
\end{equation}
This is the main equation on which the analysis of the present paper is based. Since the oblateness of isochoric surfaces
in the Sun is so small, they can be regarded as spherical surfaces
for most practical purposes.  However, from a conceptual point
of view, it is important to remember that our central equation (Equation~\ref{ch5:eq3})
refers to temperature variations over isochoric surfaces and
that is what gives the possibility of comparing our theoretical
results with observational data of the solar surface.

In our discussions, we sometimes will have to deal with
situations like the following. We may know the distribution of
$\Omega (r, \theta)$ in some region. Suppose we also know 
the temperature on some axis $\theta = \theta_0$, which means
that we would know the value of the temperature at one point on
a spherical surface.  From this, we want to find the temperature at 
other points. According to Equation~\ref{ch5:eq3}, the
temperature difference between the points $(r, \theta)$ and $(r,\theta_0)$
is given by  
\begin{equation}
\Delta T_{\theta_0} (r, \theta) = \frac{r^2 \gamma}{g}
\int_{\theta_0}^{\theta} d\theta \; T
\sin \theta \frac{\pa}{\pa z} \Omega^2.
\label{ch5:eq34}
\end{equation}
As stressed earlier by \citet{Choudhuri2021a, Choudhuri2021b},
whether the Coriolis force due to the Sun's rotation has any effect on
the convection cells depends on whether the convective turnover time is
comparable to the rotation period or not.  Numerical simulations suggest 
that convection in the deeper layers of the convection zone involves large
convection cells with long turnover times and are affected by the Coriolis force:
see Figure~1 in \citet{Brown2010} or Figure~3 in \citet{Gastine2014}. As a result,
heat transfer depends on latitude.  However, this is not the case near
the top of the convection zone, where the convection cells (the granules) are
much smaller in size and have turnover times as short as a few minutes.
The extent to which the temperature gradient $d T/ dr$ differs from the
adiabatic gradient depends on the mixing length (see, for exmaple, \citet{Kippenhahn1990}, Section~7).  In the top of the 
convection zone which is not affected by rotation, the mixing length is
independent of latitute and we expect $d T/ dr$ also to be
independent of latitude \citep{Choudhuri2021a}.  Although there must be a gradual transition
from deeper layers within which heat transport depends on the latitude to the
top layer within which this is not the case, we assume for simplicity that the
transition takes place at radius $r = r_c$ above which we
have $d T/d r$ independent of latitude.  We shall work out our model
by assuming different values of $r_c$ in the range 0.92\,--\,0.98~\Rsun.

In the simplest kind of a spherically symmetric model of the Sun, the
temperature $T$ would be function of $r$ alone and would be independent of
$\theta$.  If the heat transport depends on latitude, then that would introduce
a small variation of $T$ with $\theta$. We can write
\begin{equation}
T(r, \theta) = T (r, 0) + \Delta T (r, \theta).
\label{ch5:eq4}
\end{equation}
If $d T/ dr$ is independent of latitude in the layer above $r > r_c$, then we have
\begin{equation}
\frac{dT (r,\theta)}{dr} = \frac{dT (r,0)}{dr} \nonumber
\end{equation}
so that it follows from Equation~\ref{ch5:eq4} that
\begin{equation}
\frac{d}{dr} \Delta T (r, \theta)= 0 \nonumber
\end{equation}
in this top layer.  So we can write
\begin{equation}
\Delta T (r > r_c, \theta) = \Delta T (r_c, \theta).
\label{ch5:eq5}
\end{equation}
In principle, it would be possible to determine $T (r, \theta)$ throughout the convection zone if we have a theory of how convective heat transport varies with latitude due to the effect of the Coriolis force.  Since our understanding of this complex problem is limited, we can proceed in a different manner. Since there is general agreement that the thermal wind balance condition holds within the deeper layers of the convection zone, we assume Equation~\ref{ch5:eq3} to hold below the radius $r = r_c$.  If we know the angular velocity $\Omega (r, \theta)$ in this region, then it is straightforward to evaluate the left hand side of Equation~\ref{ch5:eq3}.  Once we have the value of the left hand side, we can carry on integration
in accordance with Equation~\ref{ch5:eq34} to determine $\Delta T (r, \theta)$ at all points within the convection zone below $ r = r_c$. Once we have the value of $\Delta T (r, \theta)$ at the radius $r = r_c$, we readily have the value of $\Delta T (r, \theta)$ at all points above this surface by using Equation~\ref{ch5:eq5}.  In other words, we can obtain $\Delta T (r, \theta)$ throughout the convection zone from the values of $\Omega (r, \theta)$ in the deeper layers below $r = r_c$, where Equation~\ref{ch5:eq3} is expected to hold.  Comparing Equation~\ref{ch5:eq34} with Equation~\ref{ch5:eq5}, it should be clear
that $\Delta T (r, \theta)$ is nothing but $\Delta T_{\theta_0} (r, \theta)$ with $\theta_0 = 0$.

As we already pointed out, there is a lack of consensus whether the thermal wind balance equation holds in the top layer of the convection zone.  It follows from Equation~\ref{ch5:eq5} that $(\pa/\pa \theta) \Delta T (r,\theta)$ does not vary with $r$ above the radial surface $r = r_c$.  On the other hand, the temperature scale height becomes very small in this top layer and the temperature falls by orders of magnitude as we move to the solar surface from $r = r_c$. As a result, the thermal wind term represented by the right hand side of Equation~\ref{ch5:eq3} in which $T$ appears in the denominator becomes very large.  We do not think that this term can be balanced by the dissipation term.  We suggest that the thermal wind balance must hold even in this top layer and the centrifugal term represented by left hand side of Equation~\ref{ch5:eq3} has to become very large to balance the thermal wind term, implying a strong variation of $\Omega^2$ along $z$. As we have the values of $\Delta T$ above $r = r_c$, we can evaluate the right hand side of Equation~\ref{ch5:eq3} easily.  Then we can use Equation~\ref{ch5:eq3} to determine how $\Omega^2$ varies within this top layer.

In a nutshell, our methodology is as follows.  We start by assuming a value of $r = r_c$ below which convective heat transport is affected by the Coriolis force and above which this is not the case.  To begin with, we need the value of $\Omega (r, \theta)$ below $r_c$, from which we can calculate the left hand side of Equation~\ref{ch5:eq3} and eventually obtain $\Delta T (r, \theta)$ throughout the solar convection zone, obtaining $\Delta T (r, \theta)$ above $r_c$ by using Equation~\ref{ch5:eq5}.  Once we have $\Delta T (r, \theta)$ above $r_c$, the right hand side of Equation~\ref{ch5:eq3} can be evaluated, which enables us to find out $\Omega (r, \theta)$ above $r_c$ from Equation~\ref{ch5:eq3}. Although we use Equation~\ref{ch5:eq3} for all our calculations, we proceed differently below and above $r = r_c$.  Below $r_c$ we calculate $\Delta T (r, \theta)$ from $\Omega (r, \theta)$ beginning with the left hand side of Equation~\ref{ch5:eq3}, whereas above $r_c$ we calculate $\Omega (r, \theta)$ from $\Delta T (r, \theta)$ beginning with the right hand side of Equation~\ref{ch5:eq3}.

\begin{figure}[!ht]
	\includegraphics[width=\columnwidth]{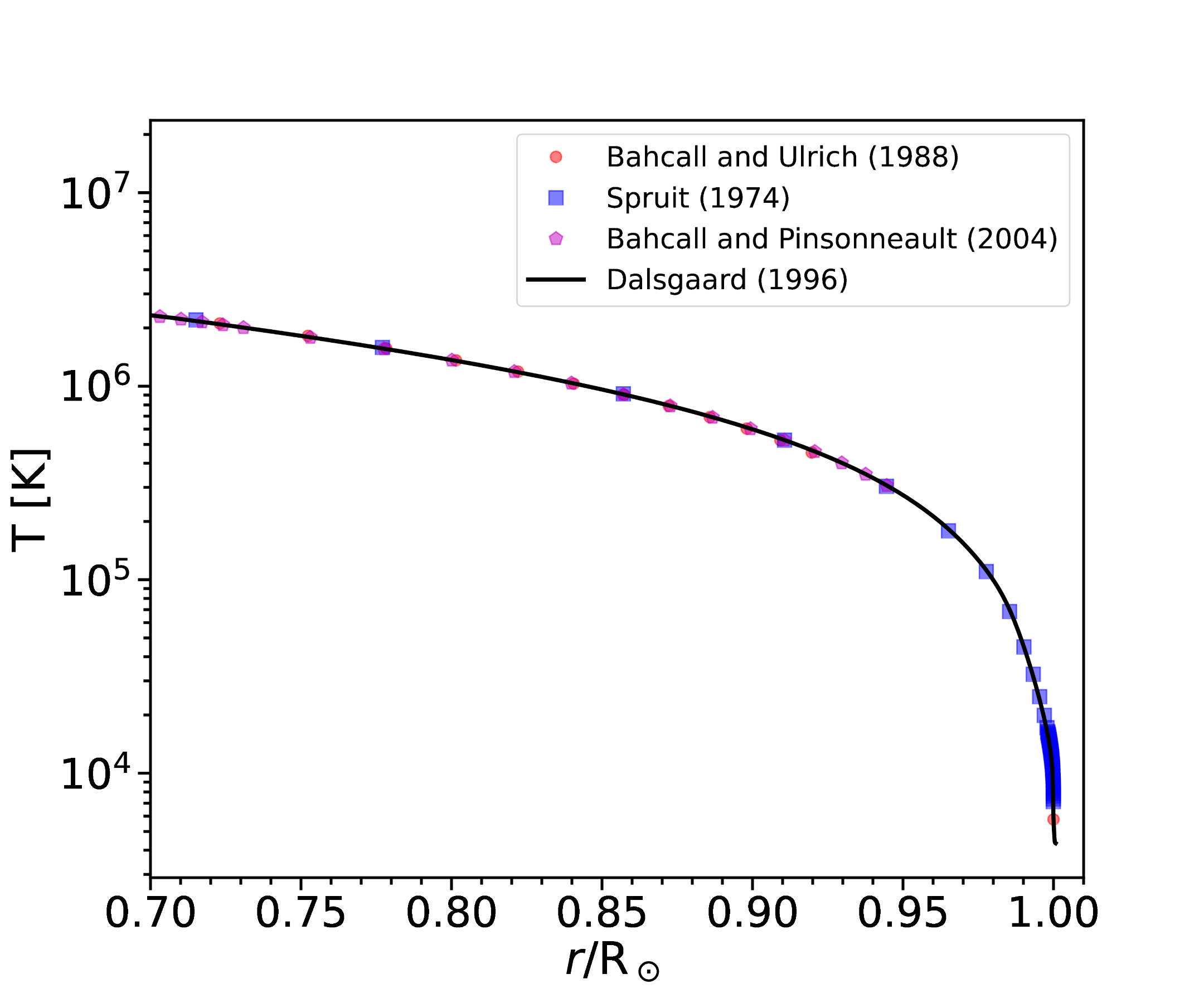}
    \caption{Variation of temperature with $r$ in different convection zone models. The curve corresponds
    to Model S used in our calculations.
    }
    \label{ch5:fig2}
\end{figure}

The earlier paper by \citet{Choudhuri2021a} presented some order-of-magnitude estimates based on the methodology outlined above.  Now we present a detailed analysis. It order to carry on this analysis, we need the values of the temperature as a function of $r$, which we can take to be $T(r, 0)$, i.e.\ temperature values on the polar axis where the effect of the Coriolis force is minimal.  Several models of the convection zone exist in the literature  \citep{Spruit1974, Bahcall1988, Dalsgaard1996, Bahcall2004}. We use what has been referred to as Model~S by \citet{Dalsgaard1996} to obtain $T$ at different values of $r$.
Actually, all the models of the convection zone give very similar $T(r)$, as can be seen in \autoref{ch5:fig2}.
We also note the sharp fall of the temperature in the outer layers of the convection zone, 
which is of crucial importance in our theory. The thermal wind term appearing in Equation~\ref{ch5:eq3} has $T$ in the denominator and becomes very large in the uppermost layers of the convection zone.  The value of the adiabatic
index $\gamma$ is taken to be 5/3 in all our calculations.  We point out that, in Model S, the value of $\gamma$ is very close to 5/3 throughout the convection zone except in the top layer 0.97\Rsun--\Rsun, where it becomes somewhat less due to the variations in the level of hydrogen ionization. We also need the values of $\Omega (r, \theta)$ below $r = r_c$ to start our calculations.  Calculations based on an analytical expression of  $\Omega (r, \theta)$ which fits helioseismology observtions reasonably well are presented in Section~\ref{ch5:empirical}.  Then Section~\ref{ch5:helio} will present calculations done with the actual helioseismology data of differential rotation $\Omega (r, \theta)$ used below $r_c$.  Calculations of both Section~\ref{ch5:empirical} and Section~\ref{ch5:helio} give the NSSL matching the observational data quite closely for appropriate values of $r_c$.  We point out that the input data of $\Omega (r, \theta)$ used in both these
sections (given by \autoref{ch5:eq6} and from helioseismology respectively) give nearly radial contours till $r_c$ without much sign of the NSSL below $r_c$.  In fact, the analytical expression of $\Omega (r, \theta)$ that we use does not incorporate the NSSL at all and gives radial contours till the solar surface.  As we get the NSSL even in this case, we can clearly rule out the possibility that there might have been some indication about the existence of the NSSL in the input data which percolated through the calculations to give the NSSL at the end.  There is no doubt that the NSSL arises primarily out of the requirement that the centrifugal term has to match the thermal wind term which has become very large in the top layers of the solar convection zone. 

It may be mentioned that \citet{Matilsky2020} calculated the temperature difference $\Delta T$ which one would get from the solar differential rotation by assuming the thermal wind balance Equation~\ref{ch5:eq3} to be valid till the top of the convection zone and plotted it in Figure~13 of their paper.  However, they did not discuss any physical significance of this. Some related issues are also discussed in a recent paper by \citet{Vasil2020}.

\section{Results Based on Analytical Expression}
\label{ch5:empirical}
\begin{figure}[!ht]
	\includegraphics[width=\columnwidth]{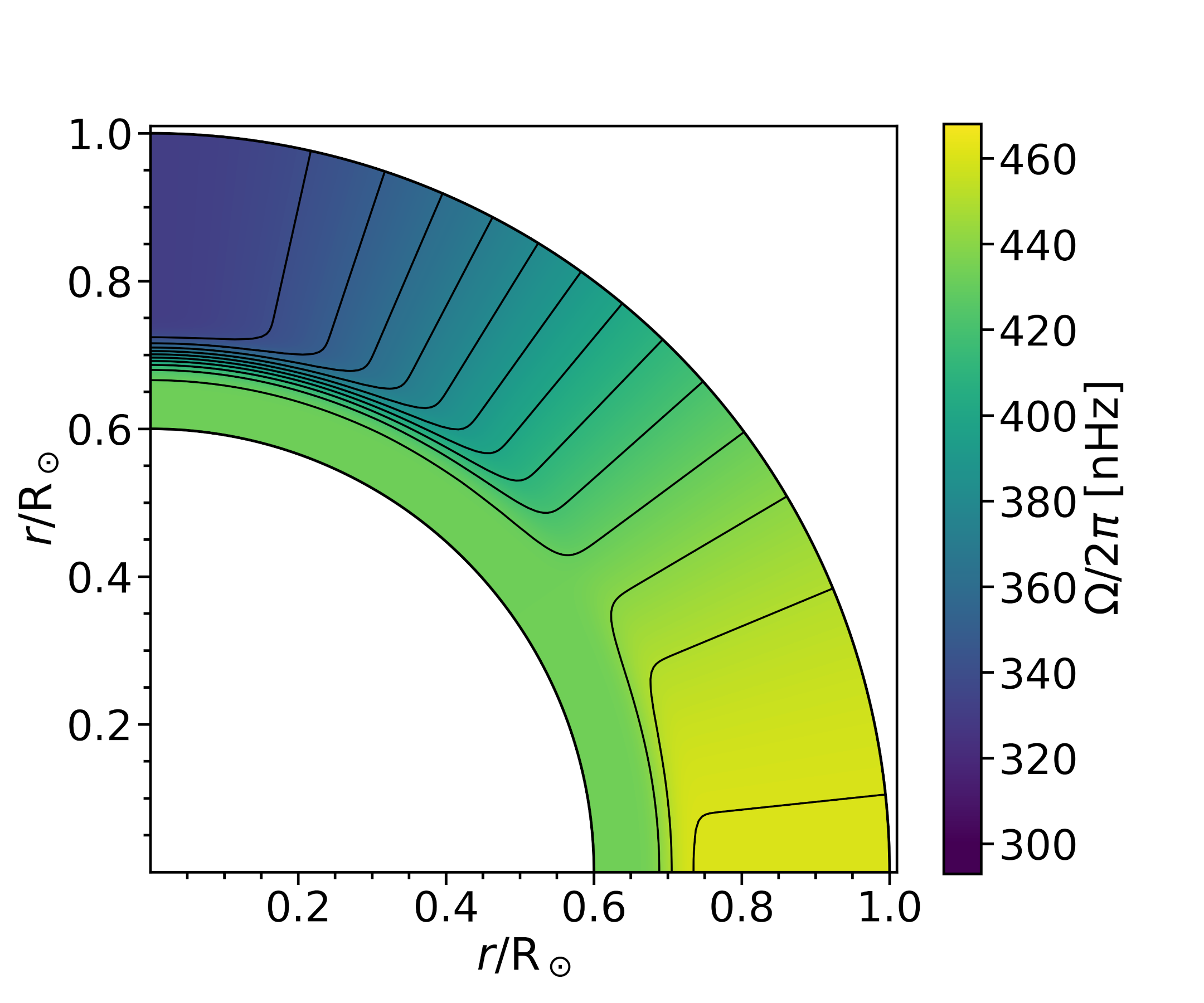}
    \caption{Rotation profile calculated by using the analytical expression (Equation~\ref{ch5:eq6}).
    The contours correspond
    to values of $\Omega$ at the interval of 10 nHz,
    the extreme contours being for the values 340 nHz
    and 460 nHz.}
    \label{ch5:fig3}
\end{figure}

As pointed out in Section~\ref{ch5:method}, we need the values of $\Omega (r,\theta)$ below $r<r_c$ to start our calculations. 
In this section, we present the results of our calculations based on the following analytical expression of $\Omega (r,\theta)$ which fits the helioseismology observations closely \citep{Schou1998, Charbonneau1999}:
\begin{equation}
    \Omega(r,\theta)= \Omega_{\rm RZ}+ \frac{1}{2}\left[1-{\rm erf}\left(\frac{r-r_t}{d_t}\right)\right] \left[\Omega_{\rm SCZ}(\theta)-\Omega_{\rm RZ}\right],
    \label{ch5:eq6}
\end{equation}
 where $r_t=0.7$\Rsun, $d_t=0.025$\Rsun, $\Omega_{\rm RZ}/2\pi=432.8$ nHz and  $\Omega_{\rm SCZ}(\theta)/2\pi=\Omega_{\rm EQ}+\alpha_2\cos^2(\theta)+\alpha_4\cos^4(\theta)$, with $\Omega_{\rm EQ}/2\pi= 460.7$ nHz, $\alpha_2/2\pi=-62.69$ nHz and $\alpha_4/2\pi=-67.13$ nHz.  \autoref{ch5:fig3} shows the rotation profile obtained from \autoref{ch5:eq6} along with contours of constant $\Omega$ (solid black lines). We note the absence of any signature of NSSL. On comparing with Figure~\ref{ch5:fig1} giving the rotation profile based on helioseismology data, we see that the analytical expression gives a
 reasonable fit to the data in the deeper layers of the convection zone.

\begin{figure}[!ht]
\centering
	\includegraphics[width=\columnwidth]{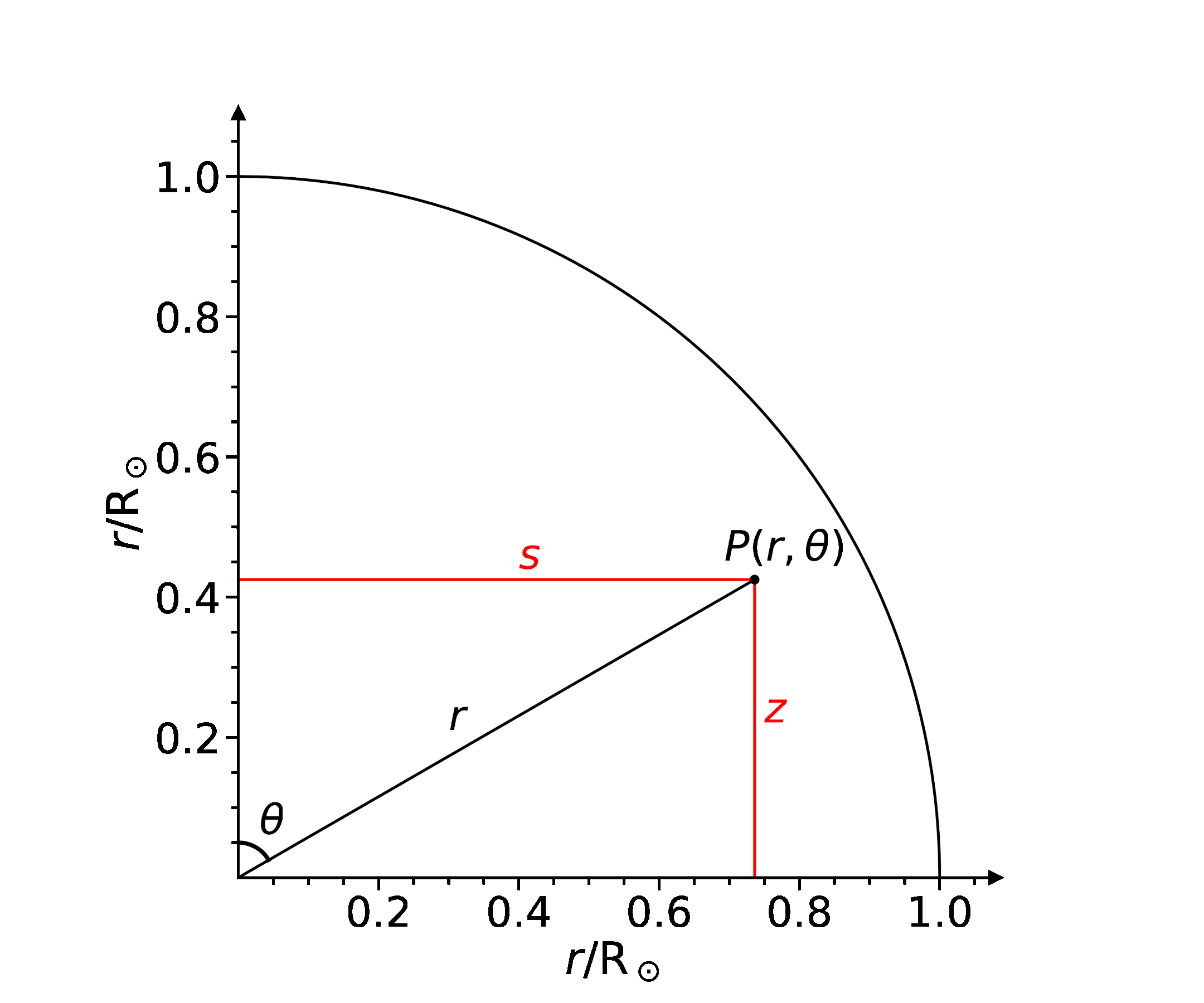}
    \caption{Relation between spherical co-ordinate system $(r, \theta)$ and the other coordinates $(s, z)$ we use.}
    \label{ch5:fig4}
\end{figure}

As explained in Section~\ref{ch5:method}, our first step is to obtain $\Delta T (r, \theta)$ for $r<r_c$ by making use of \autoref{ch5:eq3}, in which the left hand side is evaluated by using $\Omega (r, \theta)$ as given by \autoref{ch5:eq6}.
To calculate the left side of \autoref{ch5:eq3}, we need to evaluate the derivative of $\Omega^2$ along the $z$ direction. For this purpose, we use the transformation equation
\begin{equation}
    \left(\frac{\partial}{\partial z}\right)_s=\left(\frac{\partial r}{\partial z}\right)_s\frac{\partial}{\partial r}-\left(\frac{\partial \theta}{\partial z}\right)_s\frac{\partial}{\partial \theta}=\cos{\theta}\frac{\partial}{\partial r}-\frac{\sin{\theta}}{r}\frac{\partial}{\partial \theta},
    \label{ch5:eq7}
\end{equation}
on making use of $s=r\cos{\theta}$ and $z=r\sin{\theta}$ (shown in \autoref{ch5:fig4}). 
We need to choose a particular value of $r_c$. We are going to present discussions for values of $r_c$ in the range 0.92~\Rsun\,--\,0.97~\Rsun. 
We now use \autoref{ch5:eq34} to calculate $\Delta T (r,\theta)$ in the convection zone for all values of $r$ below the maximum value 0.98~\Rsun\ of $r_c$ that we consider. We calculate the numerical derivative of $\Omega^2$ with the help of the transformation \autoref{ch5:eq7} by using the first order divided difference scheme. Then we use the Runge-Kutta 4th order (RK4) method to solve \autoref{ch5:eq3}, which is equivalent to carrying on the integration in \autoref{ch5:eq34}.

\begin{figure}[!ht]
\centering
	\includegraphics[width=\columnwidth]{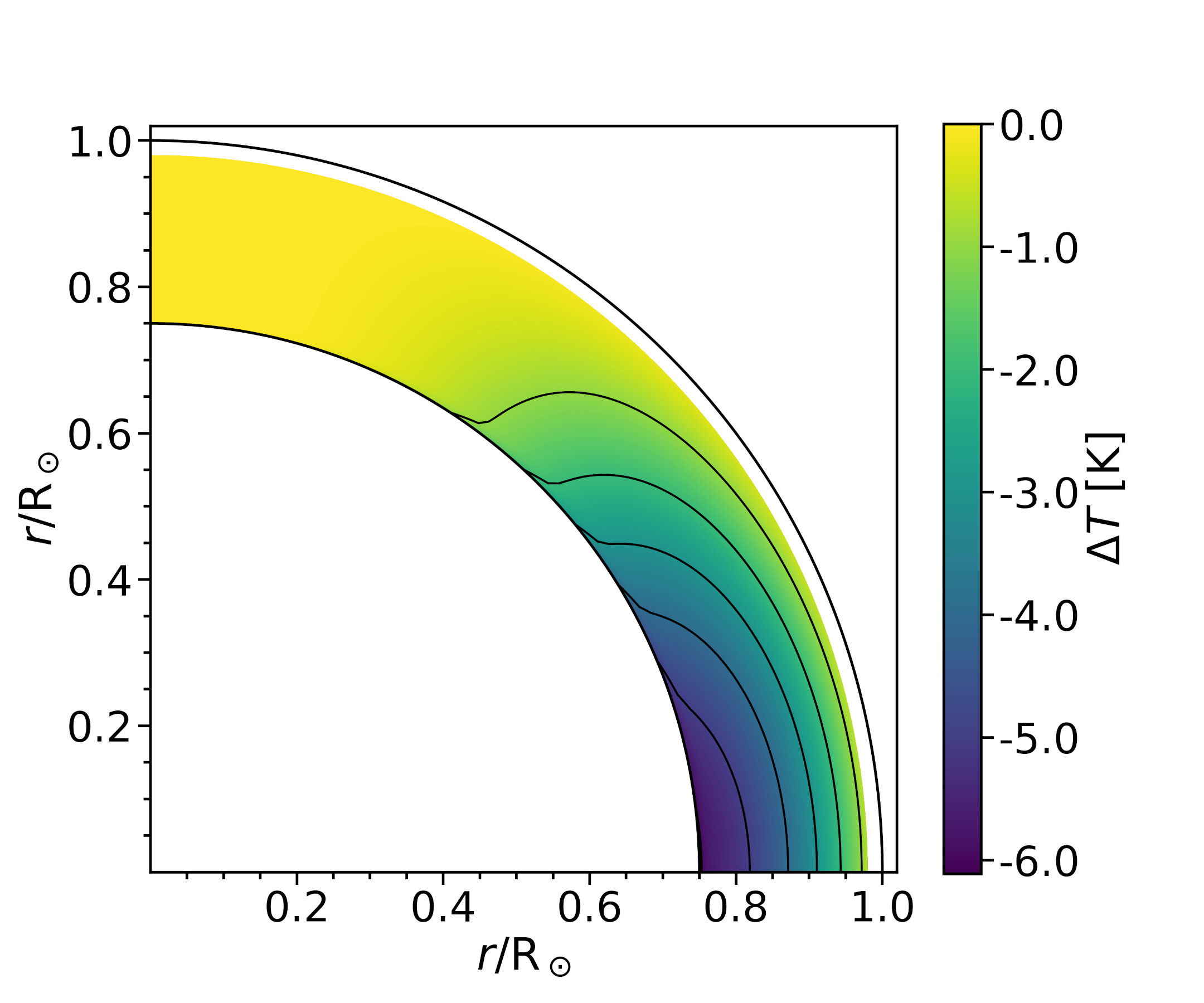}
    \caption{Profile of $\Delta T (r,\theta)$ in the solar convection zone
    below $r = 0.98$~\Rsun, which would follow from the analytical expression (\autoref{ch5:eq6})
    on taking $r_c = 0.98$~\Rsun. Contours represent the constant values of $\Delta T (r,\theta)$.}
    \label{ch5:fig5}
\end{figure}

\begin{figure}[!ht]
\centering
	\includegraphics[width=\columnwidth]{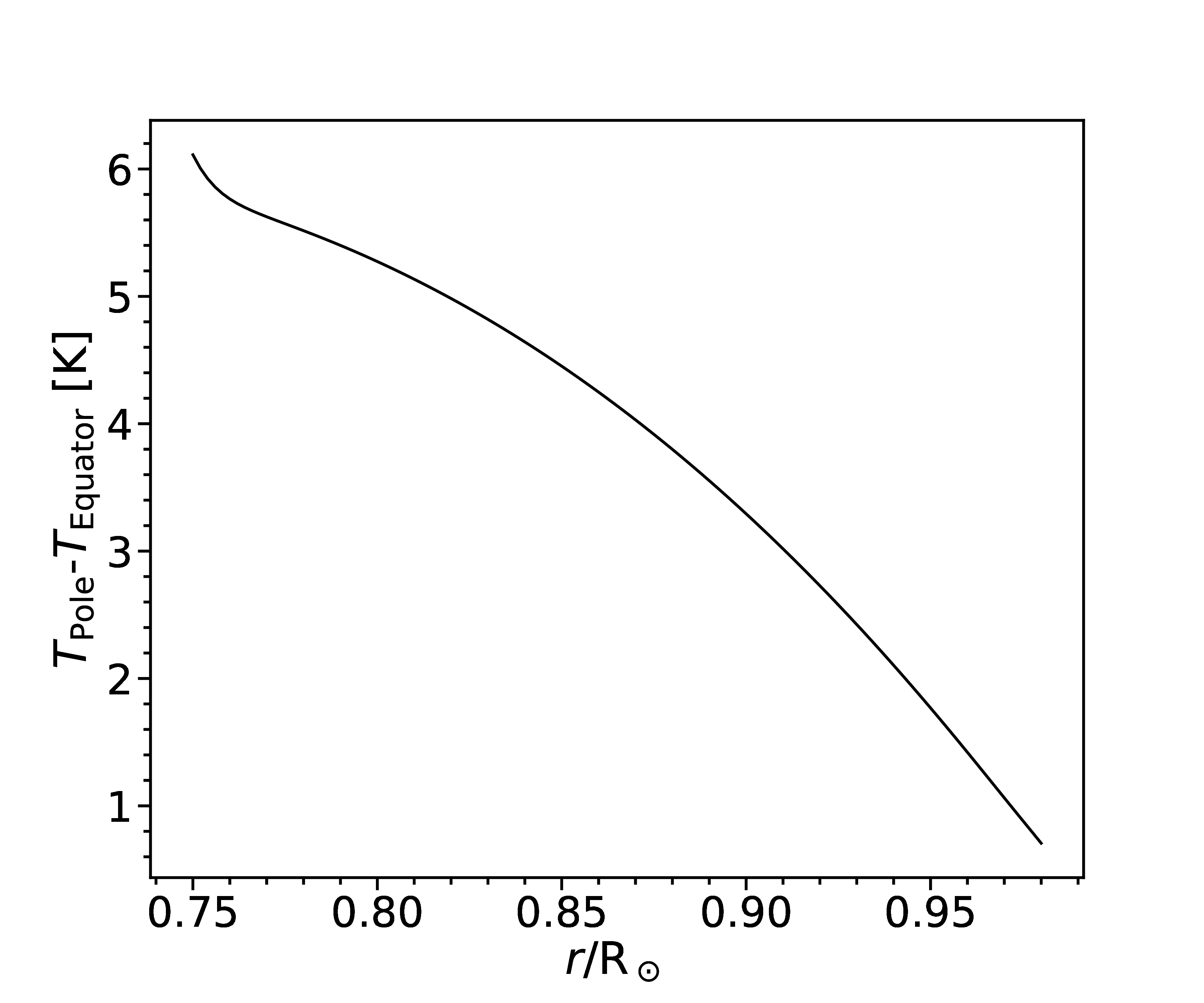}
    \caption{The pole-equator temperature difference as a function of radius, corresponding to Figure~\ref{ch5:fig5}.}
    \label{ch5:fig6}
\end{figure}

\autoref{ch5:fig5} shows the distribution of $\Delta T (r, \theta)$ in the convection zone below 0.98~\Rsun\ that would follow on assuming the thermal wind balance
and using $\Omega (r, \theta)$ given by the analytical expression \autoref{ch5:eq6}. We clearly see a decrease in  $\Delta T(r,\theta)$ with $r$ as we approach the surface. 
Now, one quantity which is of particular interest to us is the pole-equator temperature difference $\Delta T(r, \theta=0) - \Delta T(r, \theta=\pi/2)$ as a function
of $r$. Figure~\autoref{ch5:fig6} shows this pole-equator temperature difference as a function of $r$ within the convection zone for $r<0.98$\Rsun.

We shall now present our results for the NSSL by assuming different values of $r_c$ in the range 0.92\Rsun\,--\,0.97\Rsun. For a particular value of
$r_c$, we take $\Delta T (r,\theta)$ to be as given in \autoref{ch5:fig5} for $r <r_c$ and as given by \autoref{ch5:eq5} for $r >r_c$.  In this way, we obtain $\Delta T (r,\theta)$
throughout the convection zone for a chosen value of $r_c$.
We point out that the observational value
of the pole-equator temperature difference reported by \citet{Rast2008} is $\approx 2.5$ K. \autoref{ch5:fig6} shows that such a value of the pole-equator temperature difference
occurs at around $r \approx 0.92$\Rsun\ when we evaluate $\Delta T (r, \theta)$ from the analytical expression (\autoref{ch5:eq6}) of $\Omega(r, \theta)$.
This means that we have to take $r_c \approx 0.92$~\Rsun\ to get the pole-equator temperature difference at the surface which matches the 
observations of \citet{Rast2008}.

\begin{figure}[!ht]
	\includegraphics[width=\textwidth]{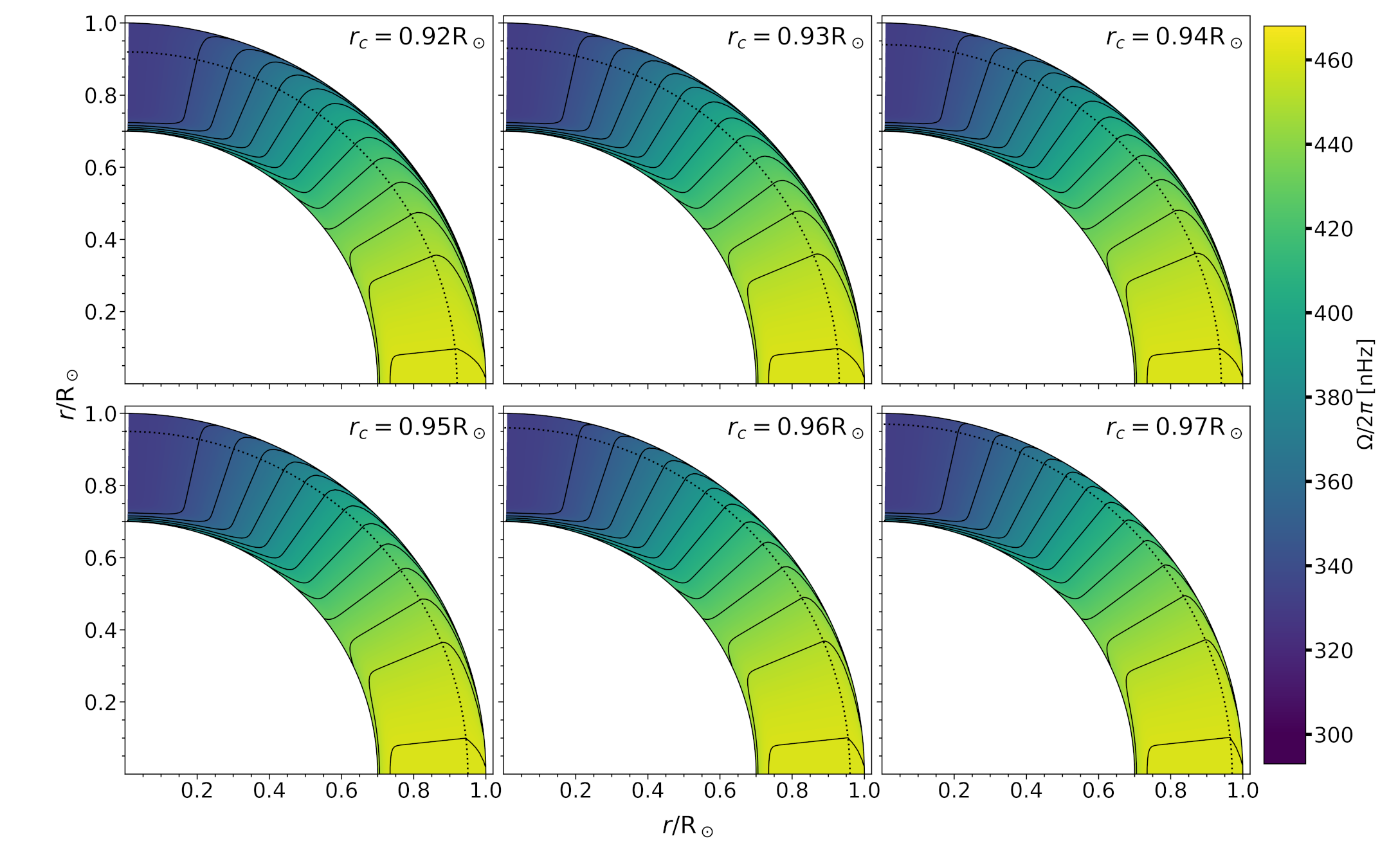}
  \caption{The profiles of $\Omega (r,\theta)$ for different values of $r = r_c$, obtained by using $\Omega (r,\theta)$ given
by \autoref{ch5:eq6} as input data for $r < r_c$ .}
    \label{ch5:fig7}
\end{figure}

Now that we have $\Delta T (r,\theta)$ throughout the convection zone including the top layer ($r\ge r_c$) for different values of $r_c$, the last step is to calculate $\Omega (r\ge r_c,\theta)$ in this top layer by assuming that the thermal wind balance holds in this layer. We have already made use of $\Omega (r < r_c,\theta)$ in the deeper layers of the convection zone, as given by \autoref{ch5:eq6}, to calculate $\Delta T (r,\theta)$ by making use of \autoref{ch5:eq34} (which is effectively
the same as \autoref{ch5:eq3}) and  \autoref{ch5:eq5}. We now use the thermal balance \autoref{ch5:eq3} in the top layer ($r\ge r_c$) in a different manner. From the value of $\Delta T (r\ge r_c,\theta)$ in this top layer, we calculate the right hand side of \autoref{ch5:eq3} and then solve \autoref{ch5:eq3} to find the distribution of $\Omega (r\ge r_c,\theta)$ in this top layer which would satisfy \autoref{ch5:eq3}. We carry on this procedure for the values $r_c = $ 0.92\,\Rsun, 0.93\,\Rsun, 0.94\,\Rsun, 0.95\,\Rsun, 0.96\,\Rsun, 0.97\,\Rsun. We can combine $\Omega (r\ge r_c,\theta)$ obtained in the top layer in this manner with $\Omega (r < r_c,\theta)$ in the deeper layers as given by \autoref{ch5:eq6}. This combination for the different values of $r_c$ which we have used are shown in \autoref{ch5:fig7}. The dotted circles in the various sub-figures indicate the values of $r_c$ for all these cases. 

The contours of constant $\Omega$ bend towards the equator in the top layers of the convection zone for all values of $r_c$ shown in \autoref{ch5:fig7}. This indicates the clear presence of the NSSL in all these cases.  We stress again that the initial input data $\Omega (r < r_c,\theta)$ which we had used in the deeper layers of the convection zone in order to start our calculations did not have the NSSL.  In fact, the analytical expression (\autoref{ch5:eq6}), which we had used for the values of $\Omega (r < r_c,\theta)$ in the deeper layers, does not give rise to the NSSL at all, as seen in \autoref{ch5:fig3}. It is thus clear that the NSSL that we see in \autoref{ch5:fig7} could not be an artifact of the input data.  The NSSL arises from the fact that the thermal wind term becomes very large in the top layers due to the falling temperature there and the centrifugal term also has to become very large to balance it.  This requirement for satisfying the thermal wind condition (\autoref{ch5:eq3}) in the top layer can only be met if there is an NSSL.  We propose this as the explanation for the existence of the NSSL in the top layer of the solar convection zone. The different sub-plots in \autoref{ch5:fig7} show that the increase in $r_c$ causes the NSSL to be confined to an increasingly narrower layer near the solar surface.   

\section{Results Based on Helioseismology Data}
\label{ch5:helio}

After presenting the results based on the analytical expression Equation~\ref{ch5:eq6} of $\Omega (r, \theta)$ in the previous section (Section~\ref{ch5:empirical}),
we now carry on exactly the same calculations based on the value of $\Omega (r, \theta)$ as given by helioseismology. We use $\Omega (r, \theta)$ averaged
over cycle~23, as supplied to us by H.M.\ Antia. The methodology that was used for
obtaining the $\Omega (r, \theta)$ profile from helioseismology data has been described
by \citet{Antia1998, Antia2008}.
 Our calculations are based on the tabulated value of temporally averaged  $\Omega (r, \theta)$ for all $r$ in the range 0.7\,\Rsun to \Rsun at steps of 0.005\,\Rsun and for all $\theta$ in the range of $2^{\circ}$ to $90^{\circ}$ ($88^{\circ}$ to $0^{\circ}$ latitude) at steps of $2^{\circ}$. The profile of $\Omega (r, \theta)$ with the contours of constant $\Omega$ (represented as black solid lines) has been shown in \autoref{ch5:fig1}.

\begin{figure}[!ht]
\centering
	\includegraphics[height=0.8\textheight]{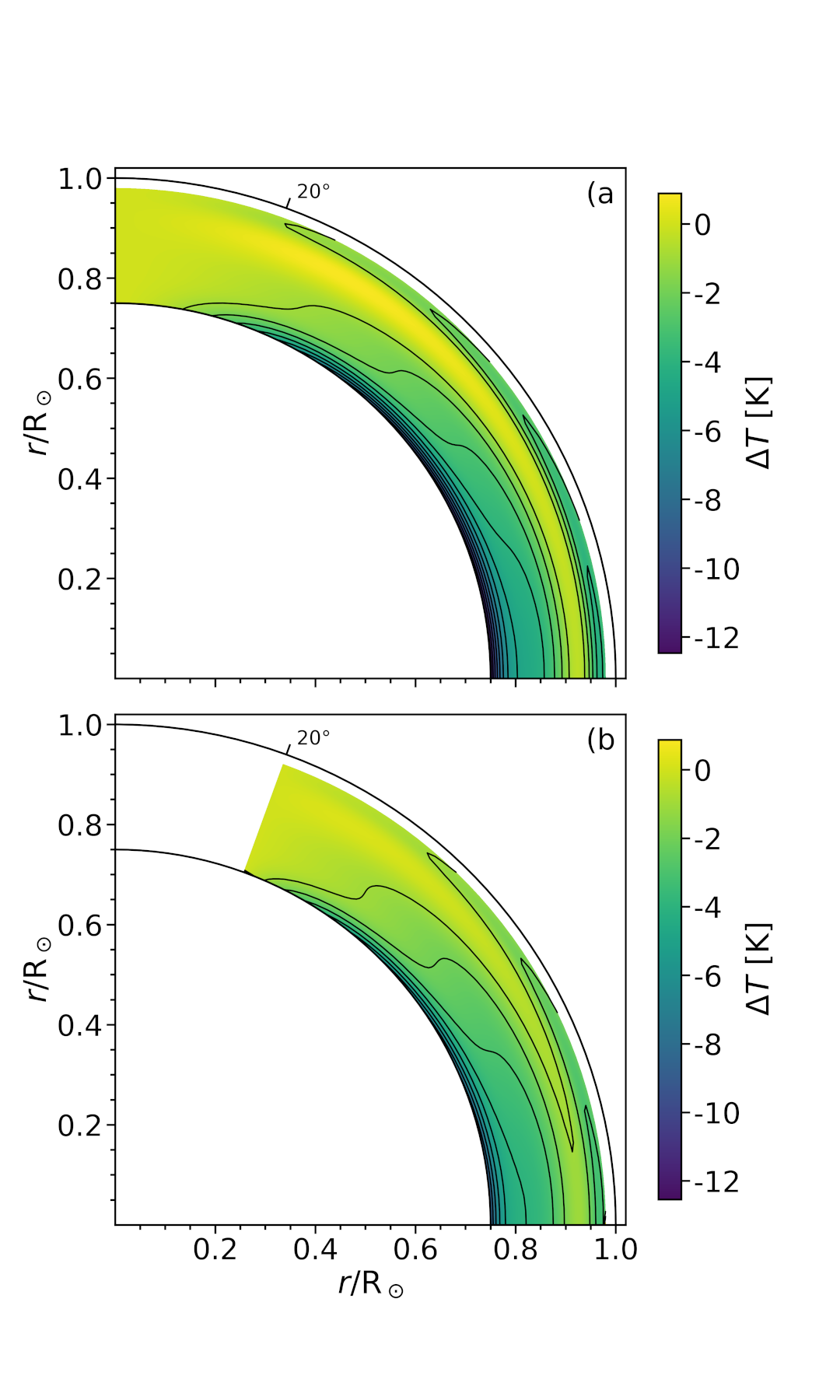}
    \caption{Profile of $\Delta T (r,\theta)$ in the solar convection zone below $r = 0.98$\Rsun, which would follow from the helioseismology data of differential rotation
    on taking $r_c = 0.98$\Rsun. Contours represent the constant values of $\Delta T (r,\theta)$.
     (a) shows the profile of 
    $\Delta T_{0^{\odot}} (r, \theta)$ and (b) the profile
    of $\Delta T_{20^{\odot}} (r, \theta)$ as defined in
    Equation~\ref{ch5:eq34}}
    \label{ch5:fig8}
\end{figure}

As in Section~\ref{ch5:empirical}, we carry on calculations for different values of $r_c$ in the range 0.93\,\Rsun\,--\,0.98\,\Rsun. For a particular value of $r_c$, we substitute the values of $\Omega (r < r_c,\theta)$ in the left hand side of \autoref{ch5:eq3} to calculate $\Delta T (r < r_c,\theta)$. The values of $\Delta T (r,\theta)$ for $r > r_c$ are again given by \autoref{ch5:eq5}.  \autoref{ch5:fig8}a shows the profile of $\Delta T(r<r_c,\theta)$ calculated for the case $r_c = 0.98$\,\Rsun. One concern we have is that the helioseismic determination of
$\Omega (r, \theta)$ has large uncertainties in the polar regions
at high latitudes and, when we use \autoref{ch5:eq34} to calculate $\Delta T (r < r_c,\theta)$, which is $\Delta T_{\theta_0} (r < r_c,\theta)$ with
$\theta_0 = 0$, we have to integrate over this region where the
value of $\Omega (r, \theta)$ is unreliable. One way of avoiding
this difficulty is to consider temperature variations only in 
regions not too close to the poles where we can trust the helioseismic
values of $\Omega(r, \theta)$. We have used \autoref{ch5:eq34} to calculate
$\Delta T_{20^{\circ}} (r < r_c,\theta)$ by avoiding the polar
region. \autoref{ch5:fig8}b shows the profile of
$\Delta T_{20^{\circ}} (r < r_c,\theta)$. Comparing the profiles
of $\Delta T (r < r_c,\theta)$ for co-latitudes higher than $20^{\circ}$ (i.e.\ latitudes lower than $70^{\circ}$)
in  \autoref{ch5:fig8}a\,--\,b, we find that various
features are in broad agreement, indicating that they are not due
to errors in $\Omega (r, \theta)$ near the polar region. 
Especially, we find an annular strip near $r = 0.925 R_{\odot}$ within which the value of $\Delta T (r < r_c,\theta)$ is close to zero in  \autoref{ch5:fig8}a.  This strip becomes
less prominent in  \autoref{ch5:fig8}b,
though it does not disappear. The reason behind this strip
is this.  On taking a careful look at \autoref{ch5:fig1}, we realize
that $\pa \Omega^2/\pa z$ just below the NSSL is close
to zero at latitudes higher than mid-latitudes and is
even positive at very high latitudes (it is usually negative within the convection zone). This explains, on
the basis of \autoref{ch5:eq34}, why we have this unusual strip even in
 \autoref{ch5:fig8}b after excluding the
polar region. \citet{Schou1998} refer to this region
at high latitudes somewhat below the surface
as ``a submerged polar jet'' and comment in Section~5.5 of
their paper that it ``is seen consistently by several independent methods''.  If this submerged polar jet is
real and not a data artifact, then what causes it is certainly an important
question.  We do not attempt to address this question in
the present paper.

\begin{figure}[!ht]
\centering
	\includegraphics[width=\columnwidth]{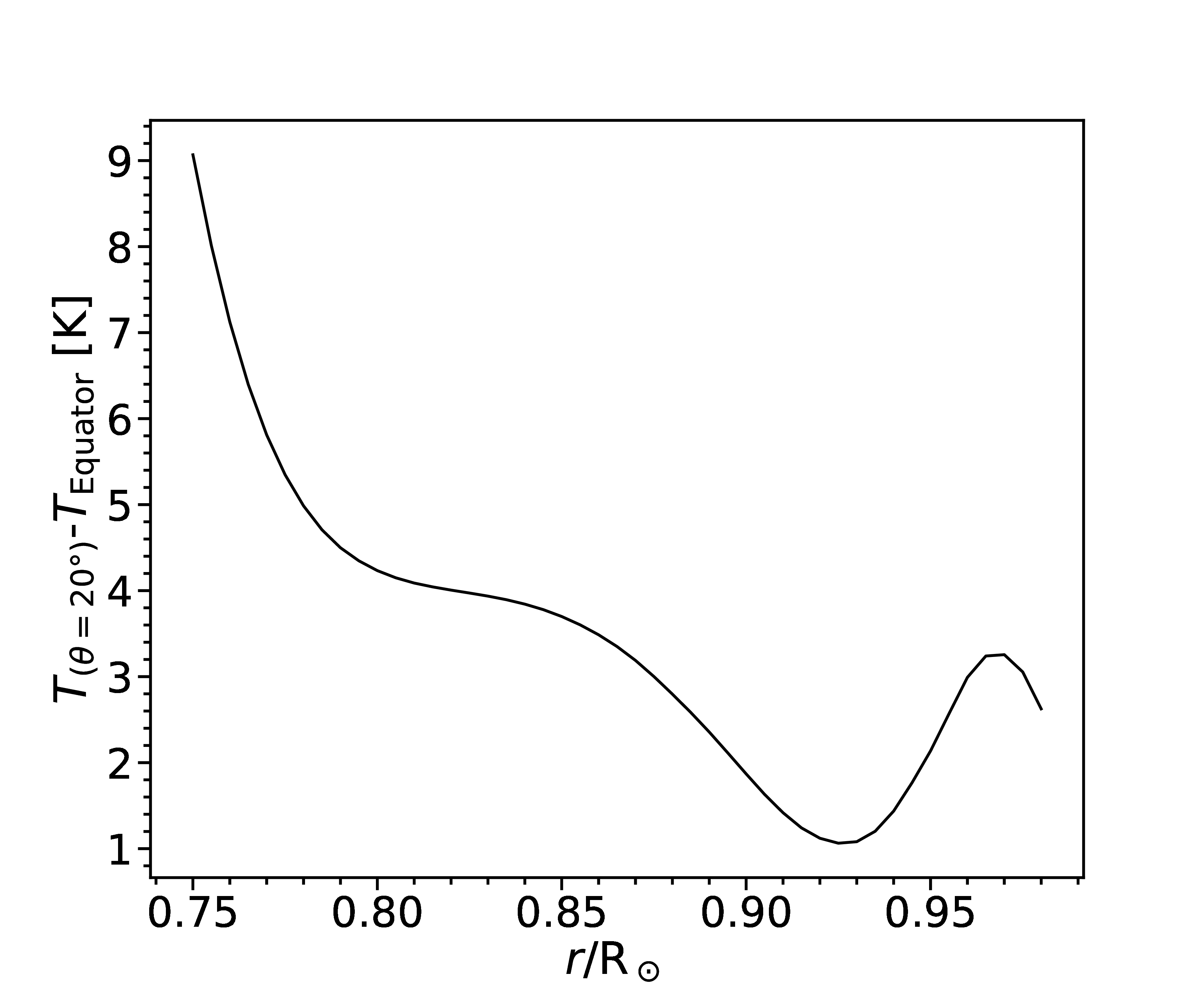}
    \caption{The temperature difference between
    the co-latitude $20^{\circ}$ and the equator
    as a function of radius, corresponding to  \autoref{ch5:fig8}.}
    \label{ch5:fig9}
\end{figure}

We now plot the temperature difference $\Delta T (r, \theta=20^{\circ}) - \Delta T (r, \theta=90^{\circ})$ between the 
co-latitude $20^{\circ}$ (i.e.\ latitude $70^{\circ}$) and the
equator as a function of $r$ in  \autoref{ch5:fig9}. It should be clear from
\autoref{ch5:eq34} that this temperature difference is given by integrating the integrand in the right hand side of \autoref{ch5:eq34}
from $\theta=20^{\circ}$ to $\theta=90^{\circ}$. In 
other words, this temperature difference is independent
of the values of $\Omega (r, \theta)$ at very high latitudes (where these values may have large uncertainties) and should be the same for both the cases shown in 
\autoref{ch5:fig8}a and \autoref{ch5:fig8}b.  We have seen in the calculations based on the analytical expression of $\Omega (r, \theta)$ in the previous section (Section~\ref{ch5:empirical}) that the pole-equator temperature difference decreased monotonically with $r$ (see  \autoref{ch5:fig6}). However,  \autoref{ch5:fig9} shows a more complicated dependence of a similar temperature difference on $r$.  We indeed find a monotonic decrease of the temperature difference with $r$ for values of $r$ lower than $\approx0.92$\Rsun. But then it starts increasing with $r$ up to $\approx 0.97$\Rsun, beyond which it decreases again. This complicated variation of the temperature difference is  connected with the submerged polar jet which continues even a little bit beyond co-latitude $20^{\circ}.$

\begin{figure}[!ht]
\centering
	\includegraphics[width=\textwidth]{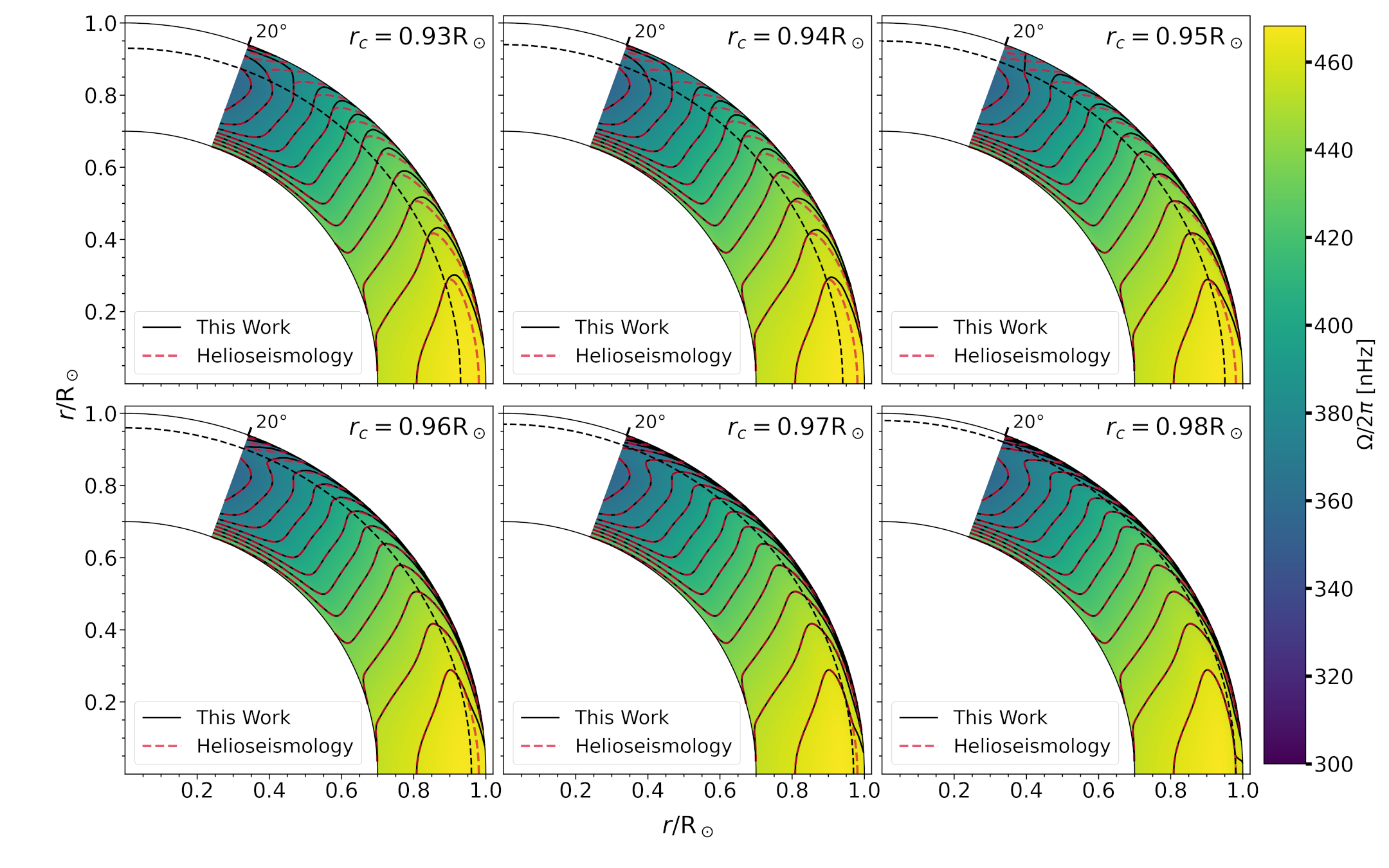}
    \caption{The profiles of $\Omega (r,\theta)$ for different values of $r = r_c$, obtained by using $\Omega (r,\theta)$ given by helioseismology as input data for $r < r_c$ .}
    \label{ch5:fig10}
\end{figure}

The final step in our analysis is exactly the same as in Section~\ref{ch5:empirical}.  Once we have $\Delta T (r, \theta)$ throughout the convection zone corresponding to different values of $r_c$ (with its value for $r \ge r_c$ being given by \autoref{ch5:eq5}), we evaluate the right hand side of \autoref{ch5:eq3} for $r\ge r_c$ and then solve \autoref{ch5:eq3} to obtain $\Omega (r\ge r_c, \theta)$ in the top layers of the convection zone. Since we have to differentiate $\Delta T (r > r_c,\theta)$  with respect to $\theta$, it does not matter whether we use
$\Delta T_{\theta_0} (r > r_c,\theta)$ with $\theta_0 = 0$ or with
$\theta_0 = 20^{\circ}$. The temperature profiles in both 
\autoref{ch5:fig8}a and \autoref{ch5:fig8}b give the same distribution
of $\Omega (r\ge r_c, \theta)$ in the top layers of the convection
zone for $\theta$ higher than $20^{\circ}$. Thus, the profile
of $\Omega (r\ge r_c, \theta)$ in the top layers of the convection
zone that we have calculated starting initially from $\Omega (r\le r_c, \theta)$
in the deeper layers of the convection is independent of the errors
in $\Omega (r\le r_c, \theta)$ in the polar region.
In \autoref{ch5:fig10}, we have shown the distribution of $\Omega (r\ge r_c, \theta)$ obtained in this way for different values of $r_c$ (represented by black dashed circles), along with $\Omega (r<r_c,\theta)$ as given by helioseismology data (same as in \autoref{ch5:fig1}). In all these cases, we clearly see the NSSL. While the input data $\Omega (r < r_c, \theta)$ used in our calculations show some indications of the NSSL for the cases $r_c = $0.97\,\Rsun, 0.98\,\Rsun, there was no sign of the NSSL 
in the input data for the cases $r_c = $0.93\,\Rsun, 0.94\,\Rsun. The fact that we get a layer just below the solar surface resembling the NSSL in all these cases strongly suggests that the NSSL arises from the requirement of the thermal wind balance with the thermal wind term becoming very large in the top layer of the convection zone. To facilitate comparison of our theoretical results with the observations, we have over-plotted in \autoref{ch5:fig10} the contours of constant $\Omega$ obtained from helioseismology observation (shown by dashed red lines).

We at last come to the question whether the theoretical results obtained on the basis of our assumption that the thermal balance equation holds in the top layers of the convection zone agree with observational data.  Comparing the solid lines indicating the theoretical results with the dashed red lines indicating helioseismology data above $r = r_c$ in \autoref{ch5:fig10}, it is evident that the agreement is very good for all values of $r$ larger than 0.96\,\Rsun. In the cases $r = $0.97\,\Rsun, 0.98\,\Rsun, the lower part of the NSSL was present in the input data and one may argue that it is not so surprising that our theoretical calculations correctly gave the structure of the NSSL in the upper layers.  However, this is clearly not case for $r = 0.96$\,\Rsun. One other aspect of the observational data we need to match is the pole-equator temperature difference. Looking at Figure~4 of \citet{Rast2008}, we find that they present measurements
up to latitudes of about $70^{\circ}$.  What they loosely
refer to as the pole-equator temperature difference (PETD)
is actually the temperature difference between $70^{\circ}$ and the equator.  If we also use the same convention,
then what is plotted in \autoref{ch5:fig9}
can be called PETD and compared directly with the results
of \citet{Rast2008}. Taking the value 2.5~K reported by \citet{Rast2008} to be the correct value, we note in \autoref{ch5:fig9} that the PETD has the value 2.5 K at $r \approx 0.96$\,\Rsun.  If this is taken to be the value of $r_c$, then PETD at the solar surface should also be 2.5 K. It is thus clear that an accurate determination of the PETD at the solar surface is extremely important and can put constraints on the appropriate value of $r_c$ to be used in theoretical calculations.  If this temperature difference is indeed 2.5 K, then we conclude that the theoretical calculations carried out with $r_c=0.96$\,\Rsun in our model are in good agreement with observational data.  This case gives a good structure of the NSSL as we see in \autoref{ch5:fig10} and the PETD at the solar surface also has the desired value 2.5 K.

\begin{figure}[!ht]
	\includegraphics[width=\columnwidth]{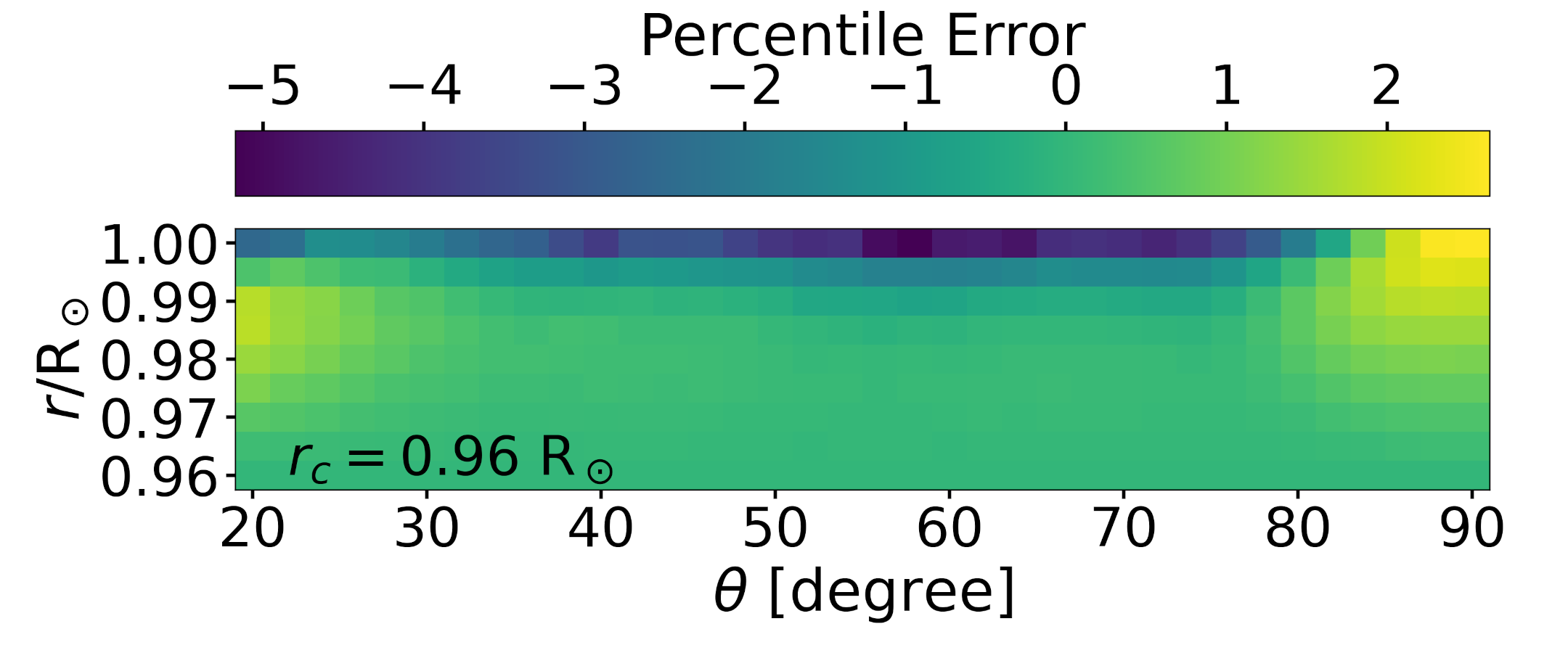}
    \caption{Distribution of percentile error $f$ for the case of $r_c=0.96$ \Rsun.}
    \label{ch5:fig11}
\end{figure}

To check quantitatively how well our theoretically
determined $\Omega$ in the NSSL
compares with observational  $\Omega_{\rm heliseismology}$, we consider the percentile error
\begin{equation}
    f = 100 \frac{\Omega - \Omega_{\rm helioseismology}}
    {\Omega_{\rm helioseismology}}
    \label{ch5:eq11}
\end{equation}
at different points. Since the observed variation of $\Omega$ within the NSSL is at the level of 5\%, we
must have $f$ considerably less than that for the fit
to be considered sufficiently good. \autoref{ch5:fig11} shows the
distribution of $f$ for the case $r_c = 0.96 $\,\Rsun  in
the top layer of the convection zone above $r = 0.96$\, \Rsun. We indeed find that $f$ is much less than
5\% within the NSSL, except in a very thin layer
above $0.995$\,\Rsun close to the solar surface. This
perhaps suggests that the thermal wind balance breaks
down in this very thin layer near the surface. The smallness
of $f$ within the NSSL below this very thin layer presumably
indicates that the thermal wind balance holds there to
a very good approximation. The root mean square (RMS) value of $f$ for the 
case 
presented in \autoref{ch5:fig9} is found to be 1.36\%. 
The RMS values of $f$ for
cases  $r_c = 0.95 $\,\Rsun and $r_c = 0.97$\,\Rsun turn
out to be 1.50\% and 2.41\% respectively.

\begin{figure}
	\includegraphics[width=\columnwidth]{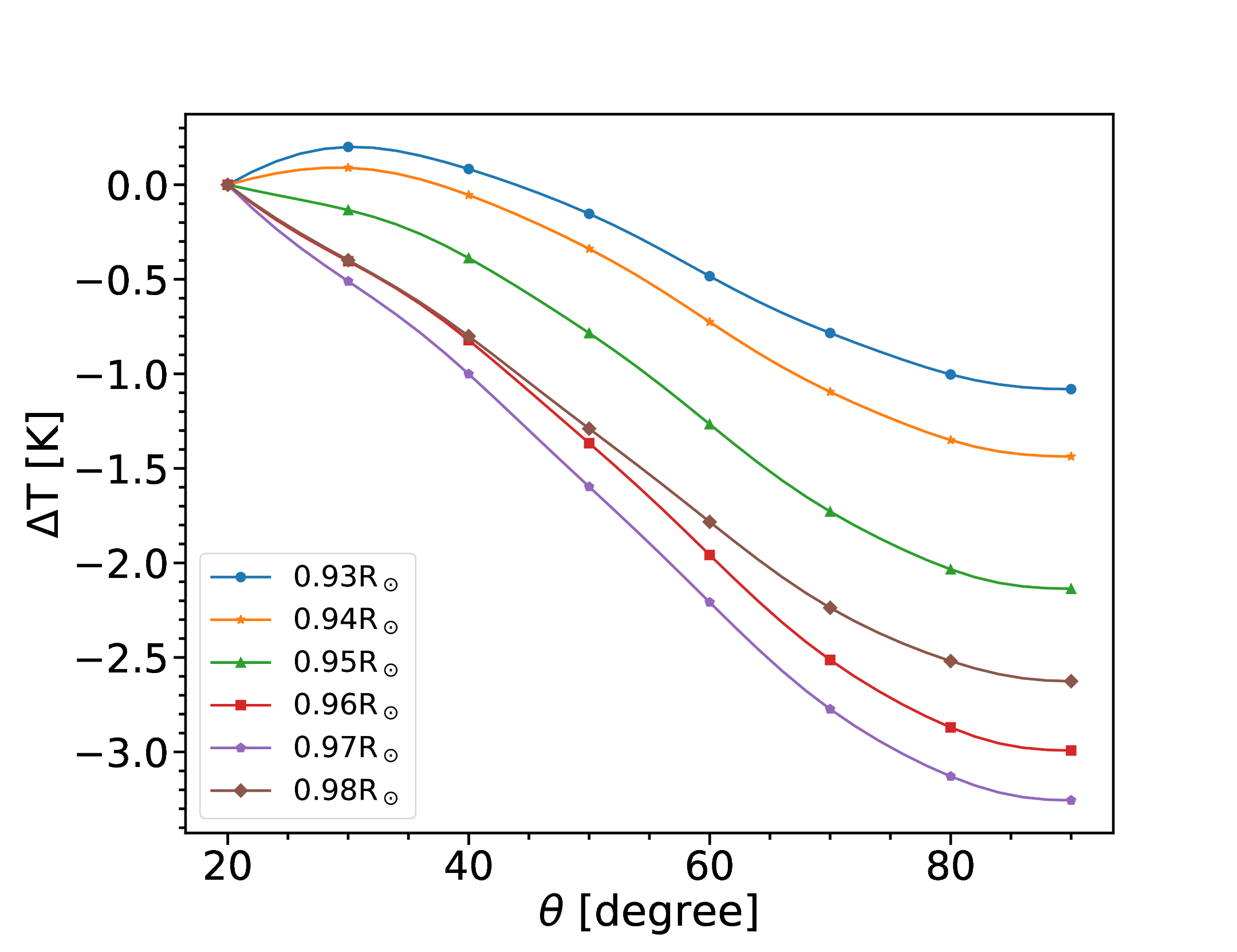}
    \caption{Variation of $\Delta T_{20^{\circ}} (R_\odot, \theta)$ on the solar surface with co-latitude $\theta$ for different values of $r_c$ (markers are plotted at every 5 data points).}
    \label{ch5:fig12}
\end{figure}

We are not aware of any independent measurements of the pole-equator temperature  difference after the work of \citet{Rast2008} done more than a decade ago. If the variation of the temperature with latitude on the solar surface is be measured more accurately by modern techniques in the future, then it will be useful to compare such observations with the theoretical results of our model. \autoref{ch5:fig12} shows how 
$\Delta T_{20^{\circ}} (R_\odot, \theta)$ varies with the co-latitude $\theta$ for different values of $r_c$.  Note that $\Delta T_{20^{\circ}} (r, \theta)$ is defined in Equation~\ref{ch5:eq34} in such a manner that its value is always zero at $\theta = 20^{\circ}$. Also, note that the curves for cases $r_c = 0.96$\,\Rsun, 0.97\,\Rsun, 0.98\,\Rsun\  do not appear in the same simple sequence as the curves for cases $r_c = 0.93$\,\Rsun, 0.94\,\Rsun, 0.95\,\Rsun . Given the complicated plot shown in \autoref{ch5:fig9}, this behaviour is not surprising.  It will be instructive to compare \autoref{ch5:fig12} with observational data when such data become available.  We remind the readers that all the results in this Section were obtained on the assumption of a sudden jump in the nature of convective heat transport at $r = r_c$. If the transition is more gradual, that may change the results slightly.  Detailed comparison between theoretical results and observational data in future may throw more light on this.

Assuming that the intensity of radiation $I$ emitted from a region of the
surface goes as $T^4$ according to the Stefan-Boltzmann law, the variation
of intensity with latitude caused by the variation of temperature with latitude would be given by
\begin{equation}
    \frac{\Delta I}{I} \approx 4 \frac{\Delta T}{T}.
\end{equation}
The pole-equator intensity difference corresponding to a temperature difference of 2.5 K would be
\begin{equation}
    \frac{\Delta I}{I} \approx 0.0016.
\end{equation}
In other words, the pole would be only about 0.16\% brighter than the
equator. The latitudinal variation of intensity measured by \citet{Rast2008}
is given in Figure~4 of their paper.  They point out that it is non-trivial
to measure this small latitudinal variation of intensity.  Apart from 
instrumental errors, the presence of polar faculae makes these measurements
difficult. However, \citet{Rast2008} claimed that their measurement of 
enhanced intensity near the polar region is a real physical effect.

\section{Conclusion}
\label{ch5:conclusion}
There are differences of opinion why the Sun has a Near-Surface Shear Layer (NSSL).  A novel explanation of the NSSL was recently proposed by \citet{Choudhuri2021a} on the basis of order-of-magnitude estimates.  We now substantiate the proposal of \citet{Choudhuri2021a} through detailed  calculations. Although it is generally agreed that the thermal wind balance equation holds within the body of the solar convection zone, whether this equation even holds in the top layers had been debated. It has been suggested that this equation breaks down in a boundary layer at the top and this may somehow give rise to the NSSL. \citet{Choudhuri2021a} argued on the other hand that the thermal wind term becomes very large in the top layers of the convection zone and the thermal wind balance equation has to hold. This means that the centrifugal term also has to be very large in the top layers to achieve the thermal wind balance, necessitating the existence of the NSSL. When we compare the final results of our
theoretical model with observational data from helioseismology, we conclude that the thermal wind
balance condition may break down only in a very thin
layer near the solar surface (having thickness of order $\approx 0.005$\,\Rsun or
$\approx 3000$ km).

The argument proposed in this paper hinges crucially on the fact that the convective cells are affected by the solar rotation within the main body of the convection zone, except in the top layers within which the convective turnover time is less than the rotation period of the Sun.  Although the transition from the layers where rotational effects are large to the layers where they are small must be a gradual transition, we simplify the calculations by assuming that this transition takes place at $r = r_c$. Since the effect of the solar rotation would make the convective heat transport in the deeper layers dependent on latitude, we expect that the temperature at a point within the convection zone will depart by an amount $\Delta T (r, \theta)$ from what we get from the standard models of the convection zone in which the effect of rotation is not taken into account.  In principle, it should be possible to calculate $\Delta T (r, \theta)$ from the theory of convective heat transport.  However, this is a formidably difficult problem in practice and we calculate $\Delta T (r, \theta)$ in the deeper layers of the convection zone from the differential rotation measured by helioseismology by assuming that the thermal wind balance condition prevails in the deep layers of the convection zone. Since the effect of rotation on the convection is negligible above $r = r_c$, we expect the radial temperature gradient $dT/dr$ to be independent of latitude in this layer, which implies that $\Delta T (r, \theta)$ does not vary with $r$ in this layer.  It is this fact, coupled with the fact that the temperature drops sharply in this top layer, which makes the thermal wind term very large in this top layer.  From the requirement that the centrifugal term also has to become large to balance the large thermal wind term in this layer, we can calculate the distribution of $\Omega (r, \theta)$ in this near-surface layer.  We have found that our calculations give a layer resembling the NSSL.

The non-variation of $\Delta T (r, \theta)$ with $r$ in the upper layers of the convection zone ensures that the pole-equator temperature difference does not vary in this layer.  The means that the value of the pole-equator temperature difference at $r_c$ gets mapped to the solar surface.  A careful measurement of the pole-equator temperature difference at the solar surface would enable us to assess the value of $r_c$ above which the convective motions are not affected much by rotation.  The value 2.5 K of the pole-equator temperature difference reported by \citet{Rast2008} led us to conclude that $r_c \approx 0.96$\Rsun.  For this value of $r_c$, the various aspects of observational data, including the structure of the NSSL, are explained very well by our theoretical model. Fairly sophisticated simulations of solar convection are now being carried on by many groups. We hope that such
simulations may also eventually be able to give an indication of the value of $r_c$ above which the effect of rotation is negligible.  
Since the measurement of the pole-equator temperature difference allows us to assess $r_c$, such a measurement can put important constraints on the simulations of solar convection. It appears that there have not been any independent measurements of the pole-equator temperature  difference after the work of \citet{Rast2008} done more than a decade ago.  We hope that other groups will undertake this measurement in the near future, since the value of this temperature difference has connections with such important issues as the nature of the solar convection and the structure of the NSSL.

The large-scale flows in the solar convection zone like the differential rotation and the meridional circulation play important roles in the flux transport dynamo model for explaining the solar cycle, which started being developed from the 1990s \citep{Wang1991,Choudhuri1995, Durney1995} and has been reviewed by several authors in the last few years \citep{Charbonneau2010, Choudhuri2011, Karak2014}.  One crucial question is whether the NSSL is important in the solar dynamo process.  One key idea in the solar dynamo models is that the toroidal magnetic field is generated by the strong differential rotation at the bottom of the solar convection zone, where the field can be stored in the stable sub-adiabatic layers below the bottom of the convection zone and can undergo amplification there. Srtands of the toroidal magnetic field eventually break out of the stable layers to rise through the convection zone due to magnetic buoyancy. Since the near-surface layer is a region of strong super-adiabatic temperature gradient which enhances magnetic buoyancy \citep{Moreno1983, Choudhuri1987}, magnetic fields are expected to rise through this layer quickly without allowing much time for shear amplification. Unless there is some mechanism to keep magnetic fields stored in the NSSL for some time, most likely the NSSL is not important for the dynamo process,
although there is not complete unanimity on this  \citep{Brandenburg2005}.
Models of the flux transport dynamo without including the NSSL give reasonable fits with observations \citep{Chatterjee2004}. 
Dynamo-generated magnetic fields, however, can react back on the large-scale flows producing temporal variations with the solar cycle \citep{Chakraborty2009, Hazra2017}. For example, the meridional circulation varies periodically with the solar cycle and modelling it requires going beyond the thermal wind balance \autoref{ch5:eq3} to include a time derivative term \citep{Hazra2017, Choudhuri2021b}. The thermal wind balance equation follows from the full equation for the meridional circulation under steady state conditions if the dissipation term can be ignored.  Presumably, the thermal wind balance \autoref{ch5:eq3} holds for the time-averaged part of large-scale flows, which has been our focus in this paper. However, there is evidence of random temporal fluctuations in the meridional circulation \citep{Karak2010, Karak2011, Choudhuri2012, Choudhuri2014, Hazra2019}, possibly indicating slight violations of the thermal wind balance equation. Since the terms involved in the thermal wind balance are much larger than the other terms in the equation of the meridional circulation (see, for example, the discussion in the Appendix of \citet{Choudhuri2021b}), even a slight imbalance between these large terms is sufficient to cause fluctuation in the meridional circulations and we believe that the violations of the thermal wind balance remain very small.

Lastly, we suggest that other solar-like stars with rotation periods similar to the Sun are likely to have similar shear layers near their surfaces, since the solar NSSL arises out of very general considerations which should hold for such stars.  The study of starspots and stellar cycles in the last few years have suggested that the solar-like stars also must have large-scale flows like the differential rotation and the meridional circulation giving rise to dynamo cycles, as in the case of the Sun \citep{Karak2014a, Choudhuri2017, Hazra2019b}.  Although asteroseismology has started giving some initial results of differential rotation in solar-like stars \citep{Benomar2018}, we are still very far for determining observationally whether other stars also have NSSL.

\clearpage{}

\clearpage{}\chapter[An Indication of BMRs Tilt Quenching in the Sun]{Magnetic field dependence of bipolar magnetic region tilts on the Sun: Indication of tilt quenching}
\label{Chap6}
\lhead{\emph{Chapter 6: An Indication of BMRs Tilt Quenching in the Sun}} \noindent

\section{Introduction}
\label{ch6:intro}
Sunspots are the regions of concentrated magnetic field observed as dark spots in white-light images. In the magnetograms, we find two regions of opposite
polarities appearing close to each other. Thus the sunspots that we see in white light image are essentially two poles of a more general feature called the Bipolar Magnetic Regions (BMR). However, the weaker BMRs produce negligible intensity contrast and hence go undetected in white light images.
In general, BMRs are tilted with respect to the equator and statistically, this tilt increases with latitude---popularly known as Joy's law \citep{Hale1908}.

The tilt is crucial for the generation of the poloidal magnetic field through the decay and dispersal of the BMRs near the solar surface, 
which is popularly known as the \bl\ process. While this was proposed in the 60s by \citet{Babcock1961} and \citet{Leighton1964}, in recent years, this process has received significant attention due to its support from observational studies \citep{DasiEspuig2010, Kitchatinov2010, MunozJaramillo2013, Priyal2014}. 
 Based on this \bl\ process, several surface flux transport models have been constructed, 
which are successful in reproducing many features of the solar surface \citep{Jiang2014a}.
Many dynamo models, including the popular flux transport dynamo models, 
have also been constructed based on this \bl\ process \citep{Leighton1969, Wang1991a, Wang1991}; see reviews \citep{Charbonneau2010, Karak2014, Choudhuri2018}.

A serious concern in these \bl\ models is the saturation of magnetic field. There
must be a nonlinear quenching to suppress the growth of magnetic field in any kinematic dynamo model such as the \bl\ ones.
In the latter models, large-scale velocities, namely, meridional flow and differential rotation are specified (broadly through observations), while the small-scale velocity is parametrized such as in the form of turbulent diffusivity. 
Therefore, the most obvious choice in these models is to include a nonlinearity in the \bl\ process. 
In all the previous \bl\ dynamo models, a magnetic field dependent quenching is included such that the poloidal field production is reduced when the toroidal magnetic field exceeds the so-called saturation field $B_0$ \citep{Charbonneau2010}. For the \bl\ process, this requires that the tilt must be reduced when the BMR field strength
exceeds a certain value; see \citet{Lemerle2017, Karak2017, Karak2018} for specific requirement of this idea. 

We believe that the BMRs are produced due to buoyant rise of the strong toroidal magnetic flux tubes from the base of the convection zone (CZ) \citep{Parker1955a}.  From the thin flux tube model, we know that during the rise of toroidal flux in the CZ, 
the Coriolis force induced by the diverging east-west velocity near the loop apex causes a tilt \citep{DSilva1993, Fan1994}.
Therefore, we expect the rise time of toroidal flux tube and thus the tilt to decrease with increase of magnetic field in the tube. 
This idea can potentially lead to a quenching in the \bl\ process. 

Although the thin flux tube model explains some observed features of BMRs, it does not capture the detailed dynamics of solar CZ. Indeed,
including the convection, \citet{Weber2011} find a significant change in the
behaviour of BMR tilt. They find the tilt to increase with the magnetic field first and then decrease in accordance with the thin flux tube model.

Using magnetogram data corresponding to 1988--2001, 
\citet{Tian2003} found a systematic variation of the BMR tilt 
with the magnetic flux content. Surprisingly,
using {\it Michelson Doppler Imager}(MDI) magnetograms during 1996 -- 2011, \citet{Stenflo2012a} did not
find any systematic variation of the BMR tilt with the magnetic flux
and they claim that their result rules out the thin flux-tube model.
However, we should not forget that the magnetic field of BMR 
also vary with the magnetic flux \citep{Tlatov2014},
and in the analysis of \citet{Stenflo2012a}, the variation of magnetic field is ignored.
Therefore, the motivation of the present Letter is first to analyse the BMRs 
based on their magnetic field strength. 
Then we shall check how the tilt changes with the magnetic field strength
and whether there is any quenching in the tilt to support the theoretical models of 
BMR formation and the \bl\ dynamo saturation.

\section{Data and Method}
\label{ch6:method}
In this work, we have used the full disk Line of Sight (LOS) magnetogram with cadence of 6 hours and Intensity Continuum (IC) with caence of 24 hours from {\it Michelson Doppler Imager} \citep[MDI: 1996--2011;][]{Scherrer1995} and {\it Helioseismic and Magnetic Imager} \citep[HMI: 2010--2018;][]{Schou2012} for identification of BMRs.

The magnetograms taken from these two instruments, give only the LOS component of magnetic field. To get the magnetic field in the direction normal to the solar surface we have corrected  for the projection effect.
The projection effect becomes more and more critical as we go towards the limb of the solar disk.Therefore, in the first step, we have restricted ourselves up to 0.9 \Rsun. Later on, to avoid the uncertainty in the magnetic field measurement we have also excluded the BMRs which have absolute mean heliographic longitude greater than 50$^\circ$ from our analysis.

\begin{figure}[!ht]
    \centering
    \includegraphics[width=\textwidth]{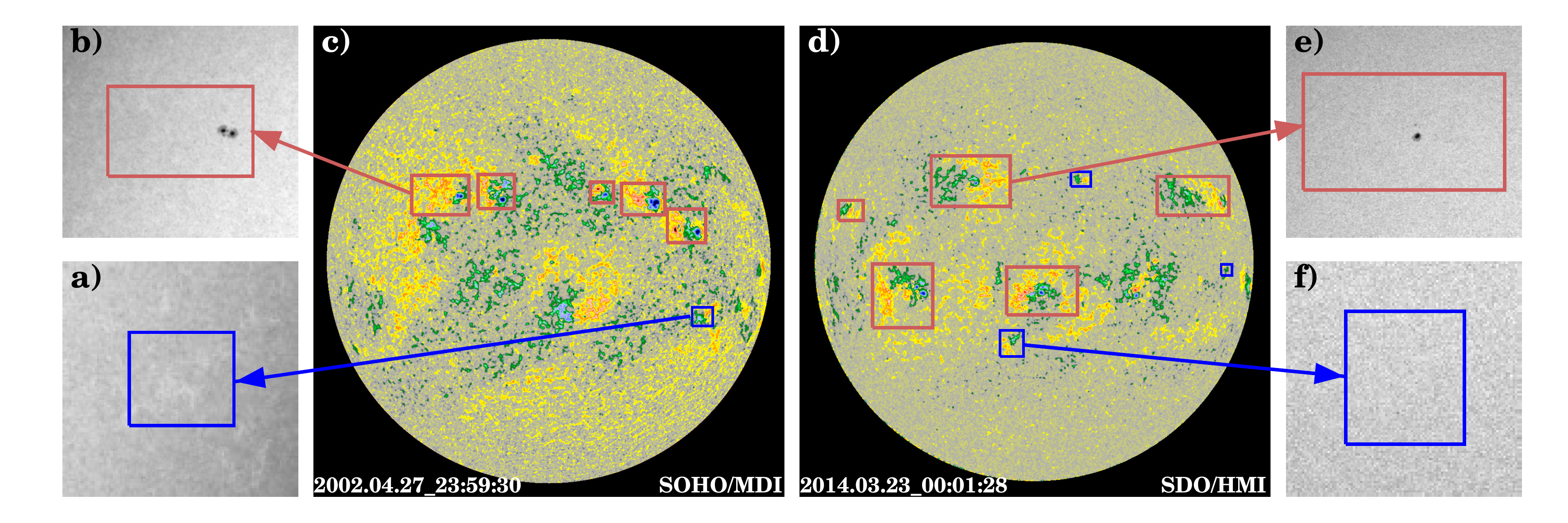}
    \caption{Representatives magnetograms of (c) MDI and (d) HMI (saturated to $\pm$1.5 kG) with BMR$_{\rm{WS}}$ (red box) and BMR$_{\rm{NS}}$ (blue). (a, b, e, and f): show IC counterparts.}
    \label{ch6:fig1}
\end{figure}

To identify BMRs, we have followed the method given in \citet{Stenflo2012a}.
So we have first applied a threshold on magnetic field strength and then a moderate flux balance condition to avoid the false detection of unipolar spot or BMR with large flux difference (see \autoref{ch6:fig1}(c--d)). 
Unlike \citet{Stenflo2012a}, we have applied a 2D Gaussian smoothing with FWHM of 3 pixels \citep{Hagenaar1999} to reduce the spatial noise, before calculating (I) heliographic coordinate, (II) magnetic flux and (III) maximum field density from detected BMRs. Since maximum magnetic field density mimics the maximum field strength, we call it as the maximum field strength $B_\mathrm{max}$. While calculating $B_\mathrm{max}$ for HMI data, we have multiplied it by a factor of 1.4 to bring two data sets on the same scale \citep{Liu2012}. Tilts of BMRs have been calculated with respect to solar E--W direction considering the spherical geometry of the Sun.

\section{Results and Discussions}
\label{ch6:res}

\begin{figure}[!ht]
\includegraphics[width=\textwidth]{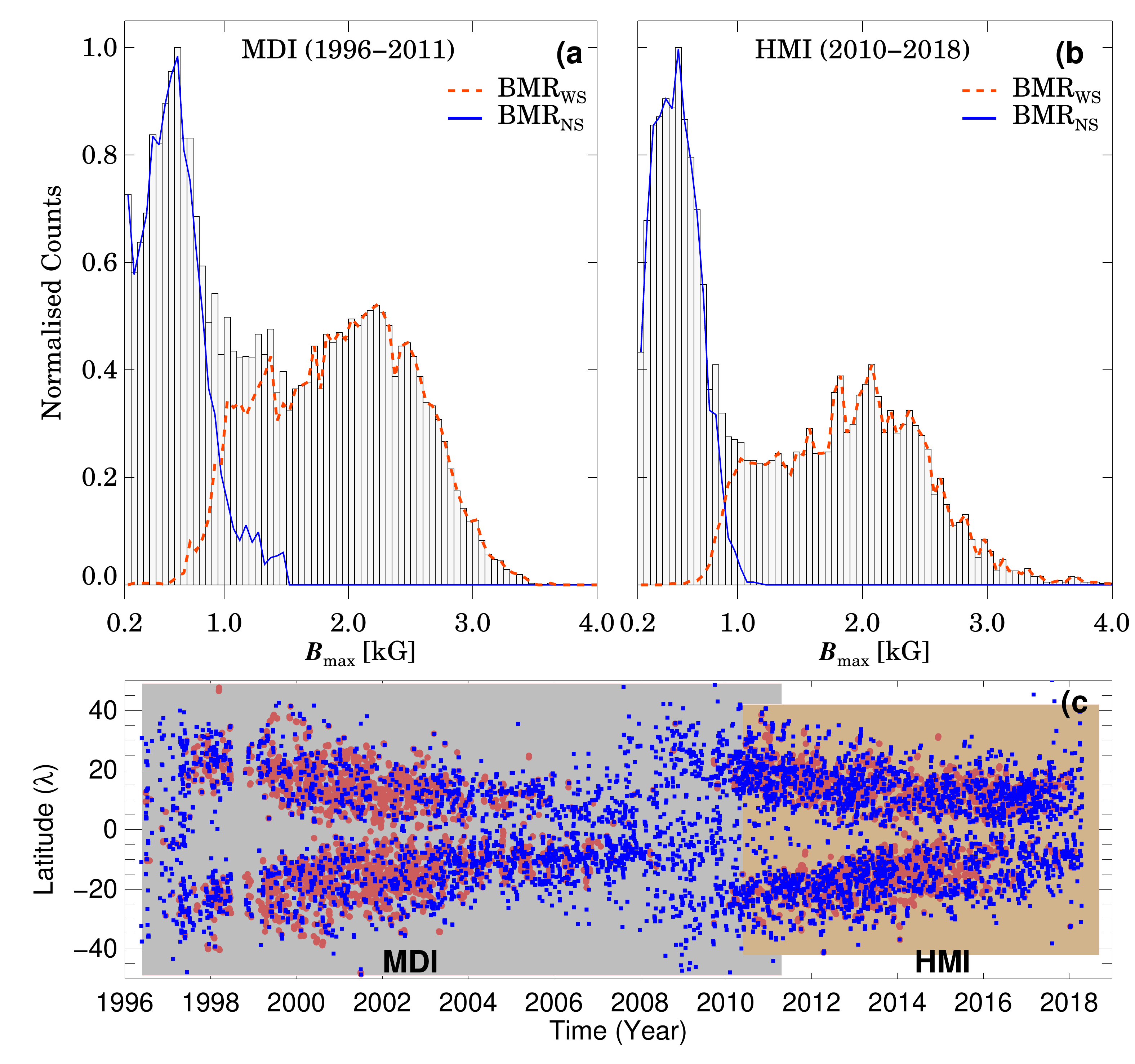}
    \caption{(a-b): Distributions of $B_{\rm max}$ in the BMRs from MDI (left panel) and HMI (right). 
             Red and blue respectively show $B_{\rm max}$ distributions of BMR$_{\rm WS}$ 
             (having counterpart in IC) and BMR$_{\rm NS}$ (no counterpart in IC).
             The vertical axes of two panes are divided by 315 and 388, respectively to bring the maxima of distributions to unity.
            Bottom: Time-latitude distribution of BMR$_{\rm WS}$ (red) and BMR$_{\rm NS}$ (blue).}
    \label{ch6:fig2}
\end{figure}

Before we explore the magnetic field dependence of BMR tilt, we first present the 
distribution of the maximum magnetic field \Bmax\ of BMRs in \autoref{ch6:fig2}(a-b). For the time being, we ignore the solid and dashed lines in these figures. We observe two well-separated peaks at around 600~G and 2100~G. These peaks are seen both in MDI and HMI data. HMI data includes the solar cycle 24, which is a relatively weak cycle and contains less number of strong field BMRs compare to weak field BMRs.
Despite the data obtained from two different instruments and two different solar cycles, we find the presence of two distinct peaks in both data sets.
These two distinct peaks remain even when we do not smooth the data or smooth with different windows.
However, as we smooth the data with a wider averaging window, these peaks tend to flatten out as well as shift slightly towards lower values. 
In the extreme limit, when we take the average magnetic field (i.e., window size equals to the BMR area), the two observed peaks 
disappear. This is expected because the magnetic field falls rapidly as we move away from the BMR center.

It appears that the whole HMI distribution is slightly shifted  to the left side and therefore the peaks appear at slightly smaller \Bmax\ than 
in MDI data. This could be due to different solar cycle, or it could be that the factor 1.4 
used to scale the HMI magnetic field is not appropriate for the entire range of \Bmax \citep{Liu2012, Pietarila2013}.
Nevertheless, these results suggest that the magnetic field distribution of BMRs is bimodal 
and possibly there are two types of BMRs having significantly different field strength. 

To understand these two peaks in our data, we analyse their  IC for the same periods. The IC images may not necessarily be simultaneous but they are near-simultaneous with a maximum time difference being 3 hours.
We find that not all BMRs have their counterparts in IC (i.e., sunspots) (\autoref{ch6:fig1}(a) and \autoref{ch6:fig1}(f)). When we say counterpart in IC, we mean whether there is any spot present in the IC on the BMR region (as identified in the magnetogram), independent of their size. It turns out that the BMRs which have their counterparts in IC (\autoref{ch6:fig1}(b) and \autoref{ch6:fig1}(f)) are having higher magnetic field. When we overplot these two distributions in \autoref{ch6:fig2}, we find that the BMRs having counterparts in IC (red/dashed line) beautifully represents the second peak at high $B_{\rm max}$ and the rest, i.e., BMRs without having a counterpart in IC (blue line), 
overlap with the first peak at the low $B_{\rm max}$.
Again we notice that in both the data sets this feature distinctly appears.
We define BMR$_{\rm WS}$ as the BMRs which have counterpart in IC, i.e., no sunspots and $B_{\rm max}$ distribution peaks at around 2~kG, while
BMR$_{\rm NS}$ as the BMRs which do not have sunspots (no counterpart in IC) and $B_{\rm max}$ distribution peaks at around 600~G.
Similar bi-modality in the maximum field distribution, have been reported in the past by \citet{Cho2015} and \citet{Tlatov2019} using sunspots and pores from SDO/HMI data. However in this work we look into the more general features, BMRs, of which sunspots and pores are part of.

Seeing the peak of BMR$_{\rm NS}$ at smaller field strength, one may conjecture that these
BMRs are produced from the small-scale magnetic field possibly originating  from the small-scale
dynamo \citep{Petrovay1993}. If this is the case, then we expect no preferred latitude distribution and no solar 
cycle variation. 
However, in \autoref{ch6:fig2}(c), we find no such evidence. Both classes of BMRs
follow similar temporal and latitudinal variations in the usual butterfly diagram. Thus, this 
result do not suggest that the origin of BMR$_{\rm NS}$ are linked to the small-scale dynamo.

\begin{figure}[!ht]
    \centering
    \includegraphics[width=\textwidth]{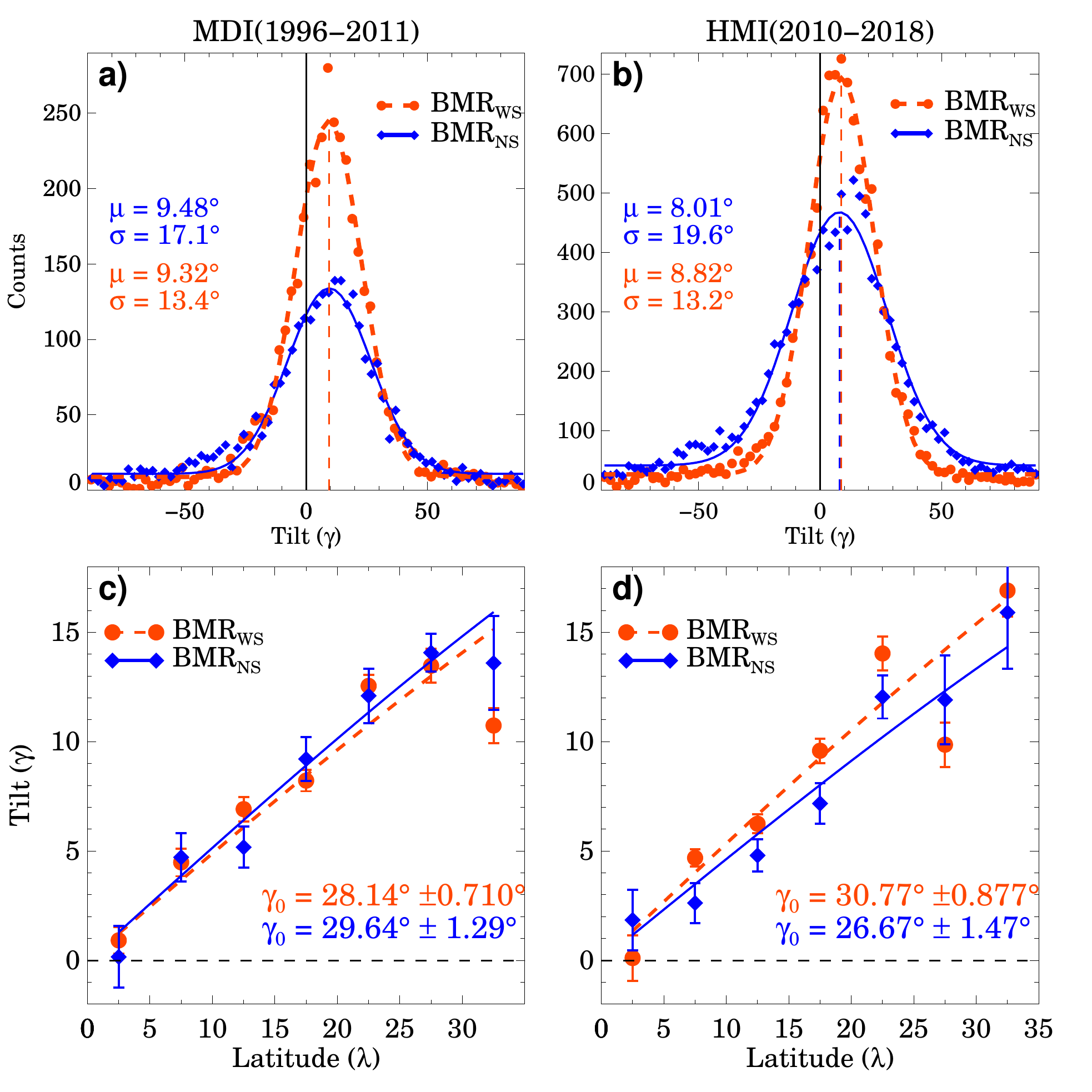}
    \caption{(a-b): Red and blue show tilt distributions of BMR$_{\rm WS}$ and BMR$_{\rm NS}$, respectively.
             Points represent the data and lines show the fitted Gaussians with parameters
             printed on the panels.
            (c-d): Mean tilt in each latitude bin as a function of the latitude. Solid and dashed lines are Joy's law ($\gamma = \gamma_0 \sin\lambda$) fits for BMR$_{\rm WS}$ and BMR$_{\rm NS}$.
            }
    \label{ch6:fig3}
\end{figure}

Now we explore the magnetic field dependence of BMR tilt. 
As we have found two distributions of BMRs, we shall first 
present the basic features of tilt of these two BMR classes separately. 
\autoref{ch6:fig3} shows the tilt distributions of these two classes of BMRs namely, BMR$_{\rm WS}$ (red) and BMR$_{\rm NS}$ (blue) in the latitude range $10^\circ$--$~30^\circ$ including both the hemisphere.
Distributions peak at non-zero tilt and show Gaussian-like behaviour, which is of course not new \citep{Wang1989, Stenflo2012a}. 
Although both distributions peak almost at the same tilt value, the distribution spreads are not identical 
and they are consistently different in two data sets. 
After fitting histograms with Gaussian profiles with mean $\mu$ and standard deviation $\sigma$, we find 
$\mu$ is around $9^{\circ}$ for both classes of BMRs and from both data sets. 
However, $\sigma$ for BMR$_{\rm WS}$ is smaller by a few degrees in both the data sets.
These results indicate that the tilt has some magnetic field dependence.

As shown in \autoref{ch6:fig3}(c-d), Joy's law slope $\gamma_0$ are 
consistently different in two classes of BMRs. 
BMR$_{\rm WS}$ has a slightly larger $\gamma_0$ in HMI data,
while in MDI data it is opposite. As MDI and HMI include data from two different times, we do not expect Joy's law trend to be identical in two data sets. Nonetheless, evidence of Joy's law in  BMR$_{\rm NS}$ further suggests that the BMR$_{\rm NS}$ class may not be originating from the small-scale magnetic field, rather they must be originating from the same large-scale 
magnetic field which produces BMR$_{\rm WS}$.

\subsection{Magnetic quenching of tilt angle}
\label{ch6:quenching}
\begin{figure}[!ht]
    \centering
    \includegraphics[width=\textwidth]{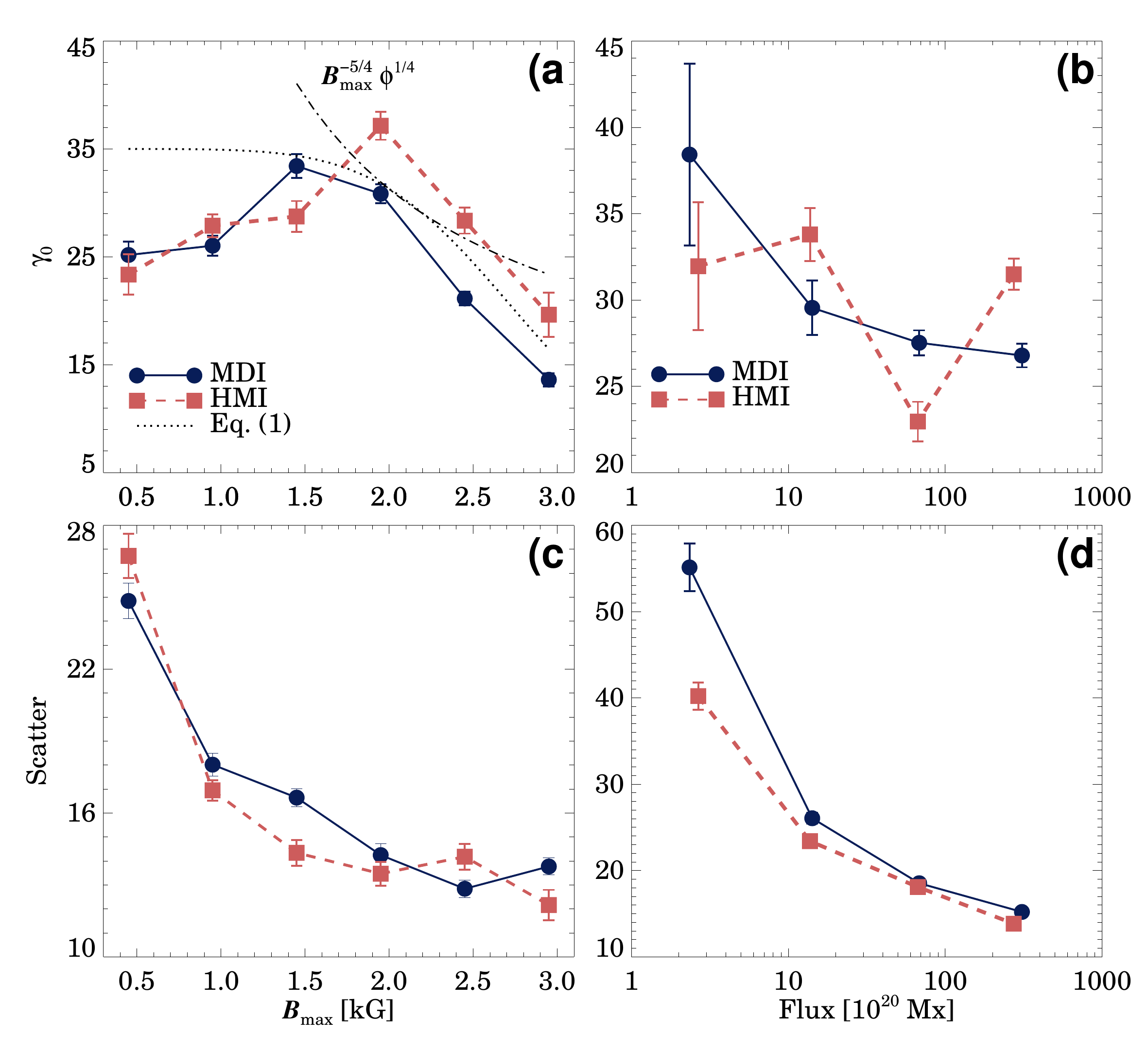}
    \caption{Magnetic field (\Bmax) dependences of: (a) Joy's law slope $\gamma_0$ and (c) the tilt scatter $\sigma$. (b) and (d) are the same as left panels but as functions of flux.
}
    \label{ch6:fig4}
\end{figure}

To quantify the magnetic field dependence of BMR tilt, we now compute 
Joy's law slope $\gamma_0$ and the scatter around the mean tilt ($\sigma$), separately in each $B_{\rm max}$ bin with bin size of 500~G. 
In \autoref{ch6:fig4}(a), we observe that for MDI data, 
$\gamma_0$ is only slightly increased 
in the small \Bmax\ range and then dropped at least by about $15^\circ$ 
in the high field values above 2~kG.
While HMI data follow a general trend, there is a significant increase
in the low-field range.
A prominent reduction of $\gamma_0$ (by about $15^\circ$) with the magnetic field strength clearly establishes the existence of BMR tilt quenching. 
We emphasize that the tilt quenching is seen when \Bmax $> 2$~kG. That is 
why in \autoref{ch6:fig3}, the mean Joy's law trend of BMR$_{\rm WS}$ is not smaller than BMR$_{\rm NS}$. It is only the strong BMRs$_{\rm WS}$ having \Bmax $> 2$~kG 
show the quenching in tilt.

We note that although the general trend of tilt quenching is seen, the results are slightly sensitive to the analysis, particularly, to the number of data.
We have checked that our results do not change when (i) taking different \Bmax\ bin, (ii) excluding data point if Joy's law fit
is not significant, and (iii) removing the data in the Joy's fitting if the BMR number is less than 50 in each latitude bin.
Further, the different beahviour of MDI and HMI always persists.
As seen in \autoref{ch6:fig4}(b), the variation with the BMR flux is monotonous for MDI data 
but not for HMI.

The indication of tilt quenching as seen in \autoref{ch6:fig4}(a) gives an observation
support of the following nonlinear quenching in the \bl\ $\alpha$
or in $\gamma_0$ routinely used to saturate the magnetic field growth in kinematic dynamo models \citep[e.g.,][]{Choudhuri1995,  Dikpati1999, Chatterjee2004, Karak2019}
\begin{equation}
f_q \propto \frac{1}{\left[1 + (\frac{B_{\rm max} }{ B_0})^n \right]} 
\end{equation}
with $n=2$ \citep[see for example, Equation 10 of][]{Karak2017}.
However, our data fits best when $n = 5.8 \pm 0.8$ (and $B_0 = 2.9 \pm 0.1$~kG with reduced-$\chi^2 = 30.9$).

Now we discuss whether our results can be connected to the theory of thin flux tube model for the BMR formation.
Based on this theory, we expect, the intense toroidal flux rises fast, and thus, the Coriolis force gets less time to induce a tilt. 
Hence, the BMR tilt is expected to decrease with the increase of the magnetic field. The thin flux tube simulations of \citet{Fan1994} predicted:
\begin{equation}
\gamma \propto \sin\lambda B_0^{-5/4}\Phi^{1/4},
\label{ch6:eq1}
\end{equation}
where $B_0$ is the initial magnetic field of the toroidal flux tube and
$\Phi$ is the flux content.
The theoretical study suggests that due to combined effects of rapid expansion, radiative cooling, and pressure buildup,
the magnetic fields of BMRs forming loops become sufficiently low as they rise towards the surface, and within a few Mm depth BMRs tends to get disconnected from their roots
\citep{Schuessler2005}. The current understanding of the whole process is 
very limited; however see \citet{Rempel2014a, Fan2014, Nelson2014}. Therefore, we do not know whether the initial magnetic field $B_0$ is related to 
the \Bmax\ that
we observe inside the BMR. 
However, if we assume that $B_0 \propto$ \Bmax, then we can make some comment on the thin flux tube model.

The BMR flux $\Phi$ is observed to vary with the
magnetic field strength \citep{Tlatov2014}. 
In our data, we find the following relation hold resonably well.
\begin{equation}
\frac{\Phi} {\angles{\Phi}} = a + b \frac{B_{\rm max}} {\angles{B_{\rm max}}} + c \left( \frac{B_{\rm max}} {\angles{B_{\rm max}}} \right) ^2+ d \left( \frac{B_{\rm max}} {\angles{B_{\rm max}}} \right) ^3,
\label{eq:eq3}
\end{equation}
where $a=-0.08 \pm 0.01$, $b=0.84 \pm 0.12$,  $c=-0.57 \pm 0.19$ and $d= 0.52 \pm 0.08$ for MDI data and $a=-0.09\pm 0.02$, $b=0.81 \pm 0.15$,
$c=-0.27 \pm 0.23$ and $d= 0.32 \pm 0.10$ for HMI. Putting this relation in \autoref{ch6:eq1}, we find that the slope of Joy's law $\gamma_0$ decreases as shown by the dashed line.
We observe that in the high-field regime, our result qualilatively supports the thin flux tube model.

In the low-field regime with $B_{\rm max} < 2$~kG 
$\gamma_0$ increases with \Bmax\ which does not fit with the thin flux tube model. However,
we should not forget that this model does not include the convection, which can affect the dynamics of the flux tube to
change the tilt through the helical convection.
By considering convection, in the thin flux tube model,
\citet{Weber2011} showed that while the general Joy's law trend is recovered,
the tilt increases with the increase of magnetic field strength first in the low
field regime, and then it reduces; see their Figure~8 and 12 \citep[also see][]{Weber2013}. 
Similar behaviour is found in our data; see \autoref{ch6:fig4}(a).

Thin-flux tube rise model also predicted that the rising-flux loops could be buffeted by the turbulent convection
during their rise in the CZ and this could cause a scatter
around the systematic tilt variations --- Joy's law \citep{Longcope1996, Longcope2002}.
When the magnetic field is strong, we expect the magnetic tension to oppose this buffeting of flux tubes and
the scatter to be less. Further, strong flux tubes rise faster (due to strong magnetic buoyancy)
and thus they get less time to be buffeted by  convection \citep{Weber2011}.
The tilt scatter computed from our data supports this idea. In \autoref{ch6:fig4}(c-d),
we see that it systematically decreases with the increase of \Bmax\ or flux.

\section{Conclusion}
\label{ch6:conclusion}
In this chapter, we have studied BMRs detected from the magnetograms of
MDI (1996--2011) and HMI (2010--2018). In both the data sets, we find that the BMR number distribution
shows a bimodal distribution when measured with respect to their maximum magnetic field \Bmax.
The first peak at low field (\Bmax$\approx600$~G) corresponds to BMRs which do not have
counterparts in IC (i.e., no sunspots),
while the second peak at high field (\Bmax$\approx 2100$~G) corresponds to BMRs
which have counterparts in IC. BMR$_{\rm NS}$ also shows a similar butterfly diagram, tilt distribution and Joy's law
as that of BMR$_{\rm WS}$. This suggests that BMR$_{\rm NS}$ are not produced from the small-scale magnetic field, rather they must be produced from the same large-scale global field which produces sunspots.
One difference between these two classes of BMRs is that the tilt scatter and
the slope of Joy's law $\gamma_0$ are smaller in BMR$_{\rm WS}$. 
However, our study does not explain why BMR show two distinct peaks in the $B_{\rm max}$ distribution, which requires further studies.

On computing the tilt in each \Bmax\ bin, we find a significant change in the BMR tilt for MDI and HMI data.
In the low \Bmax\ range, $\gamma_0$ increases with the increase of \Bmax.
However, for $B_{\rm max} > 2$~kG (which corresponds to strong sunspots), $\gamma_0$ decreases with \Bmax. These results are in qualitative agreement with the predictions of the thin flux-tube rise 
model \citep{DSilva1993,Fan1994, Caligari1995, Fan2009}and in particular the simulations 
with the convection \citep{Weber2011, Weber2013}. The reduction of tilt with the increase of the magnetic field 
in the high field regime gives a hint for the nonlinear quenching routinely used in the \bl\ type kinematic dynamo models.

We understand that the variations of BMR properties, particularly the tilt quenching with magnetic field are demonstrated in a relatively narrow range. This, however, is due to the fact that the availability of data are limited and Joy's law is a statistical relation. Furthermore, the last two cycles, during which our analyses are performed, are relatively weak, having weak BMR field strength. The highest magnetic field in our BMRs data is about 3~kG, and the magnetic quenching is expected to be more in the super-kilogauss magnetic field. Therefore, we believe that our results need to be investigated further with larger data sets, especially from stronger cycles having high-field BMRs.

\clearpage{}

\clearpage{}\chapter[The Conclusion]{The Conclusion}
\label{Chap7}
\lhead{\emph{Chapter 7: The Conclusion}} \epigraph{\itshape ``Statistics: The only science that enables different experts using the same figures to draw different conclusions.''}{-- Evan Esar
}

\noindent

Now, I came to the end of the thesis, which brings me to discuss the conclusion, novelty, and future aspects of it. In the entire thesis, I have primarily used the white light data from KoSO along with white light and magnetogram data from space-based modern instruments MDI and HMI. With the help of these data sets, I have explored the multiple observational aspects of the solar long--term variability and how it can reduce the gap between observation and the theoretical aspects of solar dynamo models. Let me start to conclude my thesis for each chapter separately.

\section{Chapter~3}

Motivated by the work of \citet[references therein]{Hathaway2013} and understanding the importance of penumbra to umbra area ratio ($q$), I have studied the long-term variation of this ratio using the historical KoSO digitized data (1923--2011). I have developed an automatic method based on Otsu technique to extract the penumbra and umbra from the sunspots. I have noted that the $q$ increases initially for area $<200\mu$Hem, but after that, it increases very slow and saturates around 5.5 to 6, which is in agreement with earlier findings. Although the bigger sunspots (Area$>200\mu$Hem) show a similar trend in both the data sets (RGO \& KoSO), the smaller sunspots (sunspot area $<100~\mu$Hem) in KoSO data does not show any systematic variation in the ratio contrary to the result reported by \citet{Hathaway2013}. It has been further verified using MDI white--light data \citep{Jha2019}. The result from RGO data raise a question about the origin of two classes of sunspots, but since KoSO data does not show any difference in both, the classes imply they are from the same source. Hence, dynamo modellers might treat them on the same footing.

\section{Chapter~4}

The significant role of solar differential rotation and its variation for the solar dynamo models; and the unavailability of an automated tracking algorithm to track the sunspots motivated me to measure it using a computerized method. I have developed an automatic way to track the sunspots from the century-long KoSO digitized white-light data and calculated the solar rotation profile. Average rotation profile measured in \cref{Chap4} show a very good agreement with the earlier results \citep{Gupta1999} with the following rotation profile when measured in degree/day,
\begin{equation}
    \Omega (\theta) = (14.381\pm 0.004)-(2.72\pm0.04)\cos^2{\theta},
\end{equation}
where $\theta$ is the co-latitude. KoSO data neither show any phase dependency nor the significant long--term variation in the solar differential rotation. However, differential rotation parameters $A$ and $B$ show the negative and positive correlation with the strength of the solar cycle. In the case of KoSO white--light data, the smaller sunspots (area $<200~\mu$Hem) gives a faster rotation rate than the bigger ones (area $>400~\mu$Hem) which is believed to be due to their different anchored depth and drag imposed by them. 

\section{Chapter~5}
The differential rotation measurement based on the Doppler velocity \cite{Howard1970} gives a slightly lower rotation rate when compared with the one measured using sunspots \citep[\cref{Chap4}, and ][]{Beck2000, Jha2021}.  This is due to the sharp change in the solar differential profile near the surface called NSSL, and the helioseismology observation has confirmed the existence of this intriguing layer. The absence of theoretical understanding of this layer made me think about its physics. Therefore, based on the thermal wind balance equation, which tells how the slight difference in temperature between solar pole and equator is balanced by centrifugal force, a theoretical model for NSSL is presented. Here the idea is the Coriolis force, arising due to solar rotation, has an effect on the convective blobs in the inner convection zone where convective turnover time ($\tau$) is comparable (or greater) than the solar rotation period ($T$) whereas it will be unaffected, near the surface ($\tau<<T$ ). Hence, near the surface the rate of fall of temperature becomes independent of latitude making the thermal wind term very large. The centrifugal term increases to balance the increase in the thermal wind term, leading to the observed NSSL. The analytical expression of $\Omega (r,\theta) $ and the observed $\Omega (r, \theta)$ gives $r_c=0.92$~\Rsun\ and 0.96~\Rsun\ respectively. Here convection zone model, which is called MODEL-S \citep{Dalsgaard1996} is used for the temperature profile. The result based on the observed profile of solar differential rotation and corresponding calculated PETD show good agreement with measured PETD in \citet{Rast2008}.

\section{Chapter~6}

Tilt quenching has been regularly used in the different types of solar dynamo models, particularly the \bl\ type dynamos. What these models were lacking was the observation support to this idea. Although the \citet{DasiEspuig2010} has used the white light data, which do not have the magnetic filed information, and given some indication of the tilt quenching but the measurement based on the magnetogram data was not available. Therefore, in \cref{Chap6} I have first developed and automatic method to identify the BMRs from the MDI (1996--2011) and HMI (2010--2018) magnetogram, which is inspired from the method described in \citet{Stenflo2012a}. Interestingly, in this work a bimodal distribution of \Bmax\ has been found where each peaks corresponds to the two classes of BMR; one which does not show the sunspot (limited by resolution and contrast) in white light (peaks around 600~G) and second which show sunspot in white light (peaks around 2100~G). The \Bmax\ dependence of amplitude of Joy's law ($\gamma_0$) shows an initial increase in $\gamma_0$ and then after around 2~kG it start decreasing and this behaviour is present in both MDI and HMI data sets. This result gives an indication of tilt quenching which need to be further verified using longer data sets with stronger solar cycle.

\section{Novelty of the Thesis}

$\bm{\Rightarrow}$ The first time, a fully automatic method is developed to extract the penumbra, and umbra area from KoSO digitized white--light data. The results obtained in this work does not show any difference in the long--term variation in penumbra to umbra area ratio as speculated based on RGO data (\cref{Chap3}).

$\bm{\Rightarrow}$ A new and automated algorithm to track the sunspots has been developed and used to track the sunspot in KoSO white--light digitized data to measure the solar differential rotation. A significant difference in measurement of solar differential rotation when bigger and smaller sunspots are used has been reported (\cref{Chap4}).

$\bm{\Rightarrow}$ A theoretical model for the NSSL based on thermal wind balance equation, which shows an excellent agreement with observation,  has been given very first time (\cref{Chap5}).

$\bm{\Rightarrow}$ The indication of tilt quenching based on the magnetogram data from MDI and HMI along with the two classes of BMRs has been found, which was not reported earlier. 

\section{Future prospects}

This thesis primarily focused on developing several new automated algorithms to extract different magnetic features as seen on the solar surface, study them, and provide input to the solar dynamo models. There is always a possibility of improving these techniques, and the recent development in computational power, Machine Learning (ML) and Artificial Intelligence (AI) could make it even more efficient and sophisticated. Here I will discuss a few possibilities that I am currently exploring and looking for.

\begin{enumerate}
    \item In \cref{Chap3}, I have used the data from KoSO and MDI, which can be further extended to data from other observatories such as the National Astronomical Observatory of Japan (NAOJ) and others. It will help us to test the algorithm with different data set to improve our statistics and consistency.
    
    \item The method developed in \cref{Chap4} only tracks the sunspots in two consecutive observations, and hence it can be modified to follow the sunspots for full-disk passage or their lifetime. It will help us get more accurate results for the solar differential rotation.
    
    \item A theoretical model of NSSL, given in \cref{Chap5}, also predict the variation of solar surface temperature from the pole to the equator, which can be verified by future observation with the help of more accurate observations.
    
    \item To understand the magnetic field dependence of tilt quenching in \cref{Chap6} I have counted each BMR multiple times as every magnetogram has been treated independently. Therefore,  to better understand the dependence of tilt on \Bmax\ in BMRs it is crucial to track all of them for disk passage or their lifetimes. I am currently involved in the work that is exploring this idea.
\end{enumerate}
\clearpage{}

\addtocontents{toc}{\vspace{2em}} 

\appendix \chapter[Temperature Variations Over Isochoric Surface]{Isochoric Surfaces within the Solar Convection Zone and Temperature Variations over them}
\label{ch5:appn}
\noindent

From the basic fluid dynamical equations, we arrive at 
Equation~\ref{ch5:eq1}
involving a differentiation of $S$ with respect to $\theta$ on a surface of constant $r$.  However, the rotation of the Sun makes the Sun slightly oblate so that the solar surface is not a surface of constant $r$.  The important question is how we connect a theory involving differentiation at constant $r$ to actual observations of the solar surface.

First of all, we argue that the oblateness of the Sun is extremely small as estimated in \citet{Kuhn2012}. One can also make a rough estimate of this oblateness from theoretical considerations. If we take $\Omega$ to be constant inside the Sun for this approximate estimate, then we can introduce an effective potential
\begin{equation}
 \Phi_{\rm eff} = - \frac{G M_{\odot}}{r} - \frac{1}{2} | {\bf \Omega} \times {\bf r} |^2
 \label{ae1}
\end{equation}
inside the convection zone, which has to be constant over isobaric surfaces (see, for example, \citet{Choudhuri1998book}, Section~9.3). If $\Delta r$ is the extension of such a surface in the equatorial 
region compared to the polar region, then we can equate $\Phi_{\rm eff}$ in the polar and equatorial regions to obtain
\begin{equation}
 - \frac{G M_{\odot}}{r} = - \frac{G M_{\odot}}{r + \Delta r} - \frac{1}{2} \Omega^2 r^2,\nonumber
\end{equation}
from which
\begin{equation}
\Delta r = \frac{1}{2} \frac{\Omega^2 r^4}{G M_{\odot}}.
\label{ae2}
\end{equation}
Since the mass contained within the convection zone is small compared to the solar mass, we can
take $M_{\odot}$ to be the total solar mass for estimating the oblateness of isobaric surfaces
within the convection zone.  Using $\Omega/2 \pi = 420$ nHz and $r = 0.85$\Rsun, we find
\begin{equation}
\Delta r = 3.3 \; {\rm km},
\label{ae3}
\end{equation}
which is less than 0.0005\% of the solar radius. The expected oblateness of the Sun is minuscule.

We now focus our attention on isochoric surfaces over which the entropy differential would
be related with the temperature differential by Equation~\ref{ch5:eq2}.
We denote the distance measured along an isochore by $l$. It is pointed out in Appendix A of \citet{Choudhuri2021a} that we approximately have
\begin{equation}
\frac{dS}{dl} = \left( \frac{\pa S}{\pa \theta}\right)_r \frac{d \theta}{d l}. 
\label{ae4}
\end{equation}
Using this relation, we can put Equation~\ref{ch5:eq1} in the form
\begin{equation}
r \sin \theta \frac{\pa}{\pa z} \Omega^2 = \frac{g}{\gamma C_V} \frac{d S}{d l}, 
\label{ae5}
\end{equation}
where $d S/ dl$ is the derivative of entropy $S$ along the isochore and we have used $d l = r d \theta$
in view of the very small oblateness of the Sun.  
We can now substitute Equation~\ref{ch5:eq2} in Equation~\ref{ae5} to obtain
\begin{equation}
r \sin \theta \frac{\pa}{\pa z} \Omega^2 = \frac{g}{\gamma T} \frac{d}{d l} \Delta T.
\label{ae6}
\end{equation}
Now, as we move along an isochore, the change in $l$ 
is accompanied by a change in $\theta$.  We have
\begin{equation}
\frac{d}{d l} \Delta T = \frac{1}{r}
\left(\frac{d}{d \theta} \Delta T \right)_{\rm isochore}.
\label{ae7}
\end{equation}
Note that here the differenetiation is with respect to
the changing value of $\theta$ along an isochore.
On substituting Equation~\ref{ae7} into Equation~\ref{ae6}, we are readily led to Equation~\ref{ch5:eq3}. 

One fact is clear from Equation~\ref{ch5:eq3}.  If the Sun did not have a thermal wind (which would require $\Omega$
not to vary along $z$ for consistency), then the temperature would be constant over the
isochoric surface.  The isochoric surface would then be a isothermal surface and hence also an
isobaric surface.  Now,
the observed solar surface is a surface at which the optical depth becomes 1, the opacity
depending on both density and temperature. In the absence of the thermal wind, the isochoric, isothermal
and isobaric surfaces would coincide and the observed surface of the Sun would be one such surface.
To obtain the temperature variation on the solar surface, we need to find the temperature difference
on isochoric surfaces.  For this purpose, we can use Equation~\ref{ch5:eq3}.

If $\Omega (r, \theta)$ is given, then
Equation~\ref{ch5:eq3} can be integrated to obtain
$\Delta T$ along an isochore.  Because of the miniscule oblateness of the isochore, we can regard the isochore
to be a spherical surface.  However, at a conceptual level, we shall keep in mind that we would be considering
$\Delta T$ on isochoirc surfaces, since this will eventually enable us to compare with observational data
on the solar surface.

We point out one other thing, which can sometimes be
a source of confusion. By the chain rule of 
differentiation, we have
\begin{equation}
\frac{d}{dl}\Delta T = \left( \frac{\partial T}{\partial r} \right)_{\theta} \frac{d r}{dl}
+ \left( \frac{\partial T}{\partial \theta} \right)_r \frac{d \theta}{dl}. 
\label{ae8}
\end{equation}
Even though $d r/ dl$ is very small, the first term on
the right hand side of Equation~\ref{ae8} cannot be
neglected with respect to the second term because $(\partial T/ \partial r)_{\theta}$ is much larger
than $(1/r)(\partial T / \partial \theta)_r$.  As a result, we cannot write down an equation for temperature
similar to Equation~\ref{ae4} for entropy. However, when
we connect the theory with observations, the observed
temperature variations of the solar surface have to be
compared with $((d/d \theta) \Delta T)_{\rm isochore}$
appearing in Equation~\ref{ch5:eq3} and not with $(\partial T / \partial \theta)_r$ which is not of much interest to us.
 \addtocontents{toc}{\vspace{2em}}  

\backmatter

\newpage

\lhead{\emph{Bibliography}}

\setstretch{0}
\bibliographystyle{spr-mp-sola}

\newpage
\thispagestyle{plain}
\setstretch{1}
\section*{Comments}
{\large {\bf $\bullet$ Referee-I}}\\
\headrule
\vspace{-0.5cm}
The data analyzed in this thesis builds upon the commendable efforts in digitizing and using the long term data from the Kodaikanal solar observatory. The thesis makes original contributions by way of automated methods of measuring sunspot features. The results from these automated methods are contrasted with earlier results regarding differential rotation that used limited data. The thesis also includes detailed theoretical work on the near-surface shear layer and the tilt quenching mechanism for sunspots. Taken together, the work in this thesis makes a number of important contributions to our understanding of the solar dynamo.

Excellent chapter layout and arrangement. I would like to especially commend the organization and readability of the introduction chapter.\\[0.5cm]

{\large {\bf $\bullet$ Referee-II}}\\
\headrule
\vspace{-0.5cm}
This thesis presents four substantial results, each Of which mieht have been worthy Of expansion into a Ph.D. Most impressive, it does not read like a disjoint ``staple'' thesis, but rather comes together as a whole, greater than the sum of its parts. The thesis is strong in both theory and observation. I was also impressed by the careful and systematic approach to the research throughout the thesis. For example, the preprocessing method set up to eliminated rogue pixels (3.2), or the efforts taken to double-check/crossvalidate the method for $q$ evaluation (3.3) or differential rotation (4.4) via other datasets (SOHO/MDI, SDO/HMI).

The opening review chapter was well organized, motivated, and informative, providing the background required for the reader to fully appreciate the rest of the thesis. It also had an excellent narrative flow, making it a pleasure to read. The following chapters on data analysis, penumbra to umbra area ratio, differential rotation, NSSL, and tilt quenching make their points clearly and are well-supported with figures and tables. The final chapter does a nice job of summarizing the results and novelty of the work and describing future plans.

\end{document}